\documentclass[sigconf, nonacm]{acmart}

\usepackage{enumitem}
\usepackage{hyperref}
\usepackage{subfig}
\usepackage{multirow}

\newcommand\vldbdoi{XX.XX/XXX.XX}
\newcommand\vldbpages{XXX-XXX}
\newcommand\vldbvolume{15}
\newcommand\vldbissue{9}
\newcommand\vldbyear{2022}
\newcommand\vldbauthors{\authors}
\newcommand\vldbtitle{\shorttitle} 
\newcommand\vldbavailabilityurl{https://github.com/alexZeakis/Embedings4ER}
\newcommand\vldbpagestyle{empty} 

\usepackage{titlesec}
\definecolor{myblue}{RGB}{31, 119, 180}
\definecolor{myorange}{RGB}{255, 127, 14}
\definecolor{mygreen}{RGB}{44, 160, 44}
\definecolor{myred}{RGB}{214, 39, 40}

\usepackage{pifont}% http://ctan.org/pkg/pifont

\begin{document}
\title{Pre-trained Embeddings for Entity Resolution:\\ An Experimental Analysis [Experiment, Analysis \& Benchmark]}

\author{Alexandros Zeakis$^{1,2}$, George Papadakis$^1$, Dimitrios Skoutas$^2$, Manolis Koubarakis$^1$}
\affiliation{%
  \institution{$^1$National and Kapodistrian University of Athens, Greece $\>\>$ \texttt{\{alzeakis,gpapadis,koubarak\}@di.uoa.gr}\\
  $^2$Athena Research Center, Greece $\>\>$ \texttt{\{azeakis,dskoutas\}@athenarc.gr}}
  \country{}
}

% \author{}
% \affiliation{%
%     \institution{National and Kapodistrian University of Athens}
%   \city{Athens}
%   \country{Greece}
% }
% \email{gpapadis@uoa.gr}

% \author{}
% \affiliation{%
%     \institution{Athena Research Center}
%   \city{Athens}
%   \country{Greece}
% }
% \email{dskoutas@athenarc.gr}

% \author{}
% \affiliation{%
%   \institution{National and Kapodistrian University of Athens}
%   \city{Athens}
%   \country{Greece}
% }
% \email{koubarak@di.uoa.gr}

\begin{abstract}
% Following the 
% recent 
% advancements in NLP, 
Many recent works on Entity Resolution (ER) 
leverage
% exploit the latest breakthroughs in 
% Deep Learning in order to maximize
Deep Learning techniques involving language models
% in combination with advanced neural network architectures,
to improve effectiveness.
% Invariably, t
% These works use language models in combination with advanced neural network architectures.
This 
% applies 
is applied
to both main steps of ER,
% : 
i.e.,
blocking and matching.
Several pre-trained embeddings have been tested, with the most popular ones being fastText and variants of the BERT model.
However, 
% they are typically combined with different deep nets, 
there is no detailed analysis of their pros and cons.
% in the context of ER.
To cover this gap, we perform a thorough experimental analysis of 12 popular language models over 17 established benchmark datasets.
% We start with their vectorization, assessing their
First, we assess their vectorization overhead for converting all input entities into dense embeddings vectors.
% Next, 
Second, we investigate their blocking performance,
% in combination with a state-of-the-art algorithm for approximate nearest neighbor search.
% We compare them with the state-of-the-art deep learning-based blocking method and perform a detailed scalability analysis.
performing a detailed scalability analysis, and comparing them with the state-of-the-art deep learning-based blocking method.
Third, we conclude with their relative performance for both supervised and unsupervised matching.
% in matching, both in supervised and unsupervised settings.
Our experimental results provide novel insights into the strengths and weaknesses of the main language models, facilitating researchers and practitioners to select the most suitable ones in practice.
\end{abstract}

\maketitle

\pagestyle{\vldbpagestyle}
\begingroup\small\noindent\raggedright\textbf{PVLDB Reference Format:}\\
\vldbauthors. \vldbtitle. PVLDB, \vldbvolume(\vldbissue): \vldbpages, \vldbyear.\\
\href{https://doi.org/\vldbdoi}{doi:\vldbdoi}
\endgroup
\begingroup
\renewcommand\thefootnote{}\footnote{\noindent
This work is licensed under the Creative Commons BY-NC-ND 4.0 International License. Visit \url{https://creativecommons.org/licenses/by-nc-nd/4.0/} to view a copy of this license. For any use beyond those covered by this license, obtain permission by emailing \href{mailto:info@vldb.org}{info@vldb.org}. Copyright is held by the owner/author(s). Publication rights licensed to the VLDB Endowment. \\
\raggedright Proceedings of the VLDB Endowment, Vol. \vldbvolume, No. \vldbissue\ %
ISSN 2150-8097. \\
\href{https://doi.org/\vldbdoi}{doi:\vldbdoi} \\
}\addtocounter{footnote}{-1}\endgroup

\ifdefempty{\vldbavailabilityurl}{}{
\vspace{.3cm}
\begingroup\small\noindent\raggedright\textbf{PVLDB Artifact Availability:}\\
The source code, data, and/or other artifacts have been made available at \url{\vldbavailabilityurl}.
\endgroup
}

\section{Introduction}

% {\color{red}Basic question: which terms are we going to use to collectively refer to all embeddings models? Language models? Transformer models? Ideally, we need a synonym, too.}

Entity Resolution (ER) is a crucial and challenging task for data integration \cite{DBLP:series/synthesis/2015Christophides}, aiming to detect the different entity profiles
% (simply called entities in the following)
that pertain to the same real-world object \cite{DBLP:journals/csur/ChristophidesEP21}. By deduplicating entity collections, ER facilitates a wide range of applications, from common data analytics tasks
% (e.g., in CRM systems)
to advanced question answering
% approaches
~\cite{DBLP:journals/pvldb/DongS13}.

Typically, ER solutions operate in two steps \cite{DBLP:books/daglib/0030287,DBLP:journals/pvldb/GetoorM12}. First, \textit{blocking} restricts the search space to the most likely matches, called \textit{candidate pairs}, through a coarse-grained approach \cite{DBLP:journals/csur/PapadakisSTP20}. This is necessary in order to tame the inherently quadratic complexity of ER
% , thus allowing 
and allow
it to scale to large volumes of data. Then, \textit{matching} performs a fine-grained processing that examines each candidate pair to decide whether it constitutes a match \cite{DBLP:series/synthesis/2021Papadakis}.

Embedding text into numeric vectors has become a very common approach in NLP and related tasks \cite{DBLP:series/synthesis/2020Pilehvar}. Earlier models adopted sparse representations, based on bag-of-words or tf-idf \cite{DBLP:books/daglib/0021593}.
% Although efficient, sparse embedding vectors cannot capture important features, such as synonyms. Hence, they
However, these can only capture syntactic and not semantic similarity.
% This limitation was addressed with the introduction of pre-trained embeddings models.
This limitation is addressed by pre-trained embeddings models.

Motivated and inspired by this, the latest breakthroughs in ER leverage language models for both blocking \cite{DBLP:journals/pvldb/Thirumuruganathan21} and matching \cite{Mudgal2018sigmod,DBLP:conf/edbt/BrunnerS20}. The following steps are typically involved in this process \cite{DBLP:journals/pvldb/EbraheemTJOT18}. First, every given entity is transformed into a dense embedding vector.
% Next, 
To perform blocking, the resulting vectors are indexed and
% then, every entity is posed as a query that retrieves its $K$ most similar candidates.
a $K$-nearest neighbor query is issued for each entity to identify candidate pairs.
% The resulting set of candidate pairs is
These are then
processed by
% the selected
a
matching algorithm, which may operate in an unsupervised or a supervised mode. In the former case, the similarity between the embeddings vectors is computed and used as edge weights in a bipartite graph, where the nodes correspond to entities and the edges connect the candidate pairs.
% Clustering is then performed in order to split the graph 
The graph is then split
into disjoint sets of nodes, such that each of them contains all entities describing the same real-world object. In supervised matching, the embeddings vectors of the candidate pairs are fed as input to a binary classifier, typically a deep neural network, that decides whether each of them is a duplicate.

\begin{figure*}[t]
\centering
\includegraphics[width=0.97\textwidth]{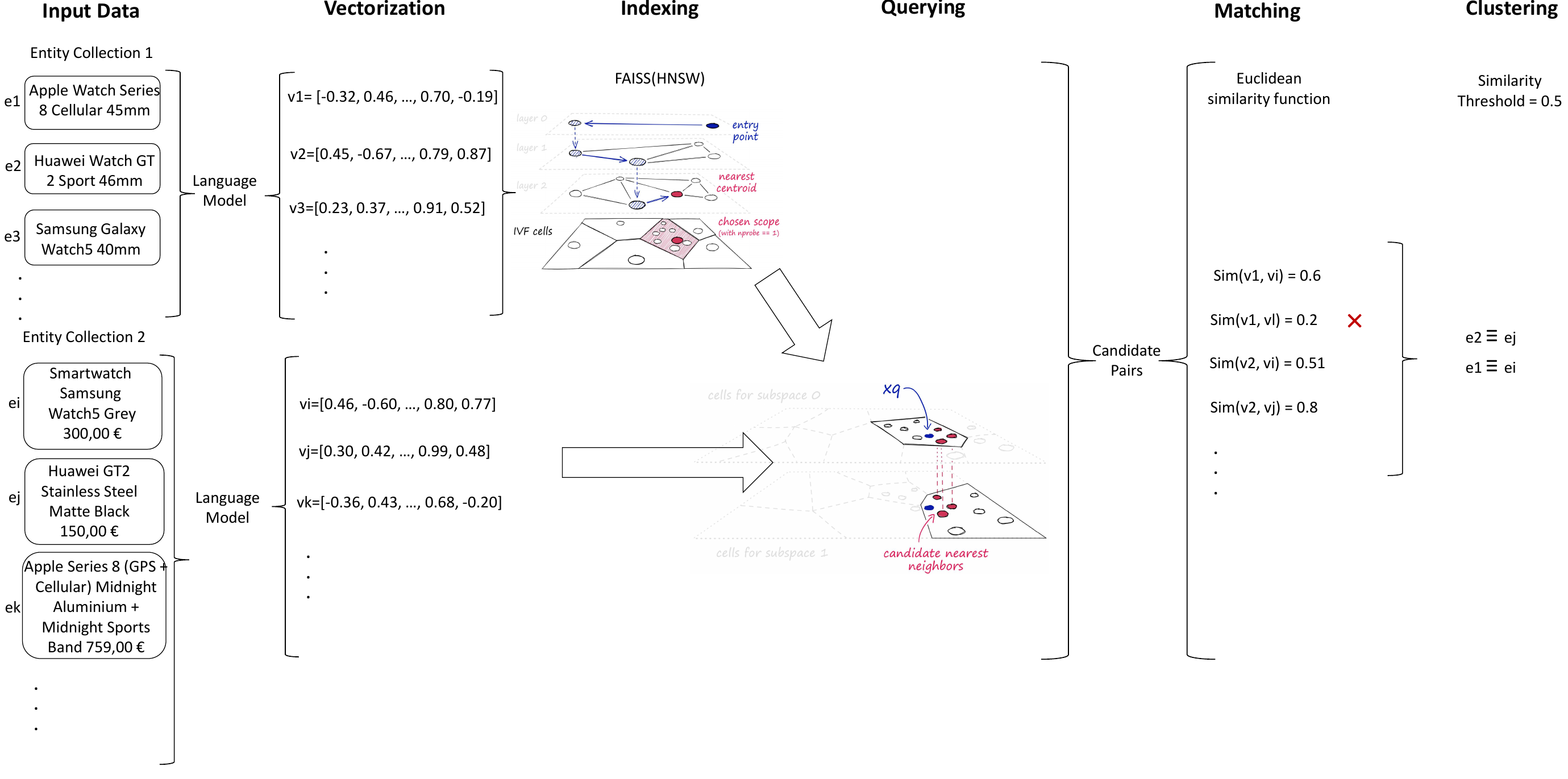}
\caption{An example illustrating the end-to-end unsupervised approach to Entity Resolution, based on language models.}
\label{fig:example}
\end{figure*}

As an example, consider the two entity collections in Figure \ref{fig:example}. They comprise data about smartwatches from shopping websites. We can see that $e_1$ matches with $_k$, $e_2$ with $e_j$ and $e_3$ with $e_k$, despite their substantially different descriptions. To use the considered language models in ER, the first step is vectorization, which in the schema-agnostic settings, converts every entity profile into a single numeric dense vector of fixed dimensionality. Entity $e_1$ is converted to vector $v_1$, entity $e_2$ to vector $v_2$ and so on and so forth. A successful language model assigns semantically similar entities to vector of low distance, as in our example. Next, FAISS(HNSW) indexes the vectors of the first entity collections, converting them to a graph structure, called Hierarchical Navigable Small-World graph, that allows for efficient searches for nearest neighbors. The entities of the second entity collection are actually posed as queries to the FAISS(HNSW) index in the next step, called querying, which retrieves the $k$ most similar vectors from the other entity collection per entity. The resulting pairs of similar vectors form the set of candidate pairs. For each pair of candidates, we estimate the (Euclidean) similarity of the corresponding vectors, associating it with a similarity score. The pairs with a scores higher than a threshold (0.5 in our example) are then processed by the Unique Mapping Clustering algorithm, which essentially associates every entity with its most similar candidate, disregarding all others. These matched entities form the output of the entire process.

Over the years, numerous language models have been used in both ER steps. Few of them have been used for blocking: GloVe \cite{DBLP:journals/pvldb/EbraheemTJOT18} and FastText \cite{DBLP:journals/pvldb/Thirumuruganathan21,DBLP:conf/wsdm/ZhangWSDFP20}. A much larger variety has been employed in matching: GloVe \cite{DBLP:journals/pvldb/EbraheemTJOT18,Mudgal2018sigmod}, FastText \cite{Mudgal2018sigmod,Nie2019cikm,DBLP:conf/aaai/Li0SZAW20,DBLP:conf/icdm/WangSWDJ20,DBLP:conf/www/ZhangNWST20,DBLP:conf/ijcai/FuHHS20,DBLP:conf/acl/YaoLDLY0LZD20} as well as BERT \cite{DBLP:journals/pvldb/PeetersB21} and its main variants, i.e., XLNet, DistilBERT, RoBERTa and AlBERT \cite{Paganelli2021edbt,DBLP:conf/edbt/BrunnerS20,DBLP:conf/www/ChenSZ20,DBLP:journals/pvldb/0001LSDT20}. However, there is no systematic analysis of their relative performance in these ER tasks.
% To cover this gap, we examine the relative performance of the main language models to answer the following research questions:
We cover this gap, answering the following research questions:

\begin{enumerate}[leftmargin=*,label=(Q\arabic*)]
    % \item What is their vectorization overhead? How does their dimensionality affect their time requirements? 
    \item How large is their vectorization overhead?
    % How does dimensionality affect execution time?
    \item What is their relative performance in blocking?
    % , in terms of both accuracy and execution time?
    % How does the custom, domain-specific vocabulary of ER datasets affect their effectiveness? 
    % {\color{blue}How robust are they to the domain-specific vocabulary of ER datasets?} {\color{red}[Dimitris: Why is this only for blocking? Doesn't it apply to matching too?]}
    % \item What is their relative effectiveness and time efficiency in both unsupervised and supervised matching? How does fine-tuning affect their performance in supervised matching?
    \item What is their relative performance in matching? How is it affected by fine-tuning in supervised matching?
    % {\color{red}[Dimitris: Doesn't this apply also in supervised matching?]}
    % \item How does fine-tuning affect their performance in supervised matching?
    % {\color{red}[Dimitris: Does fine-tuning apply perhaps more generally, and not only to supervised matching? I.e., in the sense of how sensitive each method is to the involved parameters?]}
    \item Is it possible to build high performing end-to-end ER pipelines based exclusively on pre-trained language models without providing them with labelled instances?
\end{enumerate}
% \vspace{-3pt}

To answer these questions, we perform a thorough experimental analysis,
% This suggests that no labelled dataset needs to be created for fine-tuning them. 
% Are they exclusively suitable for the schema-based or 
% We
% exclusively
% consider schema-agnostic settings, 
% which essentially represent every entity 
% where each entity is represented
% by the concatenation of all its attribute values. This
% approach inherently addresses 
% can cope with
% attribute heterogeneity (i.e., non-aligned schemata) as well as high levels of noise (e.g., misplaced values),
% while simplifying the use of language models, as no
% and does not require any
% background knowledge
% is required 
% for the data at hand.
% {\color{red}[Dimitris: Although the arguement is convincing, it may still sound like a limitation, and raise a question about that happens in the schema-aware case. Can we say something more? E.g., what about other works? If most deal with schema-agnostic, we could point this out. If many deal with schema-aware, it means this might still be worth of investigation. In that case, could we hypothesize that similar results may hold, or would someone need to do a similar experimental analysis for that scenario? In any case, perhaps it might be better to move this to a later point, where we talk in more detail about the methodology we followed, instead of pointing it out already here.]}
% In more detail, our experimental analysis involves 
involving
12 established language models: Word2Vec \cite{mikolov2013efficient, mikolov2013distributed} , GloVe \cite{pennington2014glove}, FastText 
\cite{bojanowski2017enriching}, BERT \cite{devlin2018bert}, AlBERT \cite{lan2019albert}, RoBERTa \cite{liu2019roberta}, DistilBERT \cite{sanh2019distilbert}, XLNet \cite{yang2019xlnet} as well as S-MPNet \cite{song2020mpnet}, S-GTR-T5 \cite{2020t5}, S-DistilRoBERTa and S-MiniLM \cite{wang2020minilm}. Some of them 
% have been widely used in blocking \cite{} and matching \cite{}, while others 
are applied to ER for the first time.
% {\color{red}[Dimitris: I think it would be nice here to include a summary table where the columns represent the different tasks we evaluate (vectorization, blocking, supervised matching, unsupervised matching), the rows represent each of the evaluated models, and in each cell we cite the previous works (if any) where that model has been used for that task. This table would be rather sparse, indicating that previous works have compared only some of these models and only for some of the tasks, while we compare all models for all tasks. Moreover, it would also show that some models have not been applied at all in previous works.]}
We apply them to 10 established real-world and 7 synthetic datasets, making the following contributions:

% {\color{red}[Dimitris: I think we also need some text in the intro that refers to other related existing surveys and points out that none of them does what we do.]}

% {\color{red}[Dimitris: Also, we need a paragraph that acts as a ``teaser'' for the results. Are there any useful findings, new insights, lessons learned, future directions, etc.?]}

\begin{itemize}[leftmargin=*]
    \item We organize them into a taxonomy that facilitates the understanding of their relative performance and we discuss their core aspects like their dimensionality and context awareness.
    % (i.e., static models vs dynamic ones which produce vectors according to the context) etc.
    % \item We 
    % % analytically 
    % thoroughly
    % examine the 
    % % temporal 
    % execution
    % cost of vectorization per model on GPUs.
    \item We examine the vectorization cost per model.
    %on GPU.
    % \item We compare their relative blocking performance in terms of effectiveness, time efficiency and scalability.
    \item We assess their blocking effectiveness, efficiency and scalability.
    \item We compare their relative performance for both supervised and unsupervised matching.
    \item We demonstrate that high performance can be achieved by an end-to-end solution that leverages the best language model both for blocking and matching in many datasets with either long or short textual attributes.   %The former applies to D2, D3, D8, which contain product descriptions, and D4, D9, which contain bibliographic data (i.e., publication titles and venues), whereas the rest of the datasets (including the synthetic ones of the scalability analysis) abound in short textual values. In the future, we will extend our analysis to datasets with mostly numeric attributes.
    
    % very high performance can be achieved in many datasets by an end-to-end solution that leverages the best language model both for blocking and matching without requiring any human intervention.
\end{itemize}

\section{Related Work}
\label{sec:relWork}

\noindent\textbf{Blocking.}
% Starting with blocking, 
The first approach to leverage embeddings vectors for blocking is DeepER \cite{DBLP:journals/pvldb/EbraheemTJOT18}, which uses GloVe for the vectorization of entities and hyperplane LSH for indexing and querying. AutoBlock \cite{DBLP:conf/wsdm/ZhangWSDFP20} goes beyond DeepER by leveraging FastText embeddings in combination with cross-polytope LSH. Yet, it treats blocking as a classification task, requiring a large number of labelled instances to train its bidirectional-LSTM. DeepBlocker \cite{DBLP:journals/pvldb/Thirumuruganathan21} is a generic framework for synthesizing deep learning-based blocking methods that support any language model. Most of the examined approaches, including the top-performing Auto-Encoder, leverage FastText embeddings, but support Word2Vec and GloVe too. It also considers two transformer-based models, which leverage Byte Pair Encoding, achieving slightly higher recall at the cost of much higher run-times. Given that efficiency is crucial for blocking, DeepBlocker with Auto-Encoder and FastText is the state of the art among~these~methods.
% blocking methods based on language models.

\noindent\textbf{Matching.}
Here, language models are typically used by supervised deep learning-based techniques. The first one is again DeepER \cite{DBLP:journals/pvldb/EbraheemTJOT18}, which leverages GloVe. Its generalization, DeepMatcher \cite{Mudgal2018sigmod}, is a framework for deep learning-based matching algorithms that supports the main pre-trained static models, i.e., 
% Word2Vec,
GloVe and FastText, with the last one being the predefined option. Similarly, GraphER \cite{DBLP:conf/aaai/Li0SZAW20}, Seq2SeqMatcher \cite{Nie2019cikm},  CorDEL \cite{DBLP:conf/icdm/WangSWDJ20}, MCAN \cite{DBLP:conf/www/ZhangNWST20}, HierMatcher \cite{DBLP:conf/ijcai/FuHHS20} and HIF-KAT \cite{DBLP:conf/acl/YaoLDLY0LZD20} constitute individual approaches that also leverage FastText embeddings. The reason is its character-level operation, which addresses noise in the form of misspellings, while also supporting unseen terms.
% , which are common in the domain-specific tasks of ER.

More recent works focus on BERT-based models, due to their dynamic, context-aware nature.
% , which considers the context of terms to achieve higher effectiveness. 
The first such work is EMTransformer \cite{DBLP:conf/edbt/BrunnerS20}, which 
% lays the ground for the analysis in Section \ref{sec:supMatching}. Originally, it 
considered BERT and its three main variants: XLNet, RoBERTa and DistilBERT. The same models are used by GNEM \cite{DBLP:conf/www/ChenSZ20}, which extends EMTransformer through a graph that captures the relations between all candidate pairs that are given as input to matching. GNEM also applies this idea to DeepMatcher, in combination with FastText embeddings. DITTO \cite{DBLP:journals/pvldb/0001LSDT20} extends EMTransformer by combining the BERT-based language models %(RoBERTa in particular)
with external, domain-specific information (e.g., POS tagging) and data augmentation, which provides more (synthetic) training instances. 

% On another line of research,
JointBERT \cite{DBLP:journals/pvldb/PeetersB21} goes beyond the classic binary classification definition of matching, by also supporting multi-class classification.
% As its name suggests, it relies on BERT.
% Another interesting ER problem is examined in \cite{Paganelli2021edbt}, which aims to automatically tune the configuration parameters of deep learning-based matching algorithms. To this end, it considers all BERT-based models in Section \ref{sec:bertModels}.
The problem of automatically tuning the configuration parameters of deep learning-based matching algorithms is examined in \cite{Paganelli2021edbt}, considering all BERT-based models in Section \ref{sec:bertModels}.

\textbf{Gaps.}
We observe that none of these works considers the main SentenceBERT models
% that are discussed in Section 
(cf. Section
\ref{sec:sbertModels}).
% for blocking or matching.
Moreover, no work has examined the performance of BERT models (cf. Section \ref{sec:bertModels}) on blocking. To the best of our knowledge, no work investigates the relative ER performance of the main language models in a systematic way. Closest to our work is an in depth analysis of how BERT works in the context of matching \cite{DBLP:journals/pvldb/PaganelliBBG22}, but it disregards all other models as well as the task of blocking. Surveys \cite{DBLP:journals/corr/abs-2003-07278}, books \cite{DBLP:series/synthesis/2020Pilehvar} and tutorials \cite{DBLP:journals/pvldb/Trummer22b} about language models are too generic, without any emphasis on ER, and thus, are orthogonal to our work. Our goal in this work is to cover these important gaps in the literature.

\noindent\textbf{Sentence-similarity tasks in NLP.} Semantic Textual Similarity (STS) is an important NLP task, that is mostly evaluated in Transformer Models with the STS-B task \cite{cer2017semeval} via the GLUE Benchmark \cite{wang2018glue}. Since in schema-agnostic ER all attributes in a record are concatenated into a ``sentence'', these tasks seem related. However, the characteristics of the input data differ. For instance, popular STS benchmarks contain sentences like image captions or news headlines, whereas typical ER benchmarks like the datasets considered here contain attributes such as person or product names, movie titles, addresses, etc. Concatenating such attributes does not form actual sentences, even if they are treated as such. Moreover, the task in STS is to predict a similarity score that indicates how similar the meaning of two sentences is, whereas in ER it is to decide whether two entity instances refer to the same real-world entity or not. Hence, it is not safe or straightforward to assume that language models will exhibit the same performance in ER as in STS.
\section{Language Models Used in the Evaluation}
\label{sec:approaches}

\begin{table}[]
\setlength{\tabcolsep}{1.5pt}
\small
\begin{tabular}{|c|c|c|c|c|c|}
\hline
\multicolumn{1}{|c|}{Model} & \multicolumn{1}{c|}{Dim.} & \multicolumn{1}{c|}{$|Seq.|$} & \multicolumn{1}{c|}{Param.} & 
\multicolumn{1}{c|}{Blocking} & 
\multicolumn{1}{c|}{Matching}\\
\hline
\hline
Word2Vec (WC) & 300 & - & - & \cite{DBLP:journals/pvldb/Thirumuruganathan21} & 
\cite{Mudgal2018sigmod}\\
\multirow{ 2}{*}{FastText (FT)} & \multirow{ 2}{*}{300} & - & - & \multirow{ 2}{*}{\cite{DBLP:conf/wsdm/ZhangWSDFP20,DBLP:journals/pvldb/Thirumuruganathan21}} & 
\cite{Mudgal2018sigmod,Nie2019cikm,DBLP:conf/aaai/Li0SZAW20},
 \\ & & & & &
 \cite{DBLP:conf/icdm/WangSWDJ20,DBLP:conf/www/ZhangNWST20,DBLP:conf/ijcai/FuHHS20,DBLP:conf/acl/YaoLDLY0LZD20}
\\
GloVe (GE) & 300 & - & - &  \cite{DBLP:journals/pvldb/EbraheemTJOT18,DBLP:journals/pvldb/Thirumuruganathan21} & 
\cite{DBLP:journals/pvldb/EbraheemTJOT18,Mudgal2018sigmod}\\
\hline
BERT (BT) & 768 & 100 & 110M & - &   \cite{DBLP:conf/edbt/BrunnerS20,DBLP:conf/www/ChenSZ20,DBLP:journals/pvldb/PeetersB21,DBLP:journals/pvldb/0001LSDT20,Paganelli2021edbt}  \\
AlBERT (AT) & 768 & 100 & 12M & - & \cite{Paganelli2021edbt}\\
RoBERTa (RA) & 768 & 100 & 125M & - & \cite{DBLP:conf/edbt/BrunnerS20,DBLP:conf/www/ChenSZ20,DBLP:journals/pvldb/0001LSDT20,Paganelli2021edbt}\\
DistilBERT (DT) & 768 & 100 & 66M & - & \cite{DBLP:conf/edbt/BrunnerS20,DBLP:conf/www/ChenSZ20,DBLP:journals/pvldb/0001LSDT20,Paganelli2021edbt} \\
XLNet (XT) & 768 & 100 & 110M & - & \cite{DBLP:conf/edbt/BrunnerS20,DBLP:conf/www/ChenSZ20,DBLP:journals/pvldb/0001LSDT20,Paganelli2021edbt} \\
\hline
S-MPNet (ST) & 768 & 384 & 110M & - & - \\
S-GTR-T5 (S5) & 768 & 512 & 110M & - & - \\
S-DistilRoBERTa (SA) & 768 & 512 & - & - & - \\
S-MiniLM (SM) & 384 & 256 & 22M & - & - \\
\hline
\end{tabular}
\caption{The language models used in our experiments.
% For each model, we indicate the vector dimensionality, the maximum sequence length, the number of parameters, the datasets used for training, the training task (WE:Word Encoder; SE:Sentence Encoder; MLM:Masked Language Modeling; NSP:Next Sentence Prediction; STC:Sequence \& Token Classification; QA:Question Answering; SS:Semantic Search), and ER works where it has been used.
}
\label{tb:modelCharacteristics}

\end{table}

We have used three categories of language models in our evaluation: (1) \textit{Static models}, which associate every token with a fixed embedding vector; (2) \textit{BERT-based models}, which vectorize every token based on its context; (3) \textit{Sentence-BERT models}, which associate every sequence of tokens with a context-aware embedding vector.
% \begin{enumerate}[leftmargin=*]
%     \item \textit{Static pre-trained models}, which associate every token with a fixed embedding vector.
%     \item \textit{BERT-based models}, which vectorize every token based on its context.
%     \item \textit{Sentence-BERT models}, which associate every sequence of tokens with a context-aware embedding vector.
% \end{enumerate}

For each category, we selected a representative set of language models based on the following criteria: (i) popularity in the ER and NLP literature, (ii) support for the English language, and (iii) availability of open-source implementation. Ideally, this implementation should include a documentation that facilitates the use of each model and all language models should be implemented in the same language so as to facilitate run-time comparisons. Thus, our experimental analysis relies on out-of-the-box, open-source implementations of the main language models that can be easily used by any practitioner that is not necessarily an expert in~the~field.

All static pre-trained and BERT-based models that are mentioned in Section \ref{sec:relWork} satisfy these criteria and, thus, are included in our analysis. 
% {\color{red}[Dimitris: the previous sentence seems confusing]} 
Given that none of these ER works considers an established SentenceBERT model, %In the original paper of SBERT, they implemented and evaluated BERT and RoBERTa. The following models 
we selected the top four ones from the SBERT library (\url{https://www.sbert.net/docs/pretrained_models.html}), based on their scores and overall need of resources.
The technical characteristics of the selected models are summarized in Table \ref{tb:modelCharacteristics}.
For each model, we indicate the vector dimensionality, the maximum sequence length, the number of parameters,
% the datasets used for training\footnote{GN:Google News; W:Wikipedia (en); T:Twitter; CC:Common Crawl; BC:BookCorpus},
% the training task\footnote{WE:Word Encoder; SE:Sentence Encoder; MLM:Masked Language Modeling; NSP:Next Sentence Prediction; STC:Sequence \& Token Classification; QA:Question Answering; SS:Semantic Search},
and the ER works that have used it for blocking or matching.
% {\color{blue}We observe that static models have been used for both blocking and matching, BERT-based models have only been used for matching, while Sentence-BERT models have not been used in previous works.}
% For model implementations, we used two highly popular Python packages for language models: Gensim\footnote{\url{https://radimrehurek.com/gensim}} and Hugging Face\footnote{\url{https://huggingface.co}}. The former offers Word2Vec and FastText, while the latter provides all other models. Note that for cases where multiple versions of a language model exist, we opted for the more representative and popular one. The interested reader can refer to the repository of our project for detailed links to each version of the language model that was used.
%This leaves out ELMo.
% {\color{red}Alex: Not sure why we should have a special short subsection, rather than a short sentence as conclusion to previous introduction.}
% Based on the aforementioned evolution of language models, we selected
% \begin{enumerate}
%     \item 
%     \item English language.
%     \item Popularity. Based on previous works.
%     \item Documentation
%     \item Same programming language, 
% \end{enumerate}
Below, we briefly describe the selected models per category 
% that meet these criteria below, 
in chronological order. Most models have several versions (e.g., base, large) that differ in the number of learned parameters. To ensure a fair comparison, we consider the base version of each model. We also conducted some tests with larger versions, without observing notable differences in the results.

% \subsection{Taxonomy}

% {\color{red}Check this survey: https://arxiv.org/pdf/2003.07278.pdf }

% \subsection{Introduction}

% {\color{blue}
% Note that most models have various versions that typically differ in the number of learned parameters. Given that time efficiency is crucial for ER and that our goal is to ensure a fair comparison, we consider the base (i.e., simpler and smaller) version of each model. Besides, preliminary experiments demonstrated insignificant changes in the relative performance of the considered models, when using more complex and computationally expensive versions. 
% }

\subsection{Static pre-trained models}

These models were introduced to capture semantic similarity in text, encapsulating knowledge from large corpora. They replace the traditional high-dimensional sparse vectors with low-dimensional dense ones of fixed size, which define a mathematical space, where semantically similar words tend to have low distance.

\textbf{Word2Vec} \cite{mikolov2013efficient, mikolov2013distributed} is a shallow two-layer neural network that receives a corpus as input and produces the corresponding vectors per word. It employs a local context window, as a continuous bag-of-word (order-agnostic) or a continuous skip-gram (order-aware). The latter can link words that behave similarly in a sentence, but fails to utilize the statistics of a corpus.
%\footnote{\url{https://radimrehurek.com/gensim/models/word2vec.html}} 

\textbf{GloVe} \cite{pennington2014glove} 
% offered the solution to combine 
combines 
matrix factorization, i.e., the global co-occurrence counts, with a local context window, i.e., word analogy. It is trained on large corpora, such as Wikipedia, to provide pre-trained vectors for general use.
% , thus highlighting the importance of pre-training in NLP.
Since it operates on a global dictionary, it identifies words with a specific writing and fails to detect slight modifications.
% , i.e. english-american pronunciations, plurals, etc.
%\footnote{\url{https://huggingface.co/sentence-transformers/average_word_embeddings_glove.840B.300d}}

\textbf{FastText} \cite{bojanowski2017enriching}
% does not conceive a word as a single string, but as a group of n-grams.
conceives each word as a group of n-grams instead of a single string.
It is trained to vectorize n-grams.
It then 
% and thus, each word is represented 
represents each word as the sum of its underlying n-grams. %This method led to a faster approach, since the total number of n-grams is limited.

%\footnote{\url{https://radimrehurek.com/gensim/models/fasttext.html\#gensim.models.fasttext.FastText}}

% While these models offered some solutions to find synonyms in a text and offer pre-trained word embeddings from large corpora, they provide a single vector per word, which can be restristive, based on the meaning of a word inside a sentence. This is the case of polysemy, which the next family of models solves.

\subsection{BERT-based models}
\label{sec:bertModels}

In static models, each word has a single representation, which is restrictive, since words often have multiple meanings based on their context. Context-awareness was introduced by the transformer models \cite{vaswani2017attention}, which are a natural evolution of Encoders-Decoders \cite{sutskever2014sequence}.
% , which constitute their building blocks.
The latter have many advantages, such as handling larger areas of text around a given word. Nonetheless, they cannot encapsulate any significant relationships between words. This has been fixed with the introduction of Attention \cite{bahdanau2014neural}, which facilitates the communication between each encoder / decoder, by sharing all of the corresponding hidden states and not just the last one.
% -- unlike the preceding models.
An extra optimization, suggested by the Transformer model, is the Multi-Head attention, which led each encoder to run in parallel. Another useful approach is the use of Positional Encoding for each token, which addresses polysemy, i.e., the fact that the same word has different meanings in different sentences.

% Transformer \cite{vaswani2017attention} models introduced a significant change in NLP. They are a natural evolution of Encoders-Decoders \cite{sutskever2014sequence}, since the former constitute their building blocks. Encoders-Decoders models had many advantages, such as capturing context or being able to handle larger areas of text around a given word. Nonetheless, they could not highlight any significant relationships between words, something that was fixed by the introduction of Attention \cite{bahdanau2014neural}, which facilitated the communication between each encoder / decoder, by sharing all of the corresponding hidden states and not just the last one, as done previously. An extra optimization, suggested by the Transformer model, was the Multi-Head attention, which led each encoder to run in parallel. Another useful approach was the use of Positional Encoding for each token, which solved the problem of polysemy, i.e. a word having different meaning in different sentences.

\textbf{BERT} \cite{devlin2018bert}, which stands for ``Bidirectional Encoder Representations from Transformers'', was the next major step in the evolution of the transformer models. Its main contribution is the use of multiple transformers -- only the encoder part, since it is a language representation model -- to pre-train vectors for general use. These vectors can be further fine-tuned by adding an output layer for a wide variety of tasks. BERT is trained on two tasks: masked language modeling (MLM) 
and next sentence prediction (NSP). The former is token-based, since it tries to predict a masked token based on the unmasked tokens of a sentence. NSP is sentence-based, since it receives a first sentence as input and tries to predict whether a second sentence can follow it.

\textbf{AlBERT} \cite{lan2019albert}, which stands for ``A little BERT'', is a lighter version of BERT. BERT-base comprises 12 encoders and 110M parameters with 768 hidden and equal embedding layers (cf. Table \ref{tb:modelCharacteristics}). AlBERT trains only the first encoder and then shares all its weights with the rest of the encoders. It also reduces the embedding layer by factorization to 128 layers. These reduce the total number of parameters to 12M, significantly lowering the training time.

\textbf{RoBERTa} \cite{liu2019roberta} stands for ``Robustly Optimized BERT pre-training Approach''. 
% It comes to highlight some points regarding the training of the BERT model and improve its performance. More specifically:
% Its training differs from that of BERT in the following ways:
Compared to BERT:
(1) it is trained with more data and more and bigger batches; (2) it removes the next sentence prediction objective; (3) it changes the masked tokens per epoch to make the model more robust.
% The end result 
It
typically outperforms the original BERT \cite{liu2019roberta}.

\textbf{DistilBERT} \cite{sanh2019distilbert}, which stands for ``Distillation BERT'', is a lighter version of BERT that uses distillation \cite{hinton2015distilling, romero2014fitnets}.
% on BERT. 
A second version of the original model is built, where only half of the attention layers are used -- every second layer is omitted -- and a special loss function is used in the training that 
% computes to 
compares the teacher (original BERT) with the student (DistilBERT).
% and, consequently, train the latter.

\textbf{XLNet} \cite{yang2019xlnet} tries to overcome a certain drawback of BERT: the fact that it cannot utilize the knowledge of a predicted masked token as input for a second masked token, thus making each prediction independent and possibly false. XLNet introduces a variation of the MLM task, called permutation language modeling (PLM). The goal of the new task is to permute the tokens of one sentence in all possible matters without using any masked tokens. 

\begin{table*}[t]\centering
\small
\setlength{\tabcolsep}{3.5pt}
    % \textcolor{red}{[Dimitris: Maybe keep either NVP or p?]}}
	\begin{tabular}{ | l | r | r | r | r | r | r | r | r | r | r || r | r | r | r | r | r | r |}
		\cline{2-18}
		\multicolumn{1}{c|}{}&
		\multicolumn{1}{c|}{$\mathbf{D_{1}}$} &
		\multicolumn{1}{c|}{$\mathbf{D_{2}}$} &
		\multicolumn{1}{c|}{$\mathbf{D_{3}}$} &
		\multicolumn{1}{c|}{$\mathbf{D_{4}}$} &
		\multicolumn{1}{c|}{$\mathbf{D_{5}}$} &
        \multicolumn{1}{c|}{$\mathbf{D_{6}}$} &
        \multicolumn{1}{c|}{$\mathbf{D_{7}}$} &
        \multicolumn{1}{c|}{$\mathbf{D_{8}}$} &
        \multicolumn{1}{c|}{$\mathbf{D_{9}}$} &
        \multicolumn{1}{c||}{$\mathbf{D_{10}}$} &
        \multicolumn{1}{c|}{$\mathbf{D_{s_1}}$} &
		\multicolumn{1}{c|}{$\mathbf{D_{s_2}}$} &
		\multicolumn{1}{c|}{$\mathbf{D_{s_3}}$} &
		\multicolumn{1}{c|}{$\mathbf{D_{s_4}}$} &
		\multicolumn{1}{c|}{$\mathbf{D_{s_5}}$} &
        \multicolumn{1}{c|}{$\mathbf{D_{s_6}}$} &
        \multicolumn{1}{c|}{$\mathbf{D_{s_7}}$} \\
		\hline
        \hline
        Dat${_1}$ & Rest${_1}$ & Abt & Amz & DBLP & IMDb & IMDb & TMDb & Wmt &  DBLP & IMDb & 
        \multirow{ 2}{*}{$D_{10K}$} & \multirow{ 2}{*}{$D_{50K}$} & \multirow{ 2}{*}{$D_{100K}$} &
        \multirow{ 2}{*}{$D_{200K}$} & \multirow{ 2}{*}{$D_{300K}$} & \multirow{ 2}{*}{$D_{1 }$} & \multirow{ 2}{*}{$D_{2M}$}\\ 
        Dat${_2}$ & Rest${_2}$ & Buy & GPr. & ACM & TMDb &  TVDB & TVDB & Amz & Scholar & DBP & & & & & & & \\
        \hline
        $|V_1|$ & 339 & 1,076 & 1,354 & 2,616 & 5,118 & 5,118 & 6,056 & 2,554 & 2,516 & 27,615 & \multirow{ 2}{*}{10K} & \multirow{ 2}{*}{50K} & \multirow{ 2}{*}{100K} & \multirow{ 2}{*}{200K} & \multirow{ 2}{*}{300K} & \multirow{ 2}{*}{1M} & \multirow{ 2}{*}{2M}\\
		$|V_2|$ & 2,256 & 1,076 & 3,039 & 2,294 & 6,056 & 7,810 & 7,810 & 22,074 & 61,353 & 23,182  & & & & & & & \\
        % NVP${_1}$ & 1,130 & 2,568 & 5,302 & 10,464 & 21,294 & 21,294 & 23,761 & 14,143 &  10,064 & 1.6$\cdot10^5$ \\
        % NVP${_2}$ & 7,519 & 2,308 & 9,110 & 9,162 & 23,761 & 20,902 & 20,902 & 1.14$\cdot10^5$ & 1.98$\cdot10^5$ & 8.2$\cdot10^5$ \\
        \hline
        $|A_1|$ & 7 & 3 & 4 & 4 & 13 & 13 & 30 & 6 & 4 & 4 & \multirow{ 2}{*}{12} & \multirow{ 2}{*}{12} & \multirow{ 2}{*}{12} & \multirow{ 2}{*}{12} & \multirow{ 2}{*}{12} & \multirow{ 2}{*}{12} & \multirow{ 2}{*}{12} \\
        $|A_2|$ & 7 & 3 & 4 & 4 & 30 & 9 & 9 & 6 & 4 & 7 & & & & & & & \\
% 		$|\bar{p}_1|$ & 3.33 & 2.39 & 3.92 & 4.00 & 4.16 & 4.16 & 3.92 & 5.54 & 4.00 & 5.63 \\
% 		$|\bar{p}_2|$ & 3.33 & 2.14 & 3.00 & 3.99 & 3.92 & 2.68 & 2.68 & 5.18 & 3.24 & 35.20 \\
        \hline
		$|D|$ & 89 & 1,076 & 1,104 & 2,224 & 1,968 & 1,072 & 1,095 & 853 & 2,308 & 22,863 & 8,705 & 43,071 & 85,497 & 172,403 & 257,034 & 857,538 & 1,716,102\\
		
		\hline
		$\bar{|S|}$ & 18.67 & 198.64 & 792.43 & 133.29 & 81,49 & 71.48 & 104.16 & 103.35 & 115.57 & 54.04 & 84.32 & 84.21 & 84.34 & 84.30 & 84.30 & 84.31 & 84.32\\
		
% 		$||V_1 \times V_2||$ & 7.65$\cdot10^5$ & 1.16$\cdot10^6$ & 4.11$\cdot10^6$ &  6.00$\cdot10^6$ & 3.10$\cdot10^7$ & 4.00$\cdot10^7$ & 4.73$\cdot10^7$ & 5.64$\cdot10^7$ &  1.54$\cdot10^8$ & 6.40$\cdot10^8$ \\
% 		\hline
% 		Attribute & name & name & title & title & {\color{red}title} & title & title & title & title & title \\
% 		Attribute$_2$ & phone\_number & description & description & authors & name & name & name & modelno & authors & starring/actor\_name \\
		\hline
		\multicolumn{1}{c}{}&
		\multicolumn{10}{c}{\textbf{(a) Clean-Clean ER}} &
		\multicolumn{7}{c}{\textbf{(b) Dirty ER}} \\
	\end{tabular}
	    \caption{(a) The real datasets for Clean-Clean ER, and (b) the synthetic datasets for Dirty ER, in increasing
	   % size (Cartesian product)
	    total size,
	    showing the number of entities ($|V_x|$), 
	   % NVP$_x$ for the total number of name-value pairs, 
	   % the number of
	    attributes ($|A_x|$),
    % and $|\bar{p}_x|$ for the average number of name-value pairs per entity profile in Dataset$_x$. 
    % the number of
    and duplicates ($|D|$), and the average sentence length in characters ($\bar{|S|}$). 
    % {\color{red}A-priori matching probability?? We do not need two attributes per dataset. We should just keep the best one to avoid confusing the reviewers.}. 
    % and $||V_1 \times V_2||$ for the number of pairwise comparisons executed by the brute-force approach. 
    }
    \vspace{-12pt}
	\label{tb:datasets}
\end{table*}
\begin{table}[t]\centering
    \setlength{\tabcolsep}{3pt}
    \vspace{-10pt}
	\begin{tabular}{ | l | r | r | r | r | r | r |}
	    \cline{2-7}
        \multicolumn{1}{c|}{} & \multirow{2}{*}{Dataset 1} &  \multirow{2}{*}{Dataset 2} & Total & Testing &  \multirow{2}{*}{Duplicates} & Attri- \\
        \multicolumn{1}{c|}{} &  &   & Pairs & Pairs & & butes \\
        \hline
        \hline
        $DSM_1$ & Abt & Buy & 9,575 & 1,917 & 1,028 & 3 \\ 
        $DSM_2$ & iTunes & Amazon & 539 & 110 & 132 & 8 \\
        $DSM_3$ & DBLP & ACM & 12,363 & 2,474 & 2,220 & 4 \\
        $DSM_4$ & DBLP & Scholar & 28,707 & 5,743 & 5,347 & 4 \\
        $DSM_5$ & Walmart & Amazon &  10,242 & 2,050 & 962 & 5 \\
		\hline
	\end{tabular}
	\caption{The datasets used in the Supervised Matching task.}
	\vspace{-28pt}
	\label{tb:smDatasets}
\end{table}

\subsection{SentenceBERT models}
\label{sec:sbertModels}

% While BERT-based models have many advantages and are used widely for many tasks \cite{DBLP:series/synthesis/2020Pilehvar}, they are mostly built to support token-based tasks (e.g. MLM).
BERT-based models are mostly built to support token-based tasks. Supervised tasks that need a sentence representation
% , like next sentence prediction, 
may utilize the special token [CLS], but in regression or unsupervised tasks this
% seems to
produces a computational overhead, as all pair-wise combinations need to be fed into the model. Using [CLS] or averaging the last output layer is often worse than GloVe embeddings \cite{reimers2019sentence}. SBERT \cite{reimers2019sentence} fixes this problem by suggesting a Siamese architecture, i.e. two identical models, with each one taking as input one of the sentences. This architecture produces the corresponding vectors and then evaluates the combination of the two vectors, based on the defined task, e.g., the cosine similarity of two sentences. Note that this architecture is orthogonal to the underlying BERT model.

% While BERT-based models have many advantages and they are used widely for many tasks, they are mostly built to support token-based tasks. Supervised tasks that need a sentence representation, like Next Sentence Prediction (NSP) utilize the special token [CLS], but in regression or unsupervised tasks seem to produce a computational overhead, as all pair-wise combinations need to be fed into the model. The approach to utilize [CLS] or average the output of last output layer is usually worse than GloVe embeddings. SBERT \cite{reimers2019sentence} fixes this problem by suggesting a Siamese architecture, i.e. one model considered as two, that takes as input each of the sentences, produces the corresponding vectors and then evaluates the combination of the two vectors, based on the defined task, e.g. cosine similarity of two sentences. It is clear that this method is orthogonal to the underlying BERT model. In the original paper, they implemented and evaluated BERT and RoBERTa. The following models were selected from SBERT library \footnote{{\url{https://www.sbert.net/docs/pretrained_models.html}}}, based on their scores and overall need of resources.

\textbf{S-MPNet} extends MPNet \cite{song2020mpnet}, which overcomes the drawbacks of BERT and XLNet in the MLM and PLM tasks, respectively. For the former, it solves the dependency between masked tokens predictions by permuting the tokens in a sentence. For the latter, it utilizes position information to reduce position discrepancy.

\textbf{S-GTR-T5} extends GTR \cite{ni2021large}, which is a dual encoder that encodes two pieces of text into two dense vectors respectively. This is typically used to encode a query and a document to compute their similarity for dense retrieval. GTR models are built on top of T5~\cite{2020t5}, an encoder-decoder model that aims to unify all NLP tasks under a single model.
Instead of introducing a new model, it uses existing techniques. The text-to-text transfer transformer (T5) is an encoder-decoder model, where each transformer has been structured in the same way as in BERT. The rationale is that while BERT is an encoder model, the encoder-decoder model produces good results too 
and can be used for other tasks that an encoder cannot perform (e.g., text generation). T5 is trained on a dataset called ``Colossal Clean Crawled Corpus'', containing hundreds of gigabytes of clean English text from the Web.

\textbf{S-DistilRoBERTa} applies distillation to the RoBERTa model
% , which has been through  
to produce a student, lighter model. This model is coupled with the Siamese architecture to produce the final model.
 
\textbf{S-MiniLM} extends MiniLM \cite{wang2020minilm}, which distills BERT to produce a much lighter student. Unlike DistilBERT
% , which takes one layer out of two, 
and other distillation strategies \cite{jiao2019tinybert,sun2019mobilebert}, which are bound to the architecture of the teacher layers, it mimics only the self-attention modules, which are the most important ones in the architecture.
% --- especially the last transformer's self-attention modules.
Thus, it can define the number of layers in each transformer, reducing the total number of required parameters. The distillation occurs in the pre-trained model 
% since they also want 
to avoid the computationally expensive
% performing 
fine-tuning of~the~teacher.

\section{Experimental Setup}
\label{sec:expSettings}

\subsection{Datasets}

% {\color{red}[Dimitris: For the rest?]}
% {\color{red}{Mention Supervised Matching Repo?}}

% GPU: vectorization, blocking, unsup matching, 

\noindent\textbf{Main Datasets.} Most experiments were conducted using the following ten real-world, established datasets for ER:
 $D_{1}$, which is offerred by OAEI 2010\footnote{\url{http://oaei.ontologymatching.org/2010/im}}, contains descriptions of restaurants.
 $D_2$ contains products extracted from two online retailers, Abt.com and Buy.com \cite{DBLP:journals/pvldb/KopckeTR10}.
 $D_3$ comes from the same domain, matching products from Amazon.com and the Google Base data API (Google Pr.)\cite{DBLP:journals/pvldb/KopckeTR10}.
 $D_4$ involves bibliographic data from two publication repositories, DBLP and the ACM digital library \cite{DBLP:journals/pvldb/KopckeTR10}. 
 $D_5$, $D_6$ and $D_7$ consist of three individual data sources, which comprise movie descriptions from imdb.com (IMDb) and themoviedb.org (TMDb) as well as TV shows from TheTVDB.com (TVDB) \cite{DBLP:journals/corr/abs-2101-06126}.
 $D_8$ is another dataset from the product matching domain, involving descriptions from Walmart and Amazon \cite{Mudgal2018sigmod}. 
 Similar to $D_4$, $D_9$ contains bibliographic data from DBLP and Google Scholar \cite{DBLP:journals/pvldb/KopckeTR10}.
 Finally, $D_{10}$ matches movies from IMDb and DBpedia \cite{DBLP:conf/wsdm/PapadakisINF11}, but has no overlap with the IMDb data source of $D_5$ and $D_6$.

All these datasets are publicly available through Zenodo\footnote{\url{https://zenodo.org/record/6950980}} in CSV format. Their detailed
characteristics are shown in Table~\ref{tb:datasets}(a). Note that all of them correspond to the \textit{Clean-Clean ER} task, also known as \textit{Record Linkage}, where the input comprises two individually duplicate-free, but possibly overlapping data sources and the goal is to detect the matching entities they share \cite{DBLP:series/synthesis/2021Papadakis,DBLP:series/synthesis/2015Christophides}.

\noindent\textbf{Datasets for Blocking Scalability.} To evaluate blocking scalability, we employ the datasets shown in Table~\ref{tb:datasets}(b), which are widely used in the literature for this purpose \cite{DBLP:journals/pvldb/0001SGP16,DBLP:journals/is/KenigG13,DBLP:journals/tkde/Christen12}. These correspond to \textit{Dirty ER}, a.k.a., Deduplication, where a single data source containing duplicates is given as input \cite{DBLP:series/synthesis/2021Papadakis,DBLP:series/synthesis/2015Christophides}.
They were artificially generated by Febrl \cite{DBLP:conf/kdd/Christen08a}, in the following way: clean entities were initially created by extracting real names (given and surname) and addresses (street number,
name, postcode, suburb, and state names) from frequency tables of real census data. 
Next, duplicate entities were randomly generated according to realistic error rates and types (e.g., by inserting, deleting or replacing characters or words).
In the end result, 40\% of all entities are matching with at least another one. There are at most 9 duplicates per record and up to 3 and 10 modifications per attribute value and record, resp.

\noindent\textbf{Datasets for Supervised Matching.} For this task, we used the same five datasets as in~\cite{DBLP:conf/edbt/BrunnerS20}, which are widely used in the literature~\cite{DBLP:journals/pvldb/0001LSDT20,Mudgal2018sigmod}. Their technical characteristics are reported in Table \ref{tb:smDatasets}; 60\% of all pairs form the training set, while the rest are equally split between the validation and test set. Most of them stem from the datasets in Table~\ref{tb:datasets}(a): $DSM_1$ is part of $D_2$, $DSM_3$ of $D_4$, $DSM_4$ of $D_9$ and $DSM_5$ of $D_8$. The only exception is $DSM_2$, which is not included in Table \ref{tb:datasets}(a), due to the lack of its complete groundtruth. Note also that $DSM_2$-$DSM_5$ 
% are \textit{dirty datasets} in the sense that they 
% include artificial noise in the form of missing and misplaced attribute values . The goal of these settings is 
so as to increase their difficulty \cite{DBLP:journals/pvldb/0001LSDT20}.

\subsection{Settings}

In all experiments, we  
% use the values of one or two specific attributes in each dataset. These attributes, which combine high coverage with high distinctiveness, are reported in the last two rows of Table \ref{tb:rlDatasets}. The settings that consider only specific attributes are characterized as \textit{schema-based}, while the experiments 
consider all attribute values per entity, i.e., each entity is represented by the concatenation of all its attribute values. These \textit{schema-agnostic} settings inherently address misplaced attribute values (e.g., cases where person names are associated with their profession), while exhibiting 
%and has shown to be effective for yielding 
high effectiveness both in blocking \cite{DBLP:journals/pvldb/0001SGP16,DBLP:journals/pvldb/0001APK15} and matching \cite{DBLP:conf/edbt/BrunnerS20}. 

% Schema-based experiments, which exclusively consider the values of one or two specific attributes per dataset that
% . These attributes, 
% combine high coverage with high distinctiveness, 
% are reported in the extended version of our work on GitHub.\footnote{\url{https://github.com/alexZeakis/Embeddings4ER/tree/main/extended}}
% \footnote{\url{https://github.com/alexZeakis/Embeddings4ER/tree/main/extended}}
% the last two rows of Table \ref{tb:rlDatasets}. The settings that consider only specific attributes are characterized as \textit{schema-based}, while the experiments 

All our code and datasets used in this experimental analysis are also publicly available in the above repository.
% \footnotemark[\ref{GitHub}]
% \footnote{\url{https://github.com/alexZeakis/Embeddings4ER}}.
For the language models, we used the implementations provided by two highly popular Python packages: Gensim\footnote{\url{https://radimrehurek.com/gensim}} and Hugging Face\footnote{\url{https://huggingface.co}}. The former offers Word2Vec and FastText, while the latter provides all other models.
% Note that for cases where multiple versions of a language model exist, we opted for the more representative and popular one.
% The interested reader can refer to our code and datasets repository\footnote{\url{https://github.com/alexZeakis/Embeddings4ER}} for more details.
% the repository of our project for detailed links to each version of the language model that was used.
All experiments were executed on a server with Ubuntu 20.04, AMD Ryzen Threadripper 3960X 24-Core processor, 256 GB RAM and an RTX 2080Ti~GPU.
% Our code and datasets are available on GitHub.\footnote{\url{https://github.com/alexZeakis/Embeddings4ER}}
The GPU is used where possible, i.e., for vectorization of the dynamic models, similarity score calculation in matching and blocking, nearest neighbor search in blocking, and fine-tuning of the dynamic models in supervised matching.

\subsection{Evaluated Tasks and Methodology}

We evaluate the language models in the following four tasks.

\noindent\textbf{Vectorization.} This converts every given textual entity into its embedding vector. We consider schema-agnostic ER, where each entity is represented by a ``sentence'' that is formed by concatenating all its textual attributes. For Word2Vec and GloVe, which only support word embeddings, we tokenize this sentence into words and average their vectors to obtain a single one. FastText internally splits the sentence into smaller n-grams and aggregates their embeddings into a single vector. The rest of the models generate an embedding for the entire sentence. In this task, we compare the execution time for each model.

\noindent\textbf{Blocking.} Given the vectorized entities of the input datasets, blocking produces a set of candidate pairs. For each input entity, we perform a \textit{nearest neighbor search} (NNS) to find the $k$ most similar vectors to it. For the datasets for Clean-Clean ER in Table \ref{tb:datasets}(a), we perform 
exact NNS. For each entity in the smallest of the two datasets, we compute all similarity scores and return the $k$ nearest neighbors. For the datasets for Dirty ER in Table \ref{tb:datasets}(b), which are significantly larger, we perform approximate NNS. According to the state of the art, we follow \cite{li2019approximate}, leveraging an HNSW \cite{DBLP:journals/pami/MalkovY20} index. This is a graph-based index, serving as an approximation to a Delaunay graph, but with long-range links as well, to support the small-world navigation property. To avoid the connectivity issues raised by high-degree nodes in the original NSW, HNSW introduces a multi-layer separation of links based on their degree as well as on an advanced heuristic for better selecting the neighbors per node. The resulting index offers very good querying times with the tradeoff of a large overhead in building the index. In our experiments, we used the implementation provided by FAISS.\footnote{\url{https://faiss.ai/cpp\_api/struct/structfaiss\_1\_1IndexHNSW.html}} First, we vectorize and index all input entities. Then, we query the index with every entity $e$ to retrieve its $k$ approximate nearest neighbors in terms of Euclidean distance.

\noindent\textbf{Unsupervised Matching.}
This is considered a clustering task, where each cluster of entities corresponds to a different real-world object. For two datasets in Clean-Clean ER, we model the task as bipartite graph matching, where each entity from the one dataset is matched with at most one entity from the other. To solve this, we apply \textit{Unique Mapping Clustering} (UMC) \cite{DBLP:conf/kdd/Lacoste-JulienPDKGG13}, which achieves both high effectiveness and time efficiency \cite{DBLP:conf/edbt/0001ETH22}.

The entire process is as follows. First, we calculate the similarity score between all pairs of entities using the following formula: $sim(e_i, e_j) = 1 / (1 + dist(v_i, v_j))$, where $dist$ denotes the Euclidean distance between the embedding vectors $v_i$ and $v_j$ of the entities $e_i$ and $e_j$, respectively. We do not perform blocking here to avoid its impact on the effectiveness of UMC. As execution time here we measure exclusively the run-time of UMC, i.e., assuming that all similarity scores have been computed already. UMC iterates over all pairs in descending order of similarity score,  until all entities from the smallest dataset have been matched, or there are no more pairs that exceed a given similarity threshold $\delta$. This threshold is the only configuration parameter of UMC. To fine-tune it, we consider all values in $[0.05, 0.95]$ with a step of 0.05 and select as optimal the one maximizing F-measure. In this way, our experimental analysis considers the maximum effectiveness per language model, comparing their potential. In practical settings, though, specifying the optimal similarity threshold for UMC is a non-trivial task.

\begin{figure}[t]
\centering
\includegraphics[trim=0.02cm 0.12cm 0.12cm 0.12cm, clip, width=0.4\textwidth]{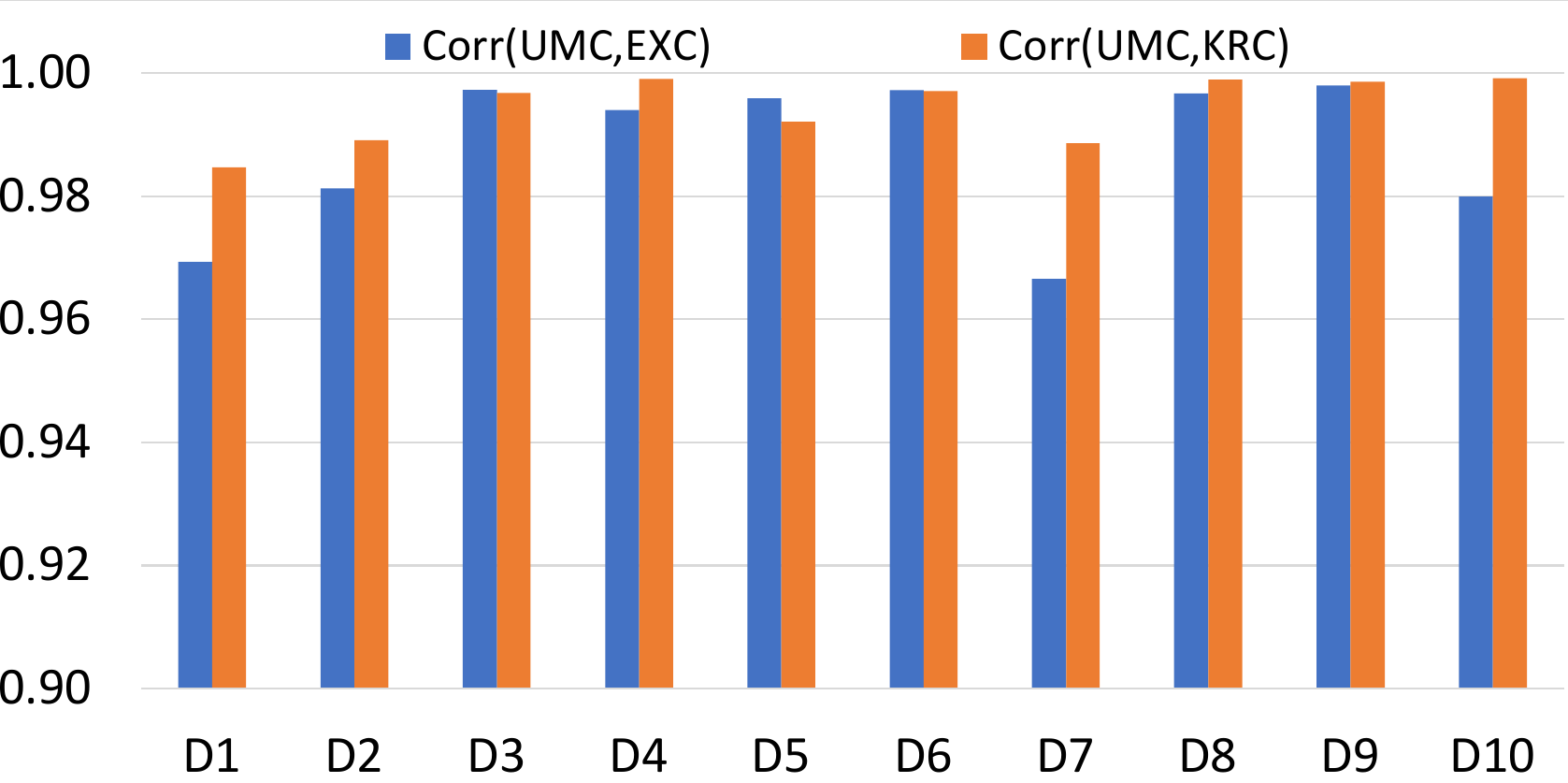}
\caption{Pearson Correlation between Unique Mapping Clustering (UMC), Exact Clustering (EXC) and Kiraly Clustering (KRC).}
\label{fig:umc_correlation}
\end{figure}

To ensure generality of our results, besides UMC we also tested two other highly-performing algorithms from \cite{DBLP:conf/edbt/0001ETH22}: \textit{Exact Clustering}, which matches two entities if they are mutually the best matches, and \textit{Kiraly Clustering}, which provides a linear time approximation to the maximum stable marriage problem. In both cases, the results exhibited very high (>0.9) Pearson correlation with those of UMC. 
% Please refer to the extended version of our work for more details$^3$.
This can be seen in Figure \ref{fig:umc_correlation}.

\noindent\textbf{Supervised Matching} This is considered a binary classification task, classifying each candidate pair as \texttt{match} or \texttt{non-match}. Typically, a training, validation and testing set are used to learn the classification model, choose its optimal configuration, and assess its performance on new instances, respectively. 

To examine the performance of BERT and SentenceBERT models,
% in the task of supervised matching, 
we combine them with \textit{EMTransformer} \cite{DBLP:conf/edbt/BrunnerS20}. We selected this approach among the open-source deep learning-based matching algorithms, because it achieves state-of-the-art performance, while relying exclusively on the embedding models. This is in contrast to DITTO \cite{DBLP:journals/pvldb/0001LSDT20}, which leverages external information, such as POS tagging. However, EMTransformer is incompatible with the static models. To address this issue, we combine them with the state-of-the-art approach for this type of models, namely DeepMatcher~\cite{Mudgal2018sigmod}.
% \cite{Mudgal2018sigmod}, which is crafted for the static models, unlike EMTransformer, which supports dynamic ones, too. % In fact, the static language models are excluded from this analysis, because their representations cannot be adjusted after training. As a result, they are incompatible with EMTransformer.
 
Note that this analysis includes models that are supported by EMTransformer or DeepMatcher with minor adjustments to the code. S-GTR-T5 and Word2Vec are thus excluded, since the existing implementations could not support them (EMTransformer cannot handle the sequence2sequence input required by the former, while the latter is not in the format required by DeepMatcher). Note also that the original implementation of EMTransformer disregards the validation set and evaluates each model directly on the testing set. However, this results in overfitting, as noted in \cite{DBLP:journals/corr/abs-2004-00584}. We modified the code so that it follows the standard approach in the literature: for each trained model, the validation set is used to check whether it maximizes F1 and this model is then applied to the testing set \cite{DBLP:journals/pvldb/0001LSDT20,Mudgal2018sigmod}. 
For this analysis, we used the five datasets shown in Table~\ref{tb:smDatasets}.

% \usepackage{multirow}

\section{Comparison on Effectiveness}
\label{sec:effectiveness}

\subsection{Blocking}

\begin{figure*}[!t]
\centering
\subfloat[Static, $k=1$]{\includegraphics[trim=2cm 0.12cm 2cm 0.12cm, clip, width=0.25\textwidth, height=40mm]{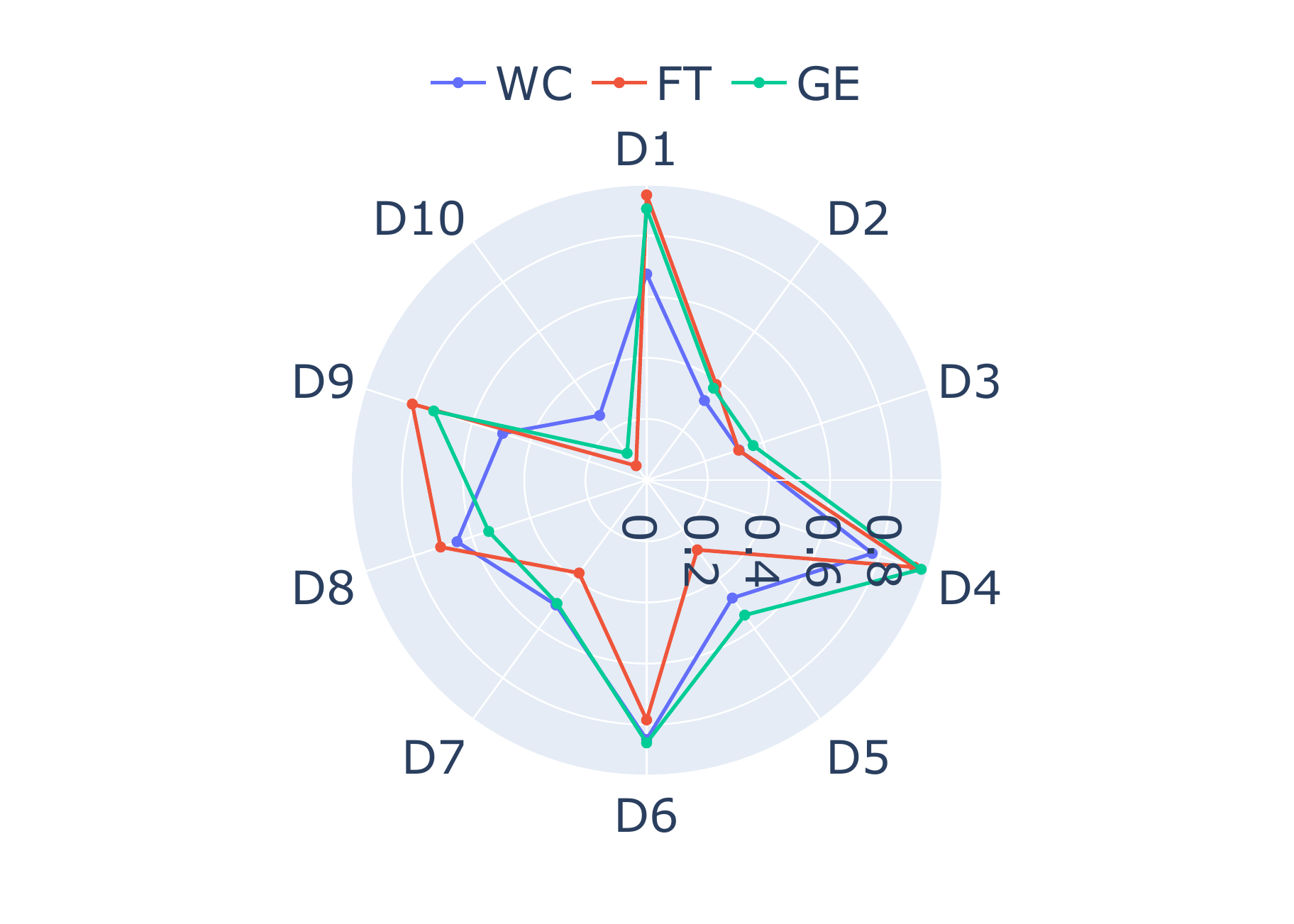}} 
\subfloat[BERT, $k=1$]{\includegraphics[trim=2cm 0.12cm 2cm 0.12cm, clip, width=0.25\textwidth, height=40mm]{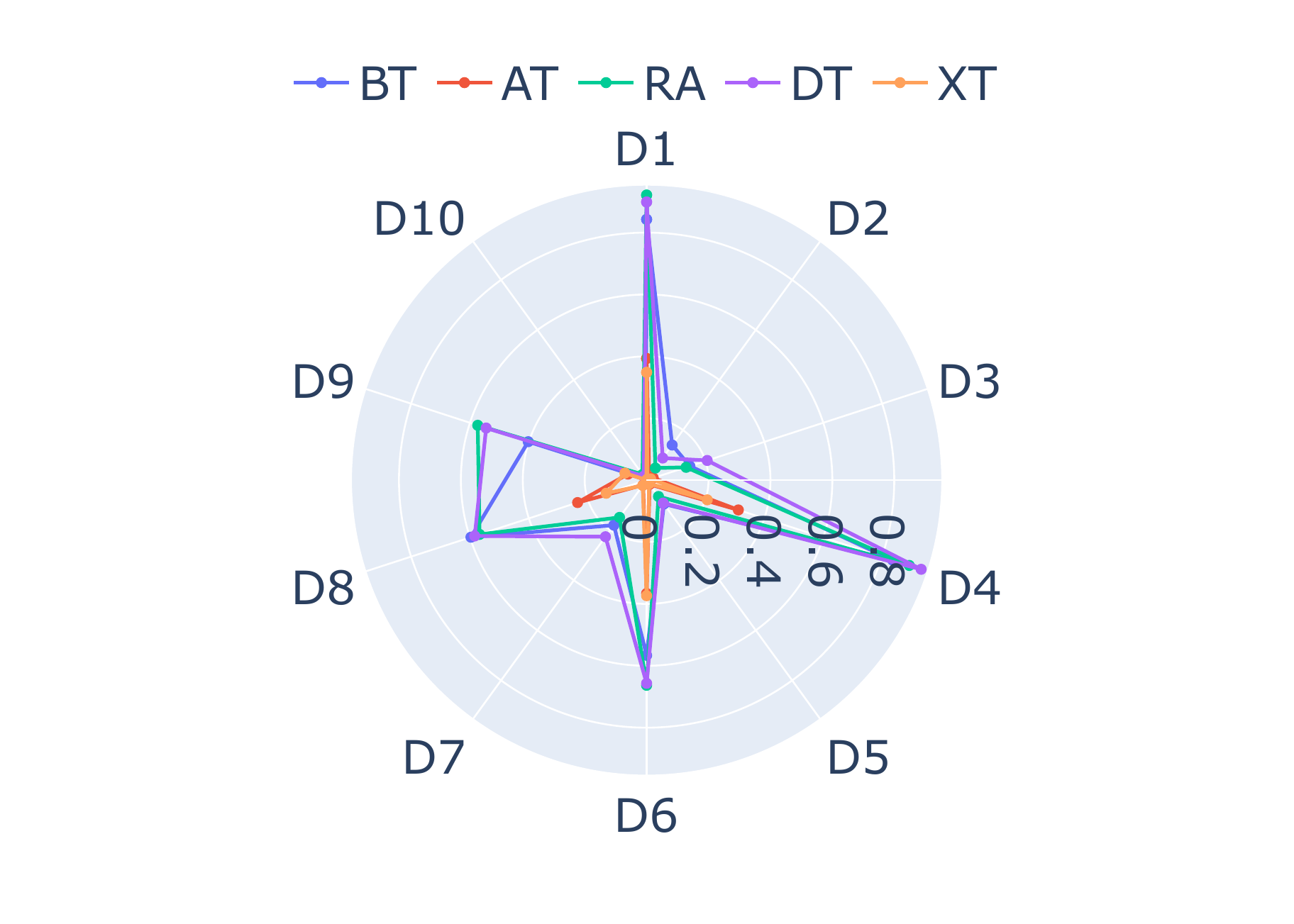}}
\subfloat[SBERT, $k=1$]{\includegraphics[trim=2cm 0.12cm 2cm 0.12cm, clip, width=0.25\textwidth, height=40mm]{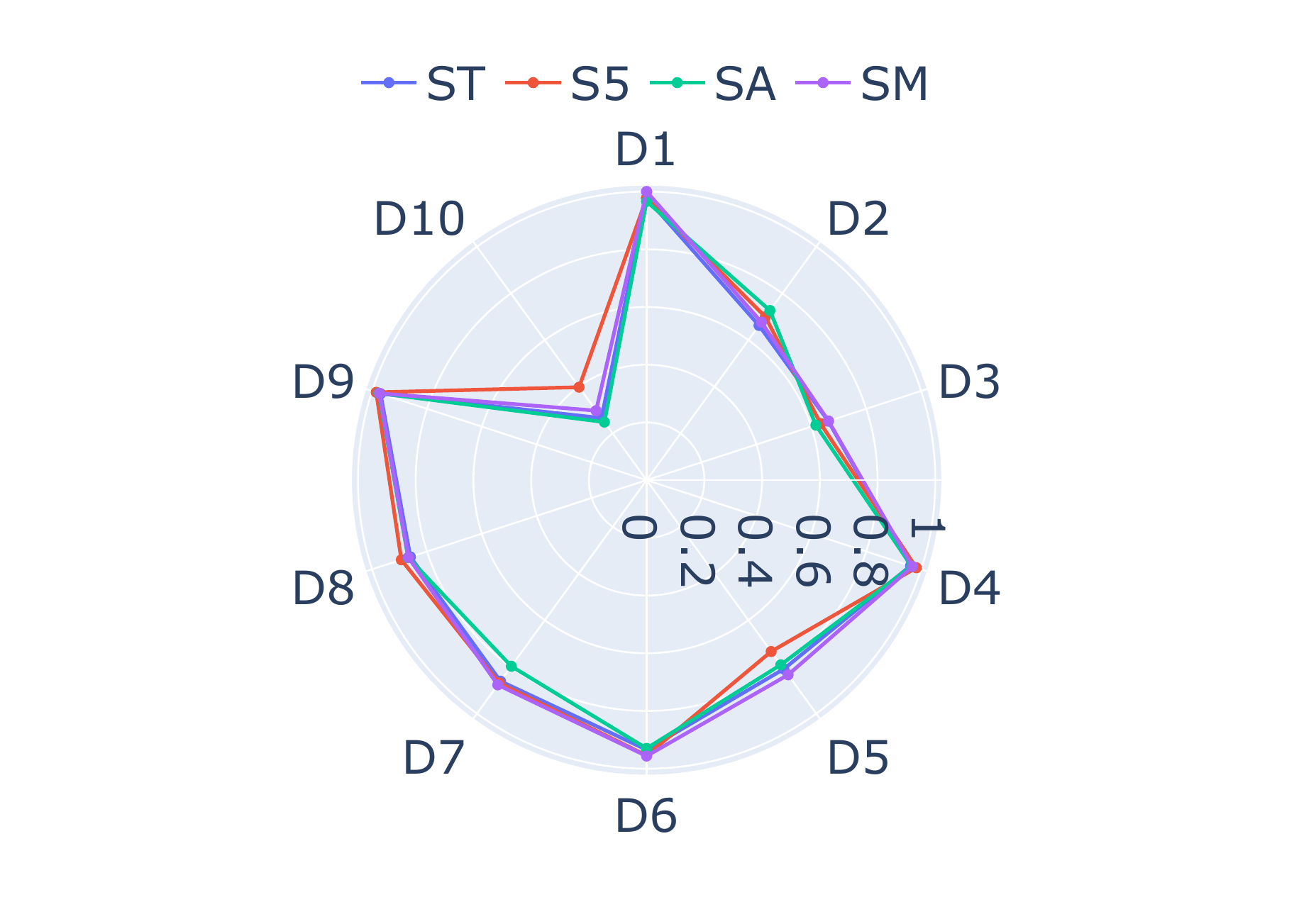}} 
\subfloat[SotA, $k=1$]{\includegraphics[trim=2cm 0.12cm 2cm 0.12cm, clip, width=0.25\textwidth, height=40mm]{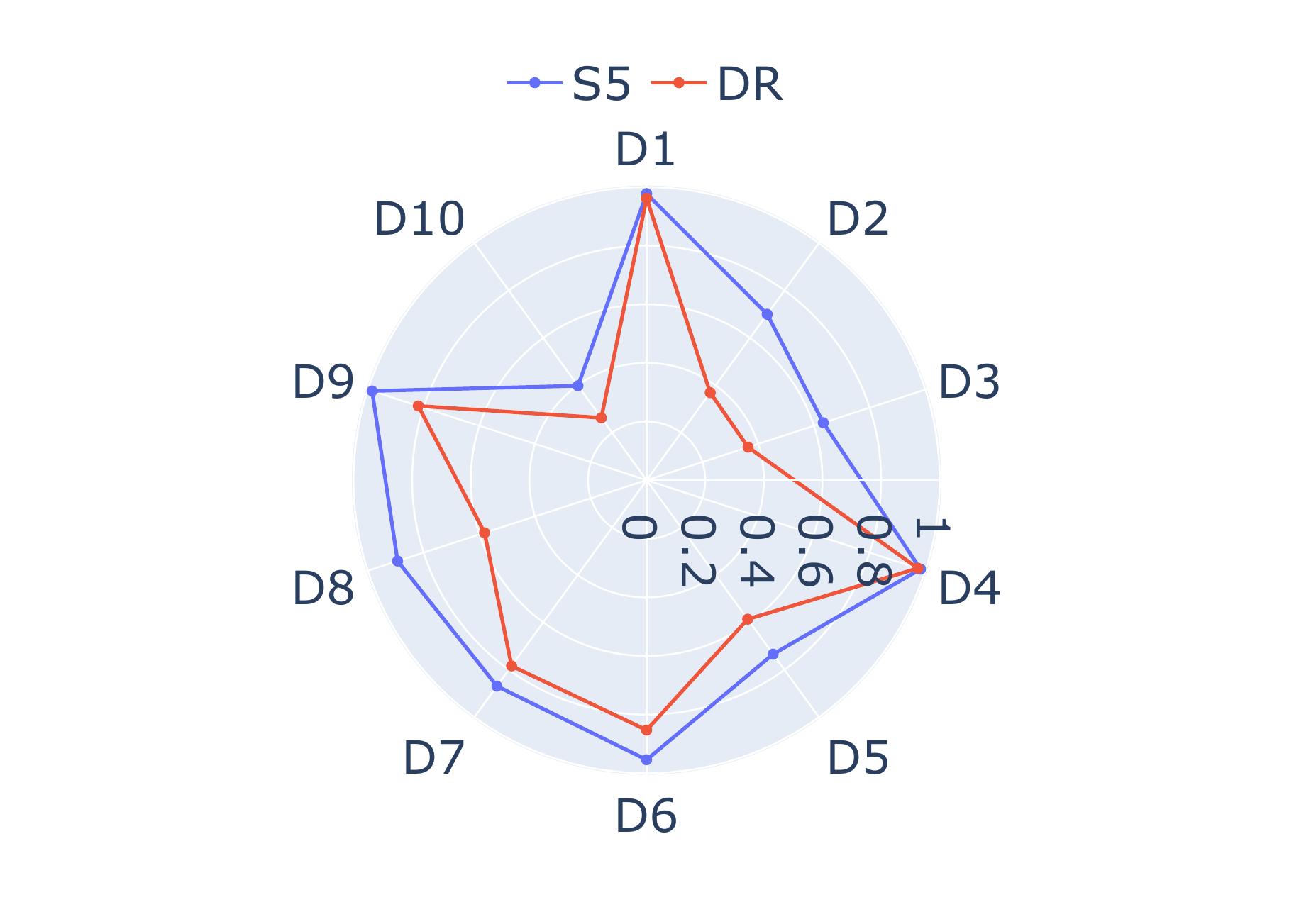}}
\newline
\subfloat[Static, $k=5$]{\includegraphics[trim=2cm 0.12cm 2cm 0.12cm, clip, width=0.25\textwidth, height=40mm]{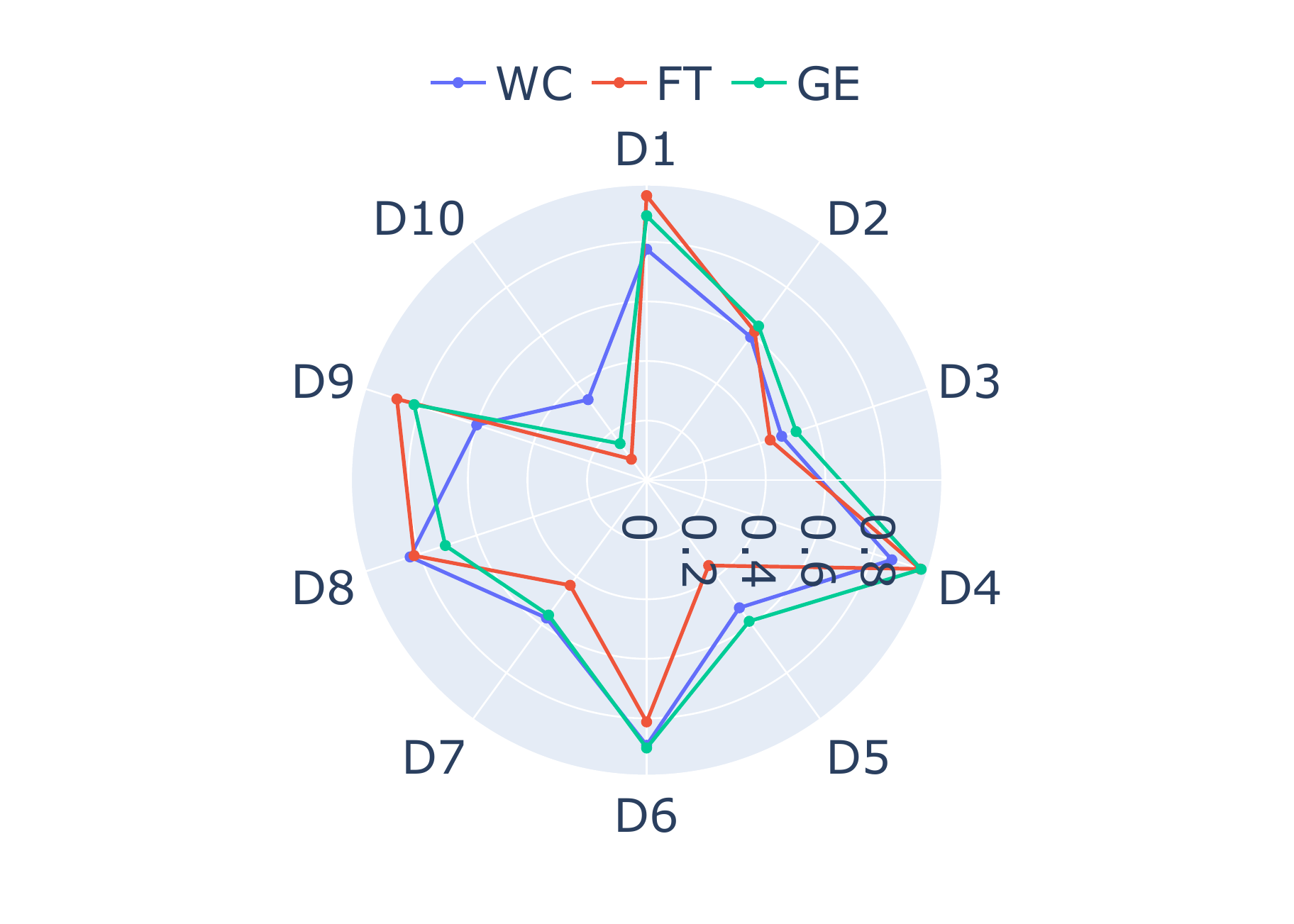}} 
\subfloat[BERT, $k=5$]{\includegraphics[trim=2cm 0.12cm 2cm 0.12cm, clip, width=0.25\textwidth, height=40mm]{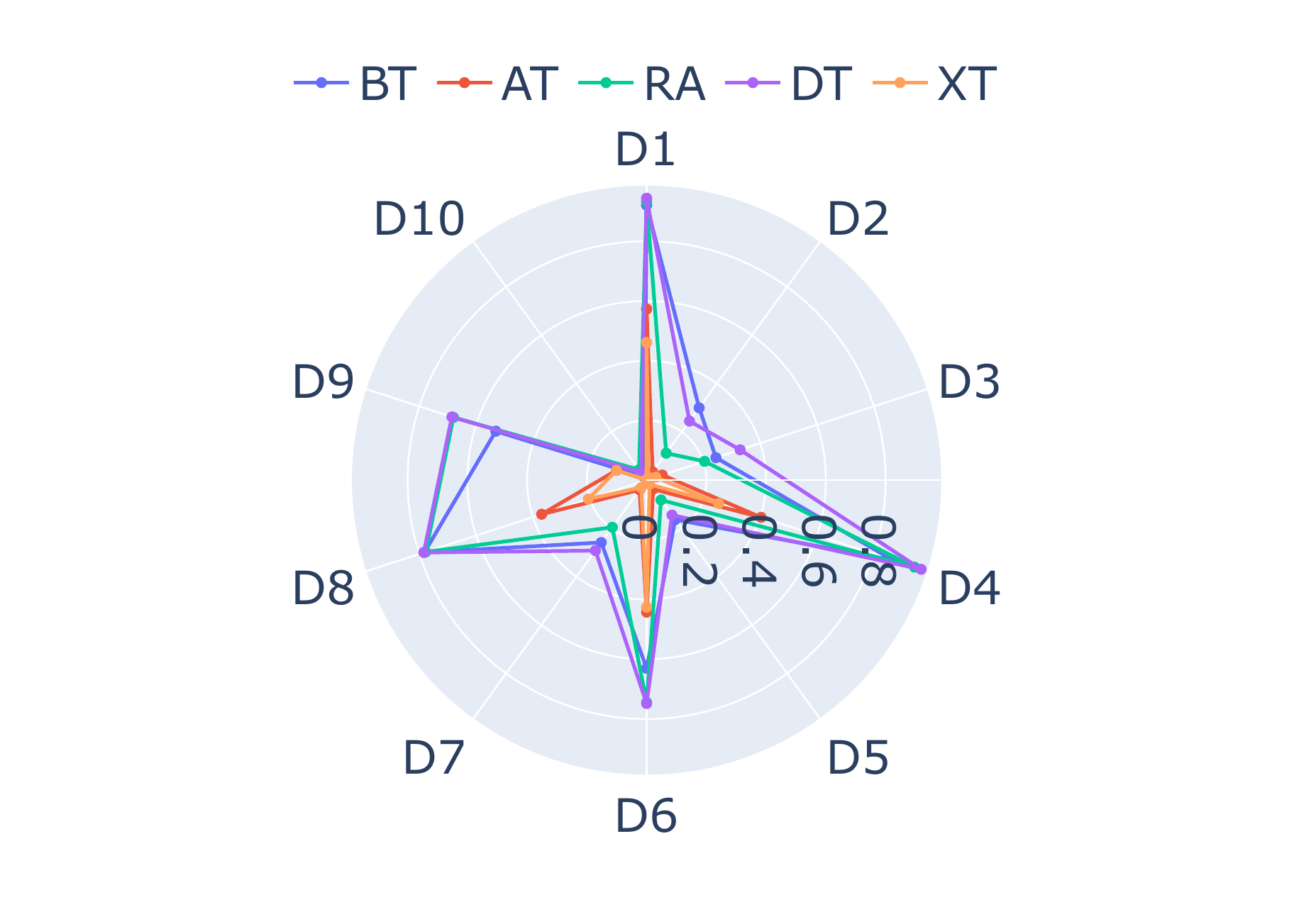}}
\subfloat[SBERT, $k=5$]{\includegraphics[trim=2cm 0.12cm 2cm 0.12cm, clip, width=0.25\textwidth, height=40mm]{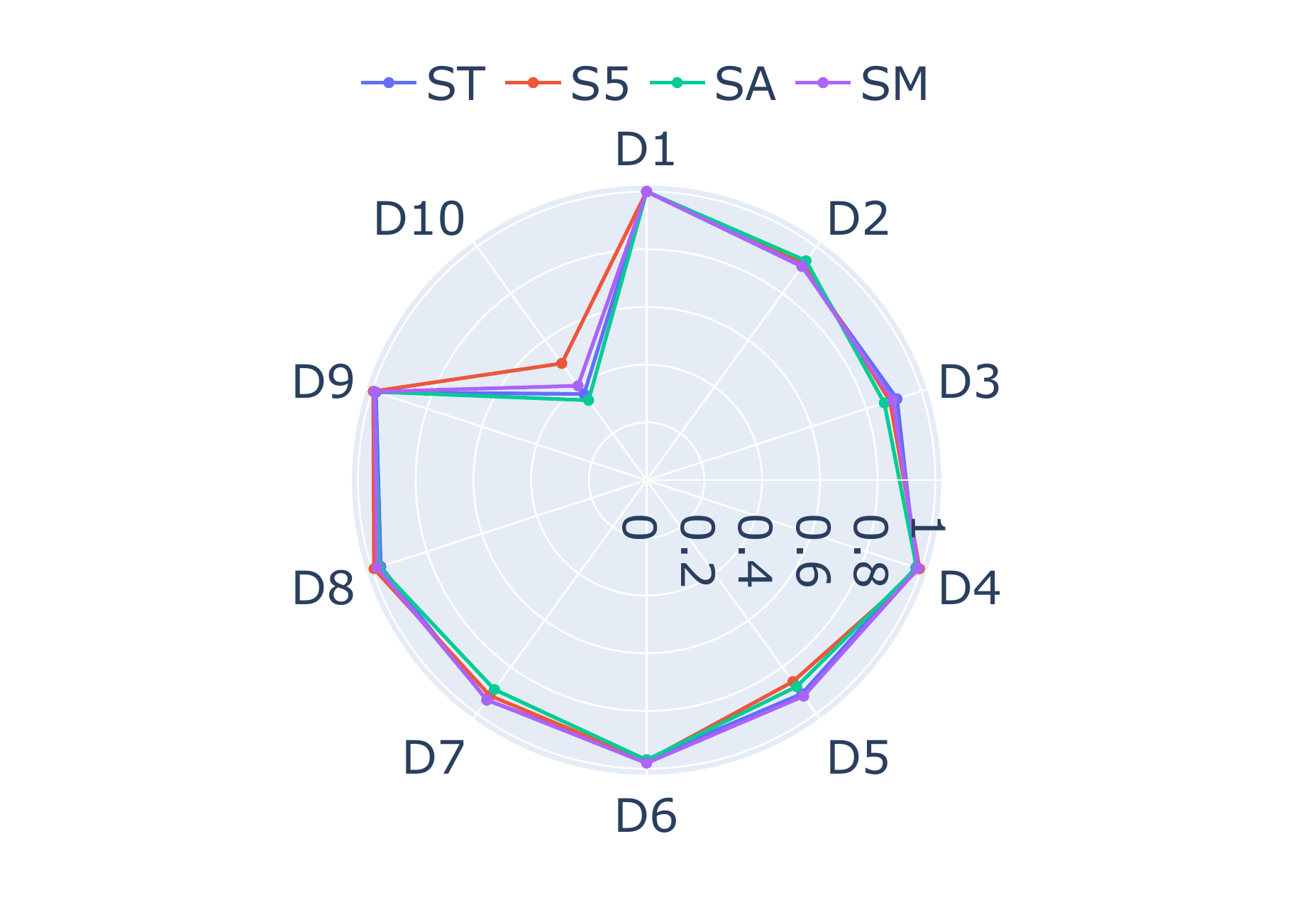}} 
\subfloat[SotA, $k=5$]{\includegraphics[trim=2cm 0.12cm 2cm 0.12cm, clip, width=0.25\textwidth, height=40mm]{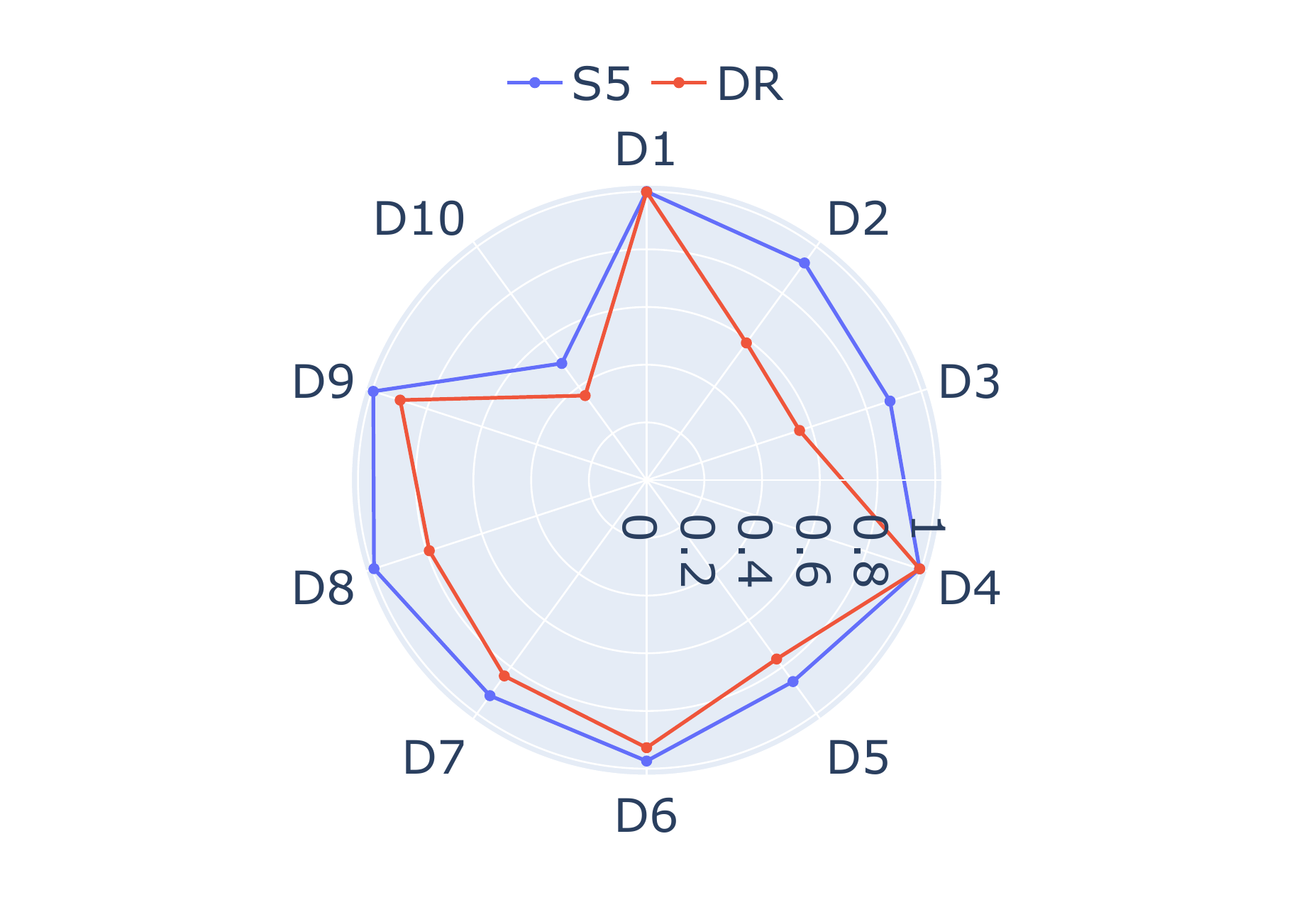}}
\newline
\subfloat[Static, $k=10$]{\includegraphics[trim=2cm 0.12cm 2cm 0.12cm, clip, width=0.25\textwidth, height=40mm]{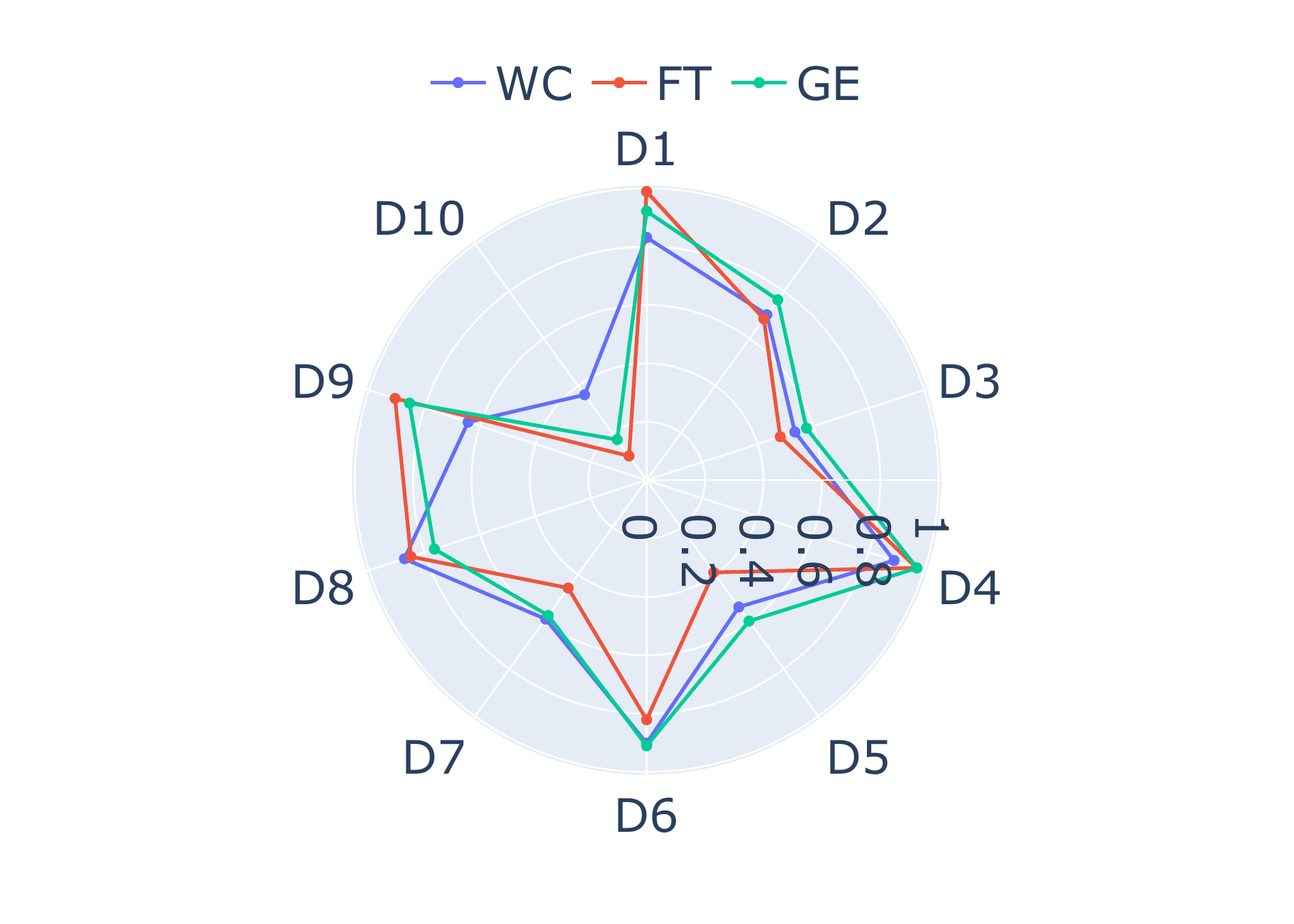}} 
\subfloat[BERT, $k=10$]{\includegraphics[trim=2cm 0.12cm 2cm 0.12cm, clip, width=0.25\textwidth, height=40mm]{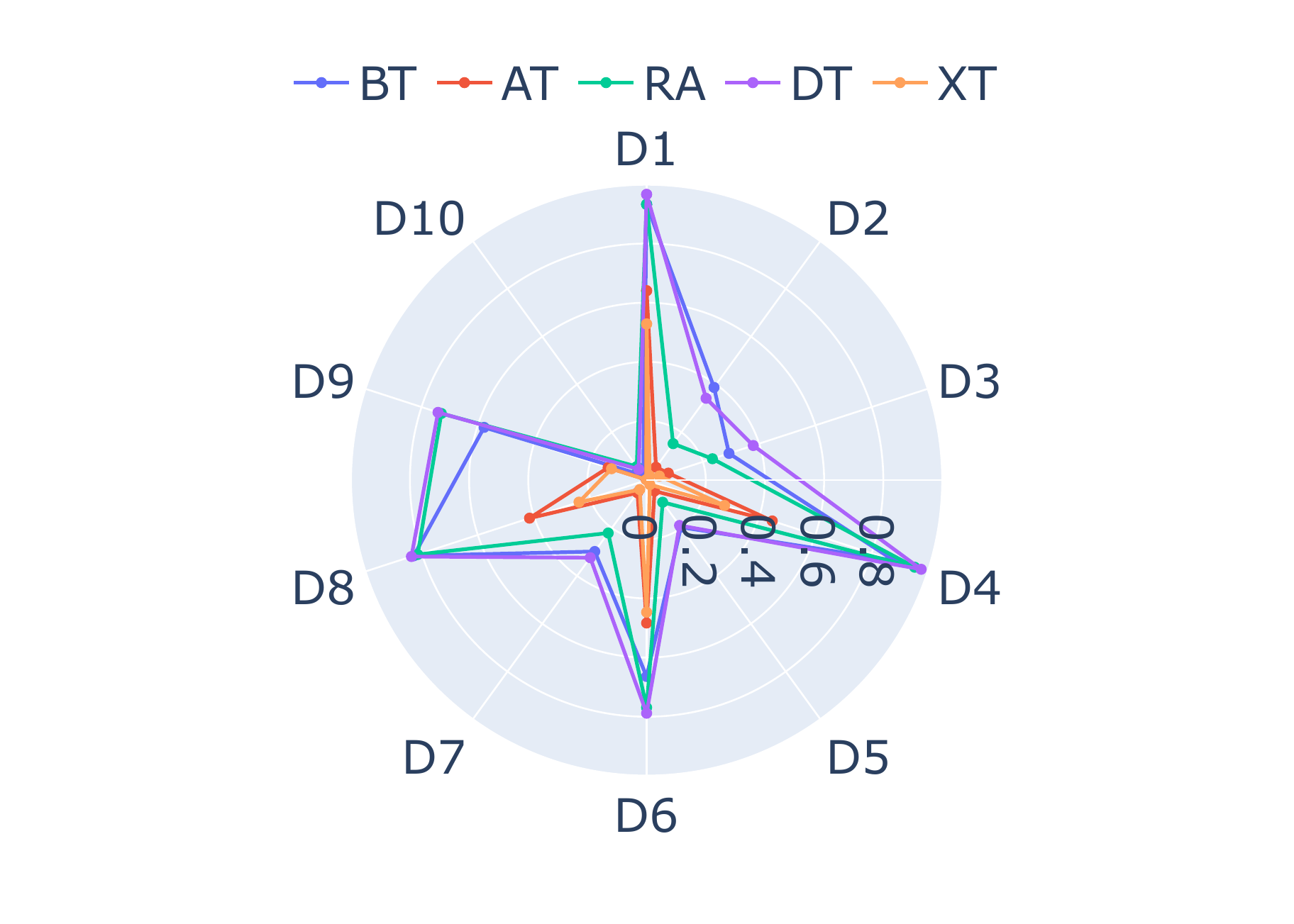}}
\subfloat[SBERT, $k=10$]{\includegraphics[trim=2cm 0.12cm 2cm 0.12cm, clip, width=0.25\textwidth, height=40mm]{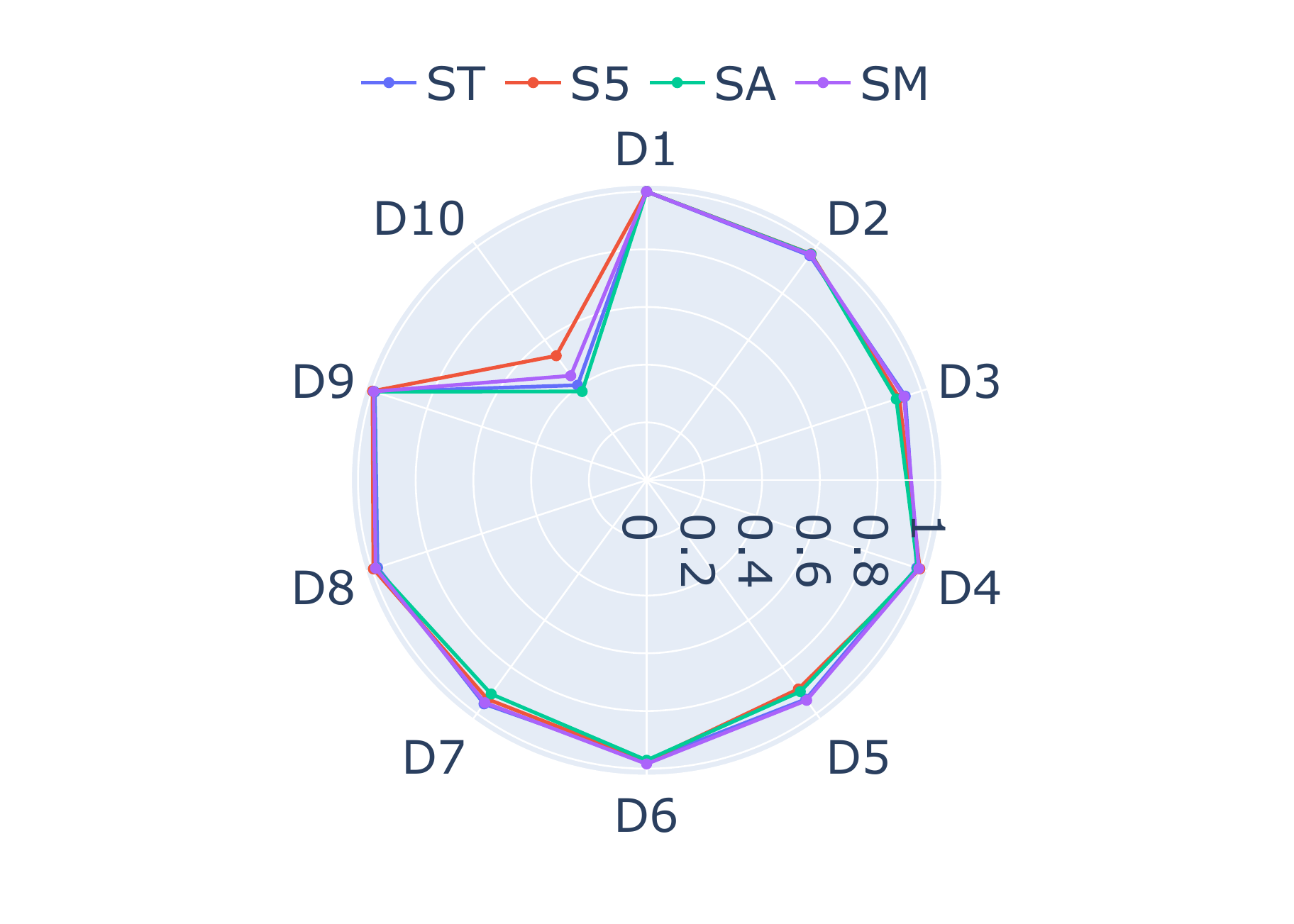}} 
\subfloat[SotA, $k=10$]{\includegraphics[trim=2cm 0.12cm 2cm 0.12cm, clip, width=0.25\textwidth, height=40mm]{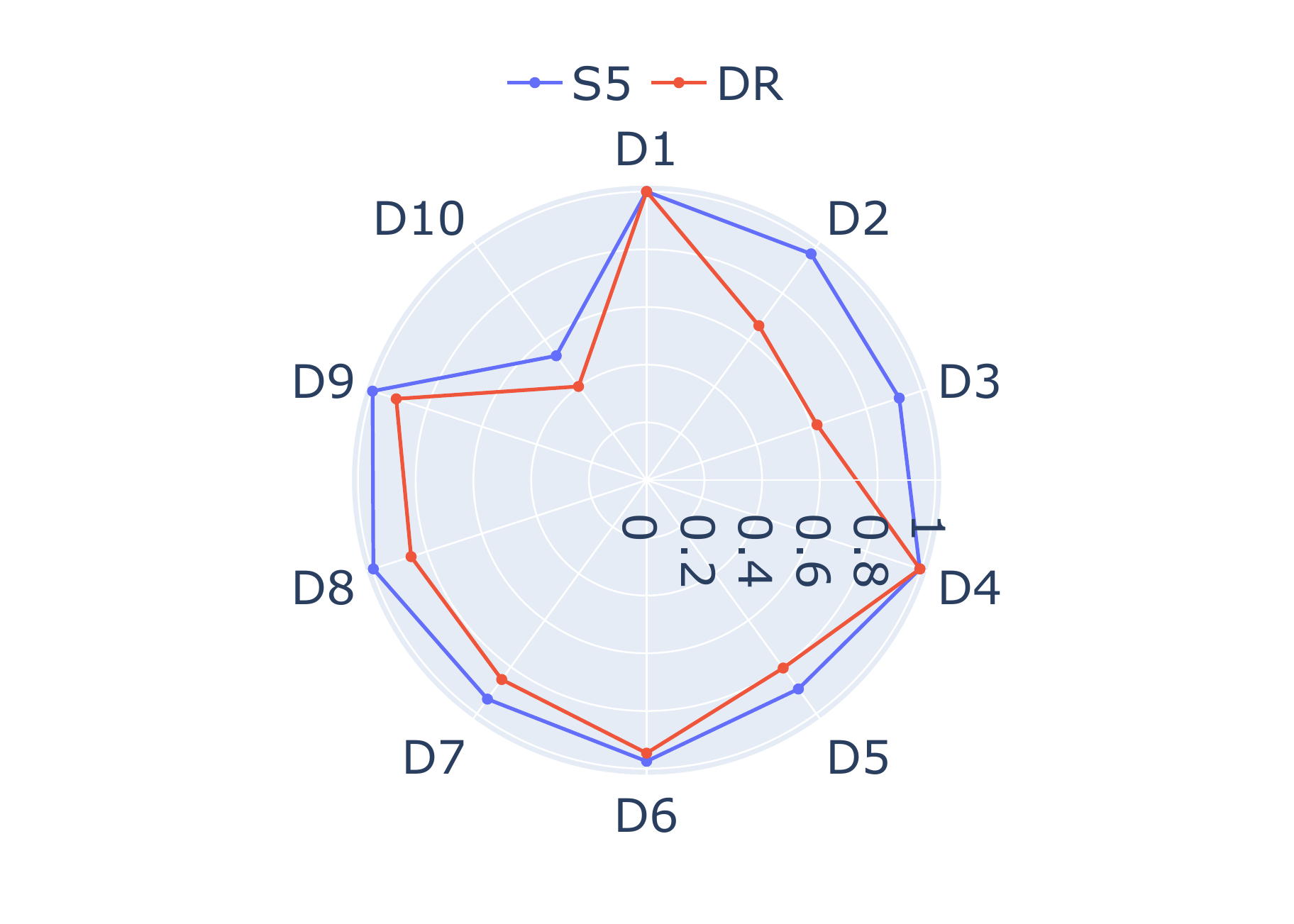}}
\newline
\caption{Blocking recall per model across all datasets in Table 
\ref{tb:datasets}(a).
Each line of plots corresponds to a value of $k \in \{1, 5, 10\}$.
}
\label{fig:blk_real}
\end{figure*}

\begin{figure}[!t]
\centering
\includegraphics[trim=0.12cm 0.12cm 0.12cm 0.12cm, clip, width=0.4\textwidth]{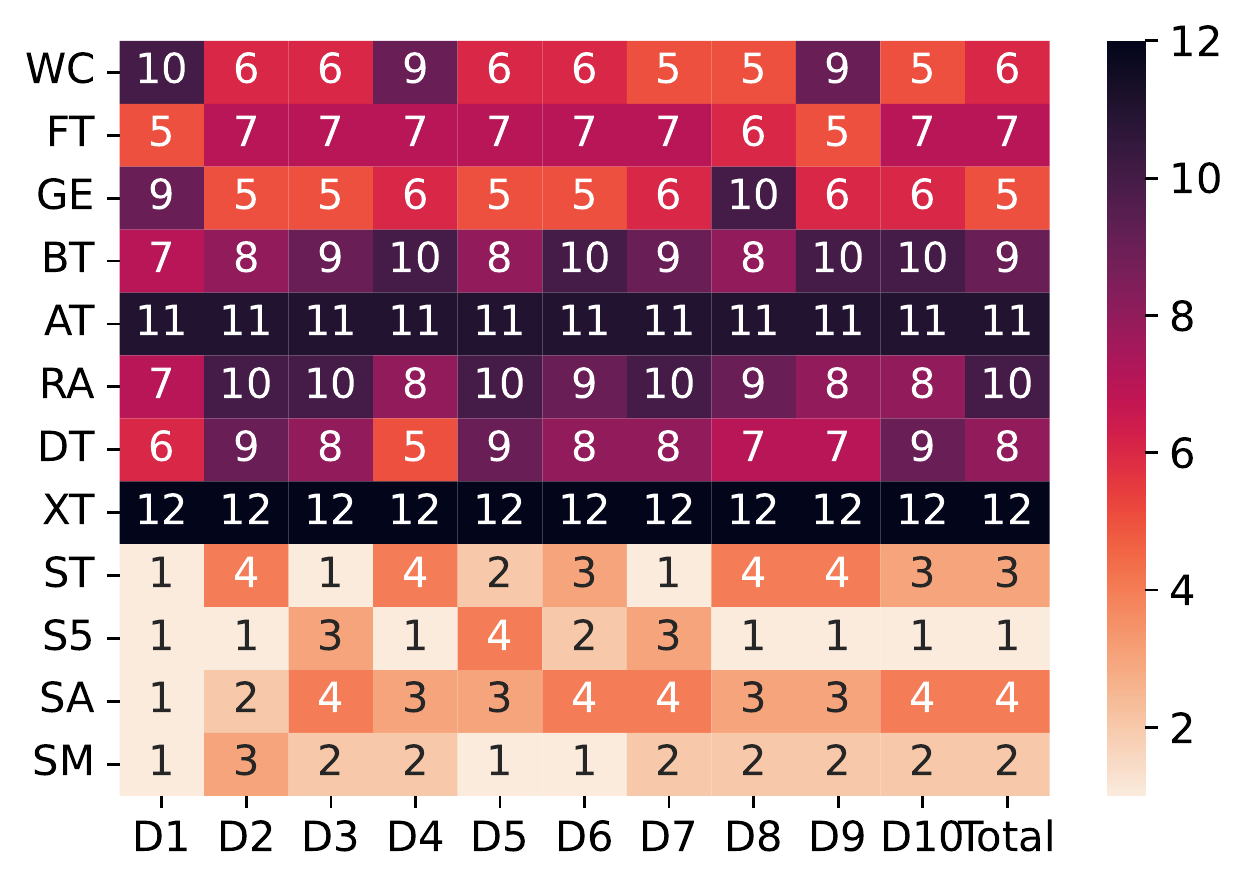}
\caption{Model ranking wrt blocking recall (lower is better).}
\label{fig:blk_heat}
\end{figure}

\begin{figure}[!t]
\centering
\includegraphics[trim=0.12cm 0.12cm 0.12cm 0.12cm, clip, width=0.44\textwidth]{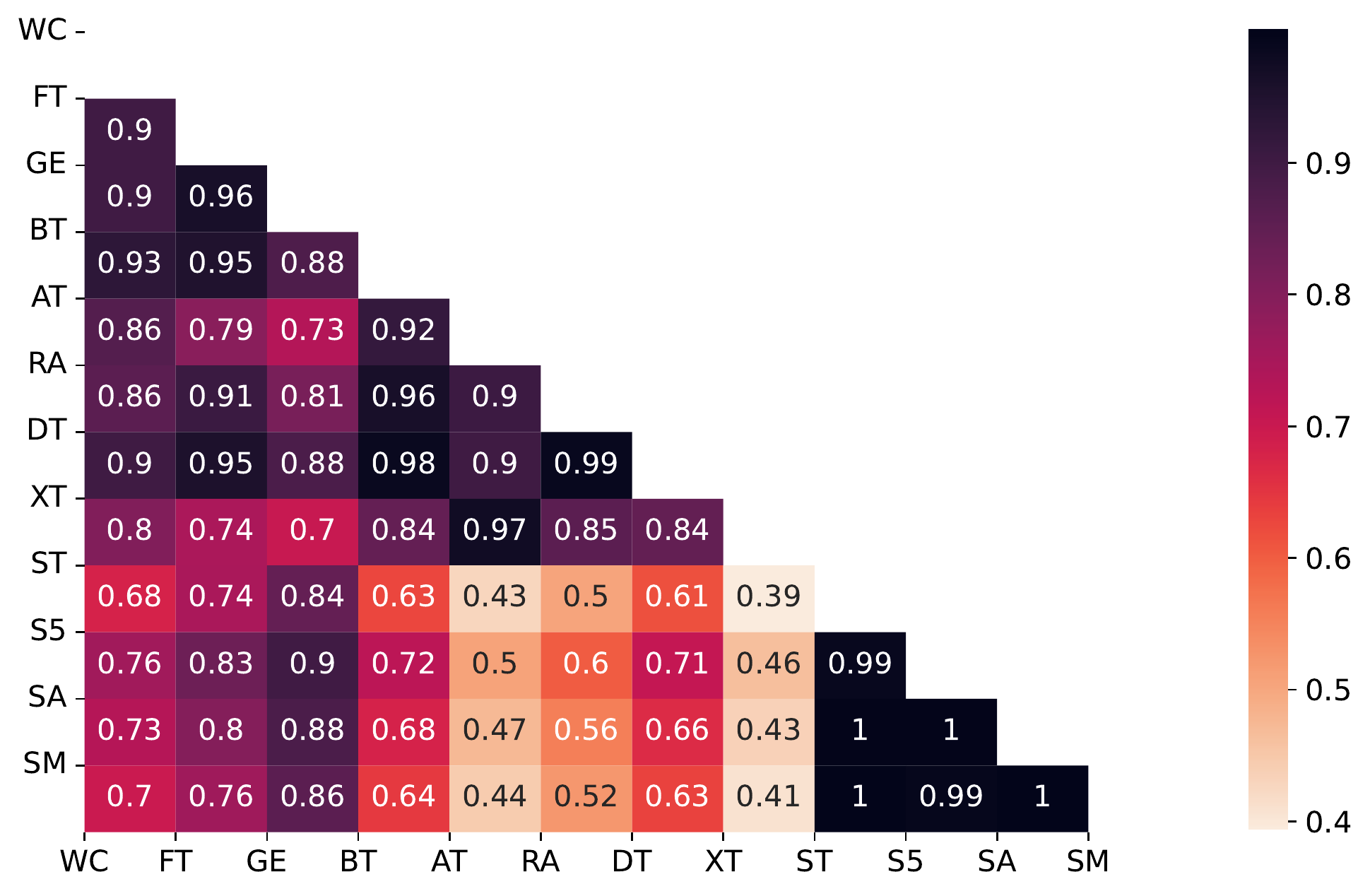}
\caption{Pearson correlation of models wrt blocking recall.}
\label{fig:blk_heat_pear}
\end{figure}

We measure the recall of the resulting candidate pairs, 
which is also known as pairs completeness \cite{DBLP:journals/is/KenigG13,DBLP:journals/tkde/Christen12,DBLP:journals/pvldb/0001APK15}.
% Pairs Quality $PQ$ (i.e., precision) 
Recall is the most critical evaluation measure for blocking, as it typically sets the upper bound for the subsequent matching step, i.e., a low blocking recall usually yields even lower matching recall, unless complex and time-consuming iterative algorithms are employed \cite{DBLP:journals/pvldb/GetoorM12,DBLP:books/daglib/0030287}. In contrast, precision is typically low after blocking, due to the large number of false positives, but significantly raises after matching. Thus, in terms of precision, all models have the same denominator (i.e. total number of candidates) and precision can be omitted, since recall and precision behave the same.

The experimental outcomes are shown in Figure \ref{fig:blk_real} 
% for each $k$ and 
for each category of models, for $k=10$.
More specifically:

In static models, GloVe is the top performer in the vast majority of cases, leaving
FastText and Word2Vec in the second and third place, respectively. On average, GloVe outperforms FastText by 19\%, except for $D_1$, $D_8$ and $D_9$, where 
FastText takes the lead. Compared to Word2Vec, GloVe's recall is higher by 12.5\%, on average, except for $D_7$ $D_8$ and $D_{10}$ (which abounds in noisy and missing values).
%This seems to confirm the results from bibliography, since GloVe was the latest and more successful model in the static category.} {\color{red}Alexandre, mporeis na valeis to citation?}

Among the BERT models, we can see that XLNet and AlBERT have very poor performance on almost all datasets. XLNet relies on permuted language modeling (PLM), which tries to model dependencies between words and phrases, instead of masked language modeling (MLM), which is adopted in BERT. This proves to be less effective in our task, where the input text is constructed by concatenating several different attributes, thus not constituting a coherent sentence.
AlBERT trains only one encoder to produce a lighter version of BERT and shares its weights with the remaining encoders. This also turns out to perform poorly here. As a result, both models suffer from poor discriminativeness, i.e., they assign low similarity scores to both matching and non-matching pairs of entities. For the same reason, albeit to a lesser extent, the same applies to 
% {\color{red} For XLNet it seems logical, since it was trained for solving PLM, rather than MLM, meaning that it searches for dependency between words and phrases, while in our datasets we have concatenated different columns, producing a possibly not coherent sentence. For AlBERT, since it trains only one encoder to produce a lighter version of BERT and shares its weights with the remaining encoders, if the task in hand can not be efficiently solved by this single encoder - and by not utilizing fine-tuning - then it is clear that it will have a very poor performance, as in our task.}
the remaining BERT models, which have a mediocre performance, with DistilBERT being the best one in all datasets.

Finally, all SentenceBERT models achieve very high recall across all datasets, but the extremely noisy and sparse $D_{10}$. The best model in this category is S-GTR-T5. This has to do with the base model that each SentenceBERT model relies to. For example, GTR-T5 trains in a dataset with more than 2B pairs, while all three other base models train on a collection of datasets that amount to 1B+ pairs. 

Finally, between groups, we can see that the point made in \cite{reimers2019sentence} holds in our task as well. BERT models, if not fine-tuned, behave worse in terms of recall than static models (GloVe) and SentenceBERT have overall the best performance. There are two main reasons for the latter: they are designed for sentence rather than word embeddings and they are trained on wider corpora.

\noindent\textbf{Summary.} 
The above patterns are summarized in Figure \ref{fig:blk_heat}, which
reports the ranking of each model per dataset with respect to recall for $k$=10, with the rightmost column indicating the average position per model.
We observe that the first four places are occupied by the SentenceBERT models, with S-GTR-T5 ranking first in most datasets. It is interesting that none of these models falls below the fourth place. here are two reasons for the superiority of SentenceBERT models: (a) They are inherently capable of transforming a sentence into an embedding vector, unlike the other two types, which are crafted for vectorizing individual tokens. (b) They encapsulate knowledge from wider corpora, while their final layer comes with reasonable weights already in its pre-trained form (unlike the BERT models).

The next three ranking positions mostly correspond to the static models, with Glove having the highest average one. Yet, FastText is the most stable one, fluctuating between positions 5 and 7, unlike the other two models, which fall up to the 10$^{th}$ place. Finally, BERT-based models are mapped to the last five positions. DistilBERT~is 

\noindent
the best one, ranked 8$^{th}$ on average, while AlBERT and XLNet are constantly ranked 11$^{th}$ and 12$^{th}$, respectively. The BERT models underperform the static ones, because they suffer from poor discriminativeness, due to the lack of fine-tuning, which guarantees their context-aware functionality. The predetermined weights in their final layer yield very low scores to most pairs of entities, regardless of whether they are matching or not. This is not true for the static models, despite their context-agnostic functionality and their lower dimensionality. 
% Please refer to the on-line extended version$^3$ of our work for more details.

These patterns suggest a high correlation in terms of effectiveness between the models that belong to the same category. Indeed, the Pearson correlation with respect to recall for $k$=10 between the models of each category is quite high ($\geq0.9)$, as shown in Figure \ref{fig:blk_heat_pear}. The correlation is equally high between the pre-trained static models and the SentenceBERT ones, due to the high performance of both categories. In contrast, the correlation between BERT-based and the other two categories is significantly lower: it fluctuates between 0.58 and 0.85 for the SentenceBERT models and between 0.69 and 0.86 for the static ones. In the latter case, DistilBERT is an exception, fluctuating between 0.84 and and 0.94, since it is the best-performing BERT model and, thus, it is closer to the static~ones. 

\begin{figure}[!t]
\centering
\subfloat[D2-Positive]{\includegraphics[trim=0.12cm 0.12cm 0.12cm 0.12cm, clip, width=0.24\textwidth]{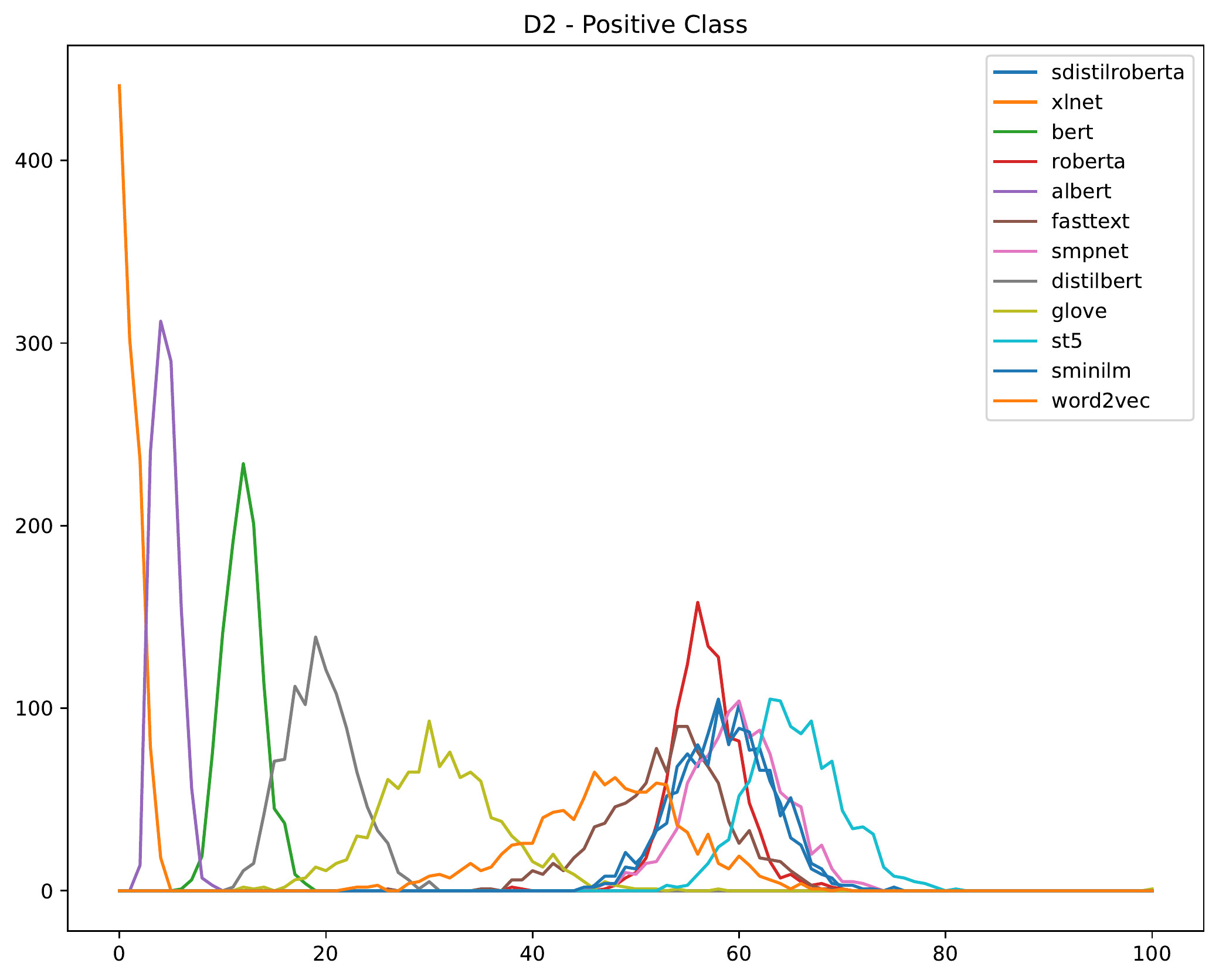}}
\subfloat[D2-Negative]{\includegraphics[trim=0.12cm 0.12cm 0.12cm 0.12cm, clip, width=0.24\textwidth]{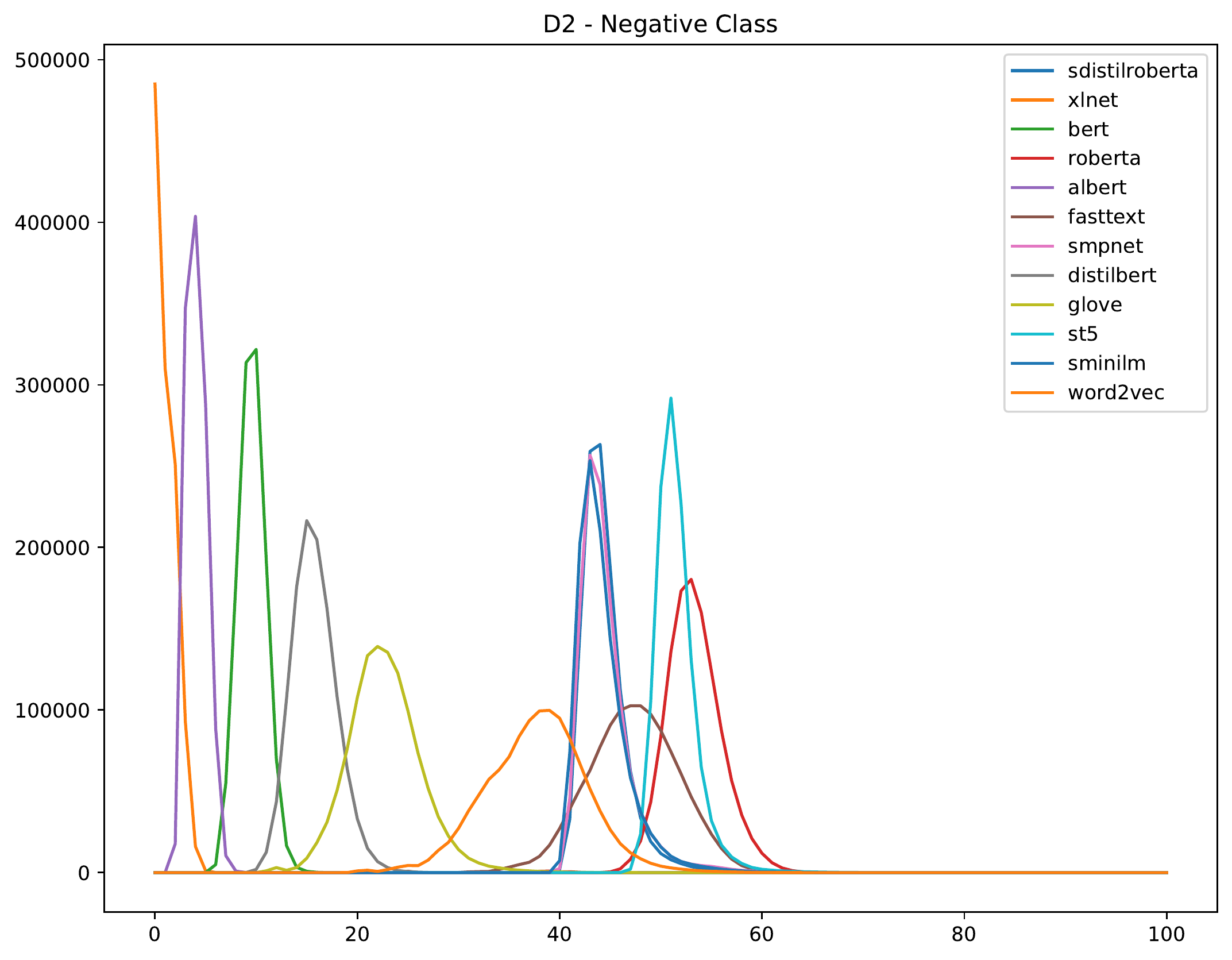}}

\subfloat[D4-Positive]{\includegraphics[trim=0.12cm 0.12cm 0.12cm 0.12cm, clip, width=0.24\textwidth]{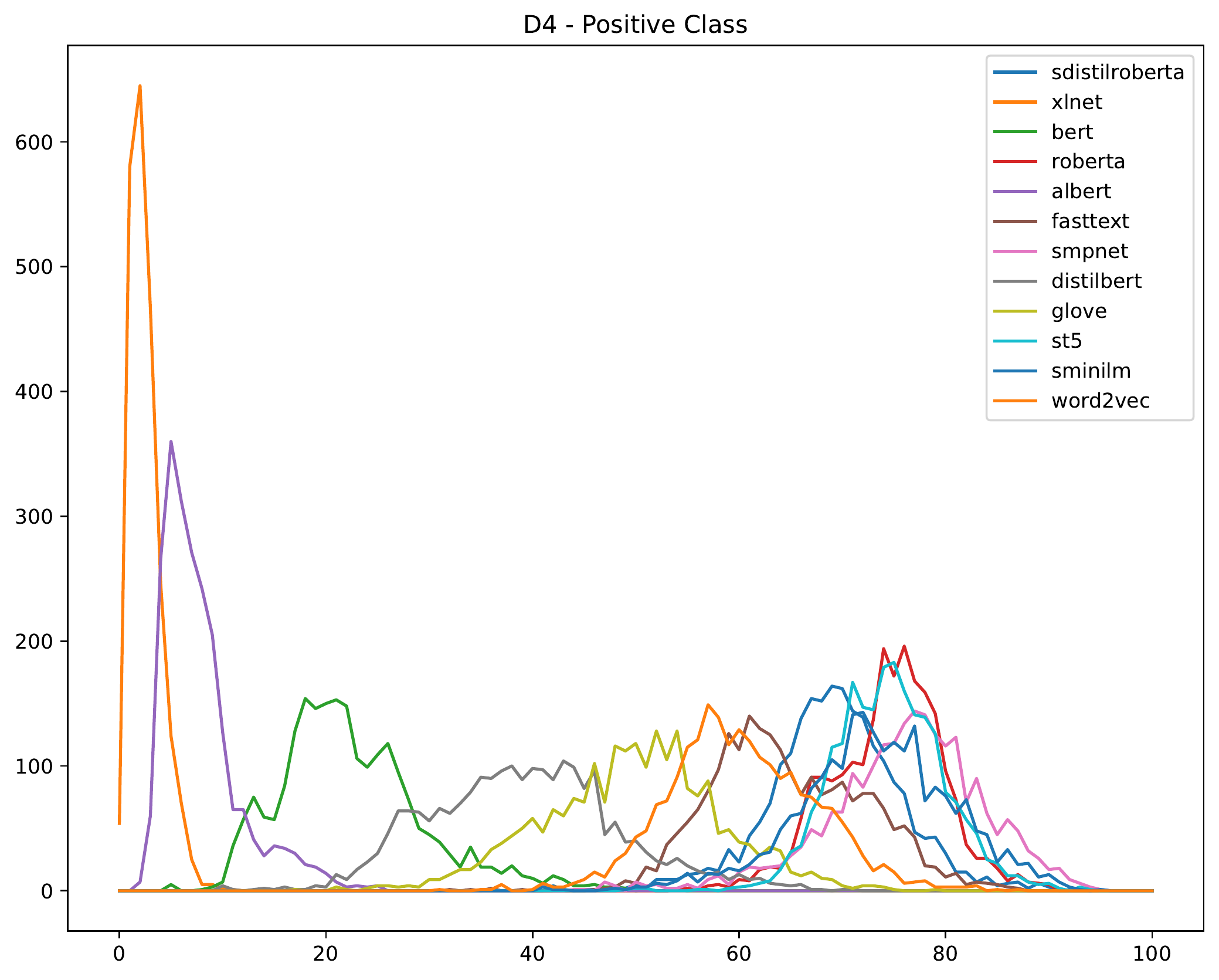}}
\subfloat[D4-Positive]{\includegraphics[trim=0.12cm 0.12cm 0.12cm 0.12cm, clip, width=0.24\textwidth]{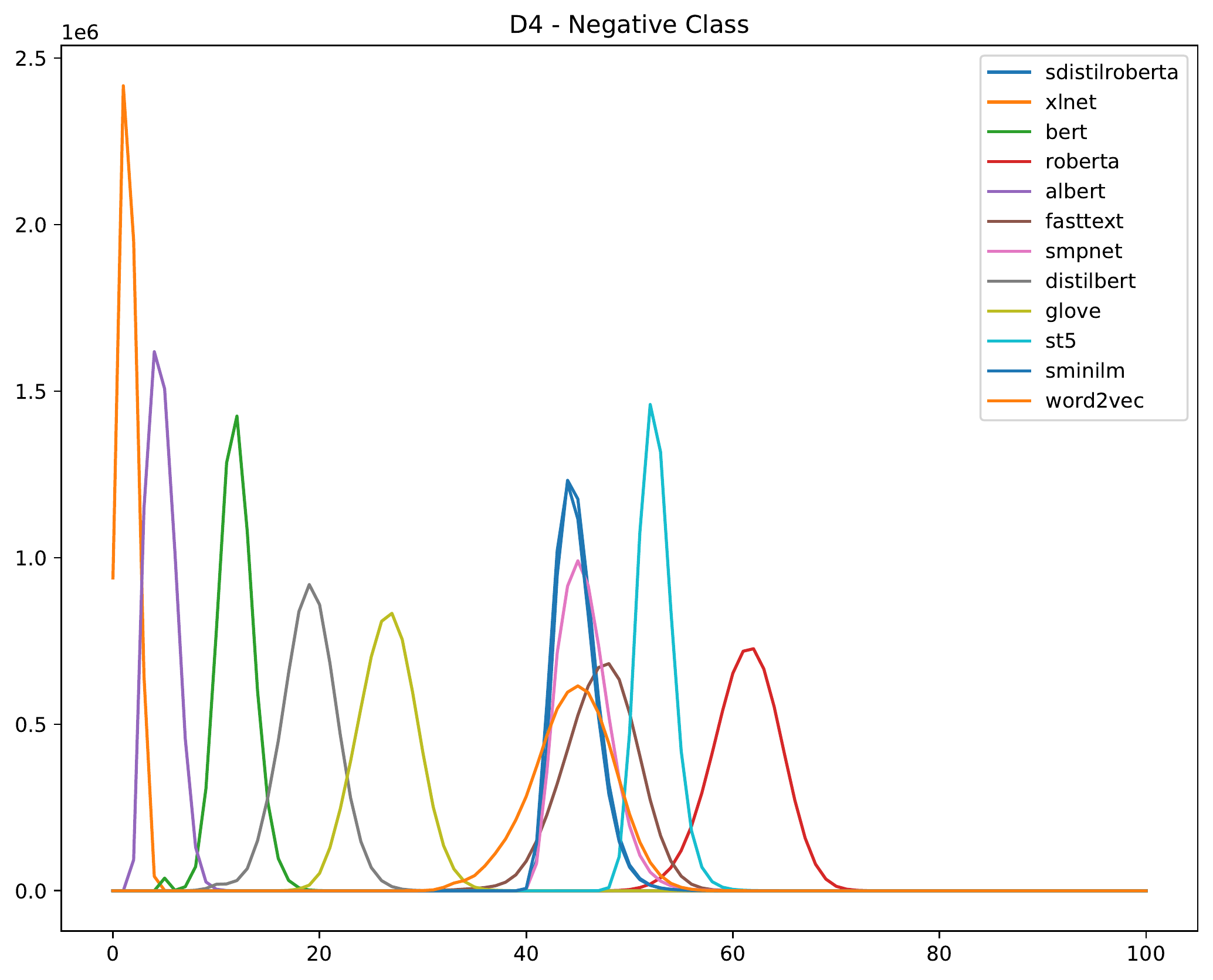}}

\caption{Discriminativeness for all models for cases D2 and D4 in the classes of Match (Positive) and Non-match (Negative).}
\label{fig:distr}
\end{figure}

To a greater or lesser extent, all BERT-based models suffer from poor discriminative power in the tasks of blocking and unsupervised matching. This can be attributed to the lack of fine-tuning, which guarantees their context-aware functionality. The predetermined weights in their final layer yield very low scores to most pairs of entities, regardless of whether they are matching or not. This is not true for the static models, despite their context-agnostic functionality and their lower dimensionality. This is shown in Figure \ref{fig:distr}, which depicts the distribution of similarity scores among datasets D2 and D4 per language model (the positive class per dataset appears in the left column and the negative one in the right). We observe high discriminativeness for SentenceBERT models, lower for the static ones, while the BERT ones fails to separate the two distributions. 

\noindent\textbf{Comparison to SotA.} The rightmost column in Figure \ref{fig:blk_real} compares the best performing language model, S-GTR-T5, with the state-of-the-art blocking approach that is based on deep learning and embeddings vectors: DeepBlocker's Auto-Encoder with FastText embeddings. 
% The relative performance of the two methods over all datasets in Table \ref{tb:datasets}(a) appears in Table \ref{tb:st5vsdeepblocker} (including vectorization time for S-GTR-T5). Note that since DeepBlocker is a stochastic approach, we take the average of 10 runs.
S-GTR-T5 consistently outperforms DeepBlocker's recall to a significant extent. The only exceptions are $D_1$ and $D_4$, where both methods achieve practically perfect recall
% -- except for $k=1$, where S-GTR-T5 is marginally better, by 3\% and 1\%, respectively. 
The reason is that $D_1$ involves a very low number of duplicate pairs in relation to the size of each data source, while $D_4$ contains relatively clean entities with long textual descriptions that are easy to match. In the remaining 8 datasets, S-GTR-T5’s recall is 15\% higher than DeepBlocker.
% In the remaining 8 datasets, the superiority of S-GTR-T5 is inversely proportional to the number of candidates per query entity, i.e., the lower $k$ is, the larger is its advantage: {\color{blue}on average, across all datasets, S-GTR-T5's recall is higher than DeepBlocker by 25\%, 17\%	and 15\% for $k=$1, 5 and 10, respectively.}

% of S-GTR-T5. 

Seemingly, this can be attributed to the FastText embeddings used by DeepBlocker. However, DeepBlocker is a comprehensive blocking method that uses FastText in a more complex way than the mere nearest neighbour search of S-GTR-T5. For example, a crucial component of DeepBlocker is self-supervision, which automatically labels a random sample of the candidate pairs in order to train its classification model. As a result, DeepBlocker is a stochastic approach, unlike S-GTR-T5, which exclusively performs nearest neighbor search. The ablation analysis in \cite{DBLP:journals/pvldb/Thirumuruganathan21} indicates that all DeepBlocker's components have a significant contribution to the final outcome. For this reason, the role of the language models is restricted and, thus, the performance of DeepBlocker exhibits a low correlation with that of FastText+NNS.

% Overall, we can conclude that S-GTR-T5 exhibits much higher blocking recall than DeepBlocker in all non-trivial cases at the cost of much higher overhead in case of small datasets and few candidates per query entity.

%S-GTR-T5 is also dominant over state-of-the-art algorithm, DeepBlocker.
%{\color{blue}Alex: Not sure about numbers, need to check this again.}

\subsubsection{Scalability.}

\begin{figure}[!t]
\centering
\subfloat{\includegraphics[width=1.0\linewidth]{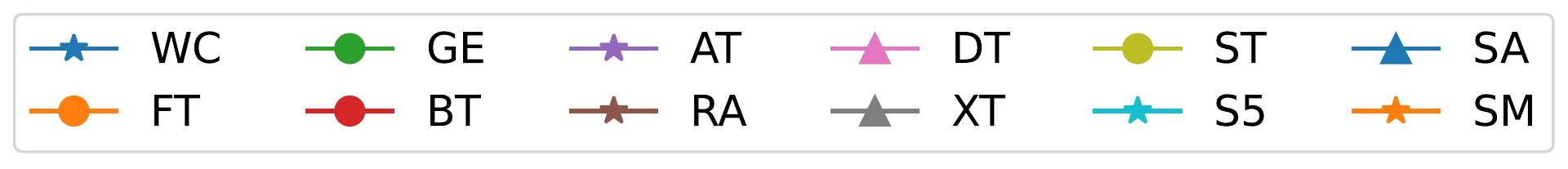}}
\setcounter{subfigure}{0}
\subfloat[Recall]{\includegraphics[trim=0.12cm 0.12cm 0.12cm 0.12cm, clip, width=0.24\textwidth]{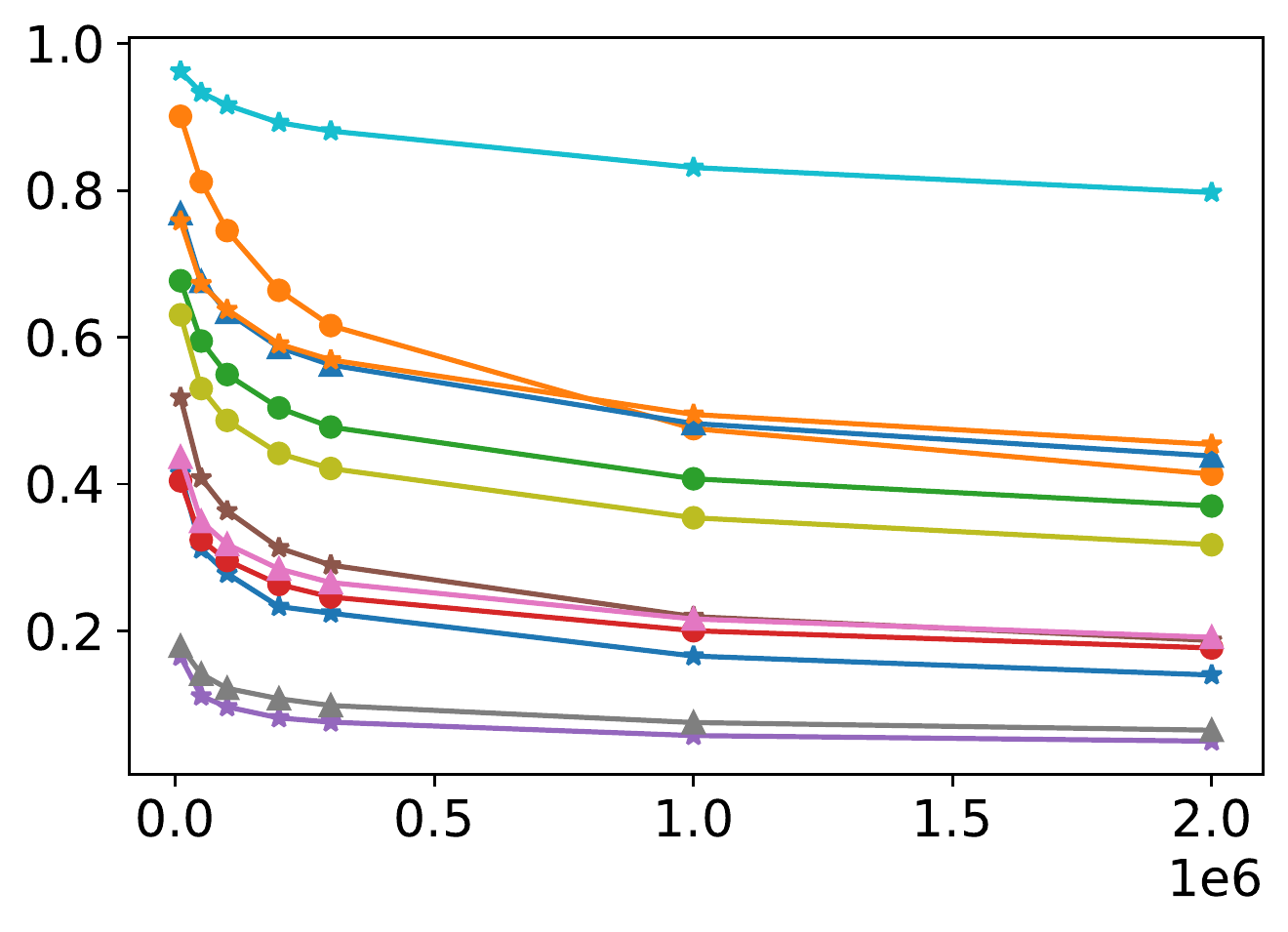}\label{subfig:blk_synth_rec_app}}
\subfloat[Precision]{\includegraphics[trim=0.12cm 0.12cm 0.12cm 0.12cm, clip, width=0.24\textwidth]{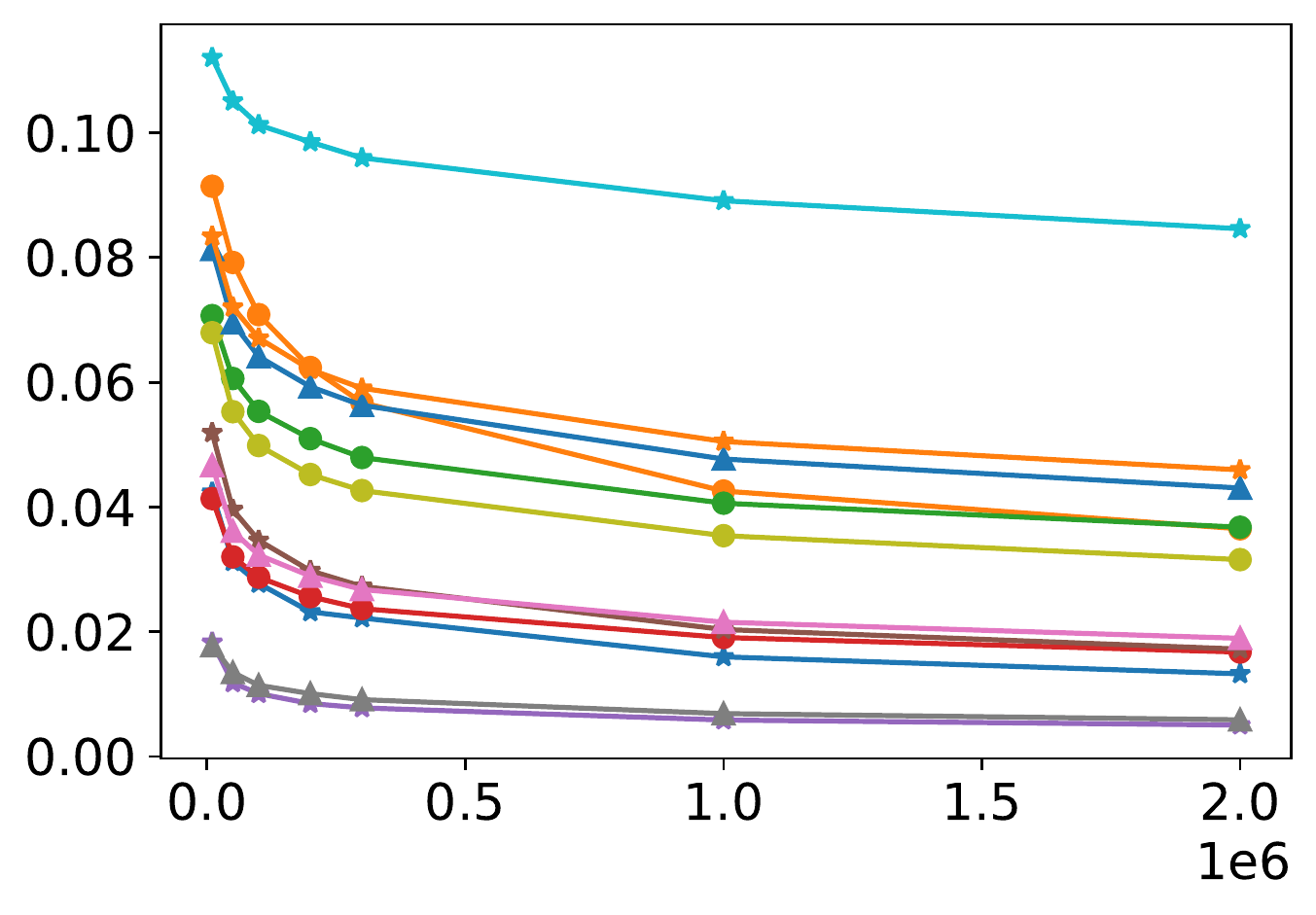}\label{subfig:blk_synth_pre_app}}

\caption{{\small Scalability effectiveness over the synthetic datasets in Table \ref{tb:datasets}(b). The horizontal axis indicates the number of input entities.}}
\label{fig:blk_synth}
\end{figure}

\begin{figure*}[!t]
\centering
\setcounter{subfigure}{0}
\subfloat[Static, Precision]{\includegraphics[trim=2cm 0.12cm 2cm 0.12cm, clip, width=0.25\textwidth, height=40mm]{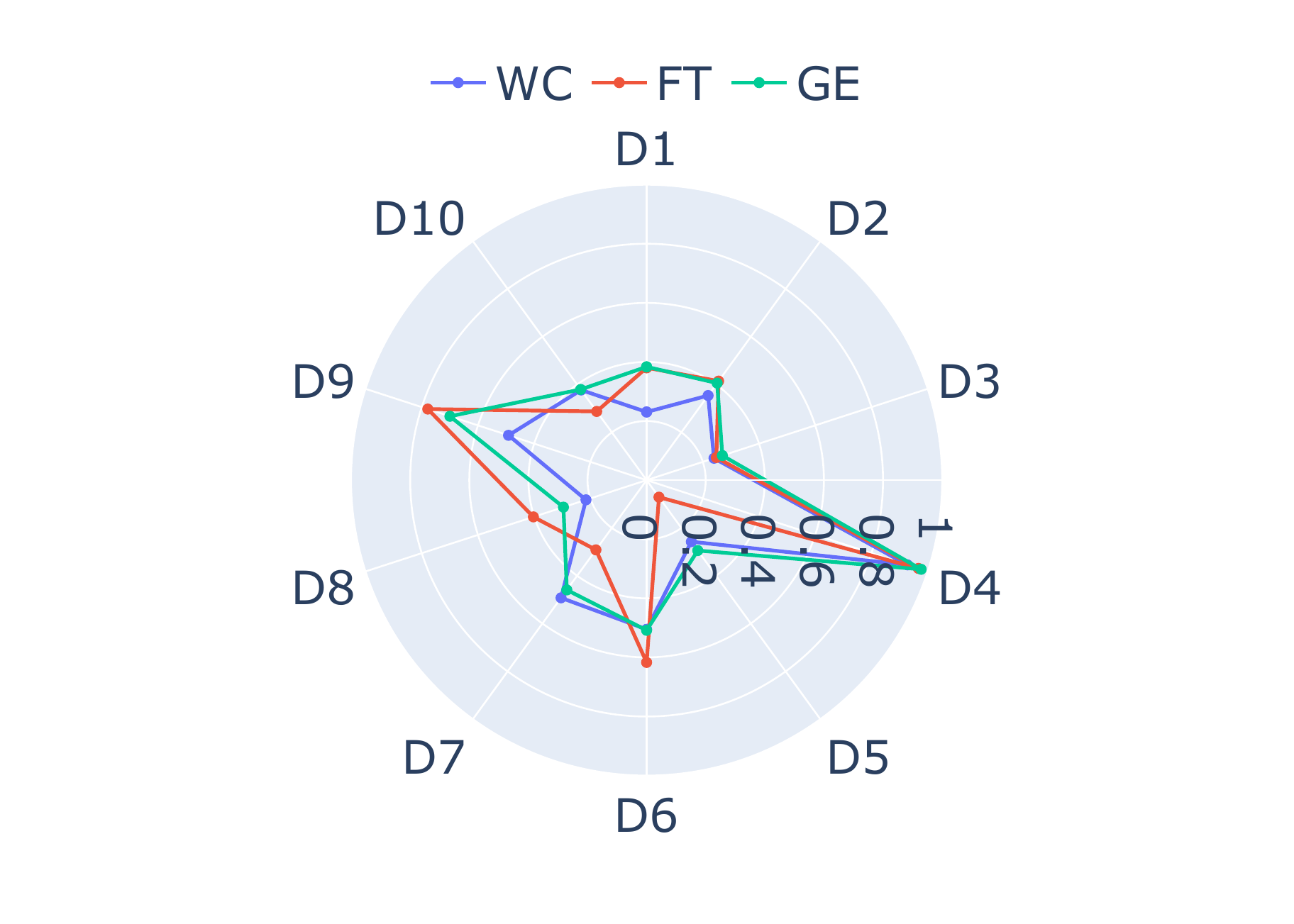}} 
\subfloat[BERT, Precision]{\includegraphics[trim=2cm 0.12cm 2cm 0.12cm, clip, width=0.25\textwidth, height=40mm]{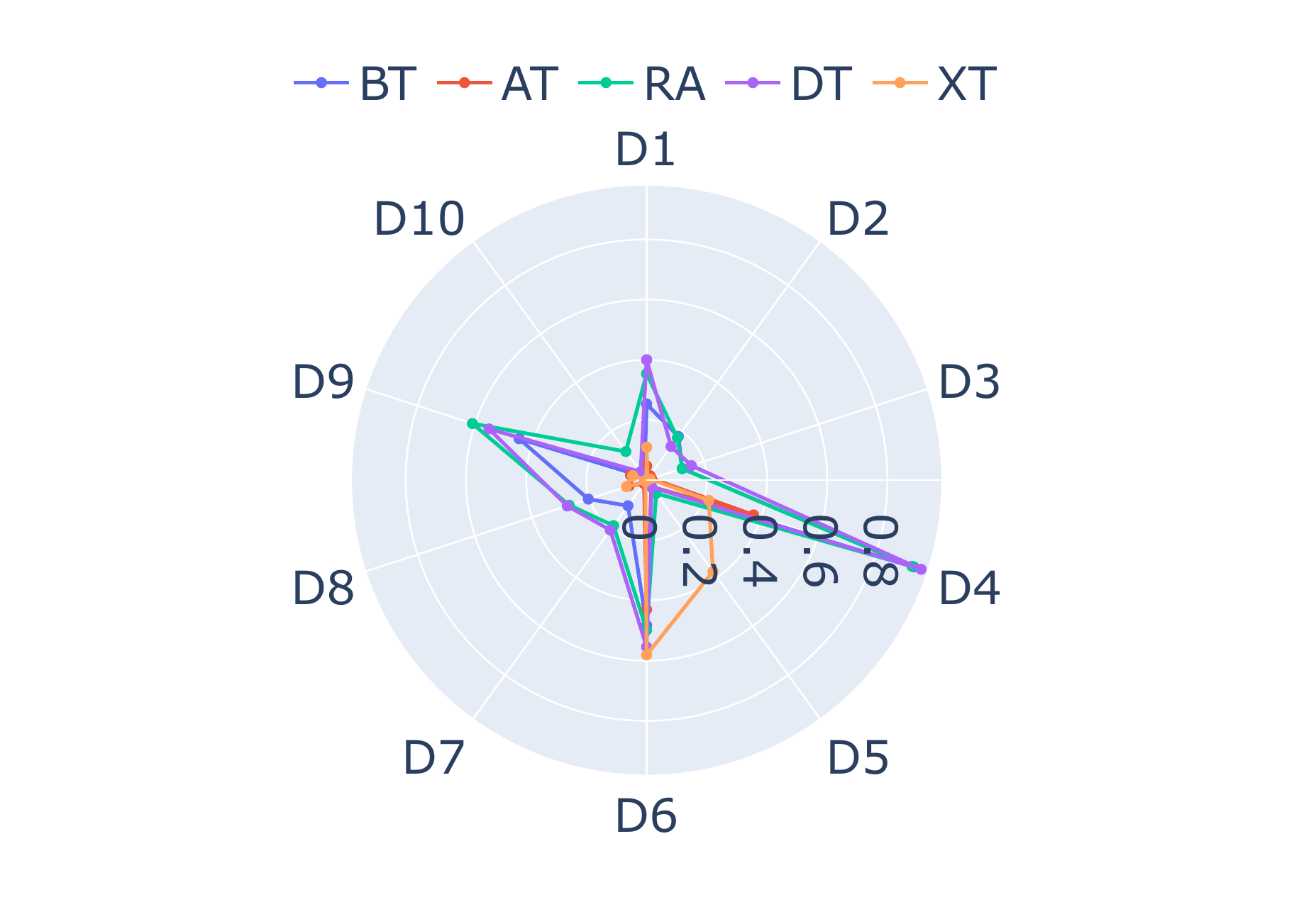}}
\subfloat[SBERT, Precision]{\includegraphics[trim=2cm 0.12cm 2cm 0.12cm, clip, width=0.25\textwidth, height=40mm]{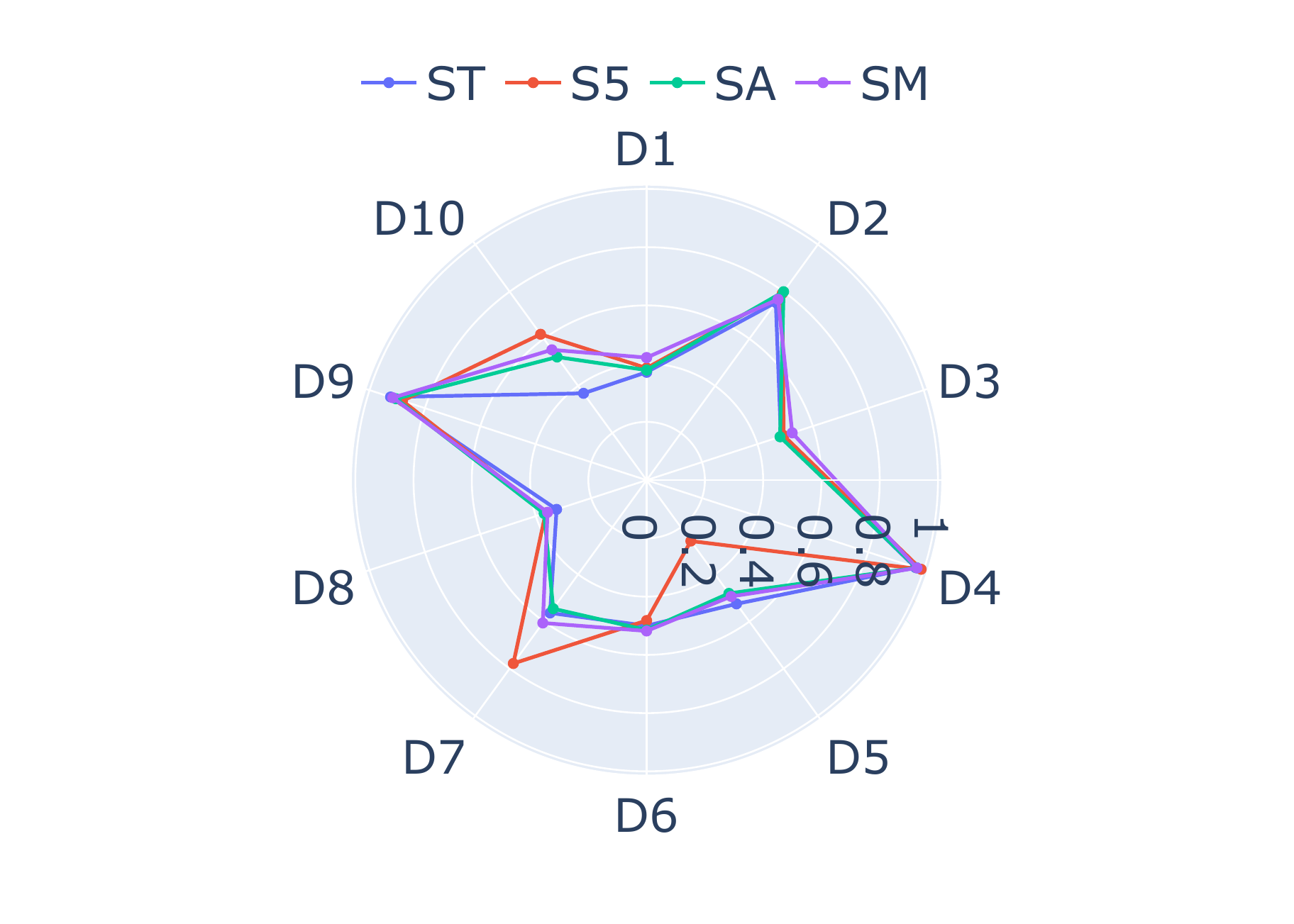}} 
\subfloat[SotA, Precision]{\includegraphics[trim=2cm 0.12cm 2cm 0.12cm, clip, width=0.25\textwidth, height=40mm]{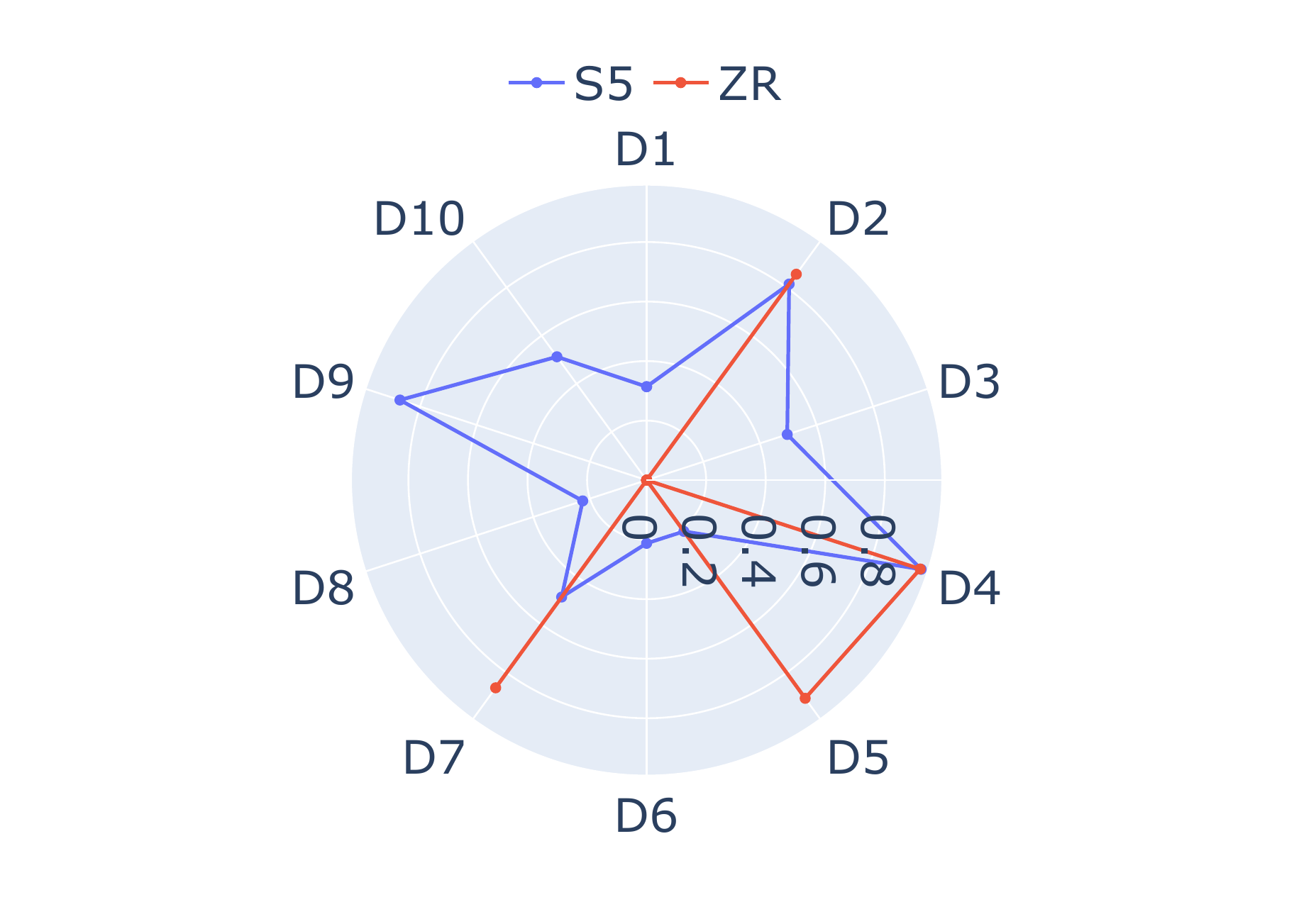}}
\newline
\subfloat[Static, Recall]{\includegraphics[trim=2cm 0.12cm 2cm 0.12cm, clip, width=0.25\textwidth, height=40mm]{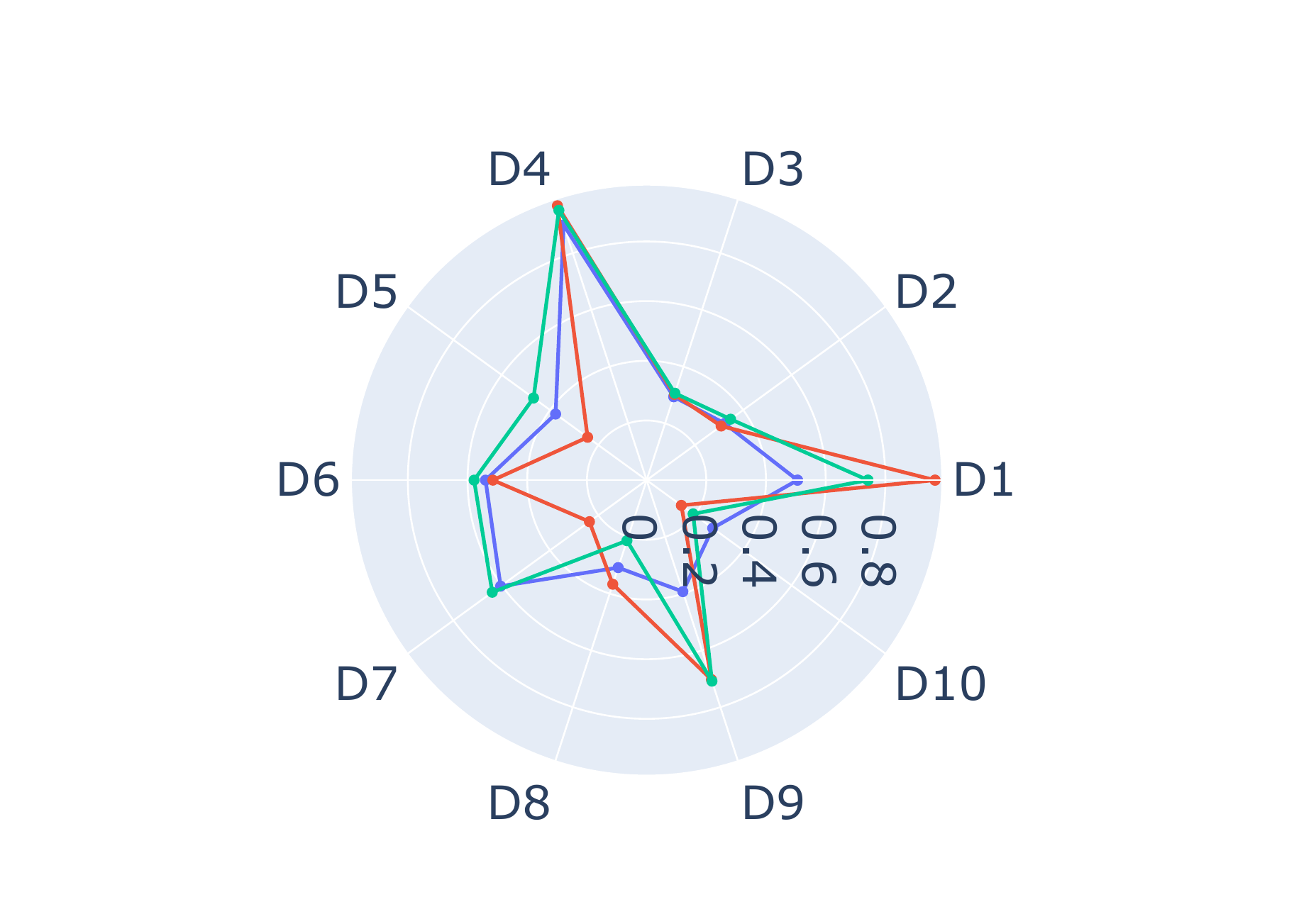}} 
\subfloat[BERT, Recall]{\includegraphics[trim=2cm 0.12cm 2cm 0.12cm, clip, width=0.25\textwidth, height=40mm]{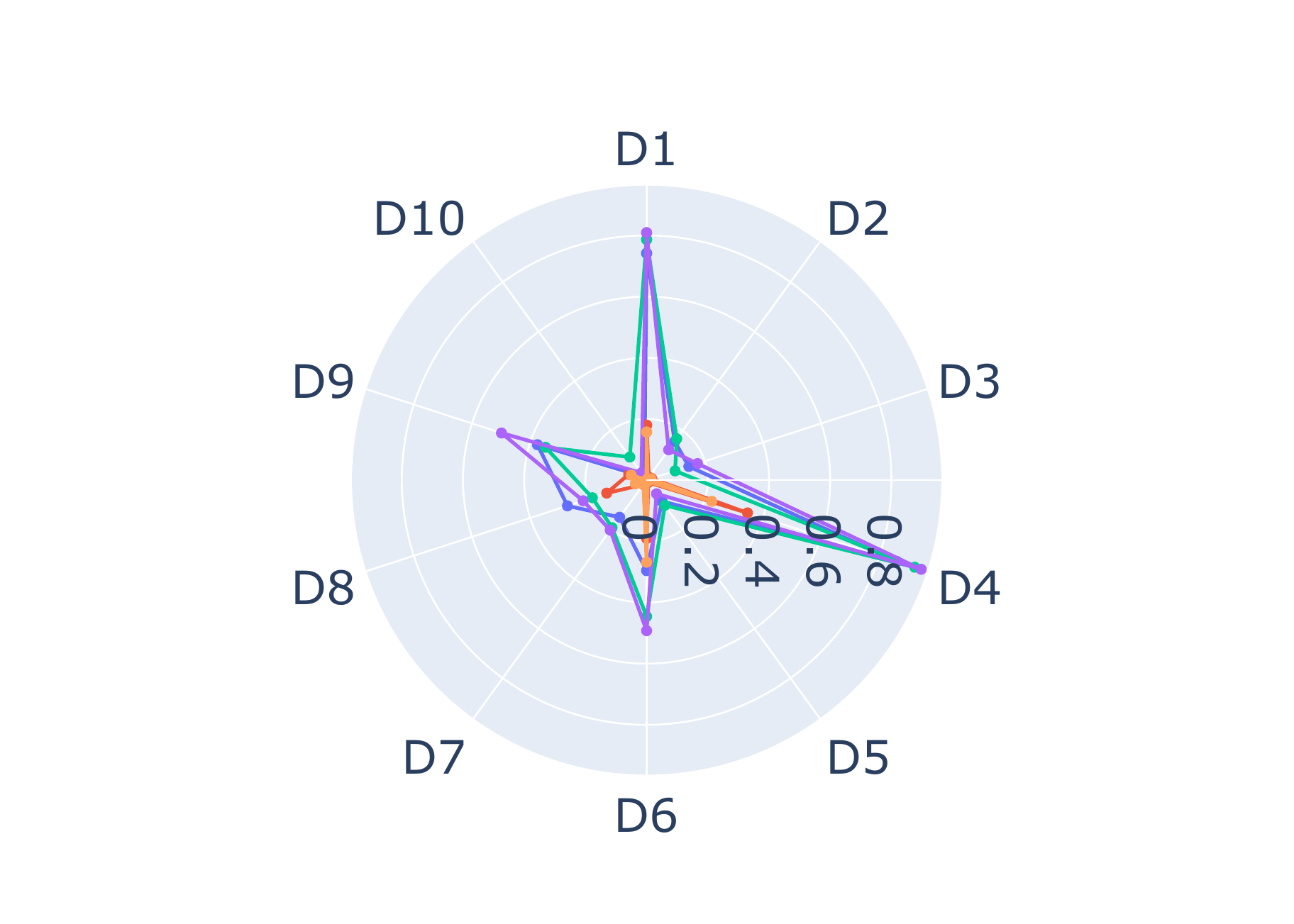}}
\subfloat[SBERT, Recall]{\includegraphics[trim=2cm 0.12cm 2cm 0.12cm, clip, width=0.25\textwidth, height=40mm]{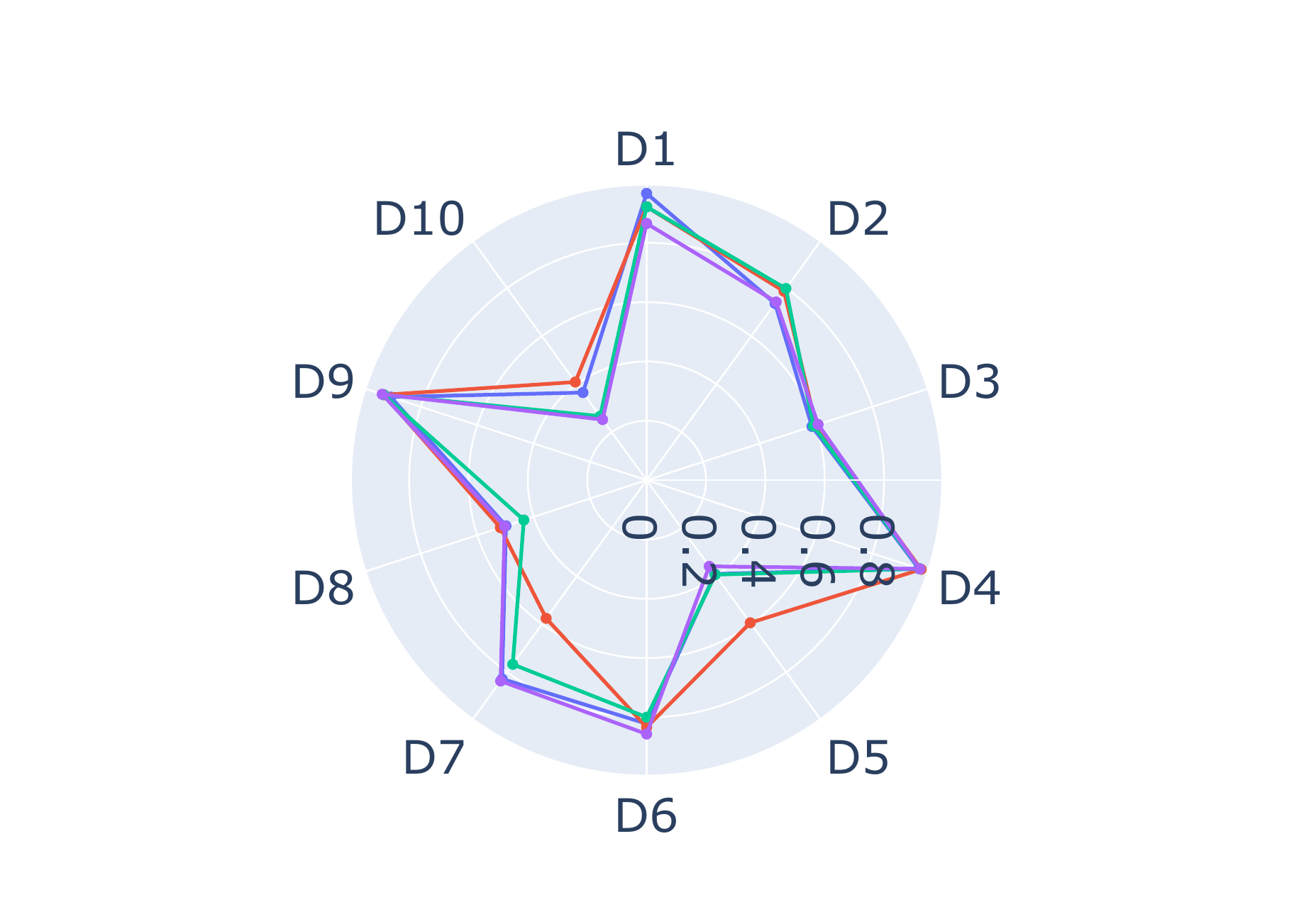}} 
\subfloat[SotA, Recall]{\includegraphics[trim=2cm 0.12cm 2cm 0.12cm, clip, width=0.25\textwidth, height=40mm]{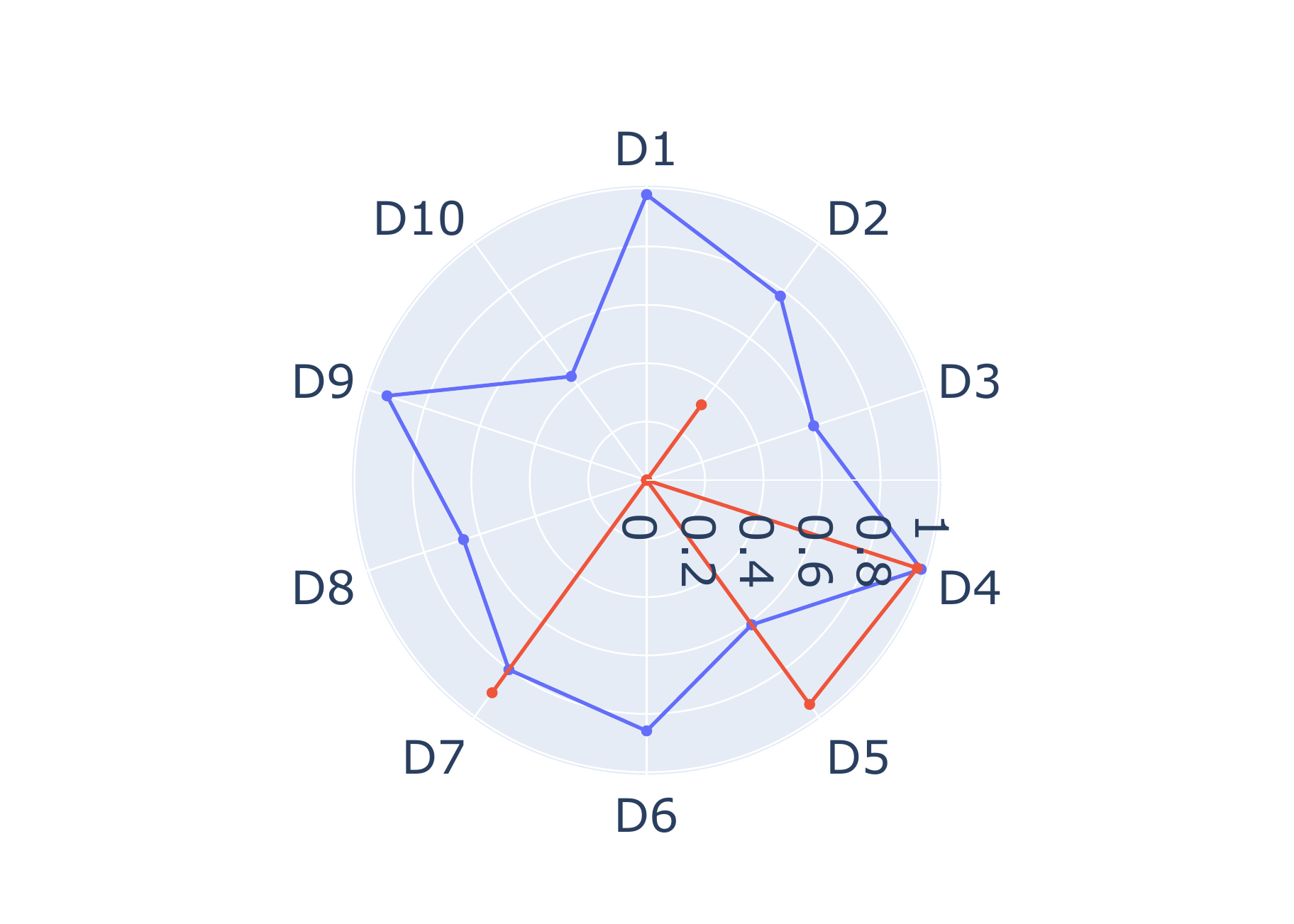}}
\newline
\subfloat[Static, F1]{\includegraphics[trim=2cm 0.12cm 2cm 0.12cm, clip, width=0.25\textwidth, height=40mm]{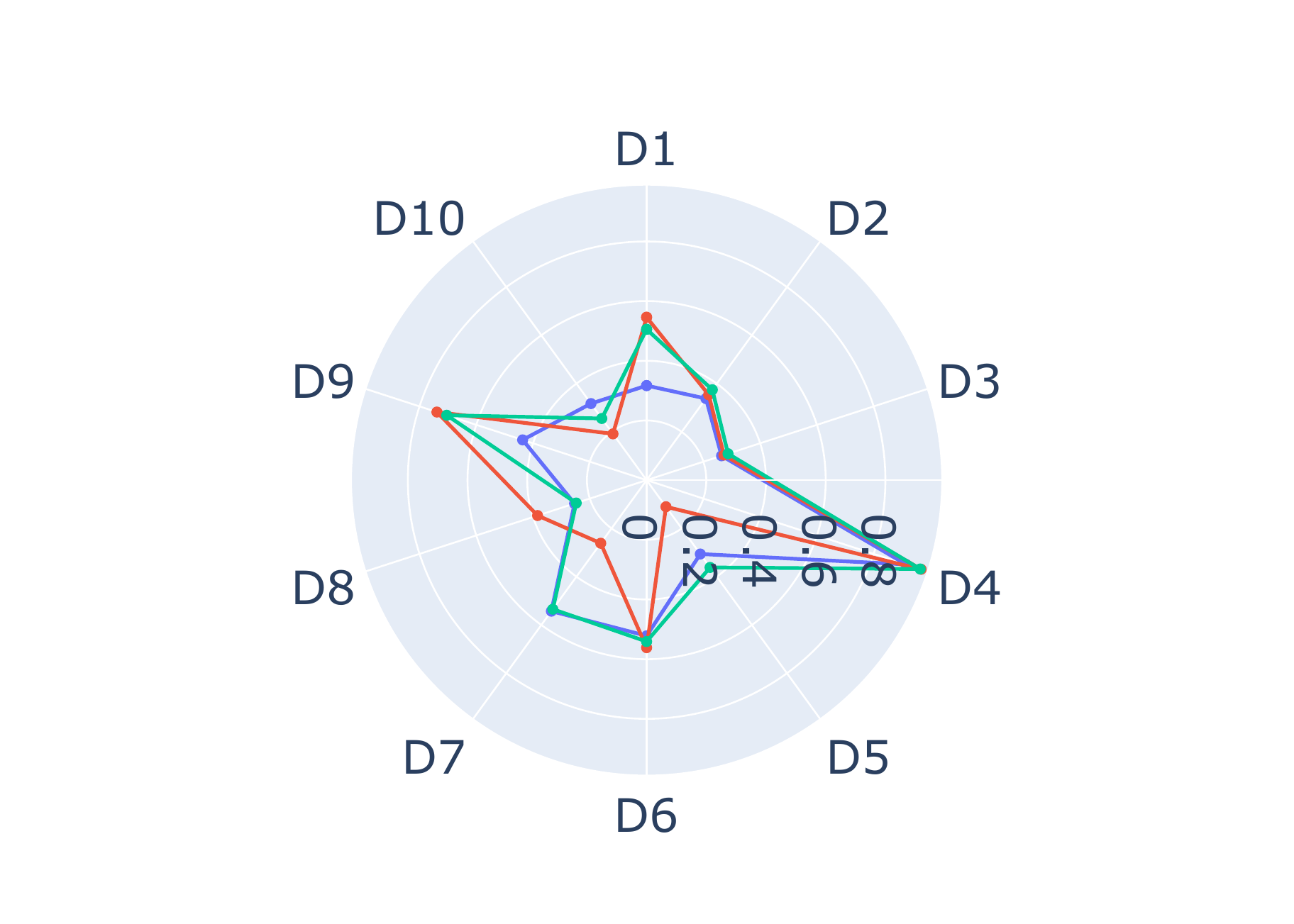}} 
\subfloat[BERT, F1]{\includegraphics[trim=2cm 0.12cm 2cm 0.12cm, clip, width=0.25\textwidth, height=40mm]{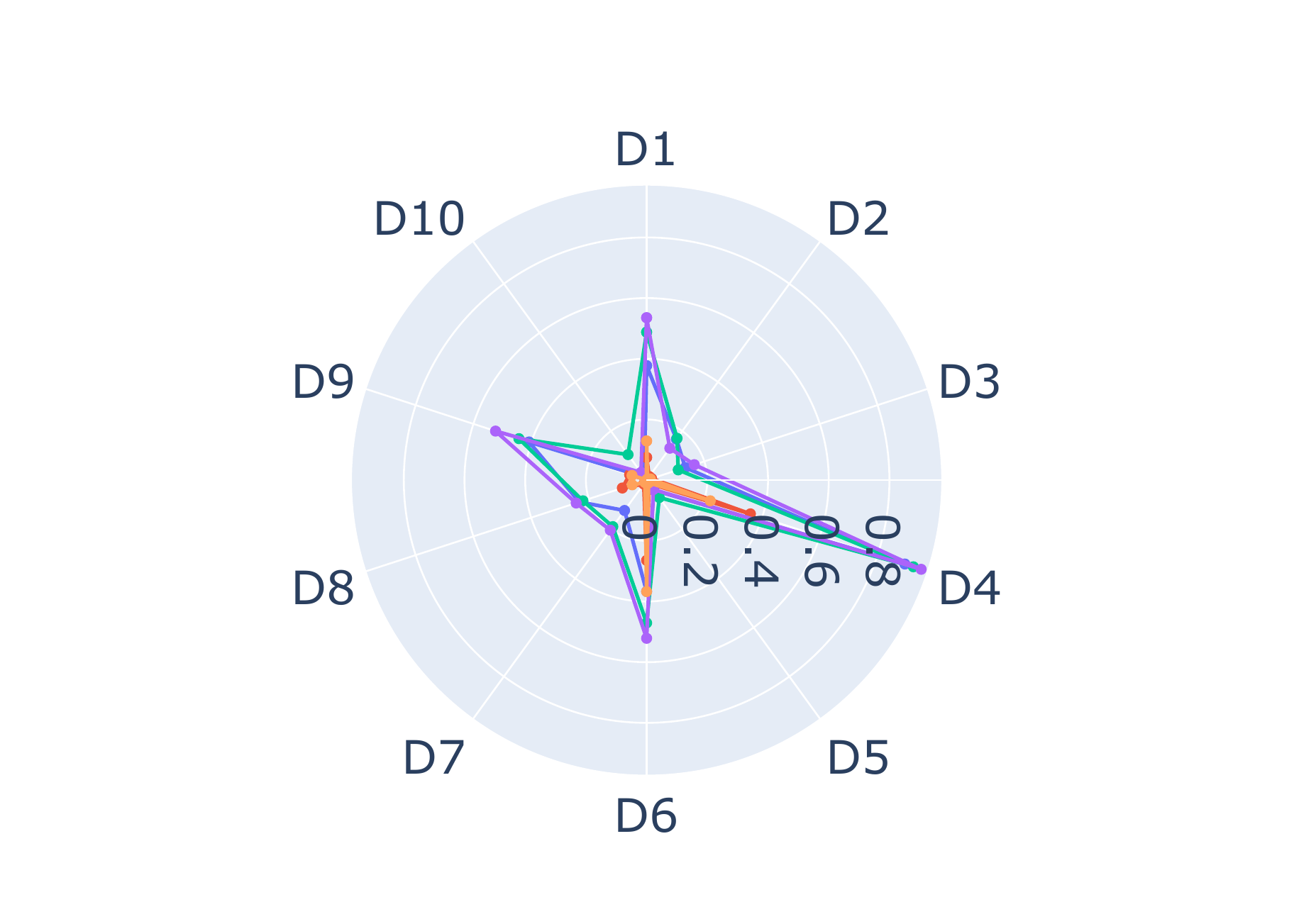}}
\subfloat[SBERT, F1]{\includegraphics[trim=2cm 0.12cm 2cm 0.12cm, clip, width=0.25\textwidth, height=40mm]{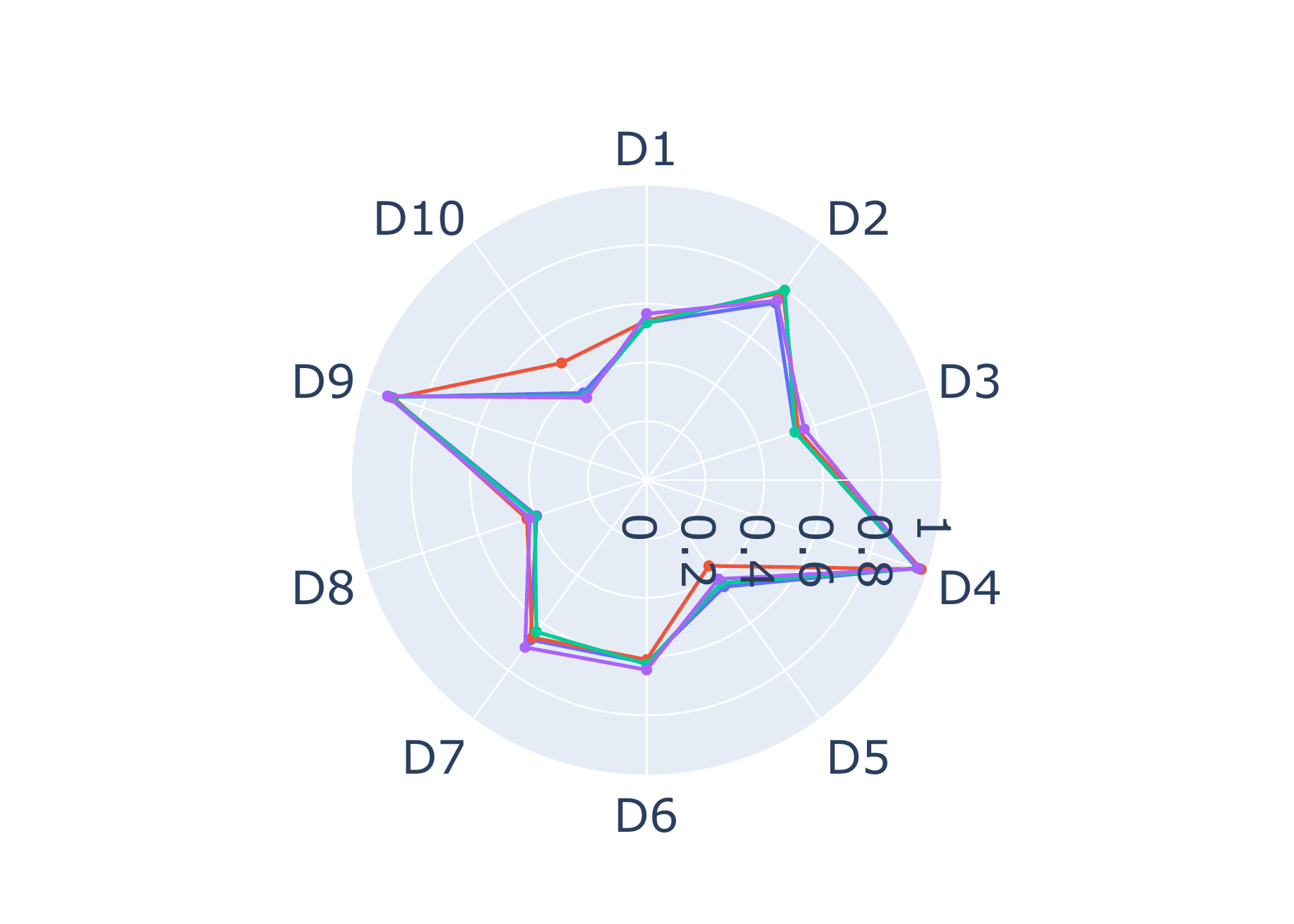}} 
\subfloat[SotA, F1]{\includegraphics[trim=2cm 0.12cm 2cm 0.12cm, clip, width=0.25\textwidth, height=40mm]{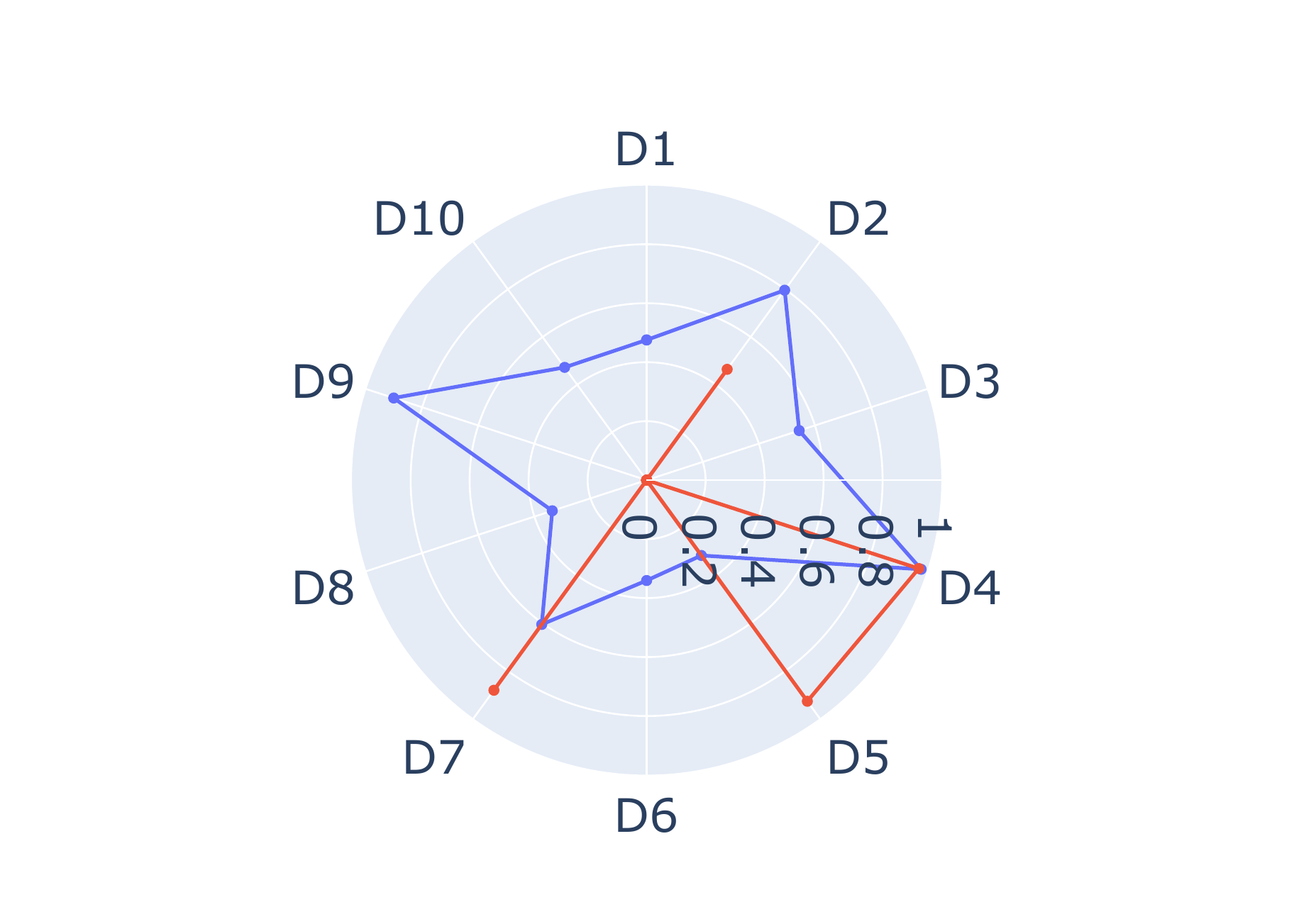}}
\newline
\caption{Precision, Recall and F1 per model across all datasets in Table 
% \caption{{\color{blue}Unsupervised Matching F-Measure per model across all datasets in Table 
\ref{tb:datasets}(a).}
\label{fig:match_unsup_rec_real}
\end{figure*}

Blocking must scale to large data volumes in order to restrict the input of matching to manageable levels, even in cases with millions of entities.
Therefore, we need to assess how well the selected language models scale as the size of the input data increases. 
To this end, we conduct an analysis using the seven datasets in Table \ref{tb:datasets}(b).
%These correspond to \textit{Dirty ER}, a.k.a., Deduplication, where a single data source containing duplicates is given as input \cite{DBLP:series/synthesis/2021Papadakis,DBLP:series/synthesis/2015Christophides}.
%Here, due to the large dataset size, we use approximate instead of exact nearest neighbor search. 
The results appear in Figure \ref{fig:blk_synth}.

Figure \ref{fig:blk_synth}(a) shows that recall consistently decreases for all models as we move from $D_{10K}$ to $D_{2M}$. This is expected, given that the number of candidates increases quadratically with the size of the input data. The best performance clearly corresponds to S-GTR-T5: its recall over $D_{2M}$ is quite satisfactory both in absolute terms (0.800) and in relation to the initial one over $D_{10K}$ (0.962), as it drops by just 17\%. 
S-GTR-T5 has been trained on a much larger and richer corpus. The advantage that this offers to S-GTR-T5 becomes much more evident when testing it on the synthetic datasets, which are more challenging due to the much larger number of candidate pairs.
The second best approach in most datasets is FastText, whose recall is reduced by 54\%, from 0.901 to 0.415. On the other extreme lie AlBERT and XLNet: their recall is lower than 0.18 across all datasets (even for $D_{10K}$) and is reduced by 2/3 over $D_{2M}$. The rest of the models fluctuate between these two extremes and can be arranged into three groups according to their performance. The best group comprises S-DistilRoberta and S-MiniLM, which start from $\sim$0.750 over the smallest dataset and end up $\sim$40\% lower at 0.450. The worst group
includes Word2Vec, DistilBERT, BERT and RoBERTa, whose recall drops by 56\%-66\%, falling far below 0.2 over $D_{2M}$. GloVe and S-MPNet lie in the middle of the first two extremes: their recall is reduced by less than 50\% and exceeds 0.32 over $D_{2M}$.

We observe almost the same patterns with respect to precision in Figure \ref{fig:blk_synth}(b). This is due to the linear relation between the two measures, as explained earlier.
Only minor variations occur in the context of Dirty ER, due to the different number of redundant candidate pairs, which are counted once (a candidate pair $<$$e_i$, $e_j$$>$ is redundant if $e_j$ is included in the nearest neighbors of $e_i$ and vice versa). Hence, there is a gradual decrease in precision for all models as the input size increases. This is larger than the decrease in recall by 2-3\% in most cases, because the denominator of recall (i.e., the number of existing duplicates) increases at a slower pace than the denominator of precision (i.e., number of distinct candidate pairs).

\subsection{Unsupervised Matching}

\begin{figure}[!t]
\centering
\includegraphics[trim=0.12cm 0.12cm 0.12cm 0.12cm, clip, width=0.4\textwidth]{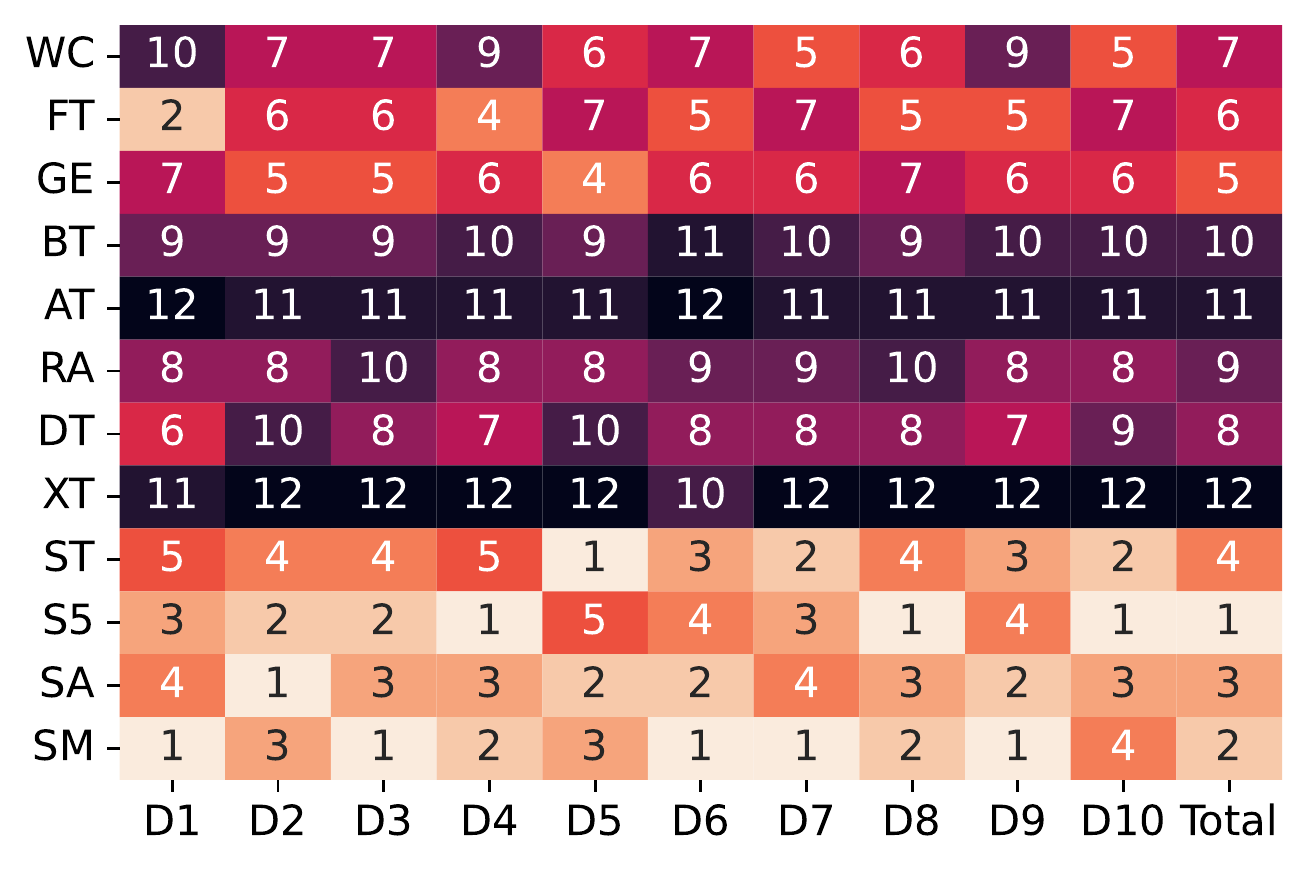}
\caption{The ranking position of each model with respect to Unsupervised Matching F1 per dataset. Lower is better.}
% \caption{Method ranking wrt blocking recall (lower is better).}
\label{fig:matching_heat}
\end{figure}

\begin{figure}[!t]
\centering
\includegraphics[trim=0.12cm 0.12cm 0.12cm 0.12cm, clip, width=0.45\textwidth]{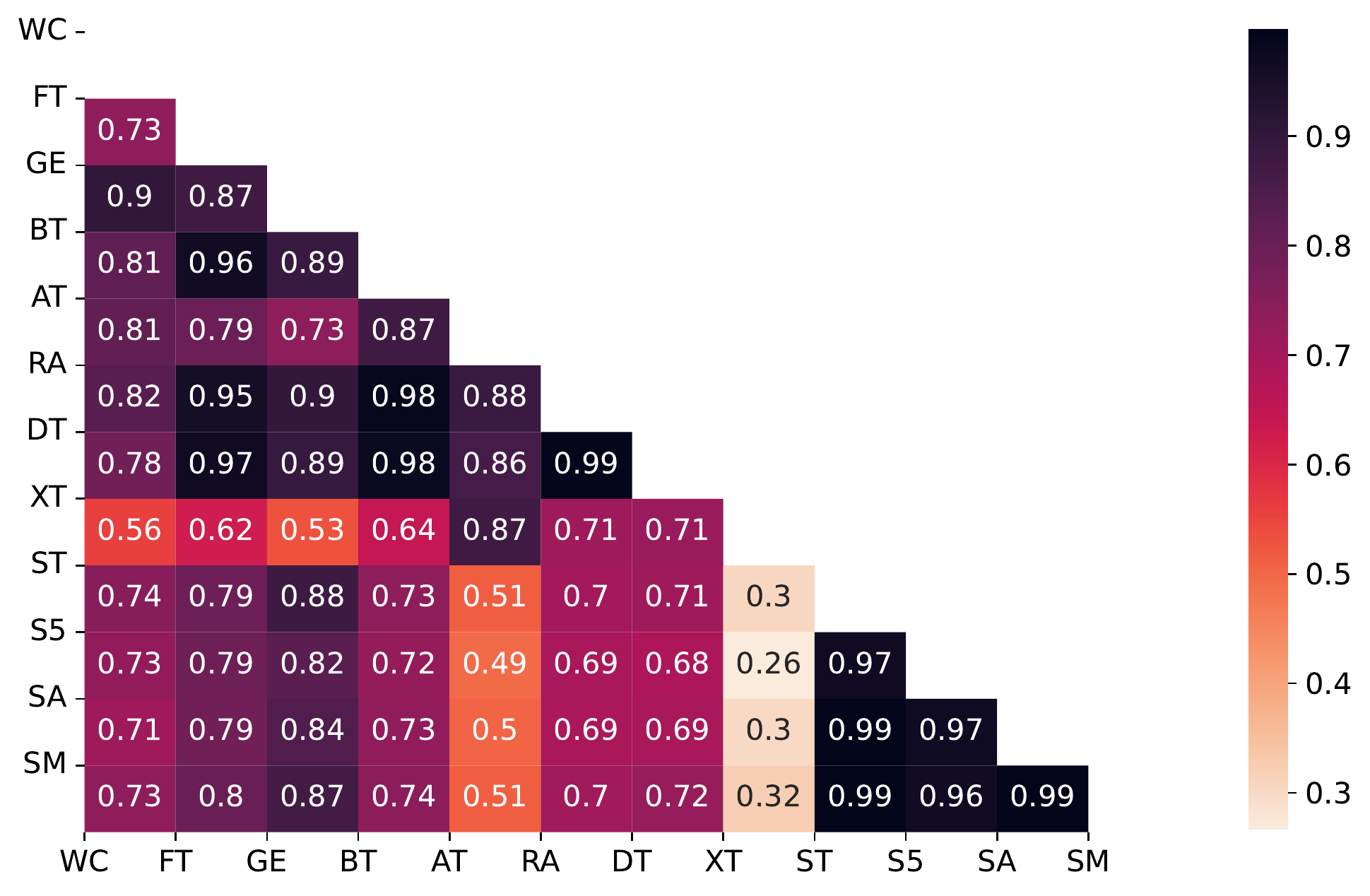}
\caption{Pearson correlation of language models with respect to Unsupervised Matching F1.}
\label{fig:match_corr}
\end{figure}

Figure \ref{fig:match_unsup_rec_real} reports the performance of all models in this task. 

In static models, based on the average distance from the maximum f-measure (F1), GloVe is the best one (31.9\%) followed by FastText (37.4\%) and Word2Vec (39.4\%). In absolute terms, their F1 remains rather low in all datasets except for the bibliographic ones: in $D_4$, it exceeds 0.9, lying very close to the SentenceBERT models, but in $D_9$, only FastText and Glove manage to surpass 0.7. The reason is that the entities from the Google Scholar are much more noisy and involve many more terminologies than $D_4$. In all other cases, their F1 falls (far) below 0.57. 

In the BERT-based models, the worst performance is consistently exhibited by XLNet and AlBERT, as their F1 does not exceed 0.37 in any dataset. On average, their F1 is almost an order of magnitude ($\sim$87\%) lower than the top one. The reason is the same as in Blocking: both were trained for a different task and cannot perform well in the task of Matching, without further fine-tuning. In the other extreme lie DistilBERT and RoBERTa, with an average distance of $\sim$55\%. Finally, BERT fluctuates between these two extremes, with an average distance from the top equal to 62\%. These three models score an acceptable F1 ($\sim$0.9) in the clean and easy $D_4$, but remain (far) below 0.54 in all other cases. 

In the SentenceBERT models, the best one is \textsf{S-GTR-T5}, achieving the highest F1 in seven out of the 10 datasets. In the remaining datasets, it is ranked second or third, lying very close to the top performer. On average, its distance from the maximum F1 is just 0.7\%, being the lowest among all models. The worst case corresponds to $D_6$, where it lies 4.9\% lower than the best method. The second best model is \textsf{S-MiniLM}, having the highest F1 in three datasets and the second lowest average distance from the maximum F1 (6.5\%). The two next best models are \textsf{S-MPNet} and \textsf{S-DistilRoBERTa}, whose average distance from the top is 10.5\%.

In absolute terms, their best performance 
corresponds to the bibliographic datasets, $D_4$ and $D_9$, where their F1 remains well over 0.9. 
The reason is that the long titles and list of authors facilitate the distinction between matching and non-matching entities. Also high ($\sim$0.8) is their F1 over $D_2$, which combines long textual descriptions with a 1-1 matching between the two data sources (i.e., every entity from the one source matches with one entity from the other). In all other datasets, their F1 ranges from 0.35 ($D_{10}$) to 0.77 ($D_7$), due to the high levels of noise they contain.

Finally, it is worth stressing that as in Blocking, S-GTR-T5 outperforms the best models of the other two types, since all other models do not surpass 0.5 in F1 -- except for the relatively clean and easy $D_4$. In contrast with Blocking, though, in Matching we can perform fine-tuning to check whether BERT models can perform better than other two types. See Section \ref{sec:supMatching} for more details. 
% This will be the object of the following section.`

\begin{figure*}[!t]
\centering
\subfloat[static]{\includegraphics[trim=2cm 0.12cm 2cm 0.12cm, clip, width=0.25\textwidth, height=40mm]{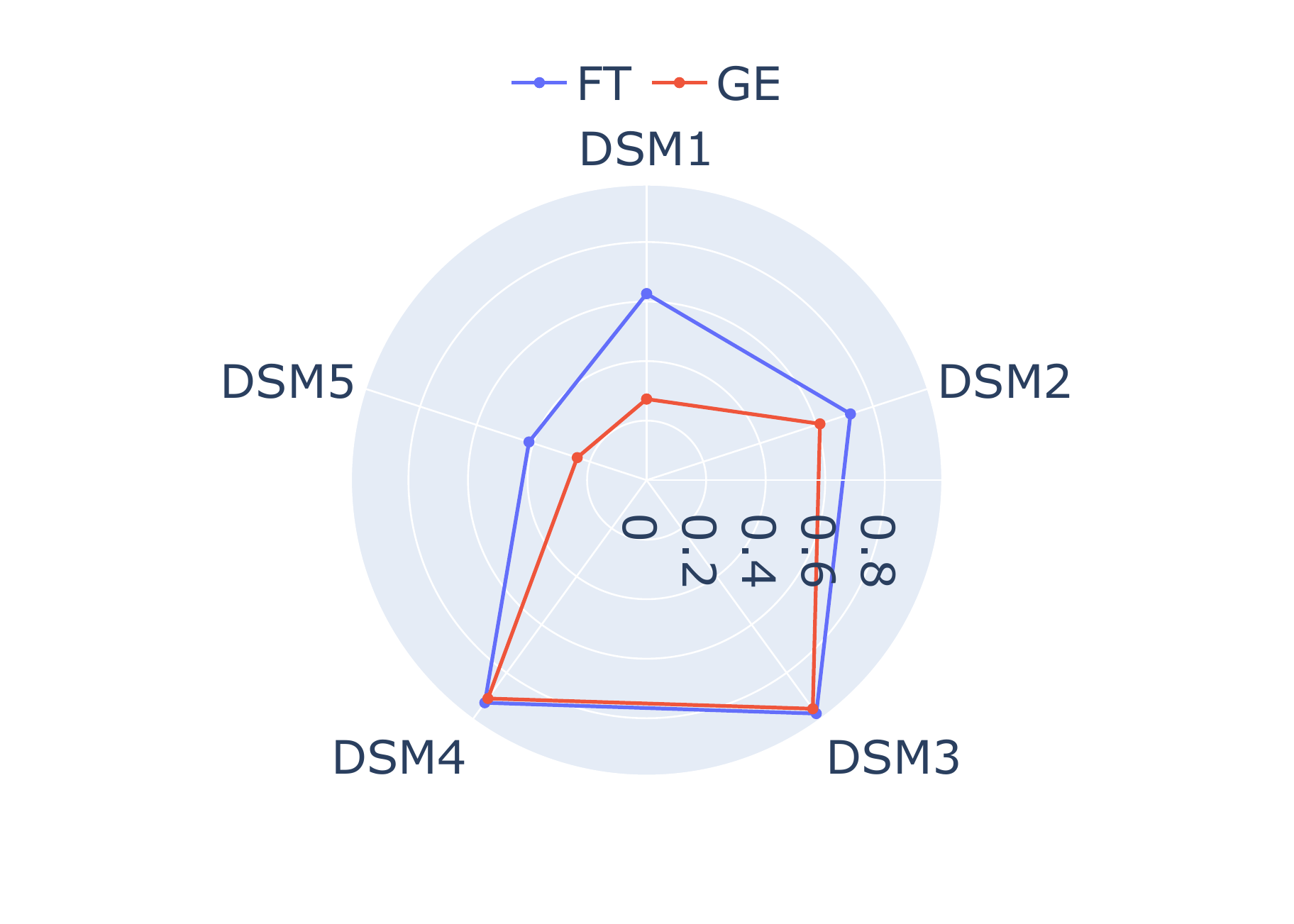}}
\subfloat[BERT]{\includegraphics[trim=2cm 0.12cm 2cm 0.12cm, clip, width=0.25\textwidth, height=40mm]{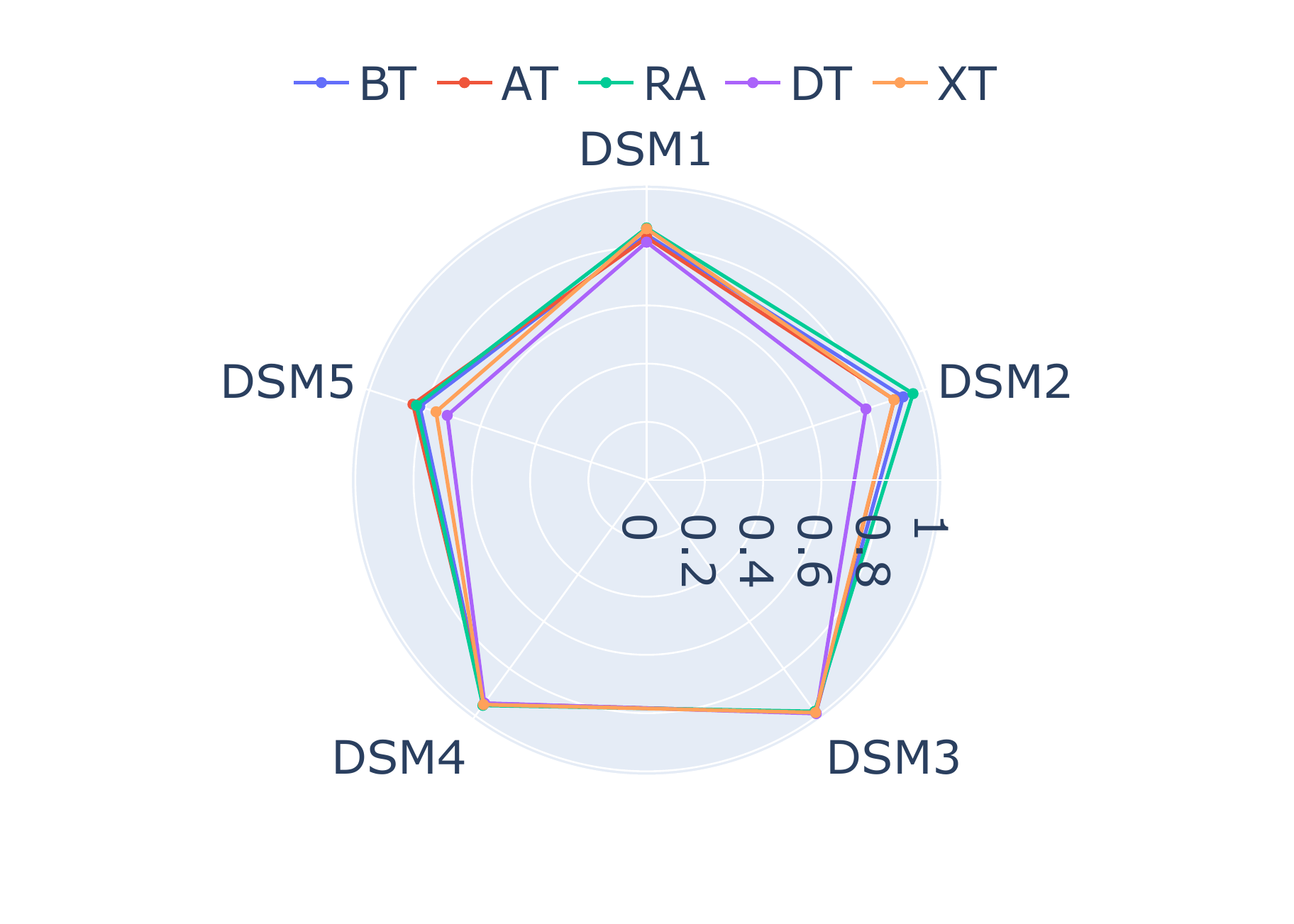}}
\subfloat[SBERT]{\includegraphics[trim=2cm 0.12cm 2cm 0.12cm, clip, width=0.25\textwidth, height=40mm]{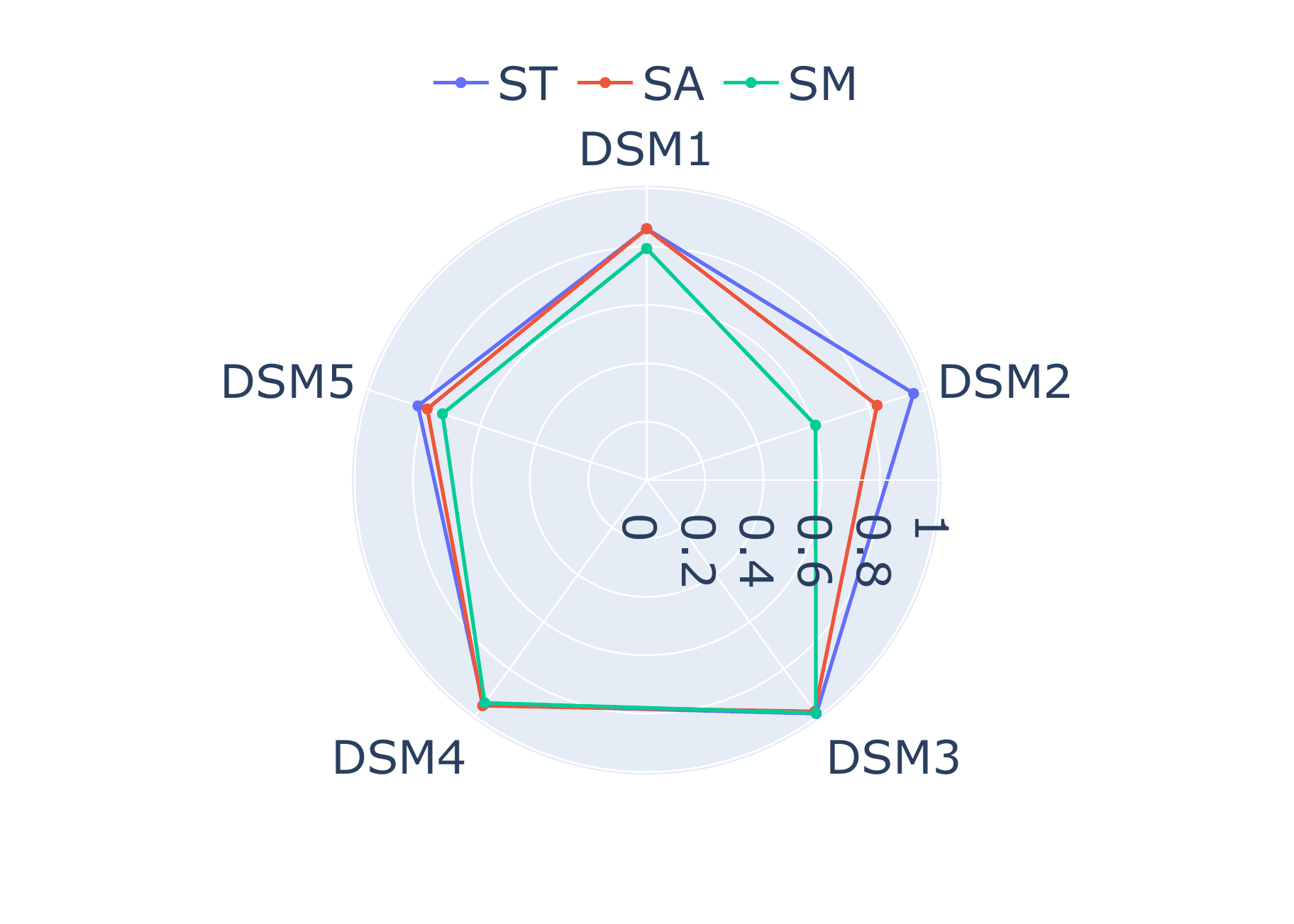}} 
\subfloat[SotA]{\includegraphics[trim=2cm 0.12cm 2cm 0.12cm, clip, width=0.25\textwidth, height=40mm]{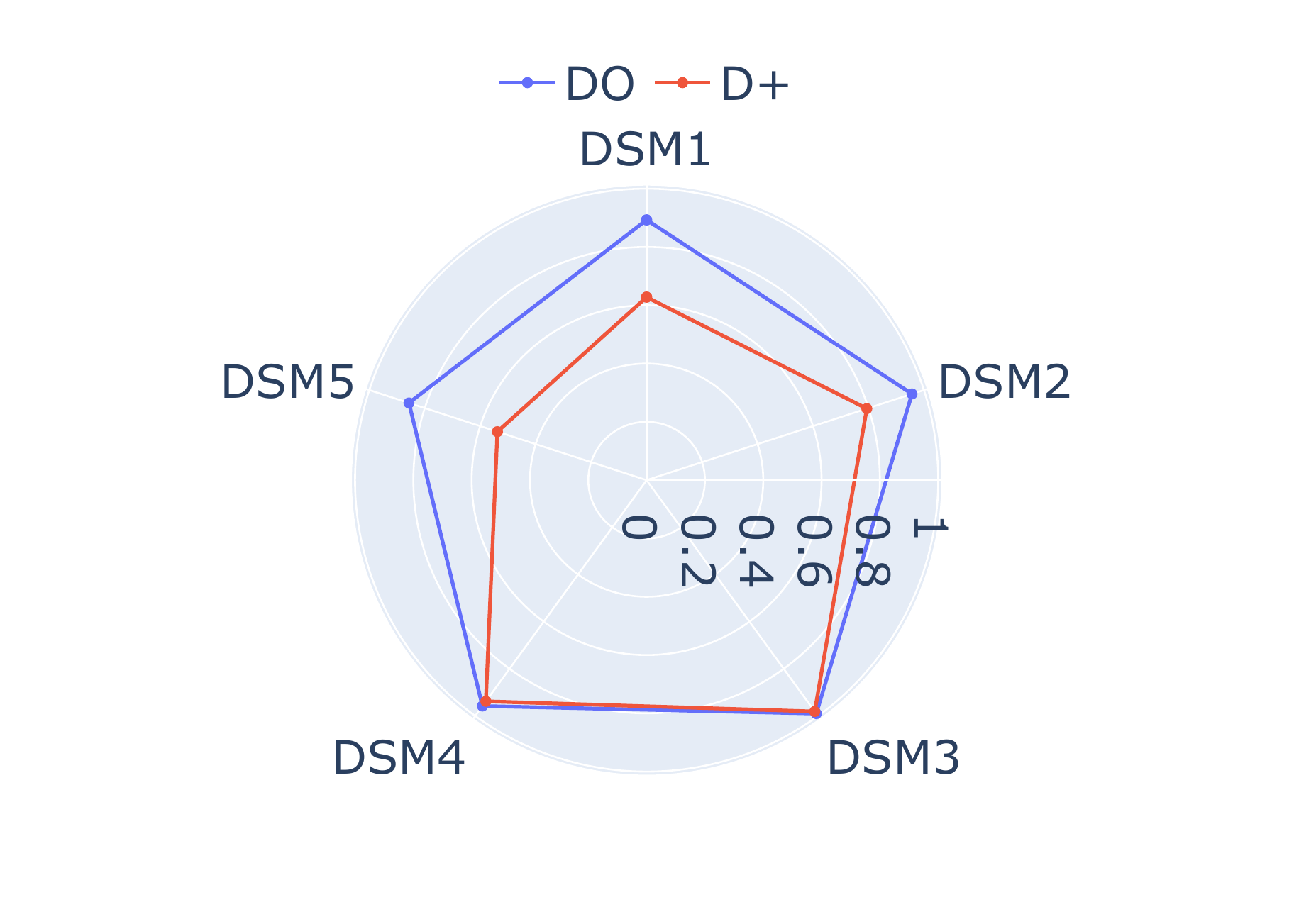}}
\newline
\caption{Supervised Matching F-Measure per model across all datasets in Table 
\ref{tb:smDatasets}.}
\label{fig:sup_matching}
\end{figure*}

\noindent\textbf{Summary.} 
Figure \ref{fig:matching_heat} summarizes the ranking position of each model per dataset.
The SentenceBERT models typically fluctuate between positions 1 and 4, the static ones between 5 and 7 and the BERT-based ones between 8 and 12. Moreover, the Pearson correlation of their F1 in Figure \ref{fig:match_corr} shows an almost perfect dependency between the SentenceBERT models and a very high one inside the group of static and BERT-based ones. The latter are weakly correlated with the SentenceBERT models. The static models have moderate correlation $(0.7-0.9)$ with the other two groups. 

Overall, the relative performance of the three model types follows the same patterns as in Blocking, due to the same root causes. The SentenceBERT models outperform all others, as they are inherently crafted for vectorizing entire ``sentences'', while they have been trained on much larger corpora. At the other extreme lie the BERT models, which lack fine-tuning, with the predefined weights in their final layer yielding low similarities for matching and non-matching pairs alike. The static models lie in the middle of these extremes, due to their context-agnostic, word-level embeddings.

\noindent\textbf{Comparison to SotA.} 
The state-of-the art in Unsupevised Matching is ZeroER \cite{wu2020zeroer}, which converts every pair of entities into a feature vector whose dimensions correspond to similarity functions. At its core lies the assumption that the resulting feature vectors are generated by a Gaussian Mixture Model with two mixture components (one for each matching category). Adaptive feature regularization is leveraged to avoid overfitting, while transitivity improves its accuracy. ZeroER uses Magellan's overlap blocking to reduce the search space to a small set of candidate pairs. 
% {\color{red}How was ZeroER configured? Is blocking used?}.

We compare ZeroER with an end-to-end framework based on the best language model for Blocking and Matching, i.e. S-GTR-T5. We actually use the above matching algorithm with the similarity threshold set to 0.5 by default, but instead of utilizing all pairs of entities, every entity of the smallest entity collection is allowed only $k$=10 candidates, produced by Blocking with exact NNS.
% {\color{red}is FAISS used or we are using exact NNS?}
% We have compared our framework with 

The relative performance of the two approaches appears in Figure \ref{fig:match_unsup_rec_real}(d). ZeroER lacks an estimated performance for half the datasets, because it did not terminate after 6 hours -- unlike S-GTR-T5, which consistently takes less than 1 minute, as shown in Table \ref{tb:sota_comparison}(b). In $D_4$, both methods have the same, almost perfect performance, due to the rather clean data and the relatively easy task. In $D_1$ and $D_2$, S-GTR-T5 outperforms ZeroER to a significant extent. ZeroER actually yields $F_1$=0 on $D_1$, because $D_1$ contains many missing and misplaced values, which cannot be supported by ZeroER's schema-based functionality (unlike the schema-agnostic settings of S-GTR-T5). $D_2$ conveys large textual values, which are also barely suitable for most similarity measures employed by ZeroER.
In contrast, $D_5$ and $D_7$ contain short attribute values that describe movies (e.g., actor names). These are ideal for the features of ZeroER, which thus achieves much higher effectiveness than S-GTR-T5, which is not crafted for rare, domain-specific (terminological) textual values.
%Nonetheless, in the occassions where it did terminate, S-GTR-T5 had a better score, with the exception of D5, where S-GTR-T5 has a very low performance.

 Overall, S-GTR-T5 performs significantly better than or at least equally well as ZeroER in most datasets, despite its parameter-free functionality, while being orders of magnitude faster (cf.~Section~\ref{sec:unsupMatching}).

\subsection{Supervised Matching}
\label{sec:supMatching}

Figures \ref{fig:sup_matching}(a)-(c) show the F1 of all models in this task 
% results of our experiment with respect to effectiveness (F1 score) 
over the datasets in Table \ref{tb:smDatasets}. We observe that the language models can be distinguished into two groups according to their context awareness. The dynamic models consistently exhibit the highest performance.
% except for the two 
% We observe that 
RoBERTa is actually the most robust model in terms of effectiveness. 
It achieves the highest F1 in most datasets, ranking first on average. In fact, its mean distance from the top F1 across all datasets amounts to just 0.5\%. It is followed in close distance by the second best model, S-MPNet, which is top performer in one dataset ($DSM_3$) and its average distance from the top F1 is 0.7\%.  
The rest of the models are sorted in the following order according to their average distance from the maximum $F_1$ per dataset: BERT, AlBERT, XLNet, S-DistilRoBERTa, S-MiniLM. 

Even the least effective dynamic model, though, is just 5\% worse than the best one, on average. The reason for this is the strong correlation between all considered models: high effectiveness for one model on a specific dataset implies similarly high effectiveness for the rest of the models. This pattern should be partially attributed to the $DSM_3$ and $DSM_4$ datasets, where the difference between the maximum and the minimum F1 is less than 1.3\%. These datasets convey relatively clean values, even though in both cases artificial noise has been inserted in the form of misplaced values. This means that the duplicate entities share multiple common tokens in the schema-agnostic settings we are considering. As a result, even classic machine learning classifiers that use string similarity measures as features achieve very high F1 ($>$0.9) in these datasets (see Magellan in \cite{DBLP:journals/pvldb/0001LSDT20,Mudgal2018sigmod}). The rest of the datasets are more challenging, due to the terminologies they involve (e.g., product names). As a result, the difference between the maximum and minimum F1 of dynamic models ranges from 4.9\% ($DSM_1$) to 
% 12.4\% ($DSM_5$) and 
17\% ($DSM_2$).

The second group of models includes the static, context-agnostic ones, which perform relatively poorly, even though they are combined with the best configuration of DeepMatcher, i.e., the state-of-the-art algorithm for this task and type of models. GloVe and FastText are actually combined with a two layer fully connected ReLU HighwayNet classifier followed by a softmax layer in the classification module in combination with a hybrid model for the attribute similarity vector module.
On average, GloVe and FastText underperform the top model per dataset by 22\% and 37\%, respectively. Only in $DSM_3$ and $DSM_4$, where all dynamic models exceed 0.95, the two static models exhibit high performance, with their F1 within 5\% of the maximum one (this is another indication about the straightforward matching task posed by the bibliographic data of these two datasets). Note that FastText consistently outperforms GloVe across all datasets, since its character-level functionality is capable of addressing the out-of-vocabulary tokens that arise, due to the domain-specific terminology of each dataset, unlike GloVe. 
% In static models we see that FastText and Glove behave moderately in three out of the five datasets.
% {\color{red}
% In BERT models, we have a very different behaviour than the one presented in Unsupervised Matching, i.e. all models perform very well in all datasets. RoBERTa seems to be the most robust of all.

% In SBERT we have similarly a very good performance for all models, with S-MPNet being the best one.

Overall, \textit{the static models underperform the dynamic ones in most cases, as reported in the literature \cite{DBLP:journals/pvldb/0001LSDT20}, while the BERT-based models match the SentenceBERT ones, unlike for the previous ER tasks, due to the fine-tuning of their last layer. SentenceBERT models also benefit from fine-tuning, but to a lesser extent, probably because the sentence representing each entity is constructed in an ad-hoc manner, lacking any cohesiveness.}
% }

\noindent\textbf{Comparison to SotA.} 
Figure \ref{fig:sup_matching}(d) depicts the performance of the state-of-the-art supervised matching algorithms that leverage static and dynamic models, DeepMatcher+ \cite{DBLP:conf/acl/KasaiQGLP19} and DITTO \cite{DBLP:journals/pvldb/0001LSDT20}. We actually consider their optimized F1 that is reported in \cite{DBLP:journals/pvldb/0001LSDT20}.

% methods we have included the performance of DITTO and  
Comparing DITTO to the dynamic models, we observe that its F1 is directly comparable to the best performing language model in each dataset. In $DSM_2$, $DSM_3$ and $DSM_4$, DITTO's F1 is lower by just $\leq$0.5\%. In $DSM_1$ and $DSM_5$, though, DITTO outperforms all language models by 3\% and 1.5\%, respectively. This should be attributed to the external knowledge and the data augmentation, whose effect is more clear when comparing DITTO to the language model at its core, i.e., RoBERTa. On average, across all datasets, the latter underperforms DITTO by 1.3\%. 

Comparing the static models to DeepMatcher+, we observe that its performance is almost identical with FastText in most datasets, because it leverages the same language model. Only in $DSM_2$ and $DSM_5$, DeepMatcher+ performs substantially better, by 9\% and 23\%, respectively. This should be attributed to its advantage over the original DeepMatcher algorithm, which DeepMatcher+ combines with transfer and active learning. Note that DeepMatcher+ consistently outperforms GloVe by 28\%, on average.
% , we observe that they outperform it in practically all cases. This is particularly true in $DSM_1$ and $DSM_5$, where the worst language model (DistilBERT) exceeds DeepMatcher+ by $\sim$20\%.}

Note also that DeepMatcher+ underperforms the dynamic models in practically all cases. This is particularly true in $DSM_1$ and $DSM_5$, where the worst dynamic model (DistilBERT) exceeds DeepMatcher+ by $\sim$20\%. This verifies the superiority of dynamic models over the static ones in supervised matching, due to their fine-tuning, which optimizes the weights of their last layer to the data at hand.

\section{Comparison on Efficiency}
\label{sec:efficiency}

\subsection{Vectorization}

\begin{table}[t]
\setlength{\tabcolsep}{2.5pt}
\small
\begin{tabular}{|l|rrr|rrrrr|rrrr|}
\cline{2-13}
\multicolumn{1}{l|}{} &   WC &    FT &   GE &    BT &    AT &    RA &    DT &    XT &    ST &    S5 &    SA &    SM \\
\midrule
\midrule
Init &  32.4 &  159.7 &  5.87 &  4.72 &  3.99 &  5.28 &  4.3 &  4.73 &  9.19 &  9.84 &  9.33 &  8.36 \\
\midrule
D1   &  0.0 &   0.2 &  1.9 &   2.6 &   2.4 &   2.3 &   1.3 &   4.0 &   1.1 &   1.1 &   0.7 &   0.5 \\
D2   &  0.1 &   1.6 &  0.2 &   3.1 &   2.4 &   2.3 &   2.2 &   3.3 &   3.4 &   3.4 &   1.8 &   0.9 \\
D3   &  0.9 &   9.6 &  0.4 &  10.1 &   6.7 &   6.3 &   8.6 &   8.3 &  10.3 &  12.4 &   5.8 &   2.3 \\
D4   &  0.2 &   2.5 &  0.3 &   5.9 &   5.2 &   5.3 &   4.2 &   7.8 &   5.1 &   5.4 &   2.8 &   1.4 \\
D5   &  0.4 &   3.8 &  0.4 &  13.6 &  13.0 &  12.9 &   8.7 &  20.3 &  10.7 &  12.1 &   6.0 &   3.2 \\
D6   &  0.6 &   5.5 &  0.5 &  15.4 &  14.3 &  13.8 &  10.4 &  21.3 &  14.9 &  17.2 &   8.2 &   3.9 \\
D7   &  0.4 &   3.6 &  0.4 &  11.9 &  11.2 &  11.4 &   8.0 &  17.7 &   9.7 &  10.4 &   5.3 &   2.8 \\
D8   &  1.0 &  10.0 &  0.8 &  28.8 &  25.0 &  24.2 &  19.5 &  38.9 &  28.5 &  27.3 &  14.9 &   6.7 \\
D9   &  2.4 &  27.7 &  1.9 &  73.4 &  66.0 &  65.5 &  49.9 &  99.9 &  58.0 &  61.5 &  31.4 &  16.0 \\
D10  &  1.1 &  10.6 &  1.2 &  51.1 &  49.1 &  47.9 &  31.6 &  78.9 &  28.9 &  30.2 &  16.5 &  10.3 \\
\bottomrule
\end{tabular}
\caption{Vectorization time in seconds per model and dataset.}
\label{tb:vectTime}
\end{table}

\noindent\textbf{Initialization.} 
The initialization time of each model is shown in the first line of Table~\ref{tb:vectTime}. It refers to the time taken to load the necessary data structures in main memory (e.g., a dictionary for the static models and a learned neural network for the dynamic ones), and is independent of the dataset used.
The static models are inefficient due to the hash table they need to load into main memory to map tokens (or character n-grams) to embedding vectors.
Regarding the dynamic models, we observe that the BERT-based ones are much faster than the SentenceBERT ones, as their average run-time is 4.7$\pm$0.8 and 8.9$\pm$0.6 seconds, respectively. This is due to the larger and more complex neural models that are used by the latter.

\noindent\textbf{Transformation.} 
The rest of Table~\ref{tb:vectTime} shows the total time required by each model to convert the entities of each dataset into dense embeddings vectors (after the initialization). Word2Vec and Glove are the fastest models by far, exhibiting the lowest processing run-times in practically all cases. Word2Vec is an order of magnitude faster than the third most efficient model per dataset, which interchangeably corresponds to FastText and S-MiniLM. Except for $D_1$, GloVe outperforms these two models by at least 6 times. In absolute terms, both Word2Vec and GloVe process the eight smallest datasets, $D_1$-$D_8$, in much less than 1 second, while requiring less than 2.5 seconds for the two larger ones.

Among the BERT-based models, DistilBERT is significantly faster than BERT, as expected, with its processing time being lower by 33\%, on average. Note, though, that it is slower than FastText by $>$50\%, on average, except for $D_3$. The next most efficient models of this category are ALBERT and RoBERTa, which outperform BERT by 11\% and 13\%, on average, respectively.
XLNet is the most time consuming BERT-based model, being slower than BERT by $\sim$30\%, on average across all datasets but $D_3$. This can be explained by the fact that $D_3$ has larger sentences, as shown in Table \ref{tb:datasets}(a).

Among the SentenceBERT models, the lowest time is achieved by S-MiniLM, which, as mentioned above, is the third fastest model together with FastText.
The second best model in this category is S-DistilRoBERTa, which is slower by 30\%, on average. Both models are faster than BERT
by 63\% and 53\%, respectively, on average.
In contrast, the quite complex learned model of S-GTR-T5 yields the highest processing time among all models in all datasets but the smallest one. S-MPNet lies between these two extremes.

\noindent\textbf{Summary.} 
The static models excel in processing time, but suffer from very high initialization time. More notably, FastText is the slowest model in all datasets, but $D_9$, where S-GTR-T5 exhibits a much higher processing time. This means that FastText's high initialization cost does not pay off in datasets with few thousand entities like those in Table \ref{tb:datasets}(a). On average, FastText requires 2 minutes per dataset. All other models vectorize all datasets in less than 1 minute, with very few exceptions: BERT over $D_9$, XLNet over $D_9$-$D_{10}$ and S-GTR-T5 over $D_8$-$D_{10}$.

\subsection{Blocking}

\begin{figure}[!t]
\centering
\subfloat[D1]{\includegraphics[trim=0.12cm 0.12cm 0.12cm 0.12cm, clip, width=0.25\textwidth, height=30mm]{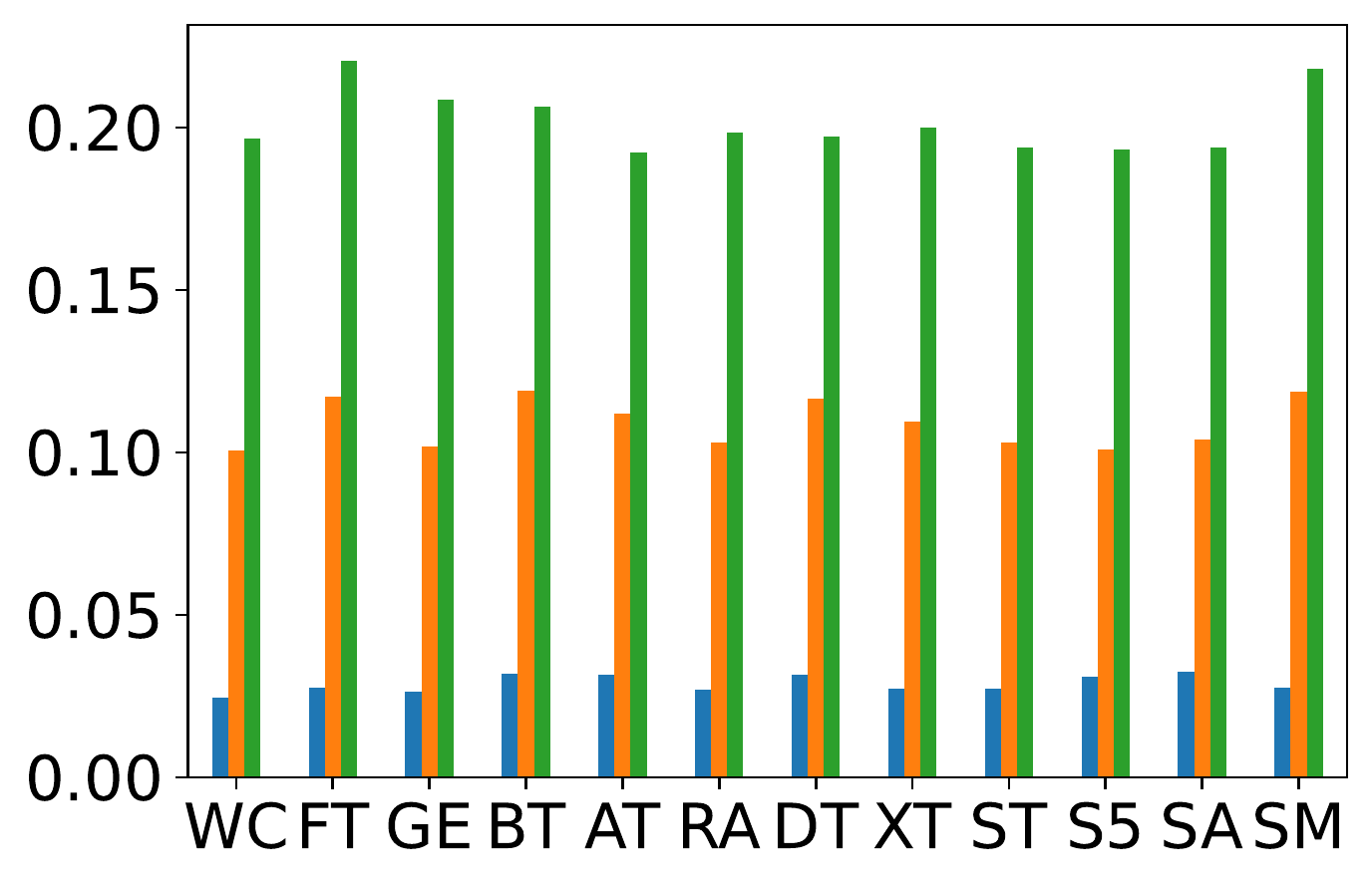}}
\subfloat[D2]{\includegraphics[trim=0.12cm 0.12cm 0.12cm 0.12cm, clip, width=0.25\textwidth, height=30mm]{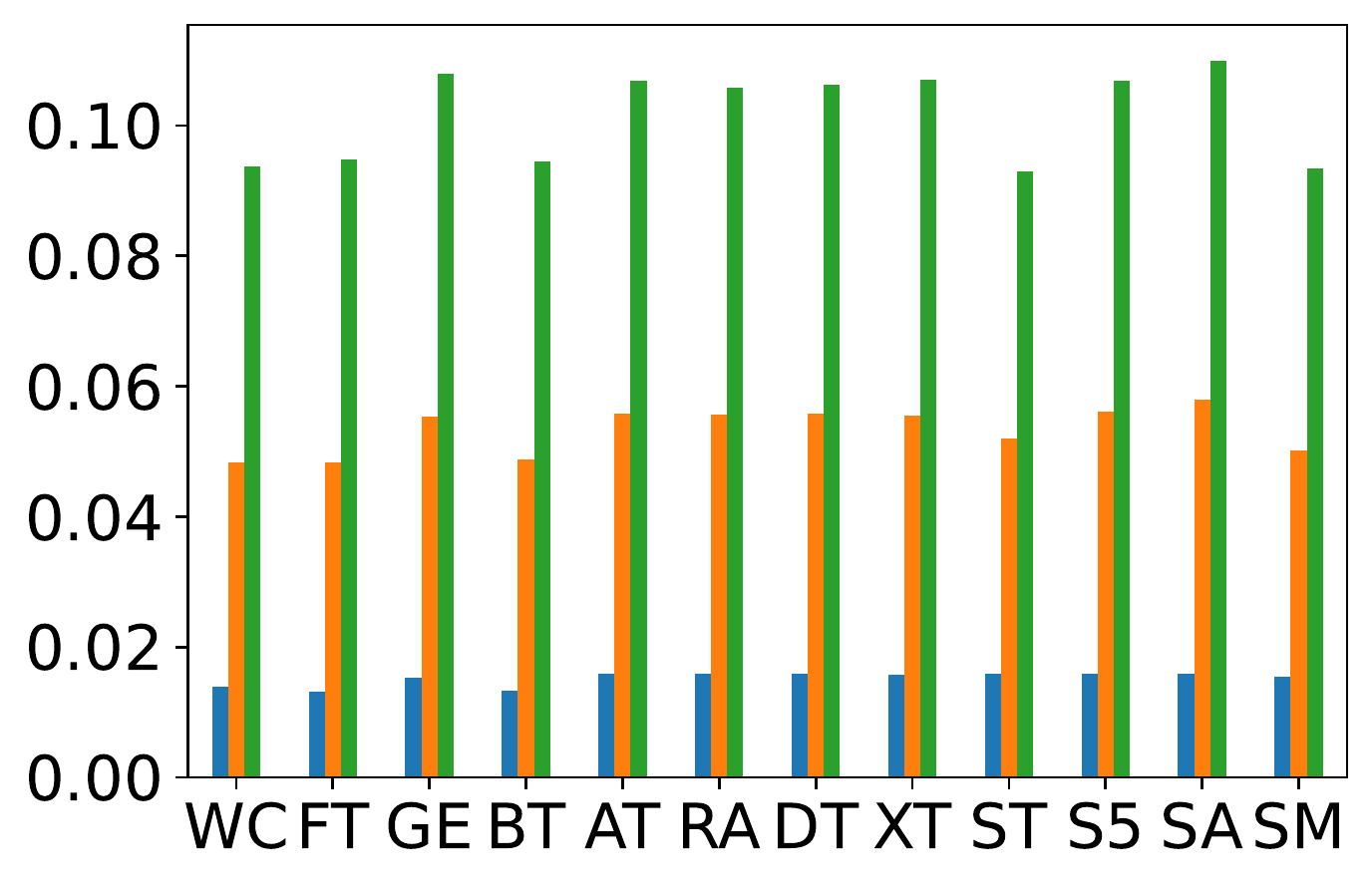}}
\newline
\subfloat[D3]{\includegraphics[trim=0.12cm 0.12cm 0.12cm 0.12cm, clip, width=0.25\textwidth, height=30mm]{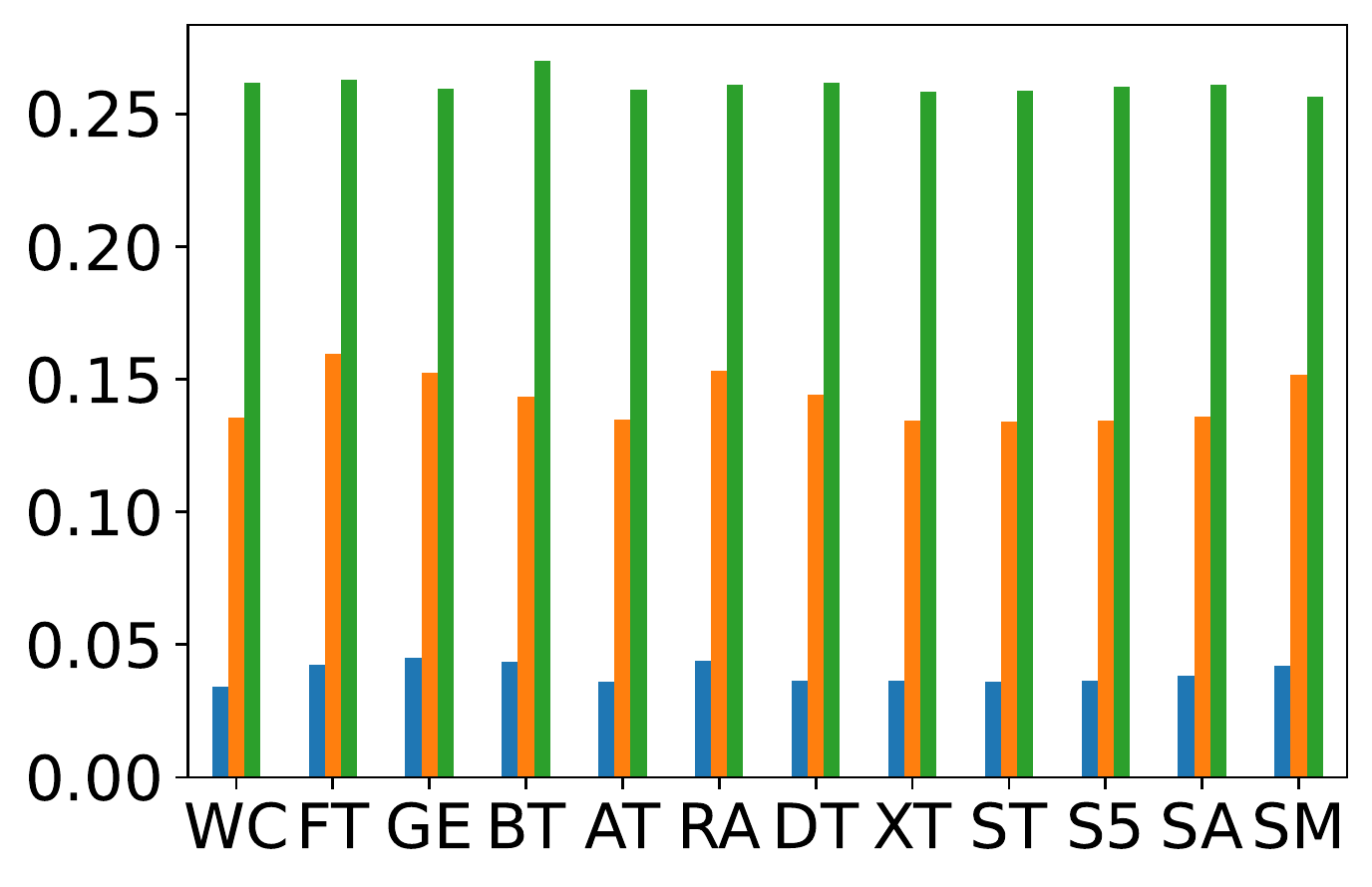}}
\subfloat[D4]{\includegraphics[trim=0.12cm 0.12cm 0.12cm 0.12cm, clip, width=0.25\textwidth, height=30mm]{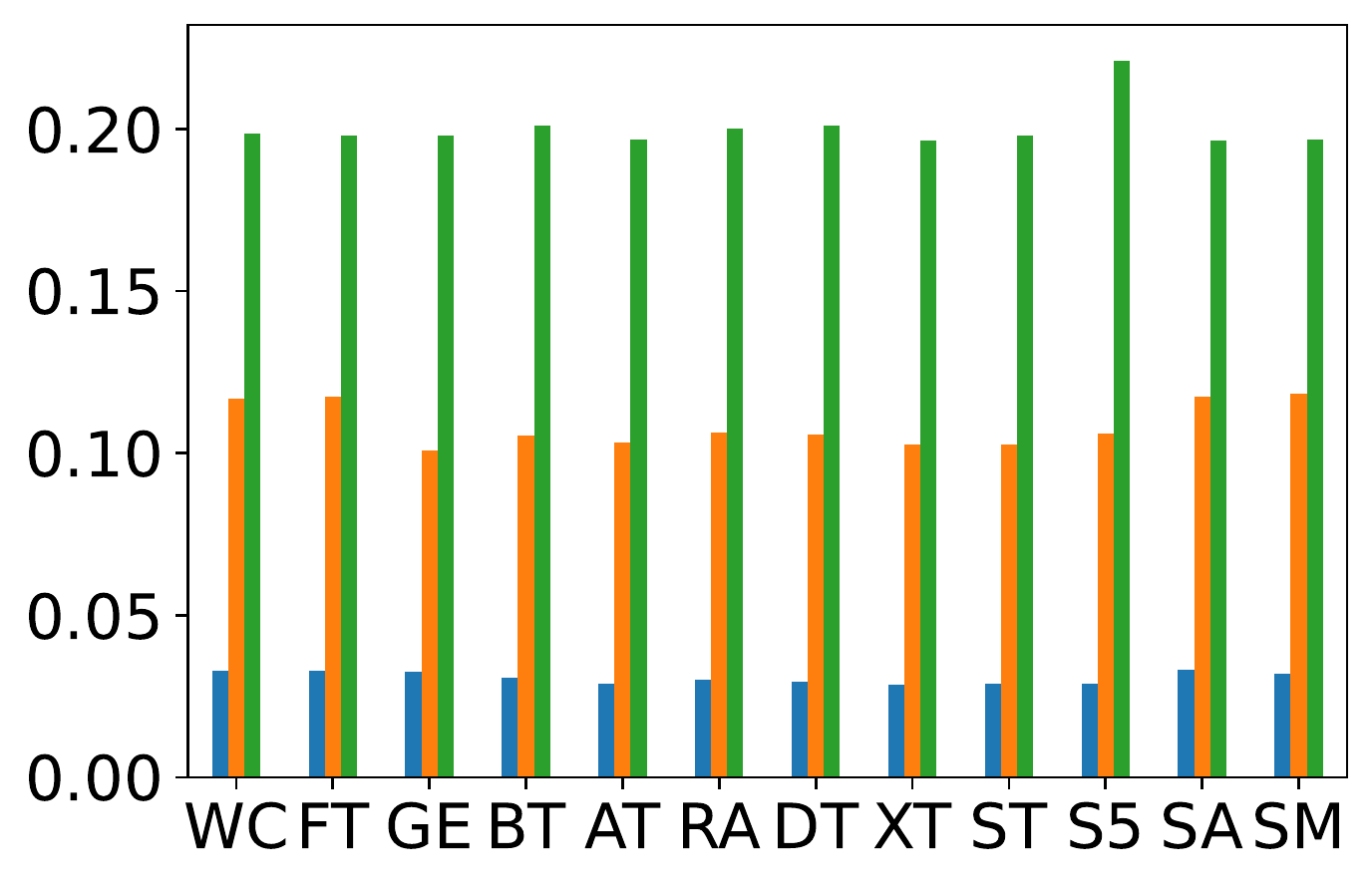}}
\newline
\subfloat[D5]{\includegraphics[trim=0.12cm 0.12cm 0.12cm 0.12cm, clip, width=0.25\textwidth, height=30mm]{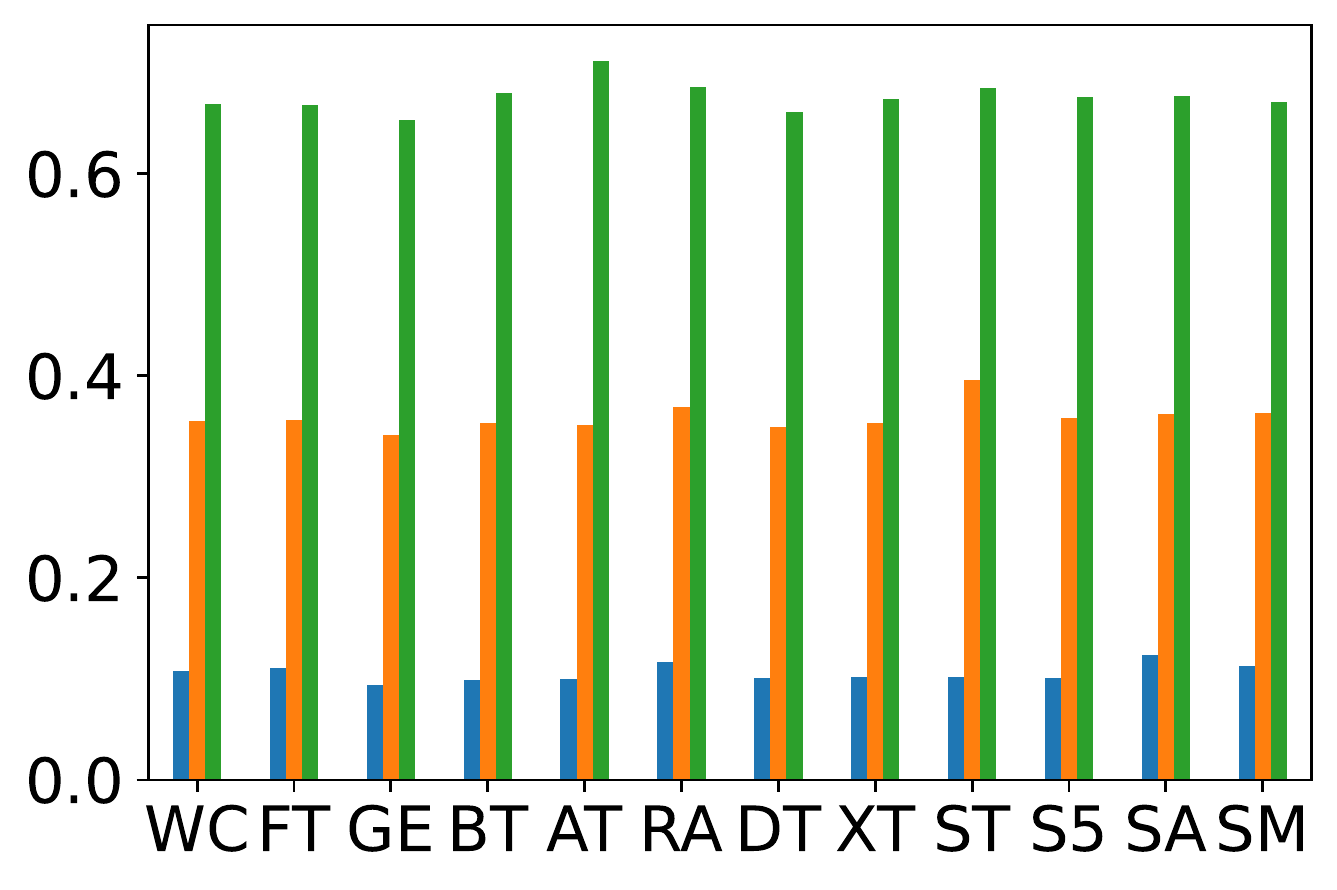}}
\subfloat[D6]{\includegraphics[trim=0.12cm 0.12cm 0.12cm 0.12cm, clip, width=0.25\textwidth, height=30mm]{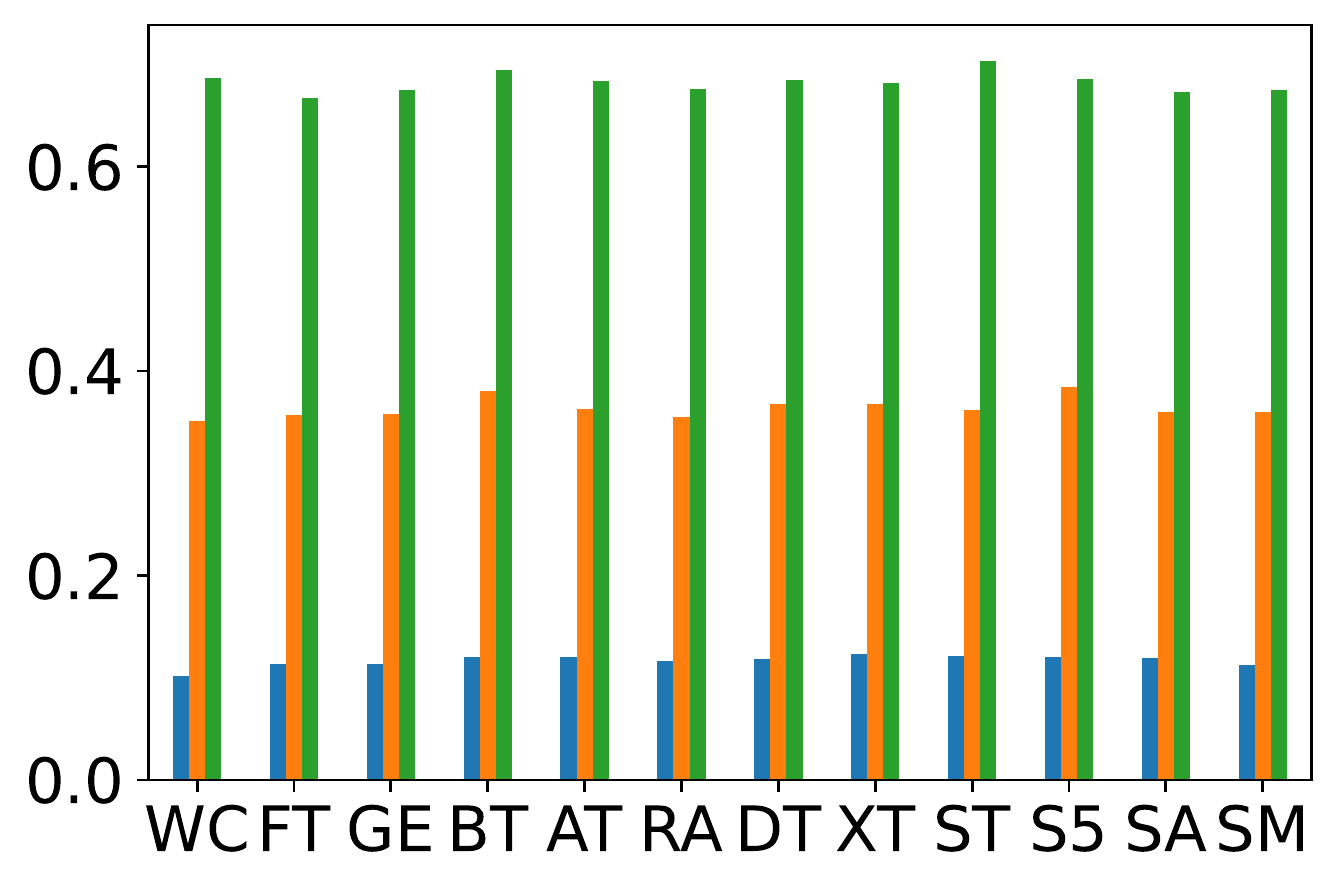}}
\newline
\subfloat[D7]{\includegraphics[trim=0.12cm 0.12cm 0.12cm 0.12cm, clip, width=0.25\textwidth, height=30mm]{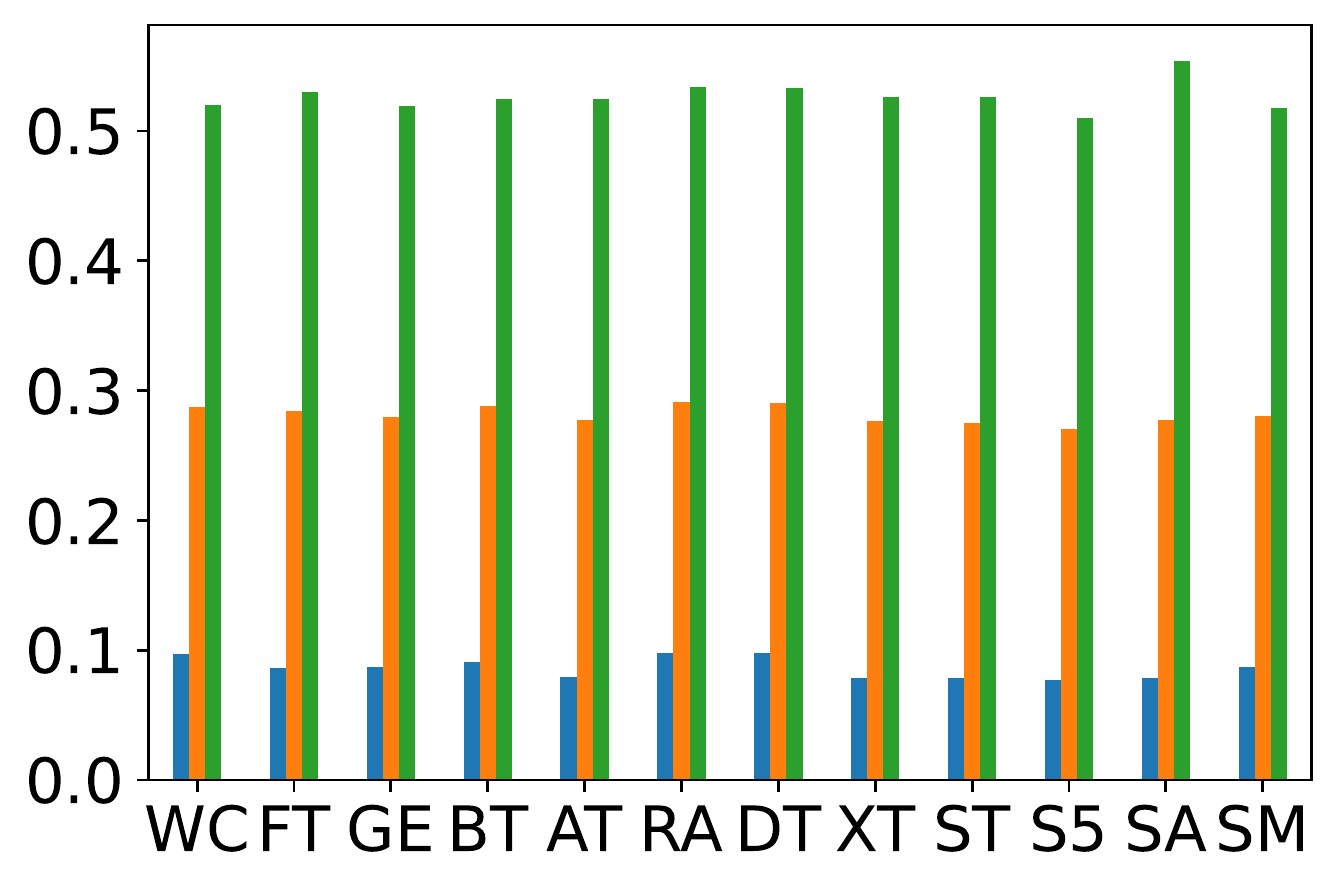}}
\subfloat[D8]{\includegraphics[trim=0.12cm 0.12cm 0.12cm 0.12cm, clip, width=0.25\textwidth, height=30mm]{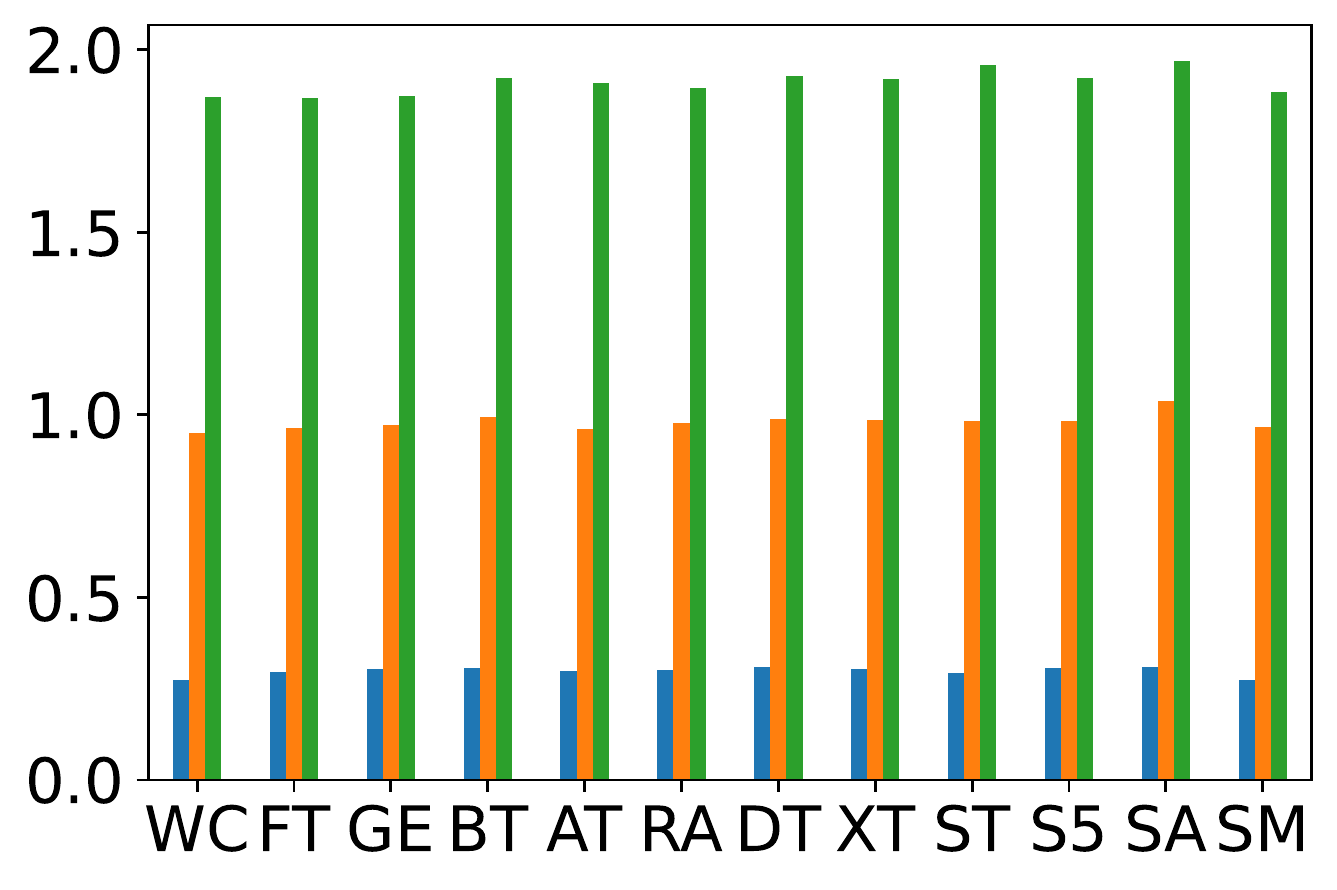}}
\newline
\subfloat[D9]{\includegraphics[trim=0.12cm 0.12cm 0.12cm 0.12cm, clip, width=0.25\textwidth, height=30mm]{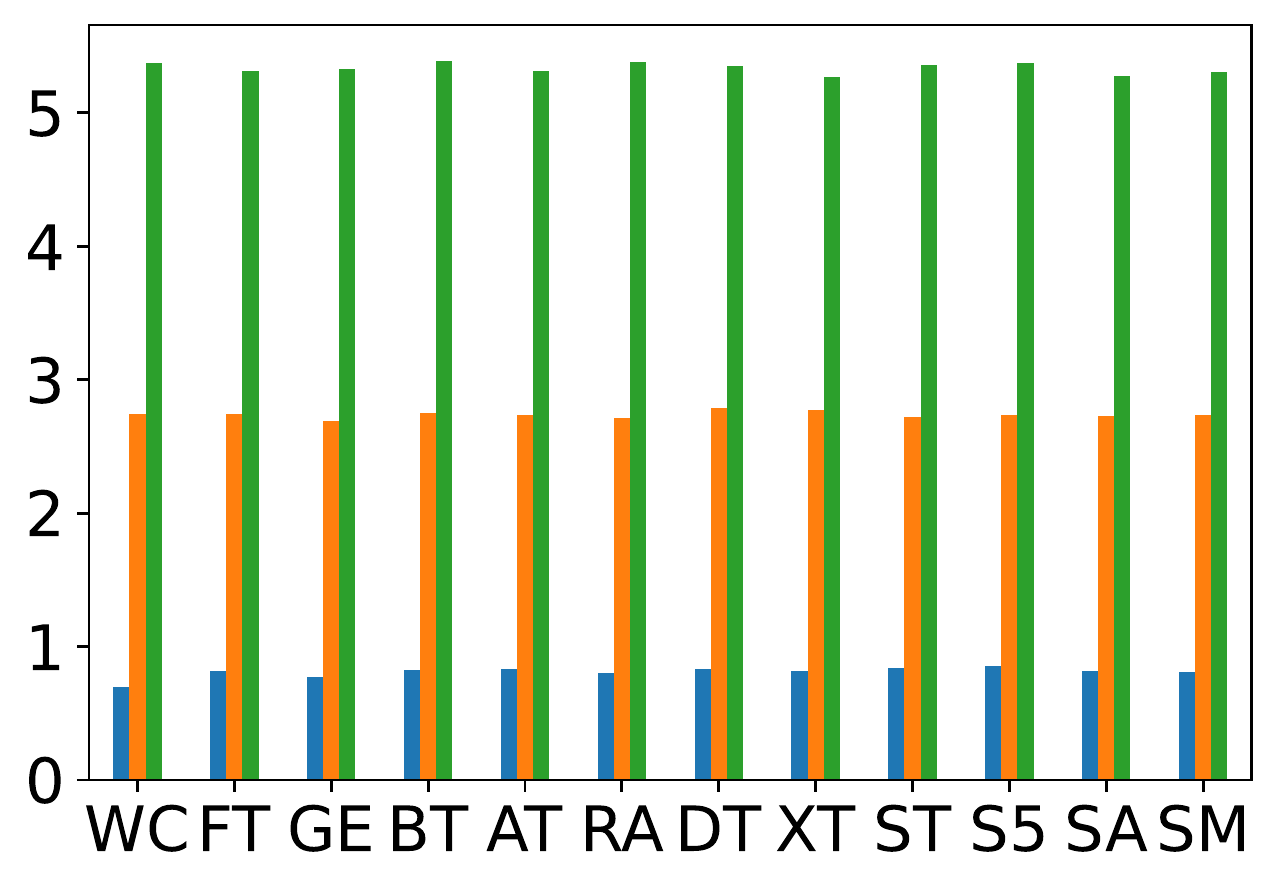}}
\subfloat[D10]{\includegraphics[trim=0.12cm 0.12cm 0.12cm 0.12cm, clip, width=0.25\textwidth, height=30mm]{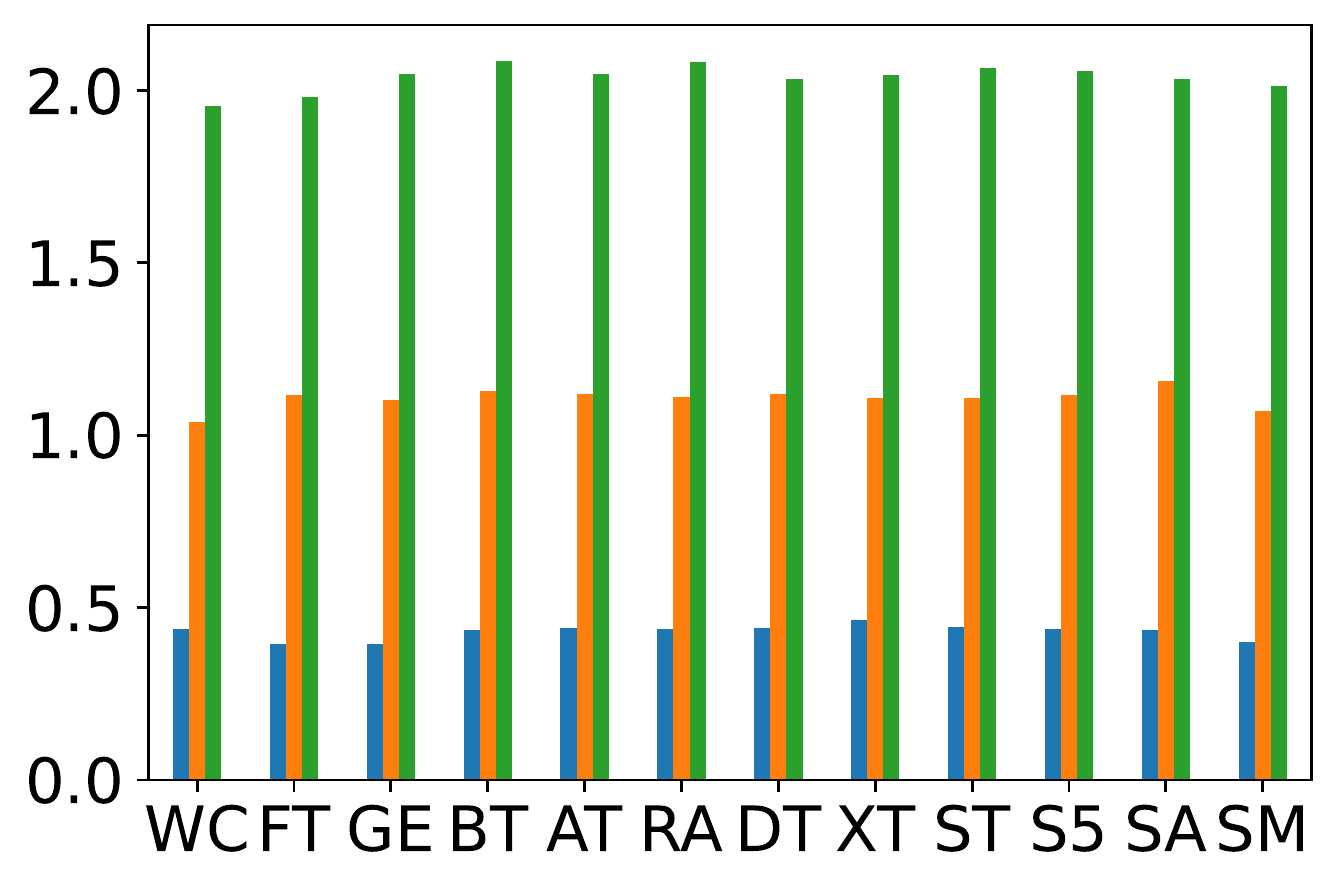}}
\caption{Blocking Time (sec) per method. There are three values for $k \in \{${\color{myblue}1}, {\color{myorange}5}, {\color{mygreen}10} $\}$.}
\label{fig:blk_time_real}
\end{figure}

% {\color{red}Do the following refer to Figure~\ref{fig:blk_time_real}?}

The execution time of blocking is very low for all models, not exceeding 0.5 seconds in most cases. The only exceptions are the two largest datasets, $D_9$ and $D_{10}$, which still require less than 2 seconds in all cases. The differences between the various models are rather insignificant.
In most cases, the lower end in these ranges corresponds to the language models with the lowest dimensionality, namely the static ones (300) as well as S-MiniLM (385), and the higher end to the rest of the models, which involve 768-dimensional vectors (see Table \ref{tb:modelCharacteristics}).
In more detail, fastText and S-GTR-T5 are consistently the fastest and slowest models, respectively.
% In figure \ref{fig:blk_time_real}, we can see the trade-off between effectiveness and efficiency. Each points holds the average time for each method across all datasets and the corresponding average recall. Since we want in each axis to achieve maximum score, we have normalized the y-axis of time by $min_{time} / time$. 
% Thus, it is clear that paying more time yields higher recall for S-GTR-T5, but an alternative and more economic choice is S-MiniLM.
However, this overhead time is negligible in comparison to the vectorization time in Table \ref{tb:vectTime}. 
% {\color{blue}Please refer to the extended version of our work$^3$ for a detailed report of run-times per model and dataset.}
% For this reason, the relative time efficiency remains the same as the one discussed in Section \ref{sec:vectorizonExp}.

% \begin{table}[t]
% \setlength{\tabcolsep}{2.5pt}
% \small
% \begin{tabular}{lrrrrrr}
% \toprule
% {} & \multicolumn{3}{l}{DeepBlocker} & \multicolumn{3}{l}{S-GTR-T5} \\
% {} &          1  &     5  &     10 &    1  &    5  &    10 \\
% \midrule
% D1  &         8.0 &    8.0 &   17.4 &  11.0 &  11.0 &  11.1 \\
% D2  &         8.9 &    8.9 &   17.3 &  13.3 &  13.3 &  13.3 \\
% D3  &        18.9 &   18.9 &   35.6 &  22.3 &  22.4 &  22.5 \\
% D4  &        12.9 &   12.8 &   27.8 &  15.3 &  15.3 &  15.5 \\
% D5  &        24.3 &   24.5 &   62.5 &  22.0 &  22.3 &  22.6 \\
% D6  &        28.1 &   27.6 &   68.3 &  27.2 &  27.4 &  27.7 \\
% D7  &        21.8 &   21.7 &   49.4 &  20.3 &  20.5 &  20.7 \\
% D8  &        46.1 &   45.8 &   46.0 &  37.4 &  38.1 &  39.1 \\
% D9  &       110.5 &  111.1 &  270.4 &  72.2 &  74.1 &  76.7 \\
% D10 &       154.0 &  149.9 &  371.4 &  40.5 &  41.2 &  42.1 \\
% \bottomrule
% \end{tabular}
% \caption{Blocking performance of DeepBlocker vs S-GTR-T5.}
% \label{tb:unsupTime}
% \end{table}

\noindent\textbf{Comparison to SotA.} Table \ref{tb:sota_comparison}(a) reports the run-times corresponding to the rightmost column in Figure \ref{fig:blk_real}. We observe that DeepBlocker is consistently faster than S-GTR-T5 for $k$=1 and $k$=5 in all datasets but $D_{10}$. The reason for the slow operation of S-GTR-T5 is its high vectorization cost, which accounts for $\sim$99\% of the overall blocking time. This explains why its run-time is practically stable per dataset across all values of $k$. This is expected, though, given that S-GTR-T5 leverages 768-dimensional embeddings vectors, compared to 300-dimensional FastText vectors of DeepBlocker. Yet, for $k$=10, DeepBlocker is faster than S-GTR-T5 only in $D_2$ and $D_3$ (by 14.4\% and 26\%, respectively). The situation is reversed in $D_1$ and $D_4$-$D_8$, where S-GTR-T5 is faster by 14.1\%, on average. The reason is that DeepBlocker does not scale well as the number of candidates per query entity increases, due to the high complexity of the deep neural network that lies at its core. Most importantly, DeepBlocker scale poorly as the size of the input data increases: in $D_{10}$, S-GTR-T5 is faster by 1.5 times for $k$$\in$$\{1, 5\}$ and 3.7 times~for~$k$=10. 
%This is further verified in Section~\ref{sec:scalability}.

\subsubsection{Scalability}

\begin{figure}[!t]
\centering
\subfloat{\includegraphics[width=1.0\linewidth]{blocking_synthetic_legend.pdf}}
\setcounter{subfigure}{0}
\subfloat[Blocking Time (sec-log)]{\includegraphics[trim=0.12cm 0.12cm 0.12cm 0.12cm, clip, width=0.24\textwidth]{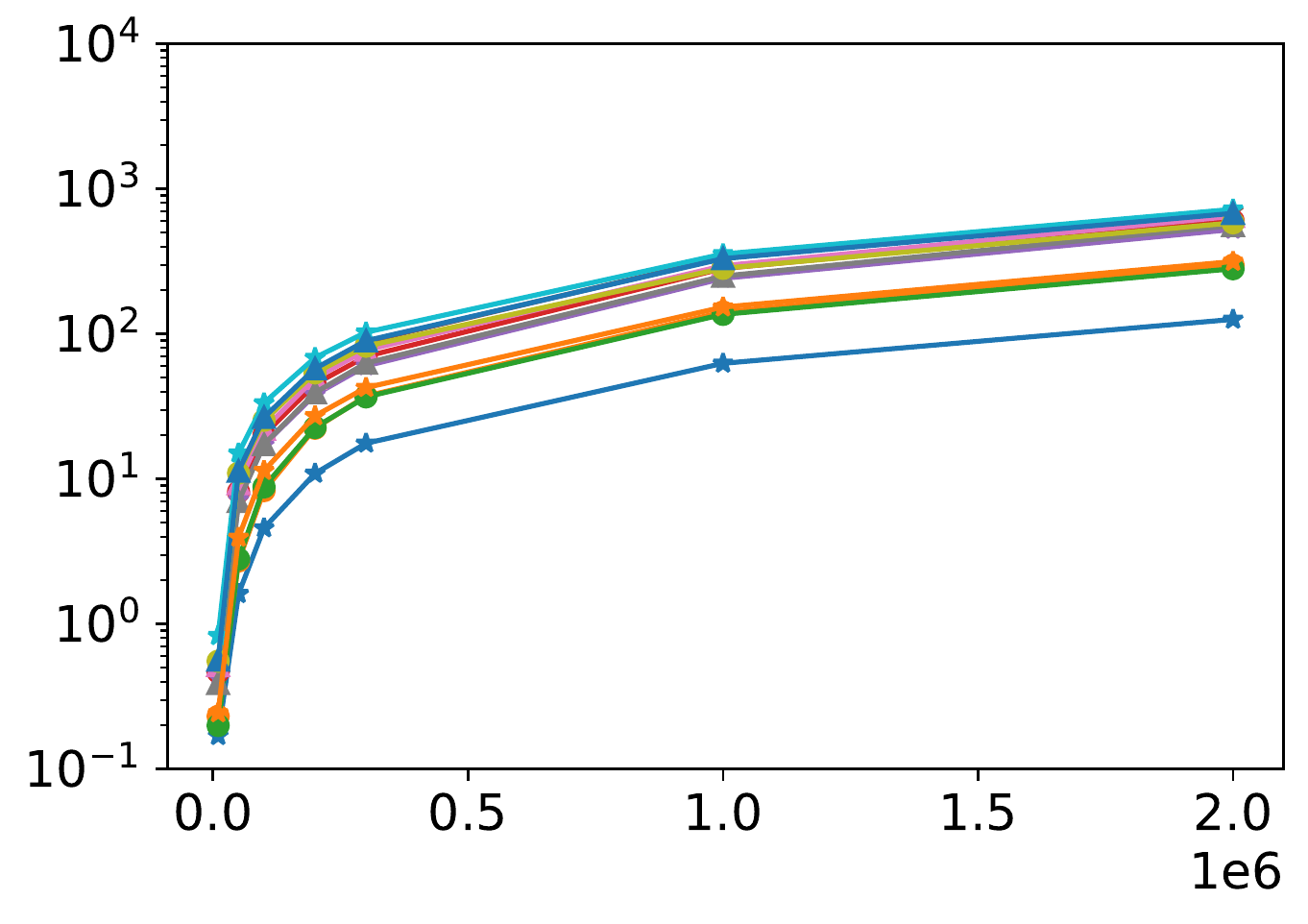}\label{subfig:blk_synth_time_app}}
\subfloat[Vectorization  Time (sec-log)]{\includegraphics[trim=0.12cm 0.12cm 0.12cm 0.12cm, clip, width=0.24\textwidth]{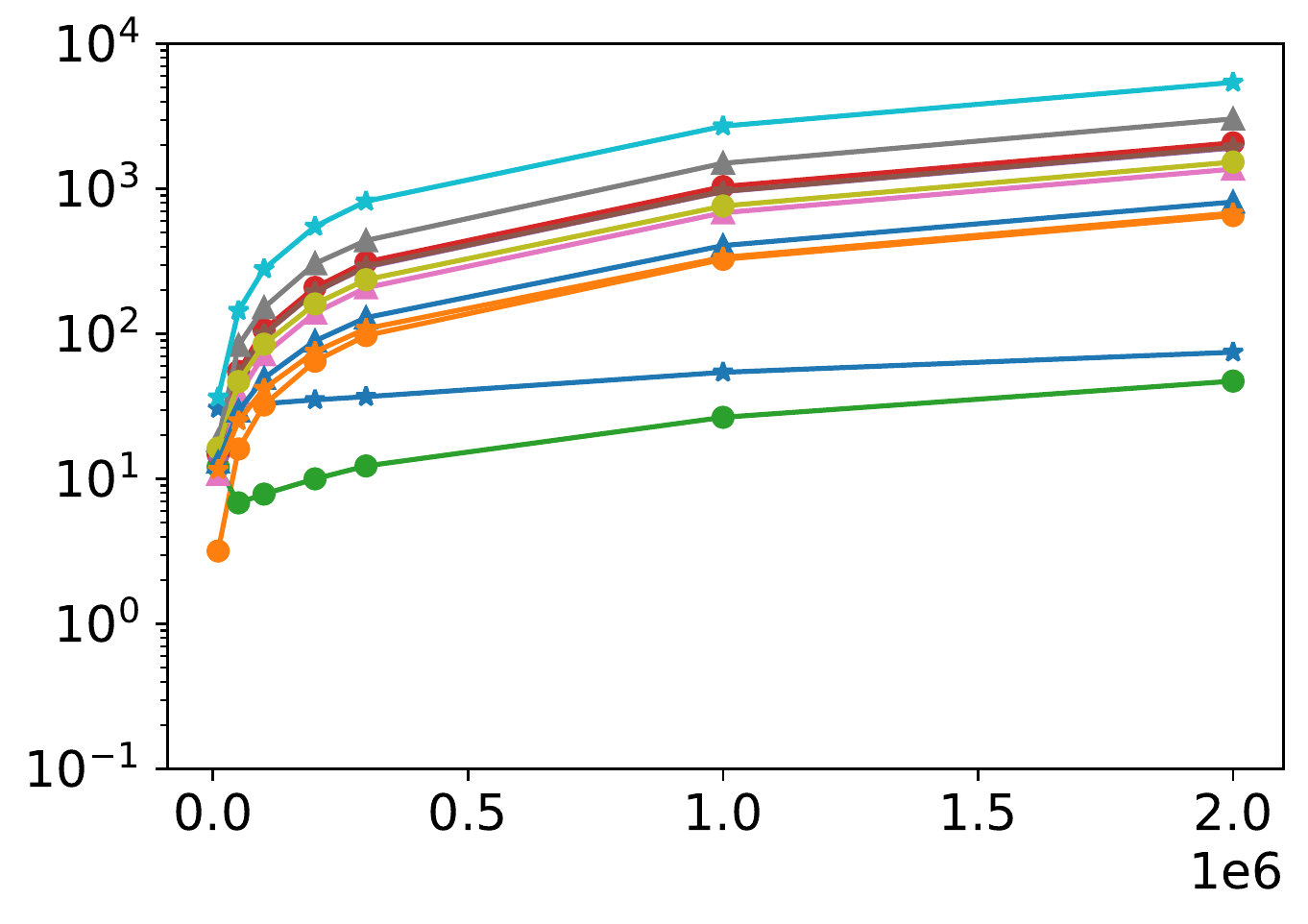}\label{subfig:vec_synth_time}}

\caption{Scalability over the synthetic datasets in Table \ref{tb:datasets}(b). The horizontal axis indicates the number of input entities.}
\label{fig:efficiency_scalability}
\end{figure}

Figure \ref{fig:efficiency_scalability}(a) shows that the blocking time of all models scales superlinearly, but subquadratically: as the number of input entities increases by 200 times from $D_{10K}$ to $D_{2M}$, the run-times increase by up to 1,435 times. With the exception of the two most efficient models, Word2Vec (742 times) and S-GTR-T5 (873 times), all other models fluctuate between 1,053 (SMPNet) and 1,435 (XLNet) times.
This is mainly attributed to the fact that FAISS(HNSW) trades high indexing time for significantly lower querying time, i.e., the former dominates the latter. During indexing, it requires complex graph-based operations that involve long paths. The larger the input size, the longer these paths get, increasing the cost of their traversal and processing superlinearly.
% Hence, the indexing time dominates the total time.

Figure \ref{fig:efficiency_scalability}(b) reports the evolution of vectorization time, which increases sublinearly with the size of the input data for all models. The increase fluctuates between 127 (DistilBERT) and 165 (XLNet) times for all BERT-based models, thus remaining far below the raise in the input size (i.e., 200 from $D_{10K}$ to $D_{2M}$). A similar behavior is exhibited by S-GTR-T5, but the rest of the SentenceBERT models achieve even better scalability, as their increase is reduced to 59 (S-MiniLM), 94 (S-MPNet) and 62 (XLNet). This should be attributed to the initialization of each model, which is independent of the input size and accounts for a large portion of the overall vectorization time of each model, as shown in Table \ref{tb:vectTime}. The larger the relative cost of initialization is, the lower is the increase in vectorization time as size of the input increase. As a result, 
% All these models are outperformed by the two most efficient static ones, 
Word2Vec and GloVe, which raise their run-times by just 2.5 and 4 times, respectively. The former actually remains practically stable up to $D_{300K}$, because its vectorization time is dominated by its high initialization time across the five smallest datasets. On the other extreme lies FastText, which is the only model that scales linearly with the size of the input data (by 205 times), due to its character-level functionality.
% (i.e., its vectorization time increases by 205 from $D_{10K}$ to $D_{2M}$).

\begin{table}[t]\centering
\small
\setlength{\tabcolsep}{3pt}
    \begin{tabular}{|l|rrr|rrr||rr|rr|}
    \cline{2-11}
    \multicolumn{1}{c|}{} &
    \multicolumn{3}{c|}{DeepBlocker} &
    \multicolumn{3}{c||}{S-GTR-T5} & 
    \multicolumn{2}{c|}{ZeroER} &
    \multicolumn{2}{c|}{S-GTR-T5} \\
    % \multicolumn{1}{c|}{} & $k=1$ & $k=5$ & $k=10$ & $k=1$ & $k=5$ & $k=10$ & $t_p$ & $t_m$ & $t_p$ & $t_m$ \\
    \multicolumn{1}{c|}{} & $k=1$ & $k=5$ & $k=10$ & $k=1$ & $k=5$ & $k=10$ & $t_p$ & $t_m$ & $t_p$ & $t_m$ \\    
    \midrule
    \midrule

D1  & 8 &  8 & 17 &  11 &  11 &  11 & 2 &  1 & 13 &  1\\
D2  & 9 &  9 & 17 &  13 &  13 &  13 & 1,728 & 71 & 14 &  3\\
D3  & 19 & 19 & 36 &  22 &  22 &  23 & - & - & 23 &  4\\
D4  & 13 & 13 & 28 &  15 &  15 &  16 & 9,595 & 113 & 16 &  9\\
D5  & 24 & 25 & 63 &  22 &  22 &  23 & 1,291 &  10 & 23 &  12\\
D6  & 28 & 28 & 68 &  27 &  27 &  28 & - & - & 28 &  16\\
D7  & 22 & 22 & 49 &  20 &  22 &  21 & 1,599 &  338 & 21 &  13\\
D8  & 46 & 46 & 46 &  37 &  38 &  39 & - & - & 38 &  8 \\
D9  & 111 & 111 & 270 & 72 & 74 &  77 & - & - & 72 &  10\\
D10 & 154 & 150 & 371 & 41 & 41 &  42 & - & - & 46 &  96 \\
		
		\hline
		% \multicolumn{1}{c}{}&
		% \multicolumn{6}{c}{\textbf{(a) Blocking}} &
		% \multicolumn{4}{c}{\textbf{(b) Unsupervised Matching}} \\
        \multicolumn{1}{c}{}&
		\multicolumn{6}{c}{\textbf{(a) Blocking}} &
		\multicolumn{4}{c}{\textbf{(b) Unsup. Matching}} \\
  %       \multicolumn{7}{c}{}&
		% \multicolumn{4}{c}{\textbf{Matching}} \\
  
	\end{tabular}
    \caption{Comparison of S-GTR-T5 with 
    % State-of-the-Art in 
    (a) DeepBlocker in Blocking, and (b) ZeroER in Unsupervised Matching. All columns are in seconds, but the rightmost one which is in milliseconds. 
    % Blocking has 3 columns corresponding to $k\in\{1,5,10\}$. 
    $t_p$ ($t_m$) stands for 
    % Unsupervised Matching has two columns, based on 
    preprocessing (matching)~time.
    % $t_p$ 
    % and $t_m$ for  time.
    }
    % $t_m$.} 
	\label{tb:sota_comparison}
\end{table}

In absolute terms, GloVe is consistently the fastest model in all datasets, but $D_{10K}$, with Word2Vec ranking second from $D_{200K}$ on. They vectorize $D_{2M}$ within 0.8 and 1.3 minutes, respectively. FastText is the second fastest model for the three smallest datasets, but converges to the fastest dynamic model, S-MiniLM, for the larger ones. They both need $\sim$11 minutes to process $D_{2M}$. On the other extreme lies S-GTR-T5, which consistently exhibits the slowest vectorization, with XLNet being the second worst model across all datasets. For $D_{2M}$, they take 90.3 and 50.6 minutes, respectively. 
%The rest of the models can be grouped in three pairs: the slowest one, which is faster only than XLNet, includes BERT and RoBERTa, which vectorize $D_{2M}$ within 34.6 and 31.9 minutes, respectively. The middle pair involves DistiBERT and S-MPNet, with the former being faster than the latter by 16\%, on average ($\sim$25 minutes over $D_{2M}$). The fastest pair comprises S-MiniLM and S-DistilRoBERTa, with the former outperforming the latter by 15\%, on average. These observations verify the conclusions drawn from Section \ref{sec:vectorizonExp} about the relative vectorization time of the considered models.

\subsection{Unsupervised Matching}
\label{sec:unsupMatching}
\begin{figure}[!t]
\centering
\subfloat[D1]{\includegraphics[trim=0.12cm 0.12cm 0.12cm 0.12cm, clip, width=0.25\textwidth, height=30mm]{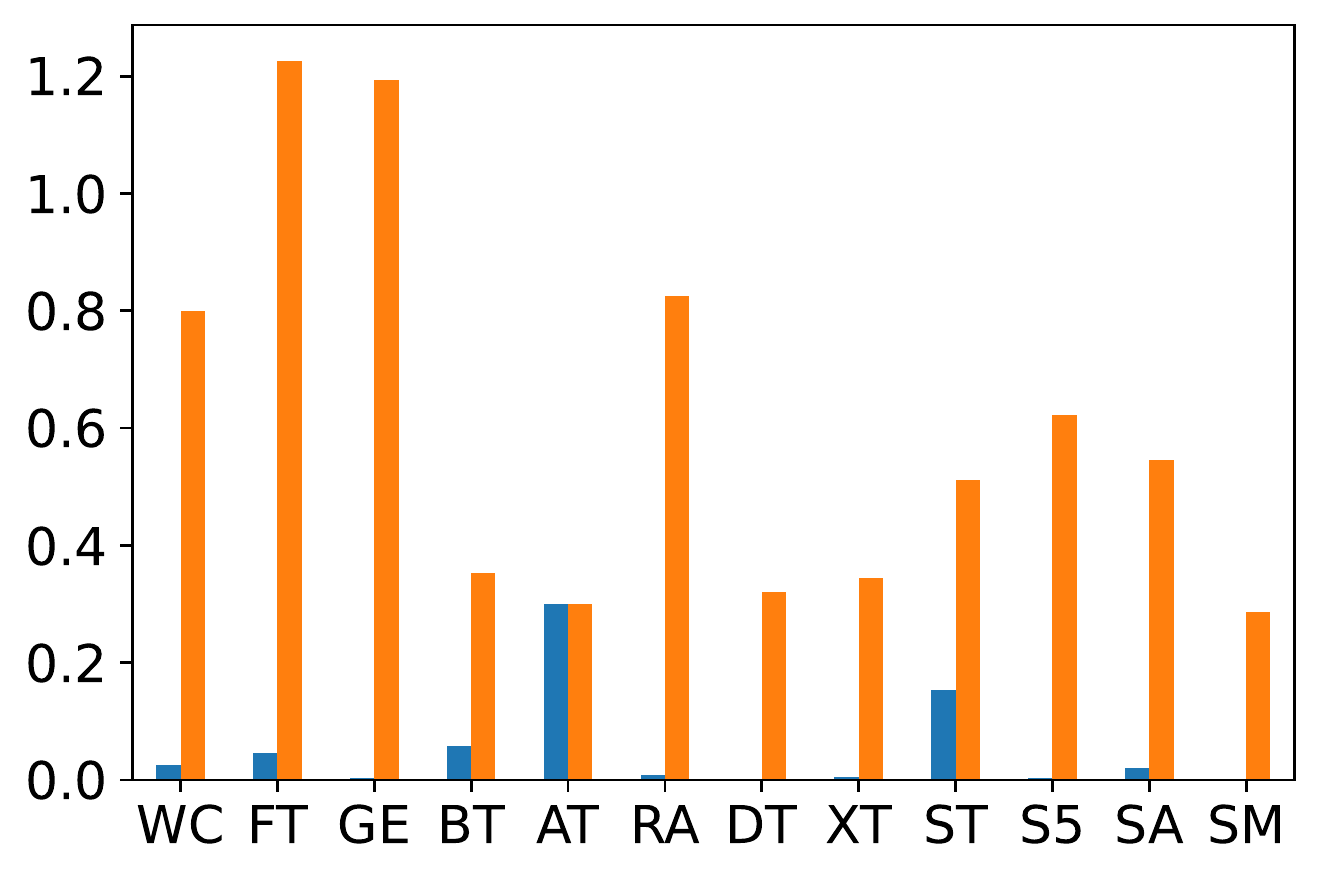}}
\subfloat[D2]{\includegraphics[trim=0.12cm 0.12cm 0.12cm 0.12cm, clip, width=0.25\textwidth, height=30mm]{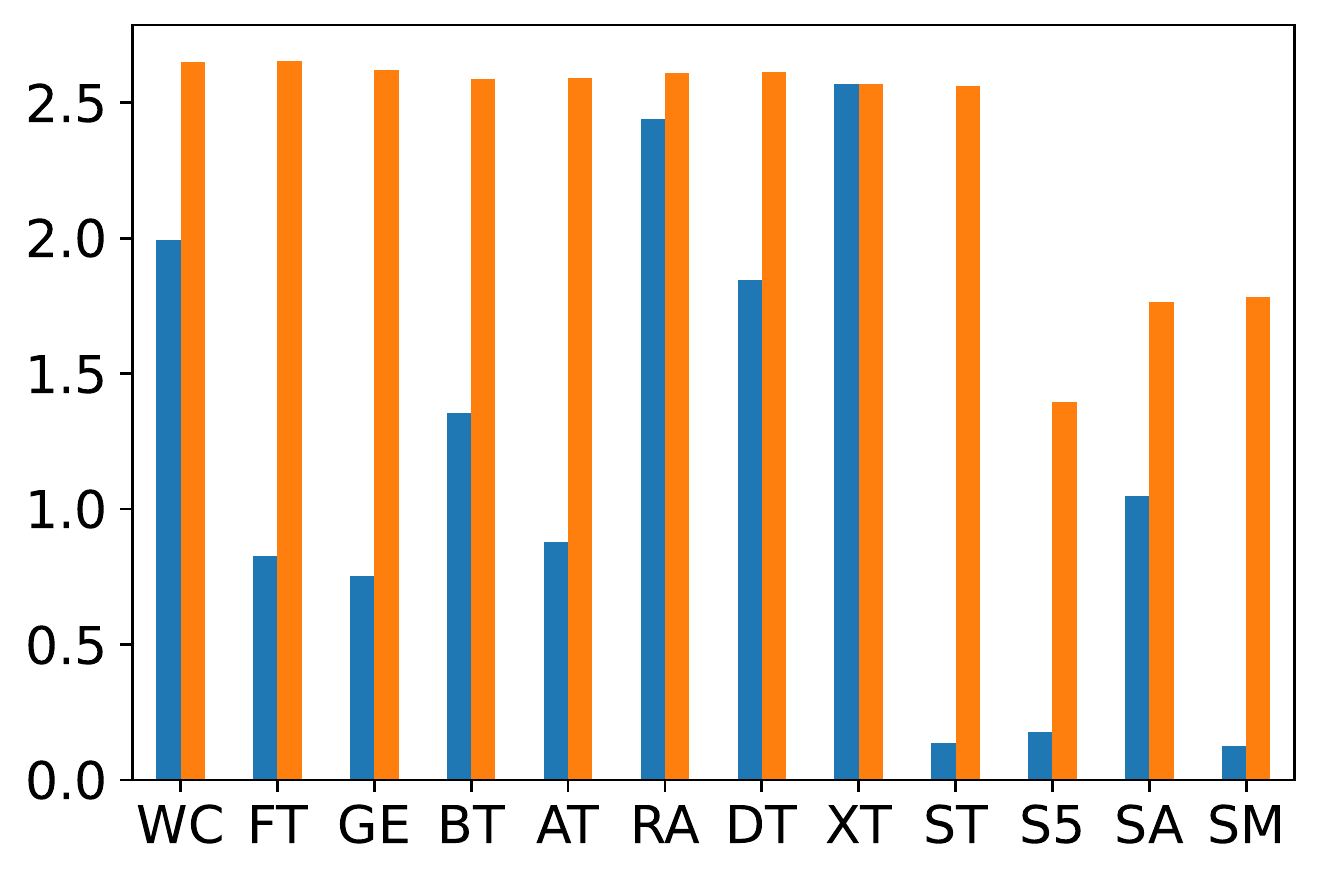}}
\newline
\subfloat[D3]{\includegraphics[trim=0.12cm 0.12cm 0.12cm 0.12cm, clip, width=0.25\textwidth, height=30mm]{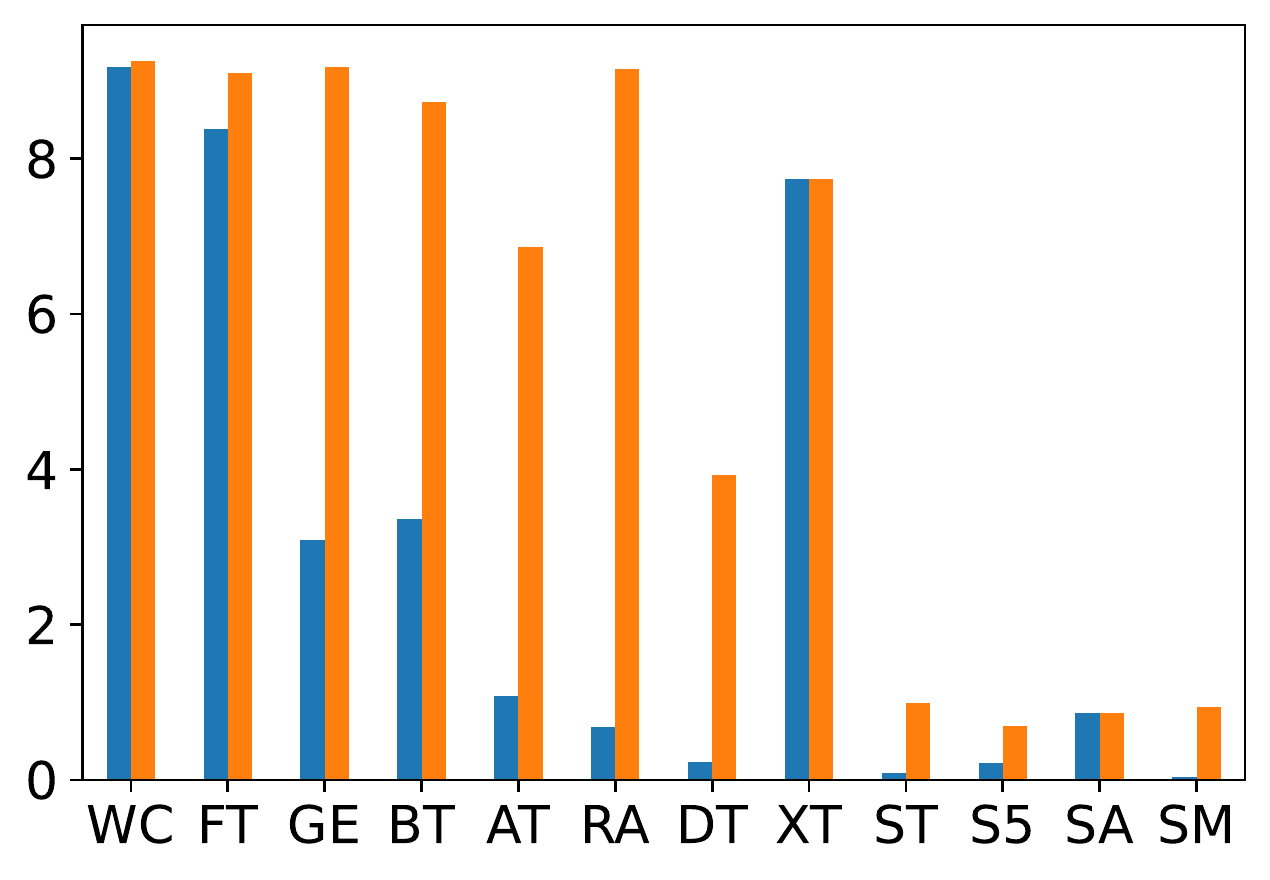}}
\subfloat[D4]{\includegraphics[trim=0.12cm 0.12cm 0.12cm 0.12cm, clip, width=0.25\textwidth, height=30mm]{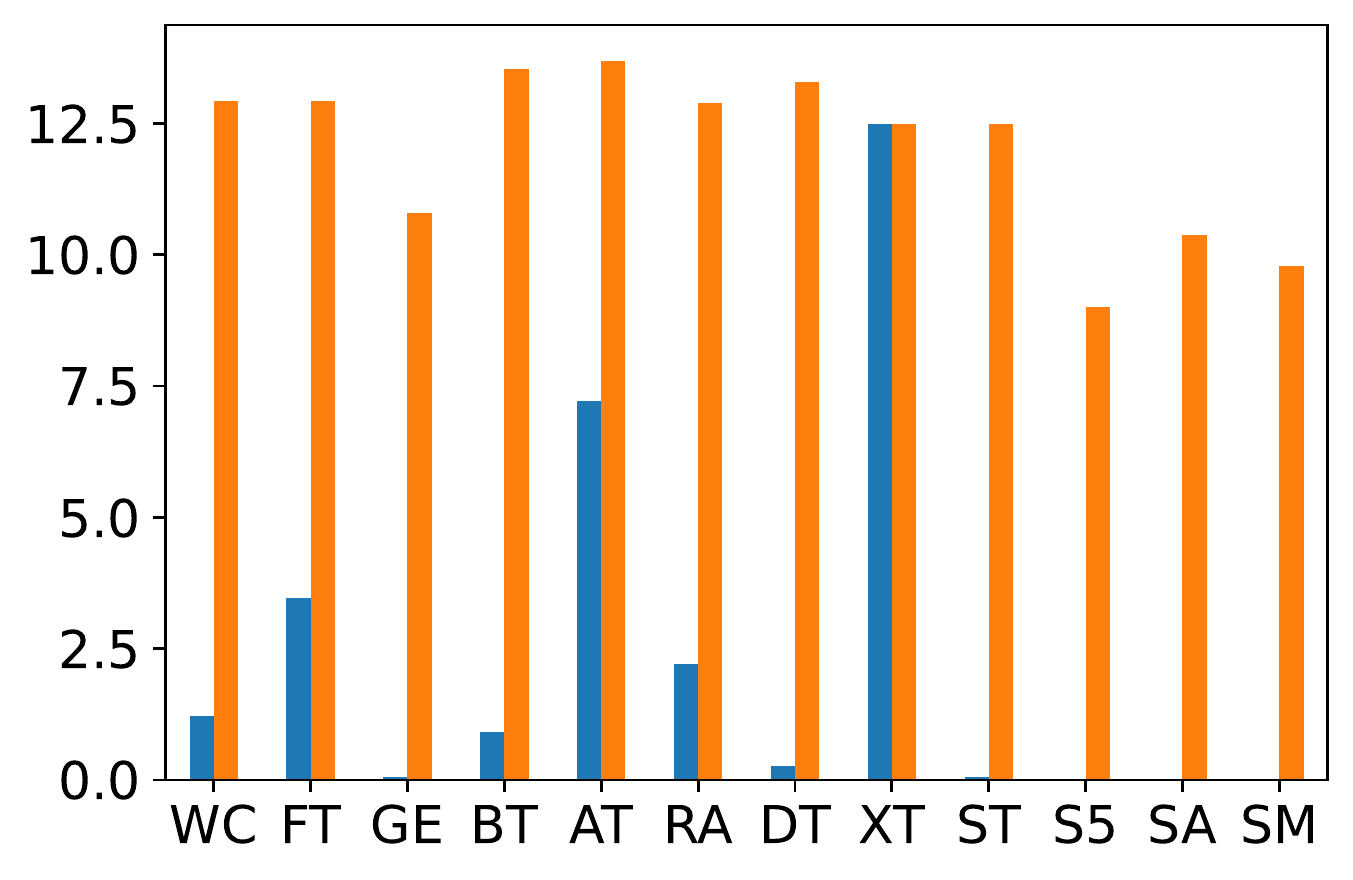}}
\newline
\subfloat[D5]{\includegraphics[trim=0.12cm 0.12cm 0.12cm 0.12cm, clip, width=0.25\textwidth, height=30mm]{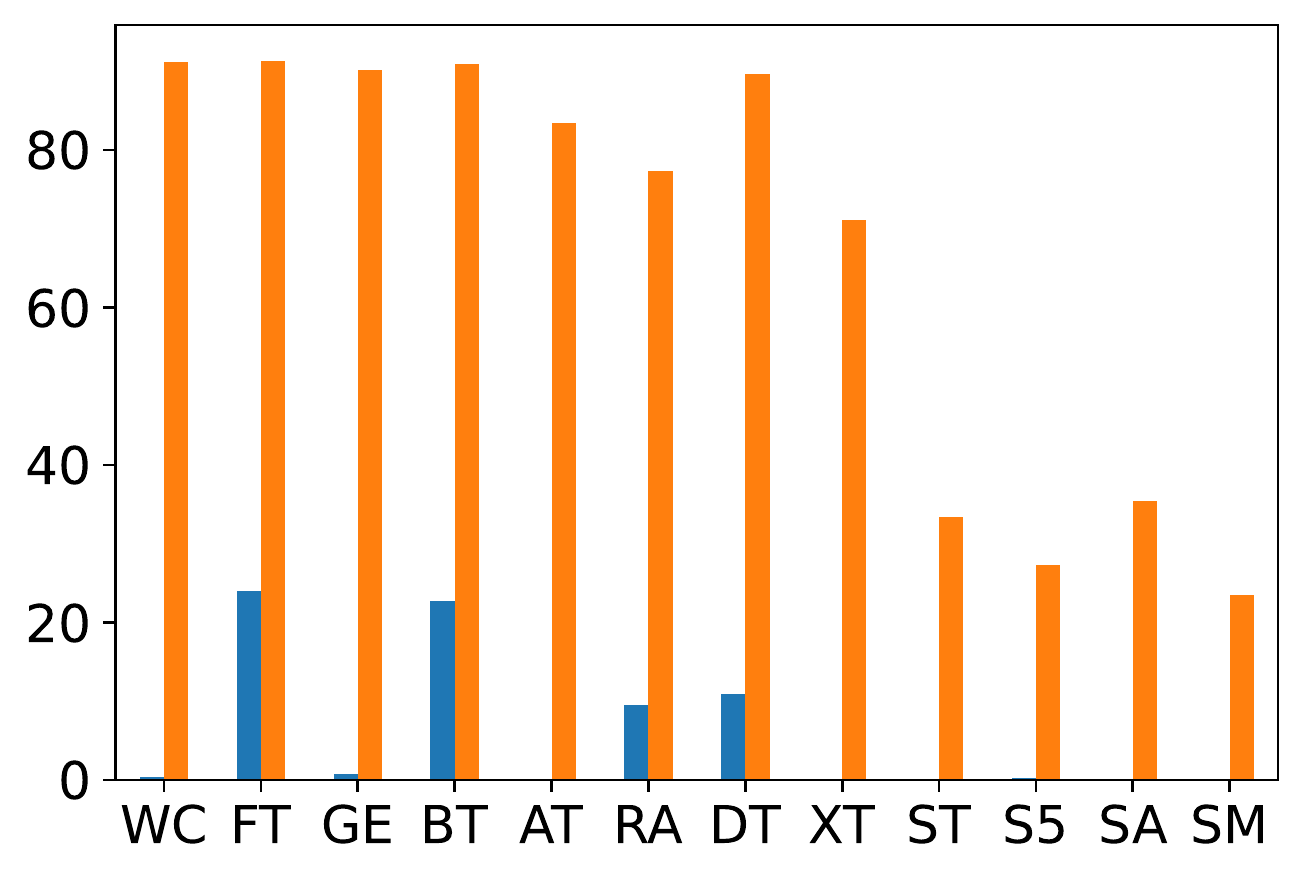}}
\subfloat[D6]{\includegraphics[trim=0.12cm 0.12cm 0.12cm 0.12cm, clip, width=0.25\textwidth, height=30mm]{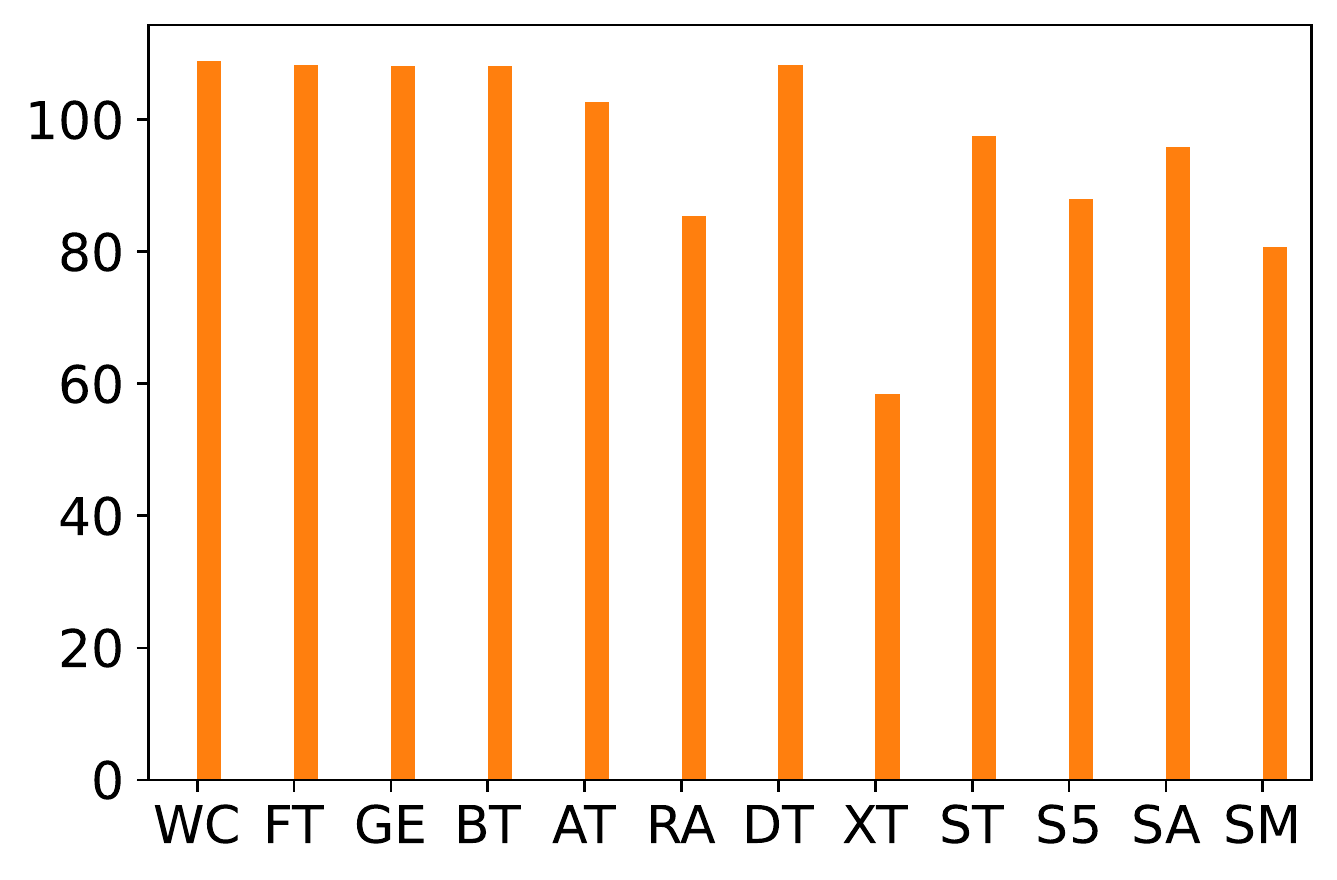}}
\newline
\subfloat[D7]{\includegraphics[trim=0.12cm 0.12cm 0.12cm 0.12cm, clip, width=0.25\textwidth, height=30mm]{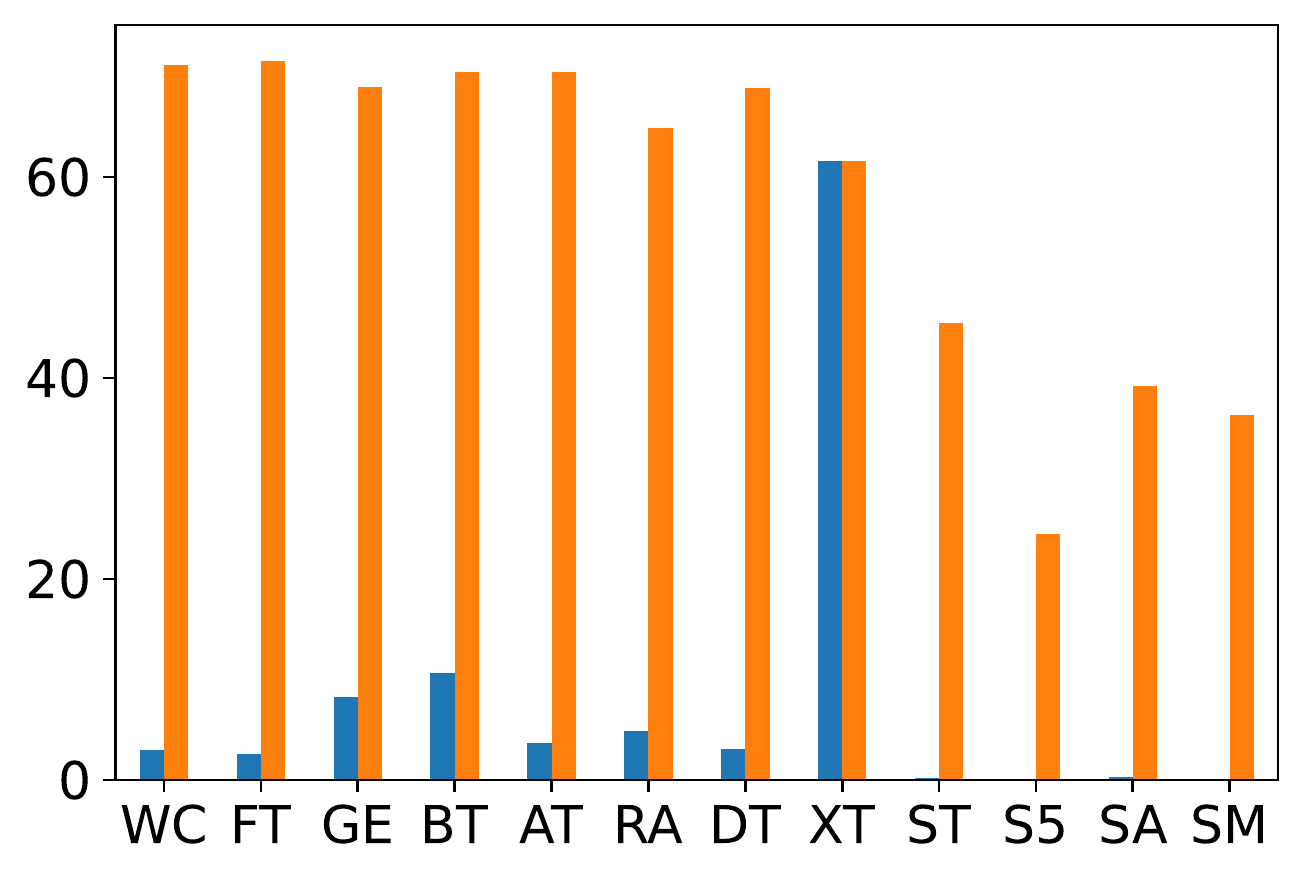}}
\subfloat[D8]{\includegraphics[trim=0.12cm 0.12cm 0.12cm 0.12cm, clip, width=0.25\textwidth, height=30mm]{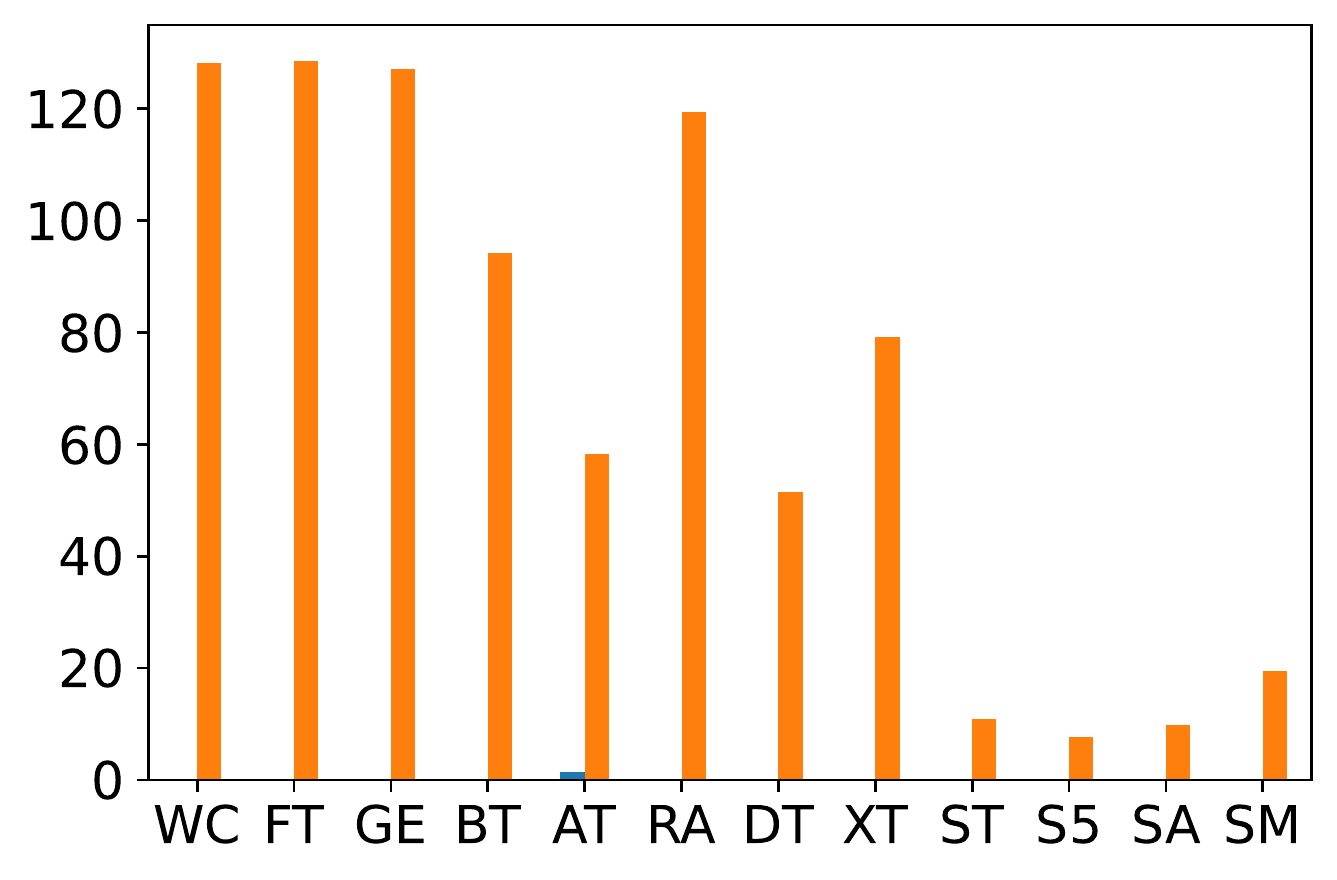}}
\newline
\subfloat[D9]{\includegraphics[trim=0.12cm 0.12cm 0.12cm 0.12cm, clip, width=0.25\textwidth, height=30mm]{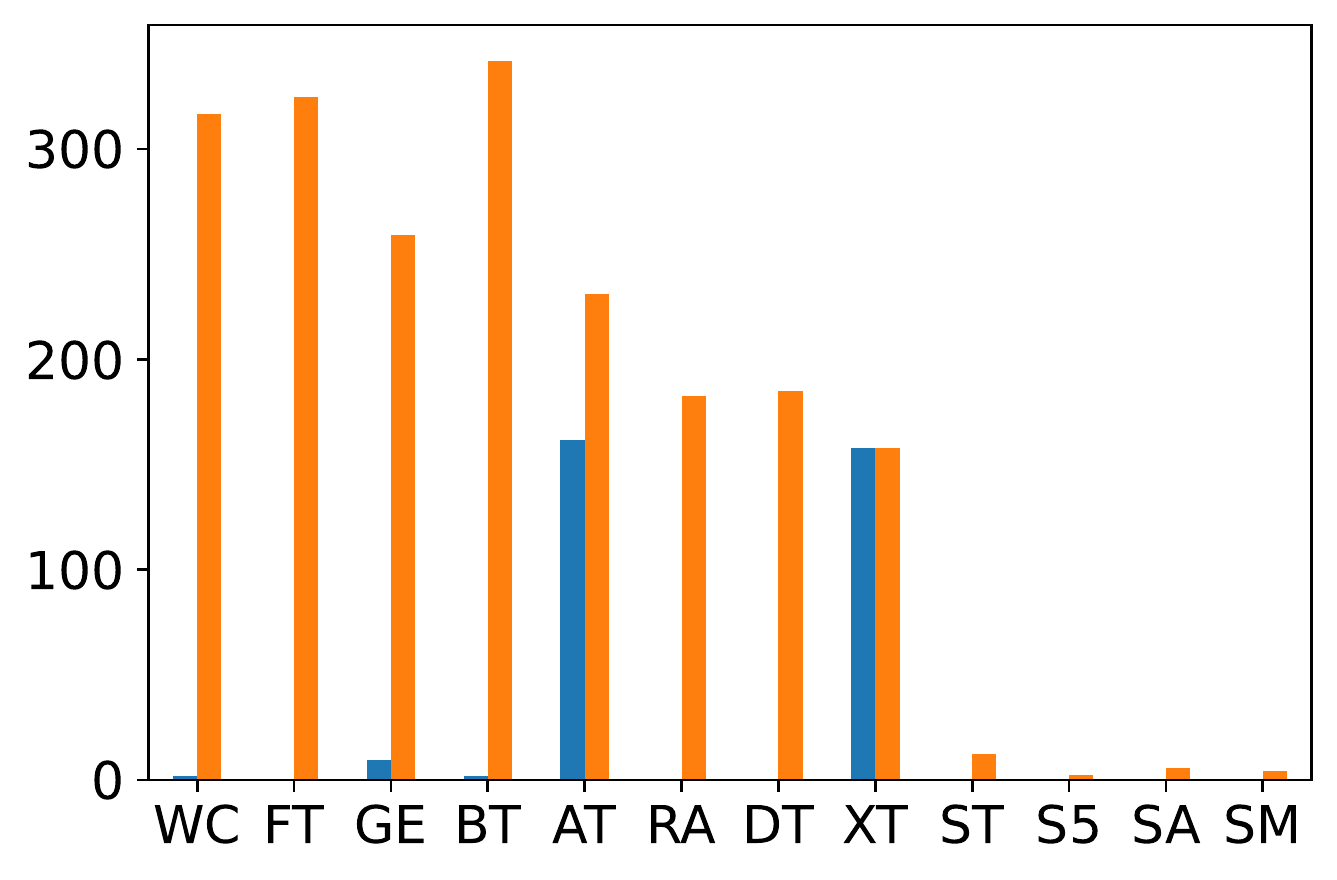}}
\subfloat[D10]{\includegraphics[trim=0.12cm 0.12cm 0.12cm 0.12cm, clip, width=0.25\textwidth, height=30mm]{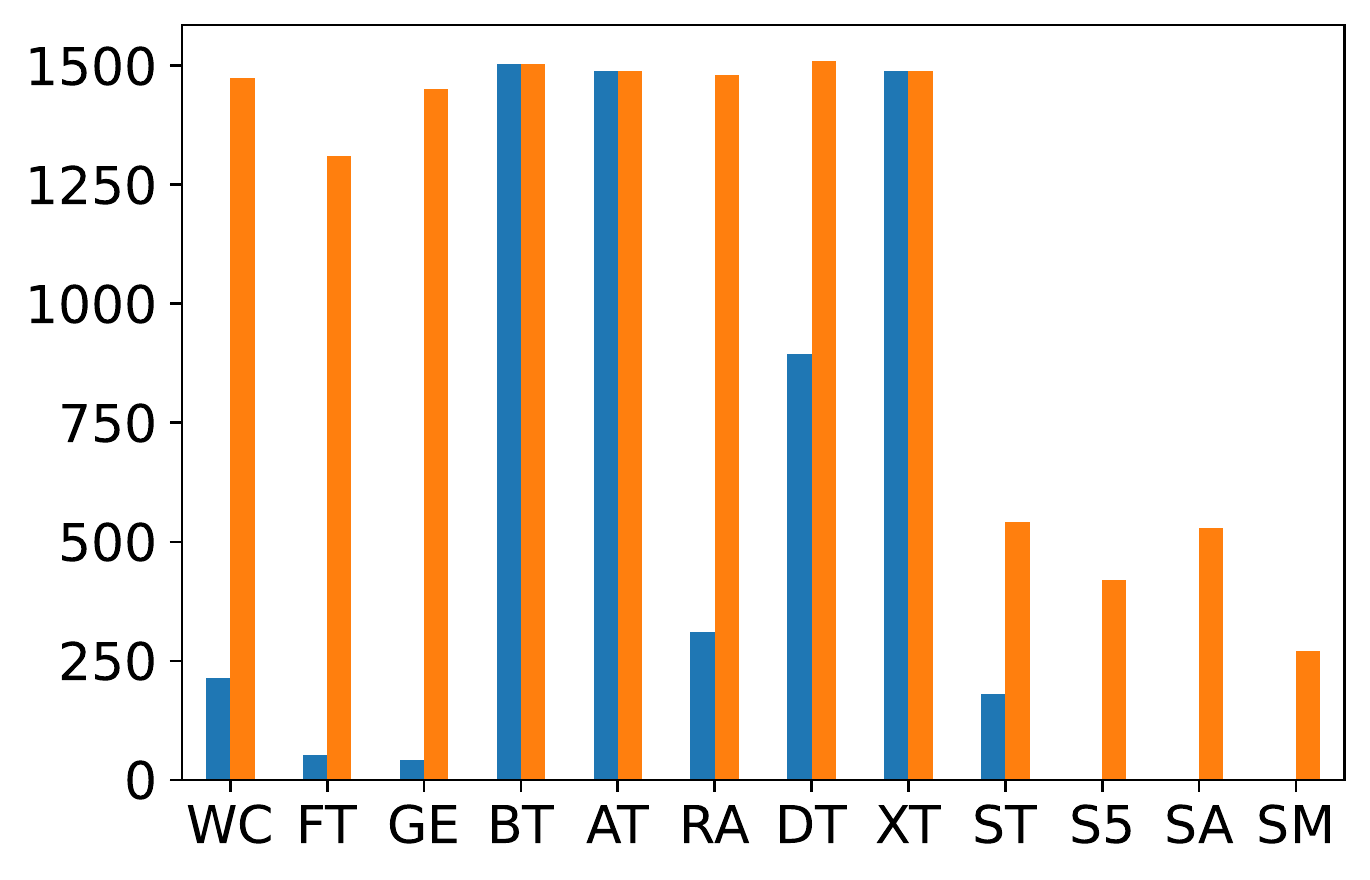}}

\caption{Unsupervised Matching Time (sec) per method. Blue is the time until the highest F1; orange is total time.}

\label{fig:unsup_match_time}
\end{figure}

% \begin{table*}[t]
% % \setlength{\tabcolsep}{2.5pt}
% \small
% \begin{tabular}{lrrrrrrrrrrrr}
% \toprule
%  &     WC &    FT &    GE &      BT &      AT &     RA &     DT &      XT &     ST &   S5 &   SA &   SM \\
% \midrule
% D1   &    0.0 &   0.0 &   0.0 &     0.1 &     0.3 &    0.0 &    0.0 &     0.0 &    0.2 &  0.0 &  0.0 &  0.0 \\
% D2   &    2.0 &   0.8 &   0.8 &     1.4 &     0.9 &    2.4 &    1.8 &     2.6 &    0.1 &  0.2 &  1.0 &  0.1 \\
% D3   &    9.2 &   8.4 &   3.1 &     3.4 &     1.1 &    0.7 &    0.2 &     7.7 &    0.1 &  0.2 &  0.9 &  0.0 \\
% D4   &    1.2 &   3.5 &   0.0 &     0.9 &     7.2 &    2.2 &    0.3 &    12.5 &    0.1 &  0.0 &  0.0 &  0.0 \\
% D5   &    0.4 &  24.0 &   0.7 &    22.7 &     0.0 &    9.5 &   10.9 &     0.0 &    0.0 &  0.3 &  0.0 &  0.0 \\
% D6   &    0.0 &   0.0 &   0.0 &     0.0 &     0.0 &    0.0 &    0.0 &     0.0 &    0.1 &  0.1 &  0.1 &  0.1 \\
% D7   &    3.0 &   2.6 &   8.2 &    10.6 &     3.7 &    4.9 &    3.1 &    61.5 &    0.2 &  0.0 &  0.3 &  0.1 \\
% D8   &    0.0 &   0.0 &   0.0 &     0.0 &     1.4 &    0.0 &    0.0 &     0.0 &    0.0 &  0.0 &  0.0 &  0.0 \\
% D9   &    1.9 &   0.7 &   9.4 &     1.7 &   161.8 &    0.1 &    0.7 &   157.8 &    0.0 &  0.1 &  0.0 &  0.1 \\
% D10  &  214.2 &  51.8 &  42.4 &  1504.4 &  1489.7 &  311.4 &  893.9 &  1488.7 &  181.5 &  0.6 &  1.1 &  0.5 \\
% \bottomrule
% \end{tabular}
% \caption{Unsupervised Matching Time (sec) per method from real data per case.}
% \end{table*}

\begin{figure}[!t]
\centering
\subfloat[D1]{\includegraphics[trim=0.12cm 0.12cm 0.12cm 0.12cm, clip, width=0.25\textwidth, height=30mm]{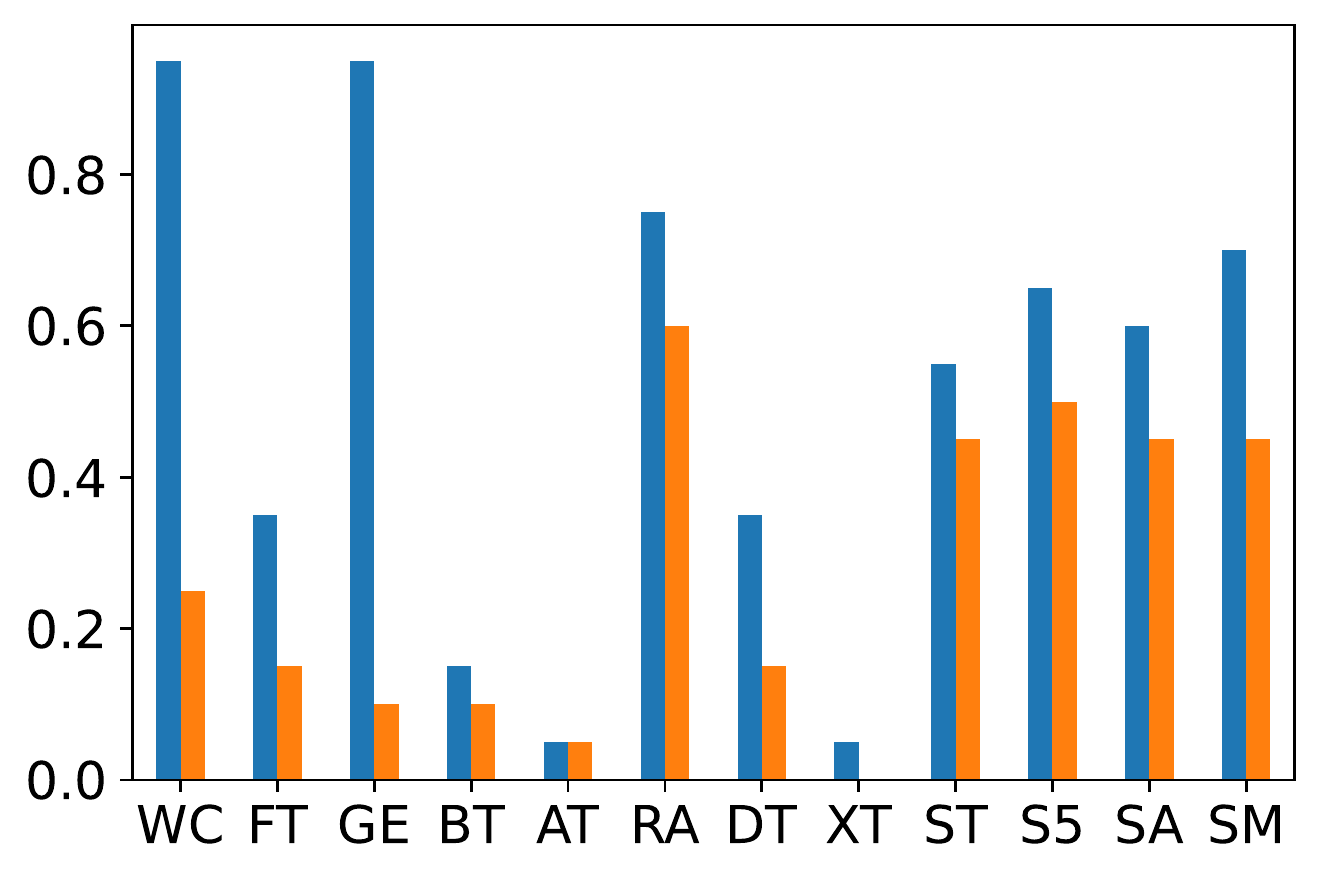}}
\subfloat[D2]{\includegraphics[trim=0.12cm 0.12cm 0.12cm 0.12cm, clip, width=0.25\textwidth, height=30mm]{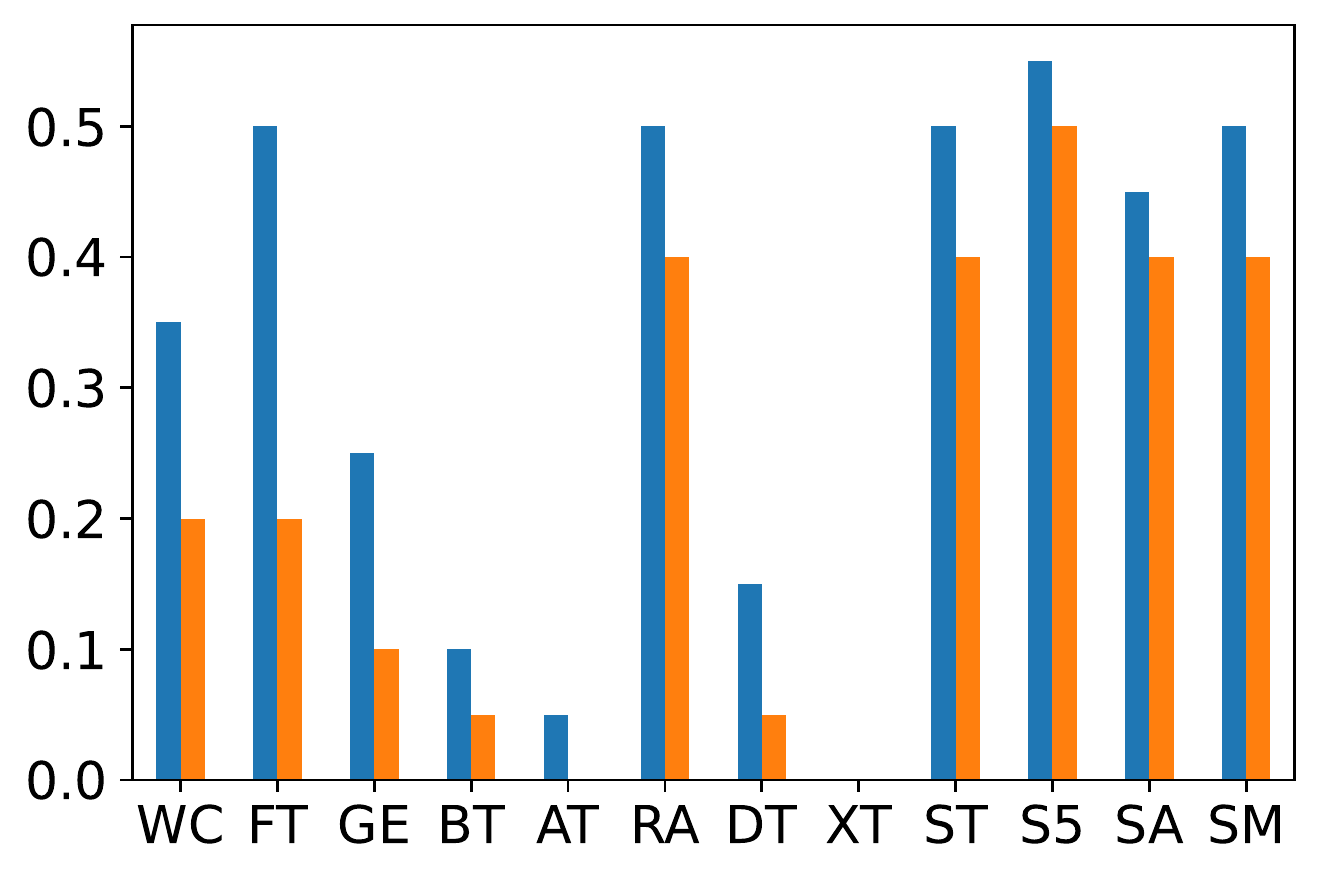}}
\newline
\subfloat[D3]{\includegraphics[trim=0.12cm 0.12cm 0.12cm 0.12cm, clip, width=0.25\textwidth, height=30mm]{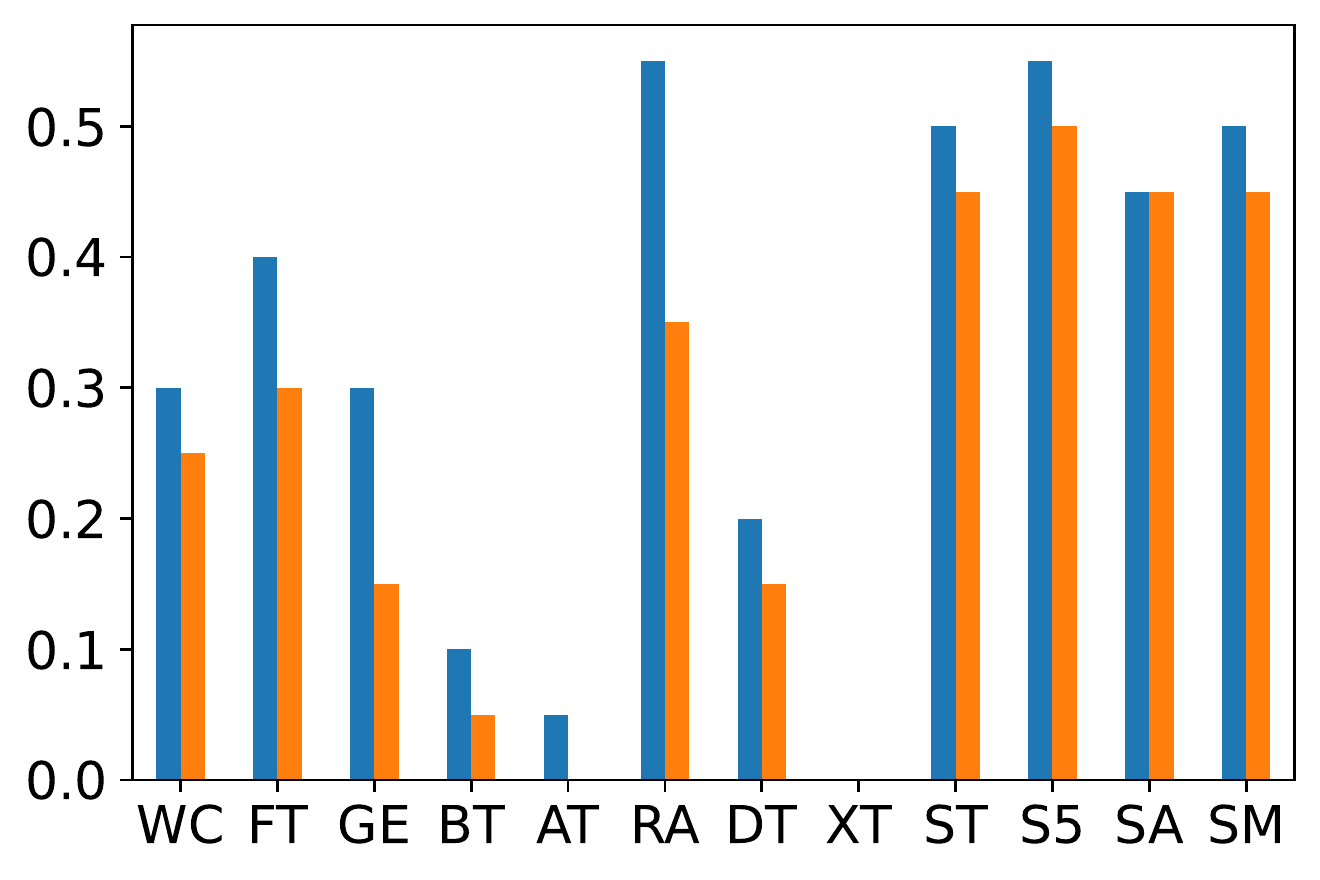}}
\subfloat[D4]{\includegraphics[trim=0.12cm 0.12cm 0.12cm 0.12cm, clip, width=0.25\textwidth, height=30mm]{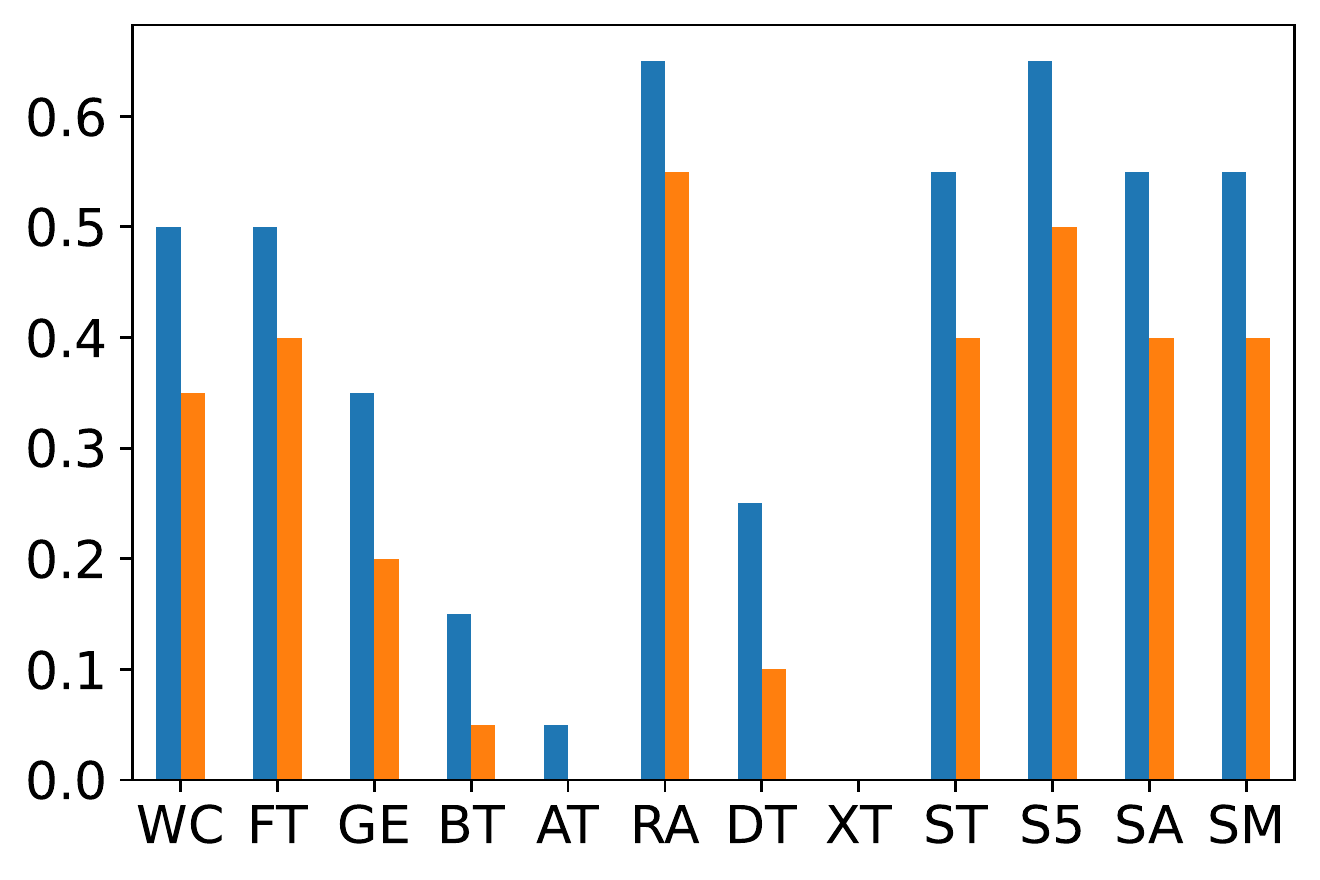}}
\newline
\subfloat[D5]{\includegraphics[trim=0.12cm 0.12cm 0.12cm 0.12cm, clip, width=0.25\textwidth, height=30mm]{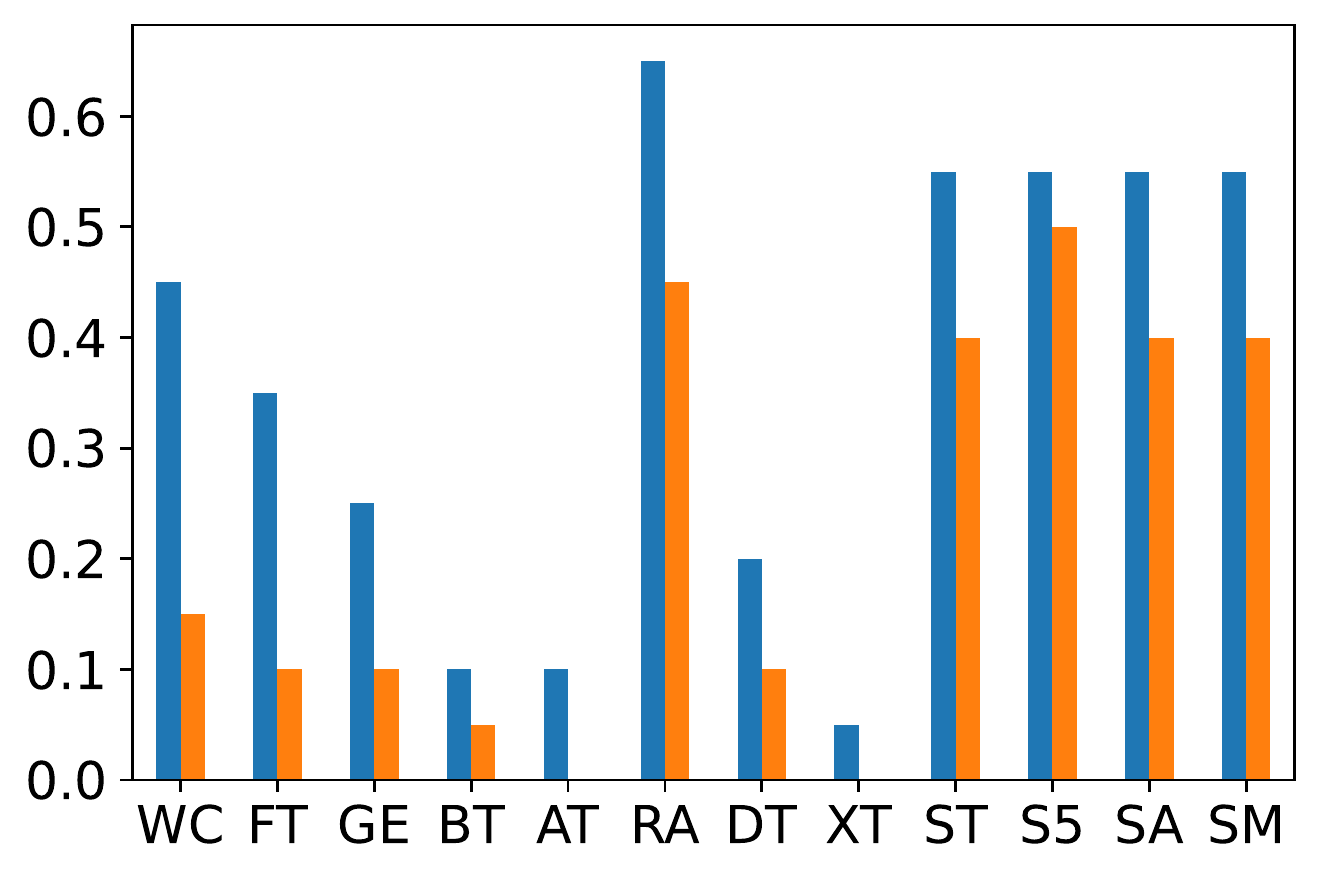}}
\subfloat[D6]{\includegraphics[trim=0.12cm 0.12cm 0.12cm 0.12cm, clip, width=0.25\textwidth, height=30mm]{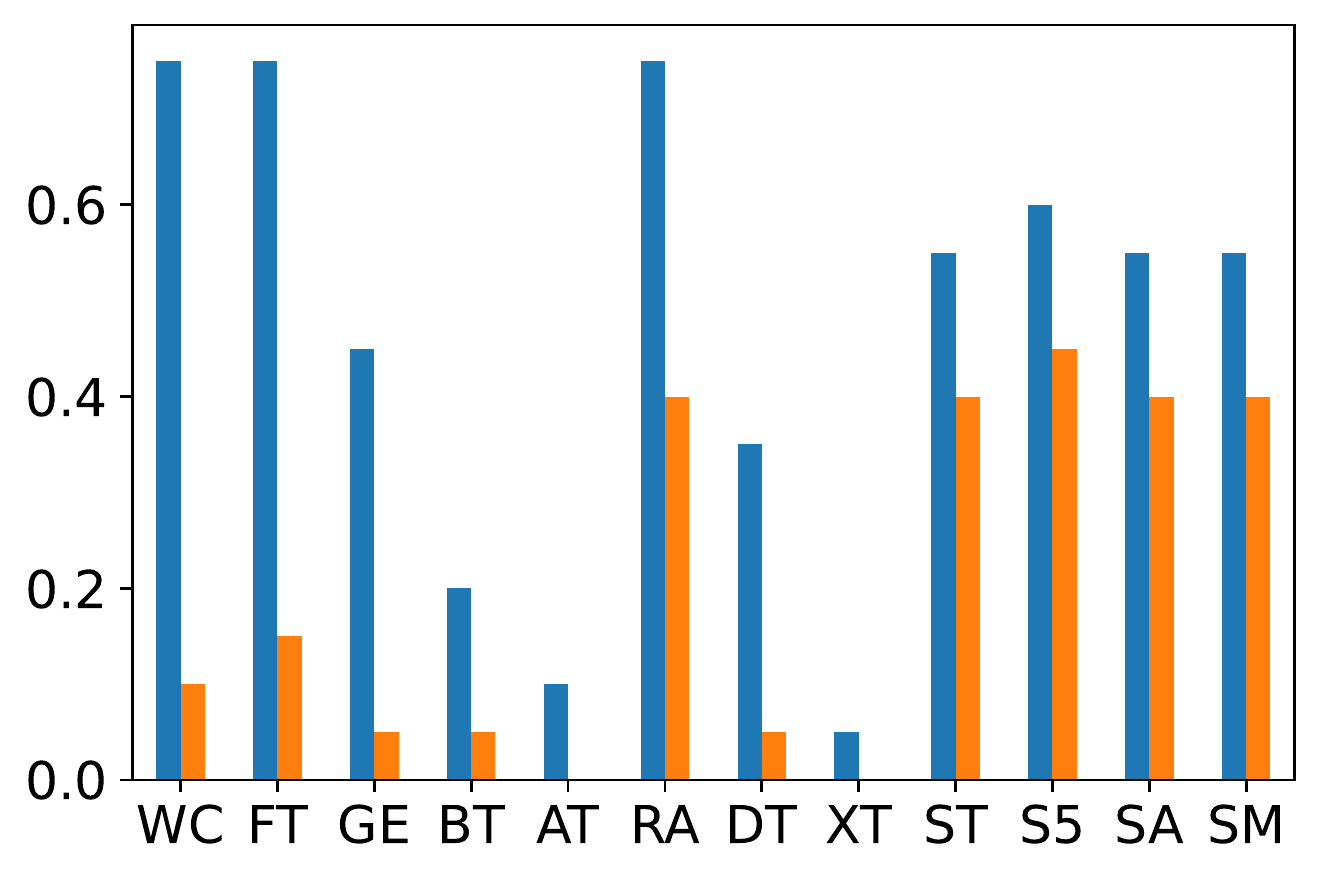}}
\newline
\subfloat[D7]{\includegraphics[trim=0.12cm 0.12cm 0.12cm 0.12cm, clip, width=0.25\textwidth, height=30mm]{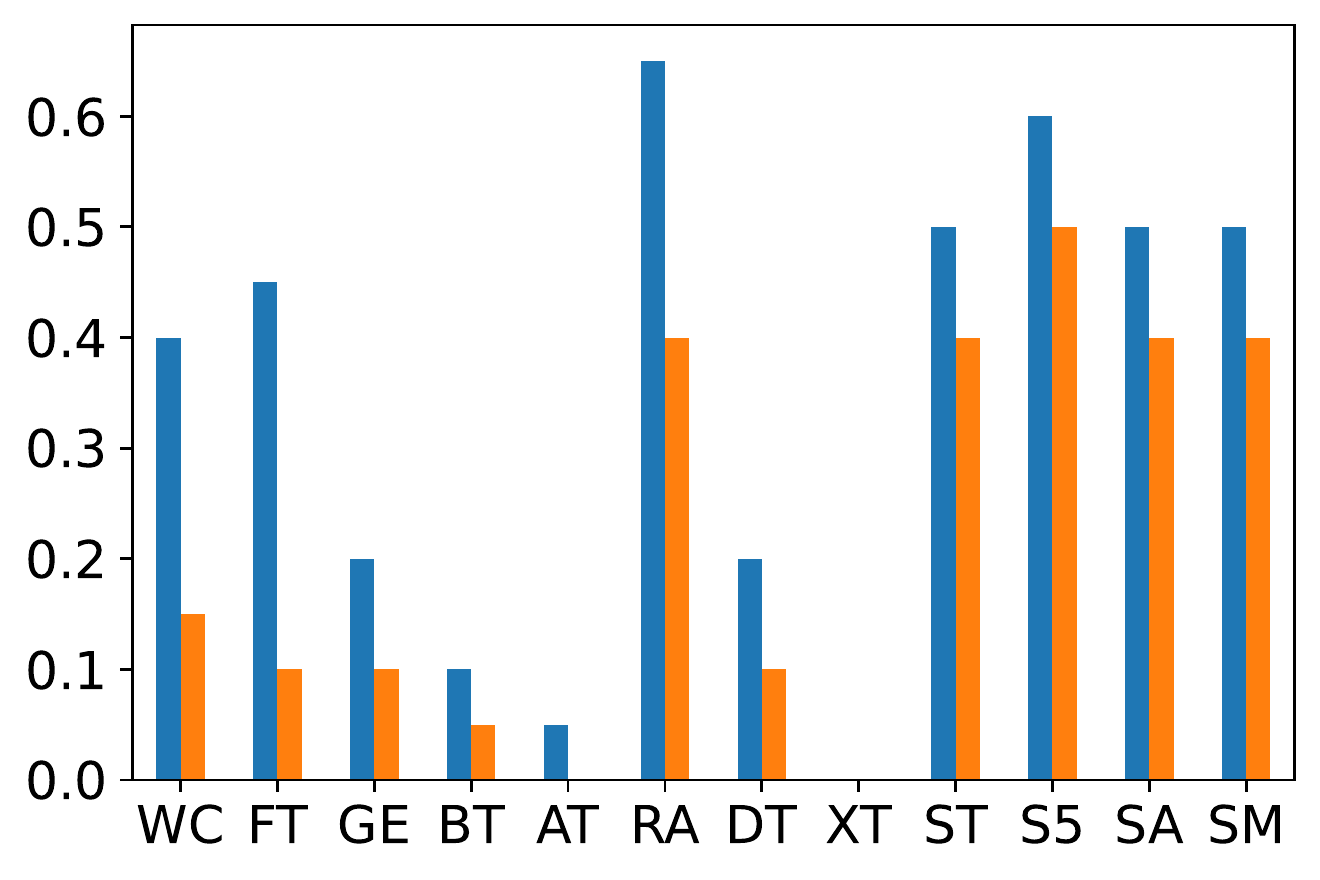}}
\subfloat[D8]{\includegraphics[trim=0.12cm 0.12cm 0.12cm 0.12cm, clip, width=0.25\textwidth, height=30mm]{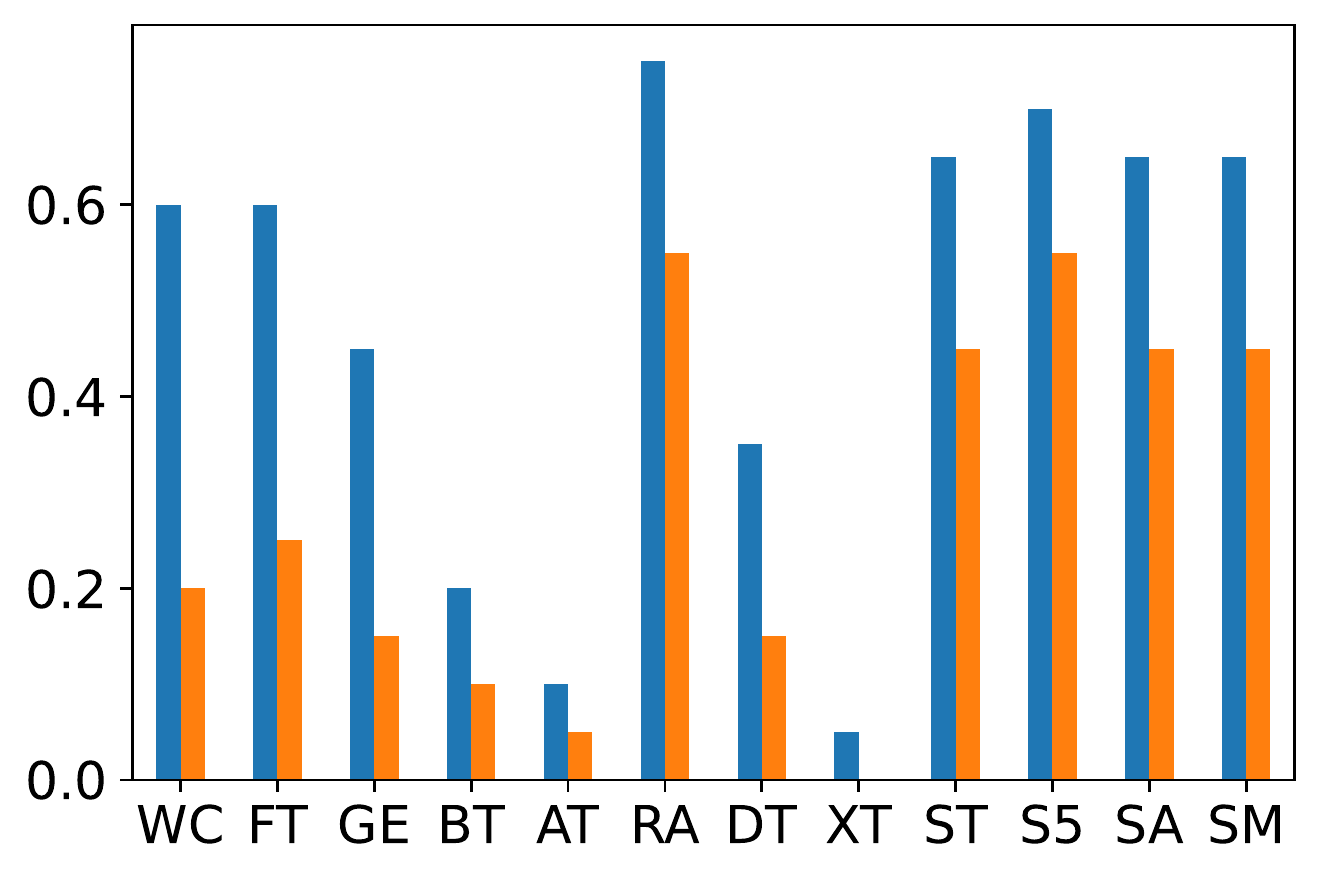}}
\newline
\subfloat[D9]{\includegraphics[trim=0.12cm 0.12cm 0.12cm 0.12cm, clip, width=0.25\textwidth, height=30mm]{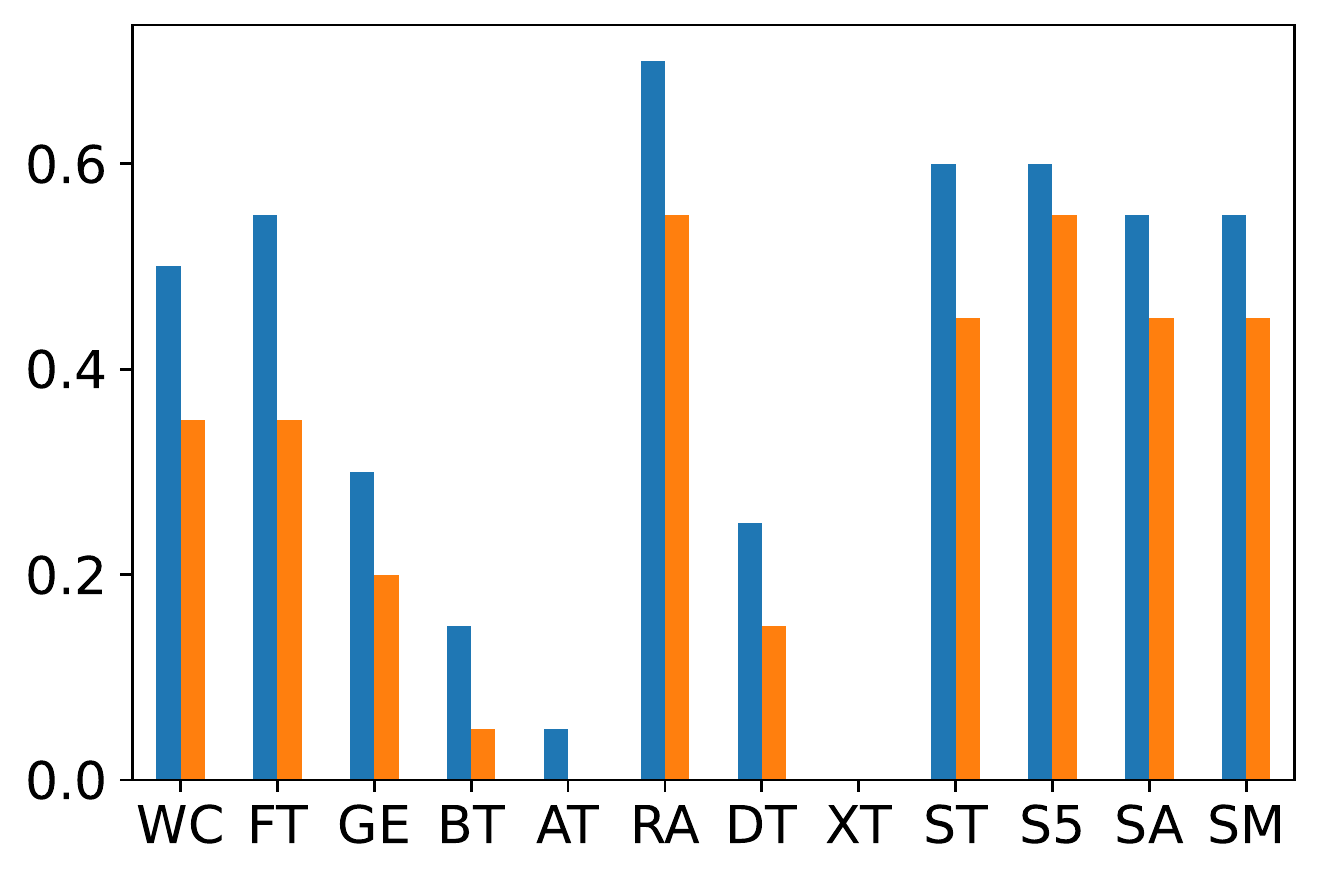}}
\subfloat[D10]{\includegraphics[trim=0.12cm 0.12cm 0.12cm 0.12cm, clip, width=0.25\textwidth, height=30mm]{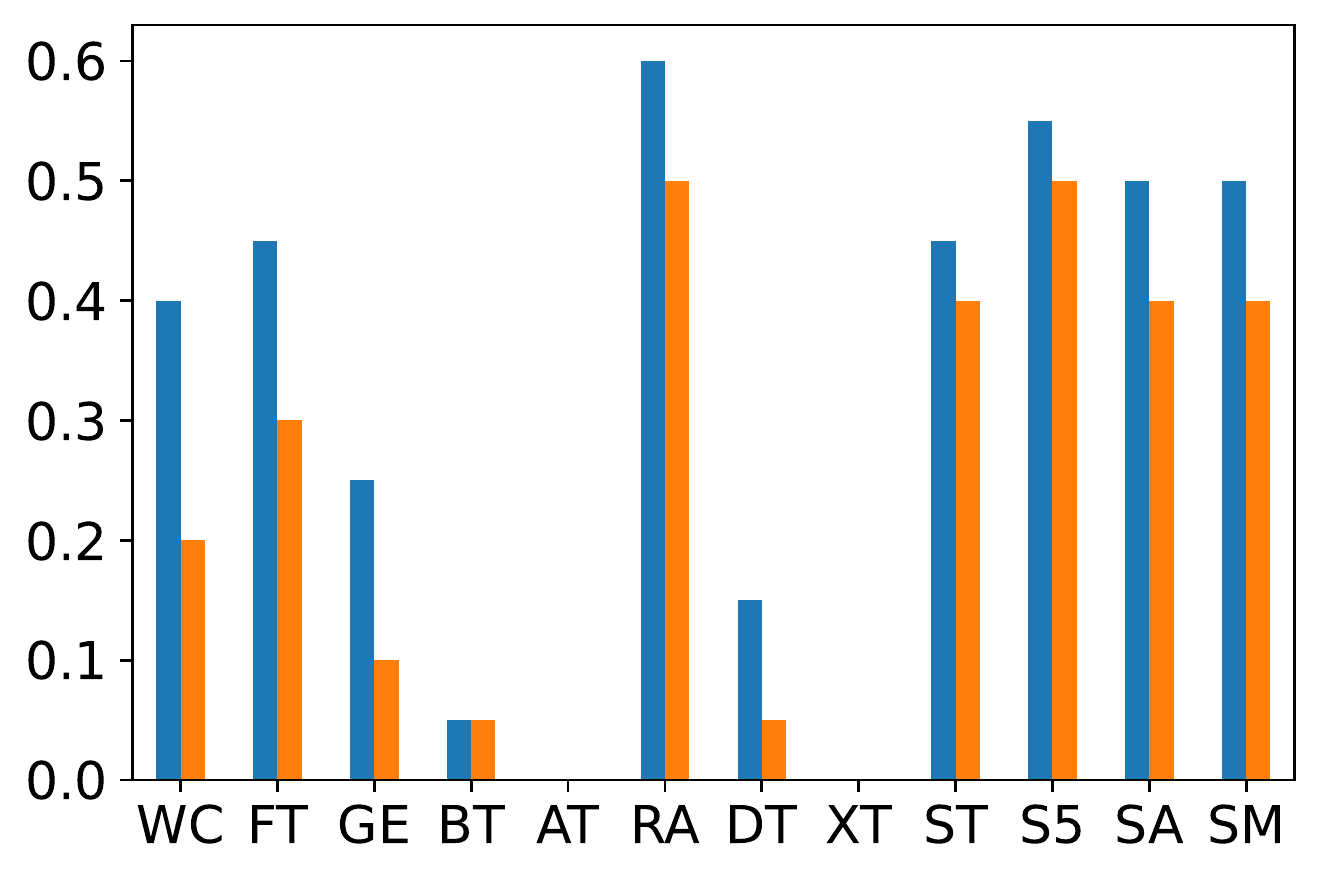}}

\caption{Unsupervised Matching $\delta$ per method. Blue is the $\delta$ where the best F1 is found and orange is the $\delta$ where the algorithm terminates.}

\label{fig:unsup_match_delta}
\end{figure}

% In Figure \ref{fig:unsup_match_time} we present the necessary time to find the best F1 mentioned in Section ?? and the total time for running the Unique Mapping Clustering algorithm. To better explain the results presented in Figure \ref{fig:unsup_match_time}, we should combine it with the results of Figure \ref{fig:unsup_match_delta}. More specifically:
% $\bullet$ It can be seen in Figure \ref{fig:unsup_match_delta} that all SBERT models terminate with high $\delta$. That means that all entities found their match with high similarity score, something that explains also the good effectiveness presented in Section ?? and that the algorithm needed a relatively small execution time, since not many pairs of similarities were examined.
% $\bullet$ In Figure \ref{fig:unsup_match_time}, XLNet and AlBERT have very high execution times to find the best F1 and when so, they also exhaust the total execution time. When that happens, they also terminate with almost $\delta=0$, meaning that we also expect to have low $F1$.
% $\bullet$ Finally, in most datasets, all models need the same time to run the Unique Mapping Clustering algorithm, but the differences occure in finding the best F1 score, i.e. which is the proper $\delta$, where F1 hits its peak value.

We define as matching time of a model the time that is required for applying the UMC algorithm to all pairwise similarities using the optimal pruning threshold. For the most effective model, S-GTR-T5, the matching time is typically much lower than half a second, except the largest dataset, $D_{10}$, where it raises to almost two seconds. This high efficiency in the context of a large number of candidate pairs should be attributed to its rather high similarity threshold, which fluctuates between 0.55 and 0.70, with a median (mean) of 0.625 (0.615), thus pruning the vast majority of pairs. Similar behavior is exhibited by the rest of the SentenceBERT models. The static models also apply UMC within few seconds in most datasets, depending on their similarity threshold. This is less frequently true for the BERT-based models, which suffer from low discriminativeness, thus yielding lower thresholds and significantly more pairs to be processed. As a result, one of the BERT-based models is the slowest one in most datasets -- typically AlBERT or XLNet.

\noindent\textbf{Comparison to SotA.} Table \ref{tb:sota_comparison}(b) demonstrates that ZeroER's run-time is dominated by its blocking time, which is higher larger than the total execution time of S-GTR-T5 in most datasets. Due to its simple functionality, the end-to-end S-GTR-T5 approach is two orders of magnitude faster than ZeroER in all datasets, but $D_1$. Its run-time is dominated by the vectorization and the indexing of the input entities, with the matching time accounting for a few milliseconds even for the largest dataset, due to blocking.

\begin{figure*}[!t]
\centering
\includegraphics[width=0.325\textwidth]{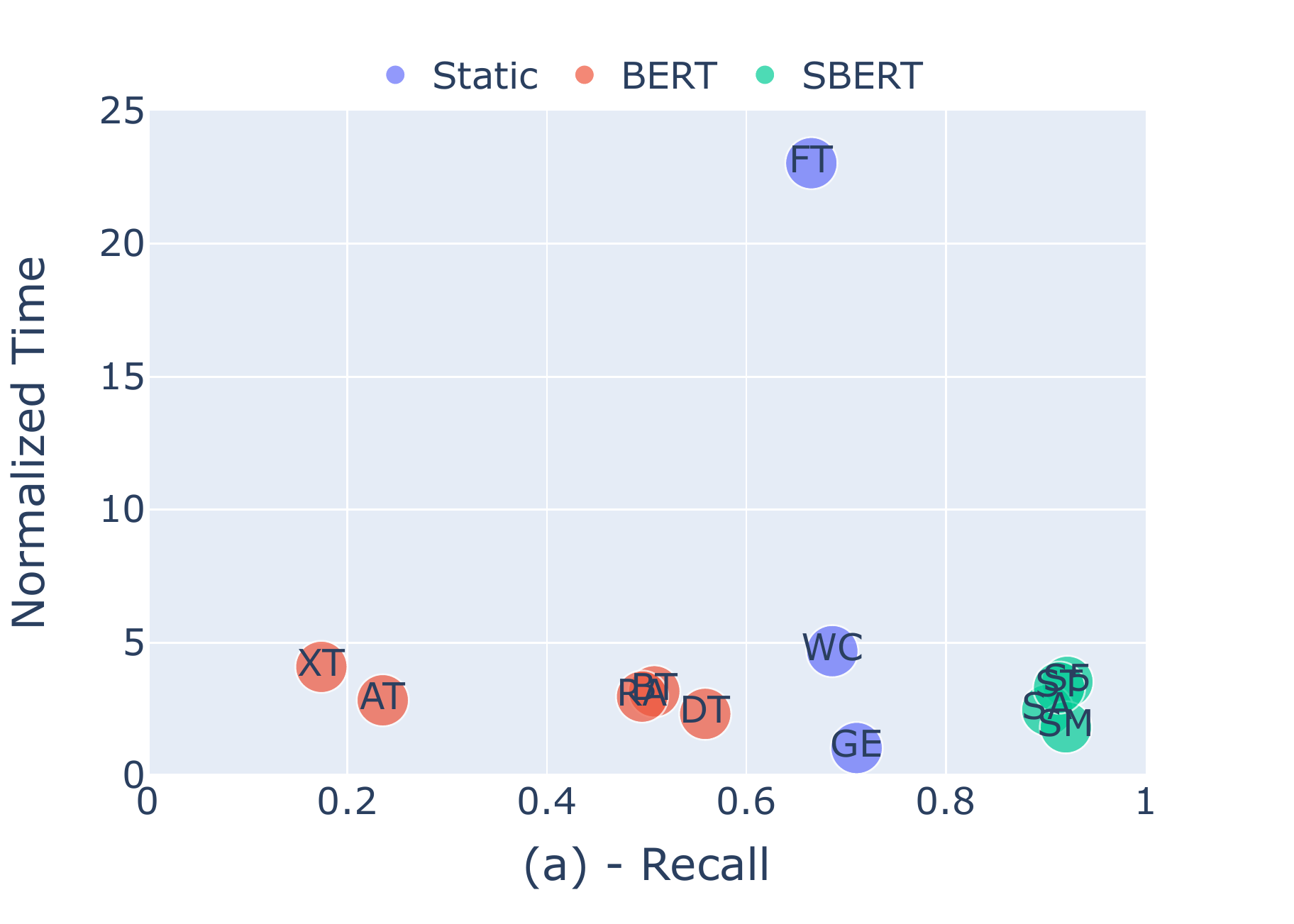}
\includegraphics[width=0.325\textwidth]{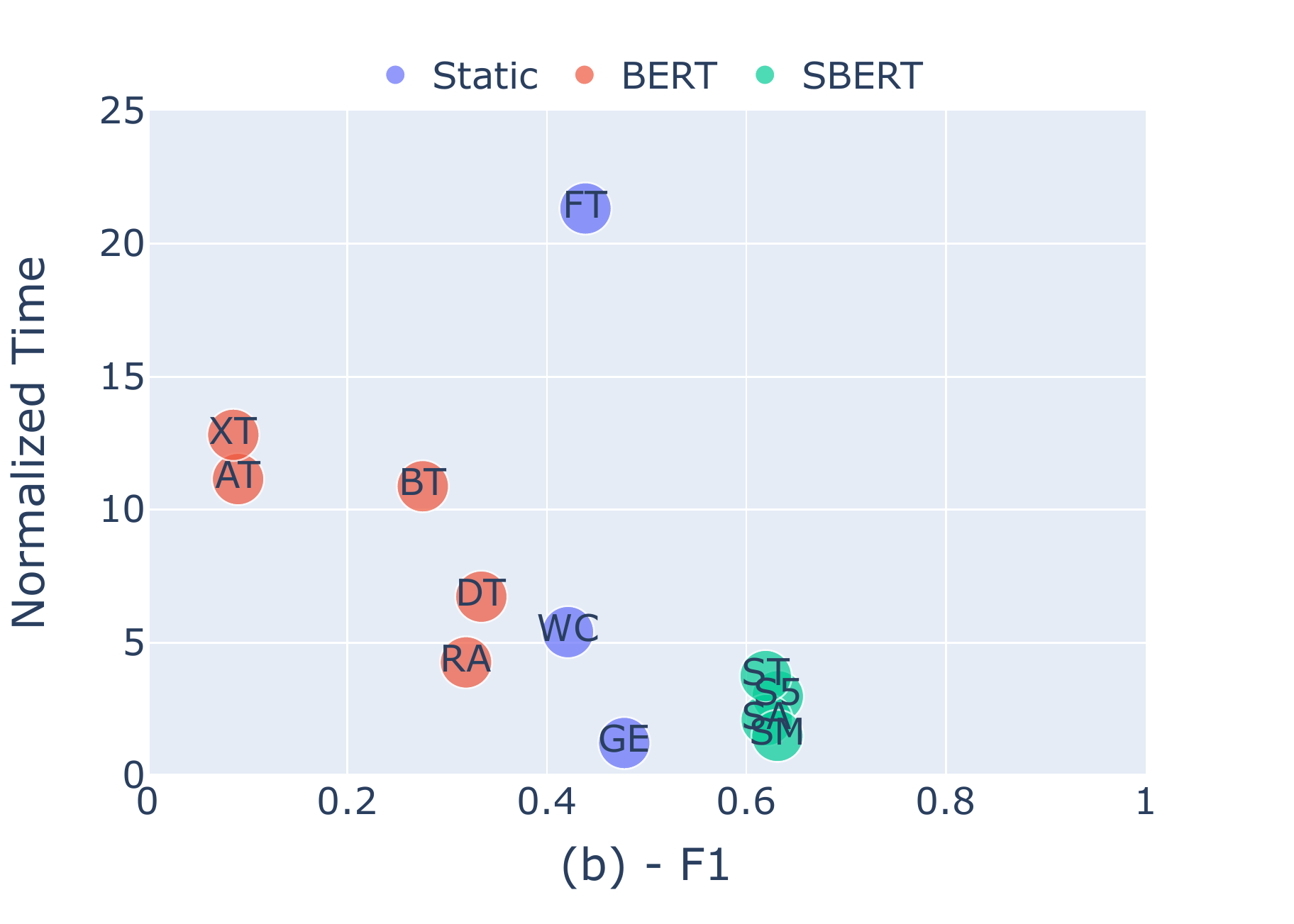}
\includegraphics[width=0.325\textwidth]{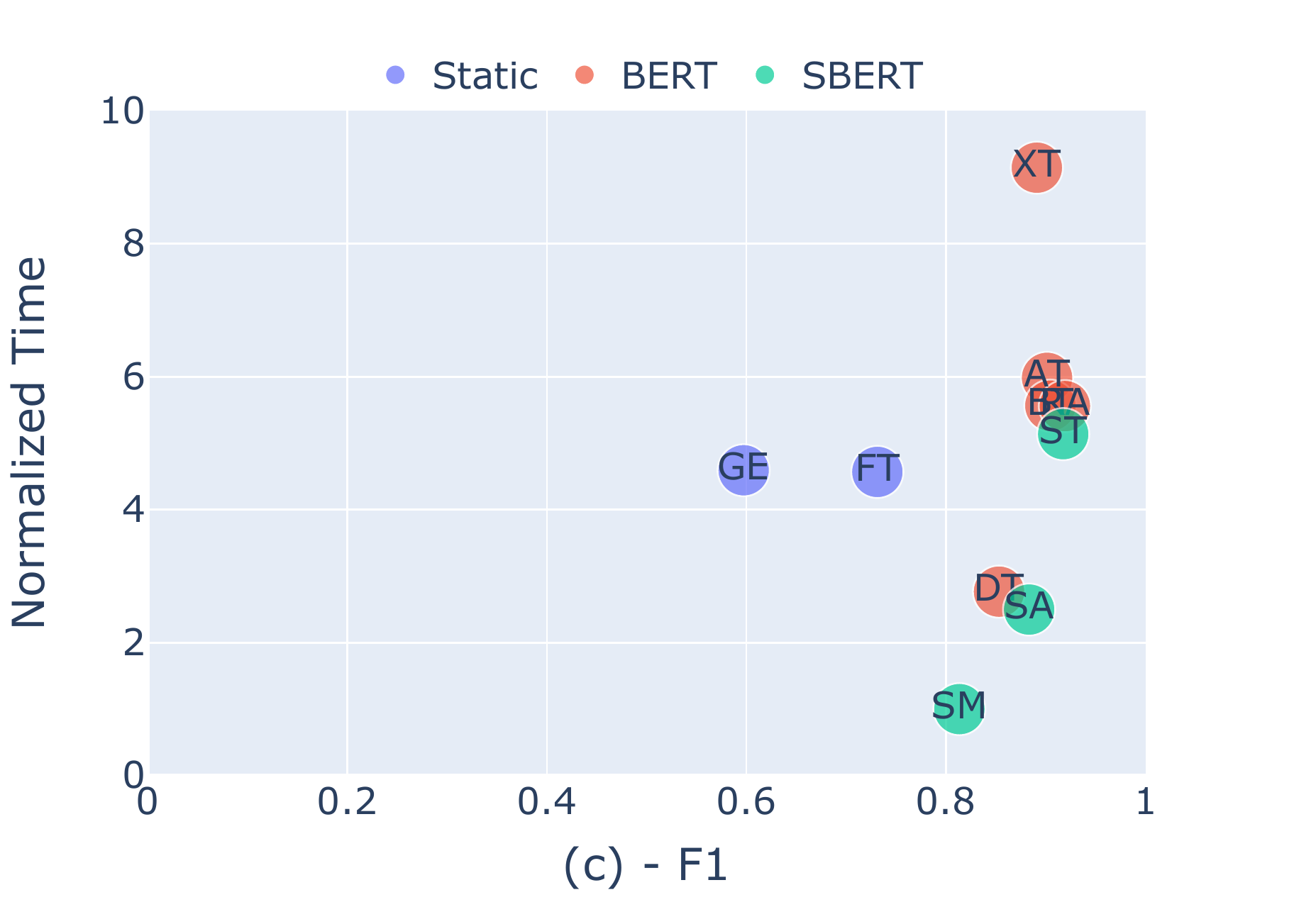}
% \subfloat[Blocking ($k=10$)]{\includegraphics[trim=0.12cm 0.12cm 2cm 0.12cm, clip, width=0.33\textwidth, height=50mm]{blocking_pareto.pdf}}
% \subfloat[Unsupervised Matching (Best F1 achievable)]{\includegraphics[trim=1.5cm 0.12cm 2cm 0.12cm, clip, width=0.33\textwidth, height=50mm]{unsup_matching_pareto.pdf}}
% \subfloat[Supervised Matching]{\includegraphics[trim=1.5cm 0.12cm 2cm 0.12cm, clip, width=0.33\textwidth, height=50mm]{sup_matching_pareto.pdf}} 
% \newline
\caption{Tradeoff between Effectiveness and Time Efficiency on average, across all datasets in Table \ref{tb:datasets}(a) for (a) Blocking with $k$=10 and (b) Unsupervised Matching (wrt best attainable F1) and all datasets in Table \ref{tb:smDatasets} for (c) Supervised Matching.}
\label{fig:tradeoff}
\end{figure*}

\subsection{Supervised Matching}

% We report the time efficiency of the language models with respect to training time ($t_t$) and testing time ($t_r$) 
% is reported 
Table \ref{tb:supMatchingTime} demonstrates that S-MPNet constitutes the best choice for applications that emphasize time efficiency at the cost of slightly lower effectiveness: its training and prediction time are consistently lower than that the top-performing model, RoBERTa, by 9\% and 7\%, respectively, on average. More significant gains in efficiency are achieved by the rest of the SentenceBERT models, S-DistilRoBERTa and S-MiniLM: they reduce RoBERTa's training and testing times by more than 50\% in all cases, while their F1 is lower by just 5\%, on average. Similarly, DistilBERT reduces RoBERTa's run-times to a half for a ~7\% reduction in F1. XLNet  underperforms RoBERTa in all respects. 
% Compared to 
XLNet is consistently the slowest by far model among all datasets,
% with respect to both training and testing time, 
thus underperforming RoBERTa in all respects. The same applies to BERT, albeit to a minor extent, i.e., $<$2\% with respect to all measures. 
% excels in both effectiveness and run-time, as it is faster by $\geq38\%$, on average. 
ALBERT achieves slightly lower training times (by 7\%) than RoBERTa at the cost of higher prediction times (by 8\%), while its F1 is lower by just 2\%, on average. Regarding the static models, on average, they are just 10\% and 17\% faster than RoBERTa with respect to training and testing time, respectively, despite their very low F1. Overall, we can conclude that \textit{the SentenceBERT models are significantly faster than the BERT-base ones, thus achieving a better trade-off between effectiveness and time efficiency.}

% The rest of the models reduce both run-times to a half, on average, at the cost of significantly lower F1 ($\ll$10\%).

% Comparing the language models to DeepMatcher+, which leverages FastText embeddings, we observe that they outperform it in practically all cases. This is particularly true in $DSM_1$ and $DSM_5$, where the worst language model (DistilBERT) exceeds DeepMatcher+ by $\sim$20\%. DITTO, on the other hand, is directly comparable to the best performing language model in each dataset. In $DSM_2$, $DSM_3$ and $DSM_4$, DITTO's F1 is lower by just $\leq$0.5\%. In $DSM_1$ and $DSM_5$, though, DITTO outperforms all language models
% by 3\% and 1.5\%, respectively. This should be attributed to the external knowledge and the data augmentation. The effect of these features is more clear when comparing DITTO to the language model that lies at its core, i.e., RoBERTa. On average, across all datasets, the latter underperforms DITTO by 1.3\%. 

\section{Discussion \& Conclusions}
\label{sec:conclusions}

Figure \ref{fig:tradeoff} summarizes our experimental results on the three ER tasks. The horizontal axis in each diagram corresponds to the effectiveness measure of the respective task, while the vertical one corresponds to the normalized run-time, which is computed by dividing the overall run-time of a model with that of the fastest one (i.e., 1 is assigned to the fastest model). The space formed by these axes illustrates the trade-off between effectiveness and time efficiency, with the top performing model lying closer to (1,1), i.e., the lower right corner. Note that for each model, we have computed its average effectiveness and normalized time across all datasets in Table \ref{tb:datasets}(a).

% In our experimental analysis we studied the performance of 12 popular language models in three ER tasks: 
% blocking, unsupervised and supervised matching. In Figure \ref{fig:tradeoff} we see the tradeoff between effectiveness and efficiency in these three tasks. Regarding efficiency, we have normalized every performance within each case and regarding effectiveness we have used the metrics mentioned in the corresponding sections earlier.

We observe that we can distinguish the ER tasks into two groups based on the relative performance of the three model types. The first group includes the unsupervised tasks, i.e., Blocking and Unsupervised Matching, where the SentenceBERT models consistently outperform the other two types to a significant extent. The reason is that 
% top performing group includes, which 
they are able to distinguish
between matching and non-matching entities without fine-tuning their top attention layers. Among them, the differences are minor, on average. Typically, though, S-GTR-T5 excels in effectiveness, but is slower, while S-MiniLM offers a better balance, trading slightly lower effectiveness for slightly higher time efficiency. The second best type includes the static
models, where GloVe clearly dominates FastText and Word2Vec, which suffer from significantly higher initialization cost, while being slightly less effective. GloVe is actually the fastest model across all datasets in both ER tasks, thus being ideal for ER applications emphasizing time efficiency at the cost of lower effectiveness. Finally, all BERT-based models consistently underperform the other two types in all respects. Their poor effectiveness stems from the lack of fine-tuning, which causes them to assign low similarity scores to both matching and non-matching pairs alike. 
%Therefore, BERT models should be avoided in unsupervised ER tasks.
% typically fail to distinguish between matching and non-matching
% entities. The reason is that they require fine-tuning, unlike the static
% models, which anyway cannot be fine-tuned. These patterns apply
% equally to the blocking and the bipartite graph matching tasks.

% Overall, S-GTR-T5 has consistent effectiveness and efficiency in all Blocking and Unsupervised Matching cases. With slightly worse performance in effectiveness, comes S-MiniLM, which has better efficiency and can also support Supervised Matching.

Different patterns are observed in Supervised Matching, where all BERT-based models excel in effectiveness. As expected, DistilBERT sacrifices effectiveness for significantly higher run-time. Note that in this case, we exclusively consider the testing/prediction time per model, because the training time constitutes an one-off cost. S-MPNet also exhibits very high effectiveness, while S-DistilRoBERTa and S-MiniLM emphasize time efficiency. The former dominates DistilBERT, while the latter is the fastest albeit the least effective dynamic model. The static models are less accurate than all dynamic models, while offering no advantage in terms of run-time. Therefore, they should be avoided in this task. Instead, any of the dynamic models can be selected, depending on requirements of the application at hand. The only exception is XLNet, which is much slower but not more effective than most dynamic models.
% differs
% Secondly, in all tasks the family of SBERT models is always best, with S-GTR-T5 being the best within. Then comes the family of static models, which is very fast and has mediocre performance and finally comes the family of BERT models. The latter have a much better performance when fine-tuned.

\begin{table}[t]
\setlength{\tabcolsep}{2.5pt}
\small
\begin{tabular}{|l|rr|rr|rr|rr|rr|}
\cline{2-11}
\multicolumn{1}{c|}{} & \multicolumn{2}{c|}{DSM1} & \multicolumn{2}{c|}{DSM2} & \multicolumn{2}{c|}{DSM3} & \multicolumn{2}{c|}{DSM4} & \multicolumn{2}{c|}{DSM5} \\
\multicolumn{1}{c|}{} &   $t_t$ & $t_e$ & $t_t$ & $t_e$ &   $t_t$ & $t_e$ &   $t_t$ & $t_e$ &   $t_t$ & $t_e$ \\
\midrule
\midrule
FT         &   851.9 &   4.8 &  100.9 &   0.6 &  1,155.0 &   6.9 &  2,527.3 &  14.8 &   882.1 &   4.7 \\
GE         &   847.0 &   4.8 &  101.4 &   0.6 &  1,157.2 &   6.9 &  2,534.3 &  14.7 &   876.9 &   4.7 \\
\midrule
BT         &  1,811.2 &  11.2 &  66.3 &   0.4 &  1,525.9 &   9.6 &  2,615.9 &  15.8 &  1,093.8 &   6.8 \\
AT         &  1,700.5 &  12.0 &  62.8 &   0.5 &  1,448.4 &  10.4 &  2,422.1 &  17.0 &  1,026.6 &   7.4 \\
RA         &  1,810.8 &  11.2 &  66.2 &   0.4 &  1,548.7 &   9.6 &  2,666.9 &  15.8 &  1,111.3 &   6.9 \\
DT         &   915.2 &   5.5 &  34.0 &   0.2 &   779.2 &   4.8 &  1,341.9 &   7.9 &   559.1 &   3.4 \\
XT         &  2,920.7 &  24.2 &  92.2 &   0.7 &  2,120.5 &  15.9 &  3,196.5 &  21.0 &  1,423.2 &  10.1 \\
\midrule
ST         &  1,667.5 &  10.5 &  63.4 &   0.4 &  1,475.8 &   8.8 &  2,549.9 &  14.4 &  1,076.8 &   6.3 \\
SA         &   828.2 &   5.0 &  31.4 &   0.2 &   716.9 &   4.3 &  1,253.7 &   7.1 &   518.1 &   3.1 \\
SM         &   406.6 &   2.1 &  13.4 &   0.1 &   299.2 &   1.7 &   521.6 &   2.7 &   216.6 &   1.2 \\
\bottomrule
\end{tabular}
\caption{Training ($t_t$) and testing ($t_e$) times in seconds of all models in Supervised Matching over the datasets in Table~\ref{tb:smDatasets}.
% The training ($t_t$) and testing ($t_e$) times are measured in minutes and seconds, respectively, {\color{red} and correspond to the average after 5 runs.
}
\label{tb:supMatchingTime}
\end{table}

% Finally, the task of Supervised Matching offers a trade-off between effectiveness and efficiency, since it needs a lot of time to train each model, but it offers a very good increase in F1. Nonetheless, the SBERT family has very good transfer learning capabilities, thus if no labeled data are available, it can offer a very good alternative with the Unsupervised Matching.

In the future, we will extent our analysis on ER datasets with numeric values.
% , rather than textual ones. 
We also intent to enhance our end-to-end, parameter- and learning-free approach to ER with SentenceBERT models, whose performance in Figure \ref{fig:match_unsup_rec_real}(d) is remarkable. 
%To further improve its F1, 
We will explore ways of replacing its default thresholds
% for blocking and matching 
with data-driven ones.

\bibliographystyle{ACM-Reference-Format}
\bibliography{references}


\begin{thebibliography}{65}

%%% the Web address.

\ifx \showCODEN    \undefined \def \showCODEN     #1{\unskip}     \fi
\ifx \showDOI      \undefined \def \showDOI       #1{#1}\fi
\ifx \showISBNx    \undefined \def \showISBNx     #1{\unskip}     \fi
\ifx \showISBNxiii \undefined \def \showISBNxiii  #1{\unskip}     \fi
\ifx \showISSN     \undefined \def \showISSN      #1{\unskip}     \fi
\ifx \showLCCN     \undefined \def \showLCCN      #1{\unskip}     \fi
\ifx \shownote     \undefined \def \shownote      #1{#1}          \fi
\ifx \showarticletitle \undefined \def \showarticletitle #1{#1}   \fi
\ifx \showURL      \undefined \def \showURL       {\relax}        \fi
% invisible to TeX
\providecommand\bibfield[2]{#2}
\providecommand\bibinfo[2]{#2}
\providecommand\natexlab[1]{#1}
\providecommand\showeprint[2][]{arXiv:#2}

\bibitem[Bahdanau et~al\mbox{.}(2014)]%
        {bahdanau2014neural}
\bibfield{author}{\bibinfo{person}{Dzmitry Bahdanau},
  \bibinfo{person}{Kyunghyun Cho}, {and} \bibinfo{person}{Yoshua Bengio}.}
  \bibinfo{year}{2014}\natexlab{}.
\newblock \showarticletitle{Neural machine translation by jointly learning to
  align and translate}.
\newblock \bibinfo{journal}{\emph{arXiv preprint arXiv:1409.0473}}
  (\bibinfo{year}{2014}).
\newblock


\bibitem[Bojanowski et~al\mbox{.}(2017)]%
        {bojanowski2017enriching}
\bibfield{author}{\bibinfo{person}{Piotr Bojanowski}, \bibinfo{person}{Edouard
  Grave}, \bibinfo{person}{Armand Joulin}, {and} \bibinfo{person}{Tomas
  Mikolov}.} \bibinfo{year}{2017}\natexlab{}.
\newblock \showarticletitle{Enriching Word Vectors with Subword Information}.
\newblock \bibinfo{journal}{\emph{Transactions of the Association for
  Computational Linguistics}}  \bibinfo{volume}{5} (\bibinfo{year}{2017}),
  \bibinfo{pages}{135--146}.
\newblock


\bibitem[Brunner and Stockinger(2020)]%
        {DBLP:conf/edbt/BrunnerS20}
\bibfield{author}{\bibinfo{person}{Ursin Brunner} {and} \bibinfo{person}{Kurt
  Stockinger}.} \bibinfo{year}{2020}\natexlab{}.
\newblock \showarticletitle{Entity Matching with Transformer Architectures -
  {A} Step Forward in Data Integration}. In \bibinfo{booktitle}{\emph{{EDBT}}}.
  \bibinfo{pages}{463--473}.
\newblock


\bibitem[Cer et~al\mbox{.}(2017)]%
        {cer2017semeval}
\bibfield{author}{\bibinfo{person}{Daniel Cer}, \bibinfo{person}{Mona Diab},
  \bibinfo{person}{Eneko Agirre}, \bibinfo{person}{Inigo Lopez-Gazpio}, {and}
  \bibinfo{person}{Lucia Specia}.} \bibinfo{year}{2017}\natexlab{}.
\newblock \showarticletitle{Semeval-2017 task 1: Semantic textual
  similarity-multilingual and cross-lingual focused evaluation}.
\newblock \bibinfo{journal}{\emph{arXiv preprint arXiv:1708.00055}}
  (\bibinfo{year}{2017}).
\newblock


\bibitem[Chen et~al\mbox{.}(2020)]%
        {DBLP:conf/www/ChenSZ20}
\bibfield{author}{\bibinfo{person}{Runjin Chen}, \bibinfo{person}{Yanyan Shen},
  {and} \bibinfo{person}{Dongxiang Zhang}.} \bibinfo{year}{2020}\natexlab{}.
\newblock \showarticletitle{{GNEM:} {A} Generic One-to-Set Neural Entity
  Matching Framework}. In \bibinfo{booktitle}{\emph{{WWW}}}.
  \bibinfo{pages}{1686--1694}.
\newblock


\bibitem[Christen({[n.\,d.]})]%
        {DBLP:conf/kdd/Christen08a}
\bibfield{author}{\bibinfo{person}{Peter Christen}.}
  \bibinfo{year}{[n.\,d.]}\natexlab{}.
\newblock \showarticletitle{Febrl -: an open source data cleaning,
  deduplication and record linkage system with a graphical user interface}. In
  \bibinfo{booktitle}{\emph{{SIGKDD}}}, \bibfield{editor}{\bibinfo{person}{Ying
  Li}, \bibinfo{person}{Bing Liu}, {and} \bibinfo{person}{Sunita Sarawagi}}
  (Eds.). \bibinfo{pages}{1065--1068}.
\newblock


\bibitem[Christen(2012a)]%
        {DBLP:books/daglib/0030287}
\bibfield{author}{\bibinfo{person}{Peter Christen}.}
  \bibinfo{year}{2012}\natexlab{a}.
\newblock \bibinfo{booktitle}{\emph{Data Matching - Concepts and Techniques for
  Record Linkage, Entity Resolution, and Duplicate Detection}}.
\newblock \bibinfo{publisher}{Springer}.
\newblock


\bibitem[Christen(2012b)]%
        {DBLP:journals/tkde/Christen12}
\bibfield{author}{\bibinfo{person}{Peter Christen}.}
  \bibinfo{year}{2012}\natexlab{b}.
\newblock \showarticletitle{A Survey of Indexing Techniques for Scalable Record
  Linkage and Deduplication}.
\newblock \bibinfo{journal}{\emph{{IEEE} Trans. Knowl. Data Eng.}}
  \bibinfo{volume}{24}, \bibinfo{number}{9} (\bibinfo{year}{2012}),
  \bibinfo{pages}{1537--1555}.
\newblock


\bibitem[Christophides et~al\mbox{.}(2021)]%
        {DBLP:journals/csur/ChristophidesEP21}
\bibfield{author}{\bibinfo{person}{Vassilis Christophides},
  \bibinfo{person}{Vasilis Efthymiou}, \bibinfo{person}{Themis Palpanas},
  \bibinfo{person}{George Papadakis}, {and} \bibinfo{person}{Kostas
  Stefanidis}.} \bibinfo{year}{2021}\natexlab{}.
\newblock \showarticletitle{An Overview of End-to-End Entity Resolution for Big
  Data}.
\newblock \bibinfo{journal}{\emph{{ACM} Comput. Surv.}} \bibinfo{volume}{53},
  \bibinfo{number}{6} (\bibinfo{year}{2021}), \bibinfo{pages}{127:1--127:42}.
\newblock


\bibitem[Christophides et~al\mbox{.}(2015)]%
        {DBLP:series/synthesis/2015Christophides}
\bibfield{author}{\bibinfo{person}{Vassilis Christophides},
  \bibinfo{person}{Vasilis Efthymiou}, {and} \bibinfo{person}{Kostas
  Stefanidis}.} \bibinfo{year}{2015}\natexlab{}.
\newblock \bibinfo{booktitle}{\emph{Entity Resolution in the Web of Data}}.
\newblock \bibinfo{publisher}{Morgan {\&} Claypool Publishers}.
\newblock


\bibitem[Devlin et~al\mbox{.}(2018)]%
        {devlin2018bert}
\bibfield{author}{\bibinfo{person}{Jacob Devlin}, \bibinfo{person}{Ming-Wei
  Chang}, \bibinfo{person}{Kenton Lee}, {and} \bibinfo{person}{Kristina
  Toutanova}.} \bibinfo{year}{2018}\natexlab{}.
\newblock \showarticletitle{Bert: Pre-training of deep bidirectional
  transformers for language understanding}.
\newblock \bibinfo{journal}{\emph{arXiv preprint arXiv:1810.04805}}
  (\bibinfo{year}{2018}).
\newblock


\bibitem[Dong and Srivastava(2013)]%
        {DBLP:journals/pvldb/DongS13}
\bibfield{author}{\bibinfo{person}{Xin~Luna Dong} {and} \bibinfo{person}{Divesh
  Srivastava}.} \bibinfo{year}{2013}\natexlab{}.
\newblock \showarticletitle{Big Data Integration}.
\newblock \bibinfo{journal}{\emph{Proc. {VLDB} Endow.}} \bibinfo{volume}{6},
  \bibinfo{number}{11} (\bibinfo{year}{2013}), \bibinfo{pages}{1188--1189}.
\newblock


\bibitem[Ebraheem et~al\mbox{.}(2018)]%
        {DBLP:journals/pvldb/EbraheemTJOT18}
\bibfield{author}{\bibinfo{person}{Muhammad Ebraheem},
  \bibinfo{person}{Saravanan Thirumuruganathan}, \bibinfo{person}{Shafiq~R.
  Joty}, \bibinfo{person}{Mourad Ouzzani}, {and} \bibinfo{person}{Nan Tang}.}
  \bibinfo{year}{2018}\natexlab{}.
\newblock \showarticletitle{Distributed Representations of Tuples for Entity
  Resolution}.
\newblock \bibinfo{journal}{\emph{Proc. {VLDB} Endow.}} \bibinfo{volume}{11},
  \bibinfo{number}{11} (\bibinfo{year}{2018}), \bibinfo{pages}{1454--1467}.
\newblock


\bibitem[Fu et~al\mbox{.}(2020)]%
        {DBLP:conf/ijcai/FuHHS20}
\bibfield{author}{\bibinfo{person}{Cheng Fu}, \bibinfo{person}{Xianpei Han},
  \bibinfo{person}{Jiaming He}, {and} \bibinfo{person}{Le Sun}.}
  \bibinfo{year}{2020}\natexlab{}.
\newblock \showarticletitle{Hierarchical Matching Network for Heterogeneous
  Entity Resolution}. In \bibinfo{booktitle}{\emph{{IJCAI}}}.
  \bibinfo{pages}{3665--3671}.
\newblock


\bibitem[Getoor and Machanavajjhala(2012)]%
        {DBLP:journals/pvldb/GetoorM12}
\bibfield{author}{\bibinfo{person}{Lise Getoor} {and} \bibinfo{person}{Ashwin
  Machanavajjhala}.} \bibinfo{year}{2012}\natexlab{}.
\newblock \showarticletitle{Entity Resolution: Theory, Practice {\&} Open
  Challenges}.
\newblock \bibinfo{journal}{\emph{Proc. {VLDB} Endow.}} \bibinfo{volume}{5},
  \bibinfo{number}{12} (\bibinfo{year}{2012}), \bibinfo{pages}{2018--2019}.
\newblock


\bibitem[Hinton et~al\mbox{.}(2015)]%
        {hinton2015distilling}
\bibfield{author}{\bibinfo{person}{Geoffrey Hinton}, \bibinfo{person}{Oriol
  Vinyals}, \bibinfo{person}{Jeff Dean}, {et~al\mbox{.}}}
  \bibinfo{year}{2015}\natexlab{}.
\newblock \showarticletitle{Distilling the knowledge in a neural network}.
\newblock \bibinfo{journal}{\emph{arXiv preprint arXiv:1503.02531}}
  \bibinfo{volume}{2}, \bibinfo{number}{7} (\bibinfo{year}{2015}).
\newblock


\bibitem[Jiao et~al\mbox{.}(2019)]%
        {jiao2019tinybert}
\bibfield{author}{\bibinfo{person}{Xiaoqi Jiao}, \bibinfo{person}{Yichun Yin},
  \bibinfo{person}{Lifeng Shang}, \bibinfo{person}{Xin Jiang},
  \bibinfo{person}{Xiao Chen}, \bibinfo{person}{Linlin Li},
  \bibinfo{person}{Fang Wang}, {and} \bibinfo{person}{Qun Liu}.}
  \bibinfo{year}{2019}\natexlab{}.
\newblock \showarticletitle{Tinybert: Distilling bert for natural language
  understanding}.
\newblock \bibinfo{journal}{\emph{arXiv preprint arXiv:1909.10351}}
  (\bibinfo{year}{2019}).
\newblock


\bibitem[Kasai et~al\mbox{.}(2019)]%
        {DBLP:conf/acl/KasaiQGLP19}
\bibfield{author}{\bibinfo{person}{Jungo Kasai}, \bibinfo{person}{Kun Qian},
  \bibinfo{person}{Sairam Gurajada}, \bibinfo{person}{Yunyao Li}, {and}
  \bibinfo{person}{Lucian Popa}.} \bibinfo{year}{2019}\natexlab{}.
\newblock \showarticletitle{Low-resource Deep Entity Resolution with Transfer
  and Active Learning}. In \bibinfo{booktitle}{\emph{{ACL}}}.
  \bibinfo{pages}{5851--5861}.
\newblock


\bibitem[Kenig and Gal(2013)]%
        {DBLP:journals/is/KenigG13}
\bibfield{author}{\bibinfo{person}{Batya Kenig} {and} \bibinfo{person}{Avigdor
  Gal}.} \bibinfo{year}{2013}\natexlab{}.
\newblock \showarticletitle{MFIBlocks: An effective blocking algorithm for
  entity resolution}.
\newblock \bibinfo{journal}{\emph{Inf. Syst.}} \bibinfo{volume}{38},
  \bibinfo{number}{6} (\bibinfo{year}{2013}), \bibinfo{pages}{908--926}.
\newblock


\bibitem[K{\"{o}}pcke et~al\mbox{.}(2010)]%
        {DBLP:journals/pvldb/KopckeTR10}
\bibfield{author}{\bibinfo{person}{Hanna K{\"{o}}pcke},
  \bibinfo{person}{Andreas Thor}, {and} \bibinfo{person}{Erhard Rahm}.}
  \bibinfo{year}{2010}\natexlab{}.
\newblock \showarticletitle{Evaluation of entity resolution approaches on
  real-world match problems}.
\newblock \bibinfo{journal}{\emph{Proc. {VLDB} Endow.}} \bibinfo{volume}{3},
  \bibinfo{number}{1} (\bibinfo{year}{2010}), \bibinfo{pages}{484--493}.
\newblock


\bibitem[Lacoste{-}Julien et~al\mbox{.}(2013)]%
        {DBLP:conf/kdd/Lacoste-JulienPDKGG13}
\bibfield{author}{\bibinfo{person}{Simon Lacoste{-}Julien},
  \bibinfo{person}{Konstantina Palla}, \bibinfo{person}{Alex Davies},
  \bibinfo{person}{Gjergji Kasneci}, \bibinfo{person}{Thore Graepel}, {and}
  \bibinfo{person}{Zoubin Ghahramani}.} \bibinfo{year}{2013}\natexlab{}.
\newblock \showarticletitle{SIGMa: simple greedy matching for aligning large
  knowledge bases}. In \bibinfo{booktitle}{\emph{{KDD}}}.
  \bibinfo{pages}{572--580}.
\newblock


\bibitem[Lan et~al\mbox{.}(2019)]%
        {lan2019albert}
\bibfield{author}{\bibinfo{person}{Zhenzhong Lan}, \bibinfo{person}{Mingda
  Chen}, \bibinfo{person}{Sebastian Goodman}, \bibinfo{person}{Kevin Gimpel},
  \bibinfo{person}{Piyush Sharma}, {and} \bibinfo{person}{Radu Soricut}.}
  \bibinfo{year}{2019}\natexlab{}.
\newblock \showarticletitle{Albert: A lite bert for self-supervised learning of
  language representations}.
\newblock \bibinfo{journal}{\emph{arXiv preprint arXiv:1909.11942}}
  (\bibinfo{year}{2019}).
\newblock


\bibitem[Li et~al\mbox{.}(2020c)]%
        {DBLP:conf/aaai/Li0SZAW20}
\bibfield{author}{\bibinfo{person}{Bing Li}, \bibinfo{person}{Wei Wang},
  \bibinfo{person}{Yifang Sun}, \bibinfo{person}{Linhan Zhang},
  \bibinfo{person}{Muhammad~Asif Ali}, {and} \bibinfo{person}{Yi Wang}.}
  \bibinfo{year}{2020}\natexlab{c}.
\newblock \showarticletitle{GraphER: Token-Centric Entity Resolution with Graph
  Convolutional Neural Networks}. In \bibinfo{booktitle}{\emph{{IAAI}}}.
  \bibinfo{pages}{8172--8179}.
\newblock


\bibitem[Li et~al\mbox{.}(2019)]%
        {li2019approximate}
\bibfield{author}{\bibinfo{person}{Wen Li}, \bibinfo{person}{Ying Zhang},
  \bibinfo{person}{Yifang Sun}, \bibinfo{person}{Wei Wang},
  \bibinfo{person}{Mingjie Li}, \bibinfo{person}{Wenjie Zhang}, {and}
  \bibinfo{person}{Xuemin Lin}.} \bibinfo{year}{2019}\natexlab{}.
\newblock \showarticletitle{Approximate nearest neighbor search on high
  dimensional data—experiments, analyses, and improvement}.
\newblock \bibinfo{journal}{\emph{IEEE Transactions on Knowledge and Data
  Engineering}} \bibinfo{volume}{32}, \bibinfo{number}{8}
  (\bibinfo{year}{2019}), \bibinfo{pages}{1475--1488}.
\newblock


\bibitem[Li et~al\mbox{.}(2020a)]%
        {DBLP:journals/pvldb/0001LSDT20}
\bibfield{author}{\bibinfo{person}{Yuliang Li}, \bibinfo{person}{Jinfeng Li},
  \bibinfo{person}{Yoshihiko Suhara}, \bibinfo{person}{AnHai Doan}, {and}
  \bibinfo{person}{Wang{-}Chiew Tan}.} \bibinfo{year}{2020}\natexlab{a}.
\newblock \showarticletitle{Deep Entity Matching with Pre-Trained Language
  Models}.
\newblock \bibinfo{journal}{\emph{Proc. {VLDB} Endow.}} \bibinfo{volume}{14},
  \bibinfo{number}{1} (\bibinfo{year}{2020}), \bibinfo{pages}{50--60}.
\newblock


\bibitem[Li et~al\mbox{.}(2020b)]%
        {DBLP:journals/corr/abs-2004-00584}
\bibfield{author}{\bibinfo{person}{Yuliang Li}, \bibinfo{person}{Jinfeng Li},
  \bibinfo{person}{Yoshihiko Suhara}, \bibinfo{person}{AnHai Doan}, {and}
  \bibinfo{person}{Wang{-}Chiew Tan}.} \bibinfo{year}{2020}\natexlab{b}.
\newblock \showarticletitle{Deep Entity Matching with Pre-Trained Language
  Models}.
\newblock \bibinfo{journal}{\emph{CoRR}}  \bibinfo{volume}{abs/2004.00584}
  (\bibinfo{year}{2020}).
\newblock
\urldef\tempurl%
\url{https://arxiv.org/abs/2004.00584}
\showURL{%
\tempurl}


\bibitem[Liu et~al\mbox{.}(2020)]%
        {DBLP:journals/corr/abs-2003-07278}
\bibfield{author}{\bibinfo{person}{Qi Liu}, \bibinfo{person}{Matt~J. Kusner},
  {and} \bibinfo{person}{Phil Blunsom}.} \bibinfo{year}{2020}\natexlab{}.
\newblock \showarticletitle{A Survey on Contextual Embeddings}.
\newblock \bibinfo{journal}{\emph{CoRR}}  \bibinfo{volume}{abs/2003.07278}
  (\bibinfo{year}{2020}).
\newblock
\urldef\tempurl%
\url{https://arxiv.org/abs/2003.07278}
\showURL{%
\tempurl}


\bibitem[Liu et~al\mbox{.}(2019)]%
        {liu2019roberta}
\bibfield{author}{\bibinfo{person}{Yinhan Liu}, \bibinfo{person}{Myle Ott},
  \bibinfo{person}{Naman Goyal}, \bibinfo{person}{Jingfei Du},
  \bibinfo{person}{Mandar Joshi}, \bibinfo{person}{Danqi Chen},
  \bibinfo{person}{Omer Levy}, \bibinfo{person}{Mike Lewis},
  \bibinfo{person}{Luke Zettlemoyer}, {and} \bibinfo{person}{Veselin
  Stoyanov}.} \bibinfo{year}{2019}\natexlab{}.
\newblock \showarticletitle{Roberta: A robustly optimized bert pretraining
  approach}.
\newblock \bibinfo{journal}{\emph{arXiv preprint arXiv:1907.11692}}
  (\bibinfo{year}{2019}).
\newblock


\bibitem[Malkov and Yashunin(2020)]%
        {DBLP:journals/pami/MalkovY20}
\bibfield{author}{\bibinfo{person}{Yury~A. Malkov} {and}
  \bibinfo{person}{Dmitry~A. Yashunin}.} \bibinfo{year}{2020}\natexlab{}.
\newblock \showarticletitle{Efficient and Robust Approximate Nearest Neighbor
  Search Using Hierarchical Navigable Small World Graphs}.
\newblock \bibinfo{journal}{\emph{{IEEE} Trans. Pattern Anal. Mach. Intell.}}
  \bibinfo{volume}{42}, \bibinfo{number}{4} (\bibinfo{year}{2020}),
  \bibinfo{pages}{824--836}.
\newblock


\bibitem[Manning et~al\mbox{.}(2008)]%
        {DBLP:books/daglib/0021593}
\bibfield{author}{\bibinfo{person}{Christopher~D. Manning},
  \bibinfo{person}{Prabhakar Raghavan}, {and} \bibinfo{person}{Hinrich
  Sch{\"{u}}tze}.} \bibinfo{year}{2008}\natexlab{}.
\newblock \bibinfo{booktitle}{\emph{Introduction to information retrieval}}.
\newblock \bibinfo{publisher}{Cambridge University Press}.
\newblock


\bibitem[Mikolov et~al\mbox{.}(2013a)]%
        {mikolov2013efficient}
\bibfield{author}{\bibinfo{person}{Tomas Mikolov}, \bibinfo{person}{Kai Chen},
  \bibinfo{person}{Greg Corrado}, {and} \bibinfo{person}{Jeffrey Dean}.}
  \bibinfo{year}{2013}\natexlab{a}.
\newblock \showarticletitle{Efficient estimation of word representations in
  vector space}.
\newblock \bibinfo{journal}{\emph{arXiv preprint arXiv:1301.3781}}
  (\bibinfo{year}{2013}).
\newblock


\bibitem[Mikolov et~al\mbox{.}(2013b)]%
        {mikolov2013distributed}
\bibfield{author}{\bibinfo{person}{Tomas Mikolov}, \bibinfo{person}{Ilya
  Sutskever}, \bibinfo{person}{Kai Chen}, \bibinfo{person}{Greg~S Corrado},
  {and} \bibinfo{person}{Jeff Dean}.} \bibinfo{year}{2013}\natexlab{b}.
\newblock \showarticletitle{Distributed representations of words and phrases
  and their compositionality}.
\newblock \bibinfo{journal}{\emph{Advances in neural information processing
  systems}}  \bibinfo{volume}{26} (\bibinfo{year}{2013}).
\newblock


\bibitem[Mudgal et~al\mbox{.}(2018)]%
        {Mudgal2018sigmod}
\bibfield{author}{\bibinfo{person}{Sidharth Mudgal}, \bibinfo{person}{Han Li},
  \bibinfo{person}{Theodoros Rekatsinas}, \bibinfo{person}{AnHai Doan},
  \bibinfo{person}{Youngchoon Park}, \bibinfo{person}{Ganesh Krishnan},
  \bibinfo{person}{Rohit Deep}, \bibinfo{person}{Esteban Arcaute}, {and}
  \bibinfo{person}{Vijay Raghavendra}.} \bibinfo{year}{2018}\natexlab{}.
\newblock \showarticletitle{Deep Learning for Entity Matching: A Design Space
  Exploration}. In \bibinfo{booktitle}{\emph{SIGMOD}}. \bibinfo{pages}{19--34}.
\newblock


\bibitem[Ni et~al\mbox{.}(2021)]%
        {ni2021large}
\bibfield{author}{\bibinfo{person}{Jianmo Ni}, \bibinfo{person}{Chen Qu},
  \bibinfo{person}{Jing Lu}, \bibinfo{person}{Zhuyun Dai},
  \bibinfo{person}{Gustavo~Hern{\'a}ndez {\'A}brego}, \bibinfo{person}{Ji Ma},
  \bibinfo{person}{Vincent~Y Zhao}, \bibinfo{person}{Yi Luan},
  \bibinfo{person}{Keith~B Hall}, \bibinfo{person}{Ming-Wei Chang},
  {et~al\mbox{.}}} \bibinfo{year}{2021}\natexlab{}.
\newblock \showarticletitle{Large dual encoders are generalizable retrievers}.
\newblock \bibinfo{journal}{\emph{arXiv preprint arXiv:2112.07899}}
  (\bibinfo{year}{2021}).
\newblock


\bibitem[Nie et~al\mbox{.}(2019)]%
        {Nie2019cikm}
\bibfield{author}{\bibinfo{person}{Hao Nie}, \bibinfo{person}{Xianpei Han},
  \bibinfo{person}{Ben He}, \bibinfo{person}{Le Sun}, \bibinfo{person}{Bo
  Chen}, \bibinfo{person}{Wei Zhang}, \bibinfo{person}{Suhui Wu}, {and}
  \bibinfo{person}{Hao Kong}.} \bibinfo{year}{2019}\natexlab{}.
\newblock \showarticletitle{Deep Sequence-to-Sequence Entity Matching for
  Heterogeneous Entity Resolution}. In \bibinfo{booktitle}{\emph{International
  Conference on Information and Knowledge Management}}.
  \bibinfo{pages}{629--638}.
\newblock


\bibitem[Obraczka et~al\mbox{.}(2021)]%
        {DBLP:journals/corr/abs-2101-06126}
\bibfield{author}{\bibinfo{person}{Daniel Obraczka}, \bibinfo{person}{Jonathan
  Schuchart}, {and} \bibinfo{person}{Erhard Rahm}.}
  \bibinfo{year}{2021}\natexlab{}.
\newblock \showarticletitle{{EAGER:} Embedding-Assisted Entity Resolution for
  Knowledge Graphs}.
\newblock \bibinfo{journal}{\emph{CoRR}}  \bibinfo{volume}{abs/2101.06126}
  (\bibinfo{year}{2021}).
\newblock


\bibitem[Paganelli et~al\mbox{.}(2022)]%
        {DBLP:journals/pvldb/PaganelliBBG22}
\bibfield{author}{\bibinfo{person}{Matteo Paganelli},
  \bibinfo{person}{Francesco~Del Buono}, \bibinfo{person}{Andrea Baraldi},
  {and} \bibinfo{person}{Francesco Guerra}.} \bibinfo{year}{2022}\natexlab{}.
\newblock \showarticletitle{Analyzing How {BERT} Performs Entity Matching}.
\newblock \bibinfo{journal}{\emph{Proc. {VLDB} Endow.}} \bibinfo{volume}{15},
  \bibinfo{number}{8} (\bibinfo{year}{2022}), \bibinfo{pages}{1726--1738}.
\newblock


\bibitem[Paganelli et~al\mbox{.}(2021)]%
        {Paganelli2021edbt}
\bibfield{author}{\bibinfo{person}{Matteo Paganelli},
  \bibinfo{person}{Francesco Del~Buono}, \bibinfo{person}{Pevarello Marco},
  \bibinfo{person}{Francesco Guerra}, {and} \bibinfo{person}{Maurizio
  Vincini}.} \bibinfo{year}{2021}\natexlab{}.
\newblock \showarticletitle{Automated machine learning for entity matching
  tasks}. In \bibinfo{booktitle}{\emph{EDBT}}.
\newblock


\bibitem[Papadakis et~al\mbox{.}(2015)]%
        {DBLP:journals/pvldb/0001APK15}
\bibfield{author}{\bibinfo{person}{George Papadakis}, \bibinfo{person}{George
  Alexiou}, \bibinfo{person}{George Papastefanatos}, {and}
  \bibinfo{person}{Georgia Koutrika}.} \bibinfo{year}{2015}\natexlab{}.
\newblock \showarticletitle{Schema-agnostic vs Schema-based Configurations for
  Blocking Methods on Homogeneous Data}.
\newblock \bibinfo{journal}{\emph{Proc. {VLDB} Endow.}} \bibinfo{volume}{9},
  \bibinfo{number}{4} (\bibinfo{year}{2015}), \bibinfo{pages}{312--323}.
\newblock


\bibitem[Papadakis et~al\mbox{.}(2022)]%
        {DBLP:conf/edbt/0001ETH22}
\bibfield{author}{\bibinfo{person}{George Papadakis}, \bibinfo{person}{Vasilis
  Efthymiou}, \bibinfo{person}{Emmanouil Thanos}, {and} \bibinfo{person}{Oktie
  Hassanzadeh}.} \bibinfo{year}{2022}\natexlab{}.
\newblock \showarticletitle{Bipartite Graph Matching Algorithms for Clean-Clean
  Entity Resolution: An Empirical Evaluation}. In
  \bibinfo{booktitle}{\emph{{EDBT}}}. \bibinfo{pages}{2:462--2:474}.
\newblock


\bibitem[Papadakis et~al\mbox{.}(2011)]%
        {DBLP:conf/wsdm/PapadakisINF11}
\bibfield{author}{\bibinfo{person}{George Papadakis},
  \bibinfo{person}{Ekaterini Ioannou}, \bibinfo{person}{Claudia
  Nieder{\'{e}}e}, {and} \bibinfo{person}{Peter Fankhauser}.}
  \bibinfo{year}{2011}\natexlab{}.
\newblock \showarticletitle{Efficient entity resolution for large heterogeneous
  information spaces}. In \bibinfo{booktitle}{\emph{{WSDM}}}.
  \bibinfo{pages}{535--544}.
\newblock


\bibitem[Papadakis et~al\mbox{.}(2021a)]%
        {DBLP:series/synthesis/2021Papadakis}
\bibfield{author}{\bibinfo{person}{George Papadakis},
  \bibinfo{person}{Ekaterini Ioannou}, \bibinfo{person}{Emanouil Thanos}, {and}
  \bibinfo{person}{Themis Palpanas}.} \bibinfo{year}{2021}\natexlab{a}.
\newblock \bibinfo{booktitle}{\emph{The Four Generations of Entity
  Resolution}}.
\newblock \bibinfo{publisher}{Morgan {\&} Claypool Publishers}.
\newblock


\bibitem[Papadakis et~al\mbox{.}(2021b)]%
        {DBLP:journals/csur/PapadakisSTP20}
\bibfield{author}{\bibinfo{person}{George Papadakis},
  \bibinfo{person}{Dimitrios Skoutas}, \bibinfo{person}{Emmanouil Thanos},
  {and} \bibinfo{person}{Themis Palpanas}.} \bibinfo{year}{2021}\natexlab{b}.
\newblock \showarticletitle{Blocking and Filtering Techniques for Entity
  Resolution: {A} Survey}.
\newblock \bibinfo{journal}{\emph{{ACM} Comput. Surv.}} \bibinfo{volume}{53},
  \bibinfo{number}{2} (\bibinfo{year}{2021}), \bibinfo{pages}{31:1--31:42}.
\newblock


\bibitem[Papadakis et~al\mbox{.}(2016)]%
        {DBLP:journals/pvldb/0001SGP16}
\bibfield{author}{\bibinfo{person}{George Papadakis}, \bibinfo{person}{Jonathan
  Svirsky}, \bibinfo{person}{Avigdor Gal}, {and} \bibinfo{person}{Themis
  Palpanas}.} \bibinfo{year}{2016}\natexlab{}.
\newblock \showarticletitle{Comparative Analysis of Approximate Blocking
  Techniques for Entity Resolution}.
\newblock \bibinfo{journal}{\emph{Proc. {VLDB} Endow.}} \bibinfo{volume}{9},
  \bibinfo{number}{9} (\bibinfo{year}{2016}), \bibinfo{pages}{684--695}.
\newblock


\bibitem[Peeters and Bizer(2021)]%
        {DBLP:journals/pvldb/PeetersB21}
\bibfield{author}{\bibinfo{person}{Ralph Peeters} {and}
  \bibinfo{person}{Christian Bizer}.} \bibinfo{year}{2021}\natexlab{}.
\newblock \showarticletitle{Dual-Objective Fine-Tuning of {BERT} for Entity
  Matching}.
\newblock \bibinfo{journal}{\emph{Proc. {VLDB} Endow.}} \bibinfo{volume}{14},
  \bibinfo{number}{10} (\bibinfo{year}{2021}), \bibinfo{pages}{1913--1921}.
\newblock


\bibitem[Pennington et~al\mbox{.}(2014)]%
        {pennington2014glove}
\bibfield{author}{\bibinfo{person}{Jeffrey Pennington},
  \bibinfo{person}{Richard Socher}, {and} \bibinfo{person}{Christopher~D
  Manning}.} \bibinfo{year}{2014}\natexlab{}.
\newblock \showarticletitle{Glove: Global vectors for word representation}. In
  \bibinfo{booktitle}{\emph{Proceedings of the 2014 conference on empirical
  methods in natural language processing (EMNLP)}}.
  \bibinfo{pages}{1532--1543}.
\newblock


\bibitem[Pilehvar and Camacho{-}Collados(2020)]%
        {DBLP:series/synthesis/2020Pilehvar}
\bibfield{author}{\bibinfo{person}{Mohammad~Taher Pilehvar} {and}
  \bibinfo{person}{Jos{\'{e}} Camacho{-}Collados}.}
  \bibinfo{year}{2020}\natexlab{}.
\newblock \bibinfo{booktitle}{\emph{Embeddings in Natural Language Processing:
  Theory and Advances in Vector Representations of Meaning}}.
\newblock \bibinfo{publisher}{Morgan {\&} Claypool Publishers}.
\newblock


\bibitem[Raffel et~al\mbox{.}(2020)]%
        {2020t5}
\bibfield{author}{\bibinfo{person}{Colin Raffel}, \bibinfo{person}{Noam
  Shazeer}, \bibinfo{person}{Adam Roberts}, \bibinfo{person}{Katherine Lee},
  \bibinfo{person}{Sharan Narang}, \bibinfo{person}{Michael Matena},
  \bibinfo{person}{Yanqi Zhou}, \bibinfo{person}{Wei Li}, {and}
  \bibinfo{person}{Peter~J. Liu}.} \bibinfo{year}{2020}\natexlab{}.
\newblock \showarticletitle{Exploring the Limits of Transfer Learning with a
  Unified Text-to-Text Transformer}.
\newblock \bibinfo{journal}{\emph{Journal of Machine Learning Research}}
  \bibinfo{volume}{21}, \bibinfo{number}{140} (\bibinfo{year}{2020}),
  \bibinfo{pages}{1--67}.
\newblock


\bibitem[Reimers and Gurevych(2019)]%
        {reimers2019sentence}
\bibfield{author}{\bibinfo{person}{Nils Reimers} {and} \bibinfo{person}{Iryna
  Gurevych}.} \bibinfo{year}{2019}\natexlab{}.
\newblock \showarticletitle{Sentence-bert: Sentence embeddings using siamese
  bert-networks}.
\newblock \bibinfo{journal}{\emph{arXiv preprint arXiv:1908.10084}}
  (\bibinfo{year}{2019}).
\newblock


\bibitem[Romero et~al\mbox{.}(2014)]%
        {romero2014fitnets}
\bibfield{author}{\bibinfo{person}{Adriana Romero}, \bibinfo{person}{Nicolas
  Ballas}, \bibinfo{person}{Samira~Ebrahimi Kahou}, \bibinfo{person}{Antoine
  Chassang}, \bibinfo{person}{Carlo Gatta}, {and} \bibinfo{person}{Yoshua
  Bengio}.} \bibinfo{year}{2014}\natexlab{}.
\newblock \showarticletitle{Fitnets: Hints for thin deep nets}.
\newblock \bibinfo{journal}{\emph{arXiv preprint arXiv:1412.6550}}
  (\bibinfo{year}{2014}).
\newblock


\bibitem[Sanh et~al\mbox{.}(2019)]%
        {sanh2019distilbert}
\bibfield{author}{\bibinfo{person}{Victor Sanh}, \bibinfo{person}{Lysandre
  Debut}, \bibinfo{person}{Julien Chaumond}, {and} \bibinfo{person}{Thomas
  Wolf}.} \bibinfo{year}{2019}\natexlab{}.
\newblock \showarticletitle{DistilBERT, a distilled version of BERT: smaller,
  faster, cheaper and lighter}.
\newblock \bibinfo{journal}{\emph{arXiv preprint arXiv:1910.01108}}
  (\bibinfo{year}{2019}).
\newblock


\bibitem[Song et~al\mbox{.}(2020)]%
        {song2020mpnet}
\bibfield{author}{\bibinfo{person}{Kaitao Song}, \bibinfo{person}{Xu Tan},
  \bibinfo{person}{Tao Qin}, \bibinfo{person}{Jianfeng Lu}, {and}
  \bibinfo{person}{Tie-Yan Liu}.} \bibinfo{year}{2020}\natexlab{}.
\newblock \showarticletitle{Mpnet: Masked and permuted pre-training for
  language understanding}.
\newblock \bibinfo{journal}{\emph{Advances in Neural Information Processing
  Systems}}  \bibinfo{volume}{33} (\bibinfo{year}{2020}),
  \bibinfo{pages}{16857--16867}.
\newblock


\bibitem[Sun et~al\mbox{.}(2019)]%
        {sun2019mobilebert}
\bibfield{author}{\bibinfo{person}{Zhiqing Sun}, \bibinfo{person}{Hongkun Yu},
  \bibinfo{person}{Xiaodan Song}, \bibinfo{person}{Renjie Liu},
  \bibinfo{person}{Yiming Yang}, {and} \bibinfo{person}{Denny Zhou}.}
  \bibinfo{year}{2019}\natexlab{}.
\newblock \showarticletitle{Mobilebert: Task-agnostic compression of bert by
  progressive knowledge transfer}.
\newblock  (\bibinfo{year}{2019}).
\newblock


\bibitem[Sutskever et~al\mbox{.}(2014)]%
        {sutskever2014sequence}
\bibfield{author}{\bibinfo{person}{Ilya Sutskever}, \bibinfo{person}{Oriol
  Vinyals}, {and} \bibinfo{person}{Quoc~V Le}.}
  \bibinfo{year}{2014}\natexlab{}.
\newblock \showarticletitle{Sequence to sequence learning with neural
  networks}.
\newblock \bibinfo{journal}{\emph{Advances in neural information processing
  systems}}  \bibinfo{volume}{27} (\bibinfo{year}{2014}).
\newblock


\bibitem[Thirumuruganathan et~al\mbox{.}(2021)]%
        {DBLP:journals/pvldb/Thirumuruganathan21}
\bibfield{author}{\bibinfo{person}{Saravanan Thirumuruganathan},
  \bibinfo{person}{Han Li}, \bibinfo{person}{Nan Tang}, \bibinfo{person}{Mourad
  Ouzzani}, \bibinfo{person}{Yash Govind}, \bibinfo{person}{Derek Paulsen},
  \bibinfo{person}{Glenn Fung}, {and} \bibinfo{person}{AnHai Doan}.}
  \bibinfo{year}{2021}\natexlab{}.
\newblock \showarticletitle{Deep Learning for Blocking in Entity Matching: {A}
  Design Space Exploration}.
\newblock \bibinfo{journal}{\emph{Proc. {VLDB} Endow.}} \bibinfo{volume}{14},
  \bibinfo{number}{11} (\bibinfo{year}{2021}), \bibinfo{pages}{2459--2472}.
\newblock


\bibitem[Trummer(2022)]%
        {DBLP:journals/pvldb/Trummer22b}
\bibfield{author}{\bibinfo{person}{Immanuel Trummer}.}
  \bibinfo{year}{2022}\natexlab{}.
\newblock \showarticletitle{From {BERT} to {GPT-3} Codex: Harnessing the
  Potential of Very Large Language Models for Data Management}.
\newblock \bibinfo{journal}{\emph{Proc. {VLDB} Endow.}} \bibinfo{volume}{15},
  \bibinfo{number}{12} (\bibinfo{year}{2022}), \bibinfo{pages}{3770--3773}.
\newblock


\bibitem[Vaswani et~al\mbox{.}(2017)]%
        {vaswani2017attention}
\bibfield{author}{\bibinfo{person}{Ashish Vaswani}, \bibinfo{person}{Noam
  Shazeer}, \bibinfo{person}{Niki Parmar}, \bibinfo{person}{Jakob Uszkoreit},
  \bibinfo{person}{Llion Jones}, \bibinfo{person}{Aidan~N Gomez},
  \bibinfo{person}{{\L}ukasz Kaiser}, {and} \bibinfo{person}{Illia
  Polosukhin}.} \bibinfo{year}{2017}\natexlab{}.
\newblock \showarticletitle{Attention is all you need}.
\newblock \bibinfo{journal}{\emph{Advances in neural information processing
  systems}}  \bibinfo{volume}{30} (\bibinfo{year}{2017}).
\newblock


\bibitem[Wang et~al\mbox{.}(2018)]%
        {wang2018glue}
\bibfield{author}{\bibinfo{person}{Alex Wang}, \bibinfo{person}{Amanpreet
  Singh}, \bibinfo{person}{Julian Michael}, \bibinfo{person}{Felix Hill},
  \bibinfo{person}{Omer Levy}, {and} \bibinfo{person}{Samuel~R Bowman}.}
  \bibinfo{year}{2018}\natexlab{}.
\newblock \showarticletitle{GLUE: A multi-task benchmark and analysis platform
  for natural language understanding}.
\newblock \bibinfo{journal}{\emph{arXiv preprint arXiv:1804.07461}}
  (\bibinfo{year}{2018}).
\newblock


\bibitem[Wang et~al\mbox{.}(2020b)]%
        {wang2020minilm}
\bibfield{author}{\bibinfo{person}{Wenhui Wang}, \bibinfo{person}{Furu Wei},
  \bibinfo{person}{Li Dong}, \bibinfo{person}{Hangbo Bao}, \bibinfo{person}{Nan
  Yang}, {and} \bibinfo{person}{Ming Zhou}.} \bibinfo{year}{2020}\natexlab{b}.
\newblock \showarticletitle{Minilm: Deep self-attention distillation for
  task-agnostic compression of pre-trained transformers}.
\newblock \bibinfo{journal}{\emph{Advances in Neural Information Processing
  Systems}}  \bibinfo{volume}{33} (\bibinfo{year}{2020}),
  \bibinfo{pages}{5776--5788}.
\newblock


\bibitem[Wang et~al\mbox{.}(2020a)]%
        {DBLP:conf/icdm/WangSWDJ20}
\bibfield{author}{\bibinfo{person}{Zhengyang Wang}, \bibinfo{person}{Bunyamin
  Sisman}, \bibinfo{person}{Hao Wei}, \bibinfo{person}{Xin~Luna Dong}, {and}
  \bibinfo{person}{Shuiwang Ji}.} \bibinfo{year}{2020}\natexlab{a}.
\newblock \showarticletitle{CorDEL: {A} Contrastive Deep Learning Approach for
  Entity Linkage}. In \bibinfo{booktitle}{\emph{{ICDM}}}.
  \bibinfo{pages}{1322--1327}.
\newblock


\bibitem[Wu et~al\mbox{.}(2020)]%
        {wu2020zeroer}
\bibfield{author}{\bibinfo{person}{Renzhi Wu}, \bibinfo{person}{Sanya Chaba},
  \bibinfo{person}{Saurabh Sawlani}, \bibinfo{person}{Xu Chu}, {and}
  \bibinfo{person}{Saravanan Thirumuruganathan}.}
  \bibinfo{year}{2020}\natexlab{}.
\newblock \showarticletitle{Zeroer: Entity resolution using zero labeled
  examples}. In \bibinfo{booktitle}{\emph{Proceedings of the 2020 ACM SIGMOD
  International Conference on Management of Data}}.
  \bibinfo{pages}{1149--1164}.
\newblock


\bibitem[Yang et~al\mbox{.}(2019)]%
        {yang2019xlnet}
\bibfield{author}{\bibinfo{person}{Zhilin Yang}, \bibinfo{person}{Zihang Dai},
  \bibinfo{person}{Yiming Yang}, \bibinfo{person}{Jaime Carbonell},
  \bibinfo{person}{Russ~R Salakhutdinov}, {and} \bibinfo{person}{Quoc~V Le}.}
  \bibinfo{year}{2019}\natexlab{}.
\newblock \showarticletitle{Xlnet: Generalized autoregressive pretraining for
  language understanding}.
\newblock \bibinfo{journal}{\emph{Advances in neural information processing
  systems}}  \bibinfo{volume}{32} (\bibinfo{year}{2019}).
\newblock


\bibitem[Yao et~al\mbox{.}(2021)]%
        {DBLP:conf/acl/YaoLDLY0LZD20}
\bibfield{author}{\bibinfo{person}{Zijun Yao}, \bibinfo{person}{Chengjiang Li},
  \bibinfo{person}{Tiansi Dong}, \bibinfo{person}{Xin Lv},
  \bibinfo{person}{Jifan Yu}, \bibinfo{person}{Lei Hou},
  \bibinfo{person}{Juanzi Li}, \bibinfo{person}{Yichi Zhang}, {and}
  \bibinfo{person}{Zelin Dai}.} \bibinfo{year}{2021}\natexlab{}.
\newblock \showarticletitle{Interpretable and Low-Resource Entity Matching via
  Decoupling Feature Learning from Decision Making}. In
  \bibinfo{booktitle}{\emph{{ACL/IJCNLP}}}. \bibinfo{pages}{2770--2781}.
\newblock


\bibitem[Zhang et~al\mbox{.}(2020a)]%
        {DBLP:conf/www/ZhangNWST20}
\bibfield{author}{\bibinfo{person}{Dongxiang Zhang}, \bibinfo{person}{Yuyang
  Nie}, \bibinfo{person}{Sai Wu}, \bibinfo{person}{Yanyan Shen}, {and}
  \bibinfo{person}{Kian{-}Lee Tan}.} \bibinfo{year}{2020}\natexlab{a}.
\newblock \showarticletitle{Multi-Context Attention for Entity Matching}. In
  \bibinfo{booktitle}{\emph{{WWW}}}. \bibinfo{pages}{2634--2640}.
\newblock


\bibitem[Zhang et~al\mbox{.}(2020b)]%
        {DBLP:conf/wsdm/ZhangWSDFP20}
\bibfield{author}{\bibinfo{person}{Wei Zhang}, \bibinfo{person}{Hao Wei},
  \bibinfo{person}{Bunyamin Sisman}, \bibinfo{person}{Xin~Luna Dong},
  \bibinfo{person}{Christos Faloutsos}, {and} \bibinfo{person}{David Page}.}
  \bibinfo{year}{2020}\natexlab{b}.
\newblock \showarticletitle{AutoBlock: {A} Hands-off Blocking Framework for
  Entity Matching}. In \bibinfo{booktitle}{\emph{{WSDM}}}.
  \bibinfo{pages}{744--752}.
\newblock


\end{thebibliography}

\section{Appendix I: Schema-based Experiments}

% \begin{table}

% \begin{tabular}{lrrrrr}
% \toprule
% {} &  Blocking (sec) &  Matching (sec) &  Precision &  Recall &    F1 \\
% Dataset &                 &                 &            &         &       \\
% \midrule
% D1      &            2.49 &            0.88 &       0.80 &    0.44 &  0.57 \\
% D2      &         1914.52 &           30.22 &       0.66 &    0.52 &  0.58 \\
% D3      &        67727.01 &           79.09 &       0.48 &    0.31 &  0.38 \\
% D4      &          616.72 &           72.36 &       1.00 &    0.98 &  0.99 \\
% D5      &          161.12 &           12.70 &       0.97 &    0.94 &  0.95 \\
% D6      &         1900.63 &           90.56 &       0.58 &    0.87 &  0.69 \\
% D7      &          135.28 &           16.46 &       0.94 &    0.96 &  0.95 \\
% D8      &         3059.40 &         1094.41 &       0.50 &    0.83 &  0.62 \\
% D9      &         1422.08 &          129.40 &       0.96 &    0.89 &  0.92 \\
% D10     &         2079.65 &          154.57 &       0.97 &    0.78 &  0.86 \\
% \bottomrule
% \end{tabular}

% \caption{Statistics for ZeroER, Schema-Based}
% \end{table}

\begin{figure}[!t]
\centering
\includegraphics[trim=0.12cm 0.12cm 0.12cm 0.12cm, clip, width=0.4\textwidth]{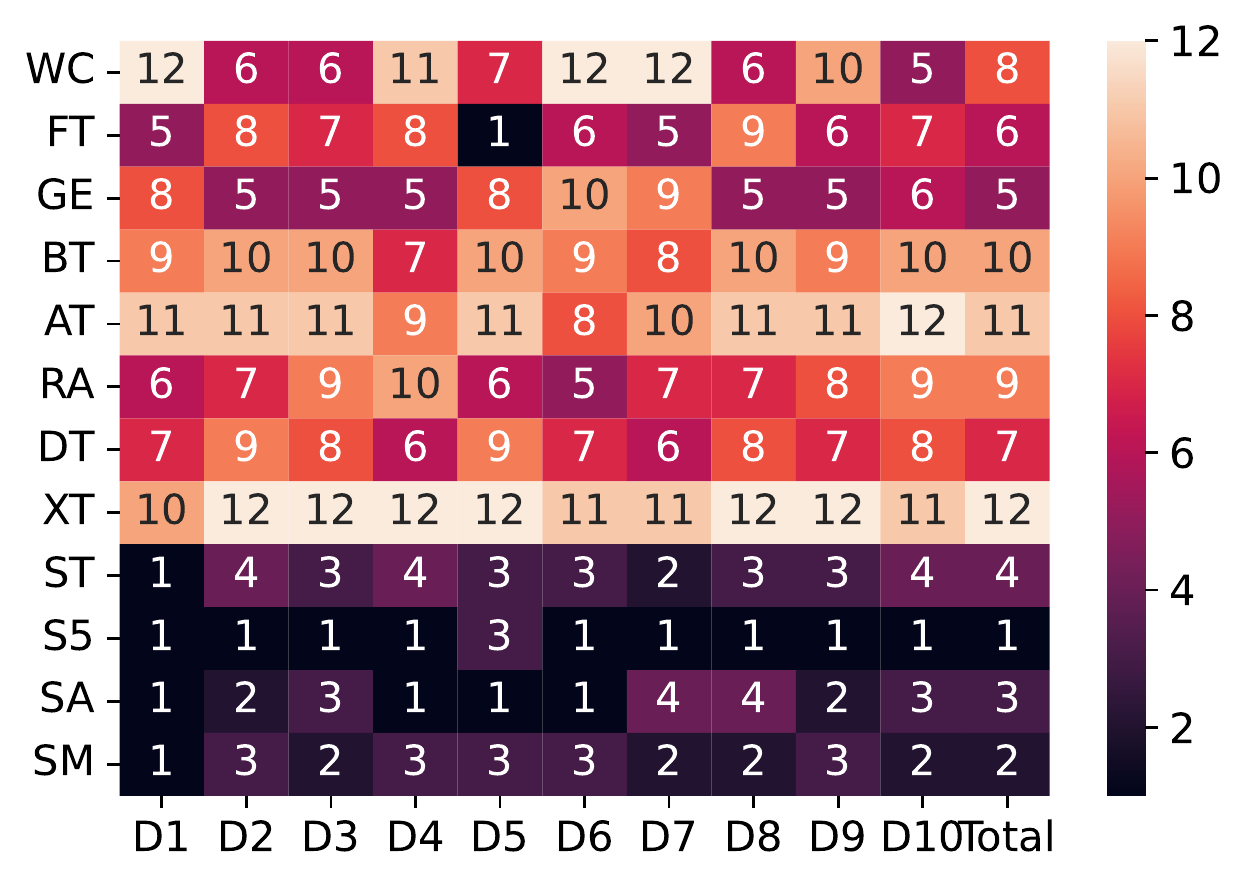}
% \caption{The ranking position of each method with respect to blocking recall per dataset. Lower is better.}
\caption{Method ranking wrt blocking recall (lower is better) (Schema-Based).}
\label{fig:blk_heat_sch}
\end{figure}

\begin{figure}[!t]
\centering
\includegraphics[trim=0.12cm 0.12cm 0.12cm 0.12cm, clip, width=0.47\textwidth]{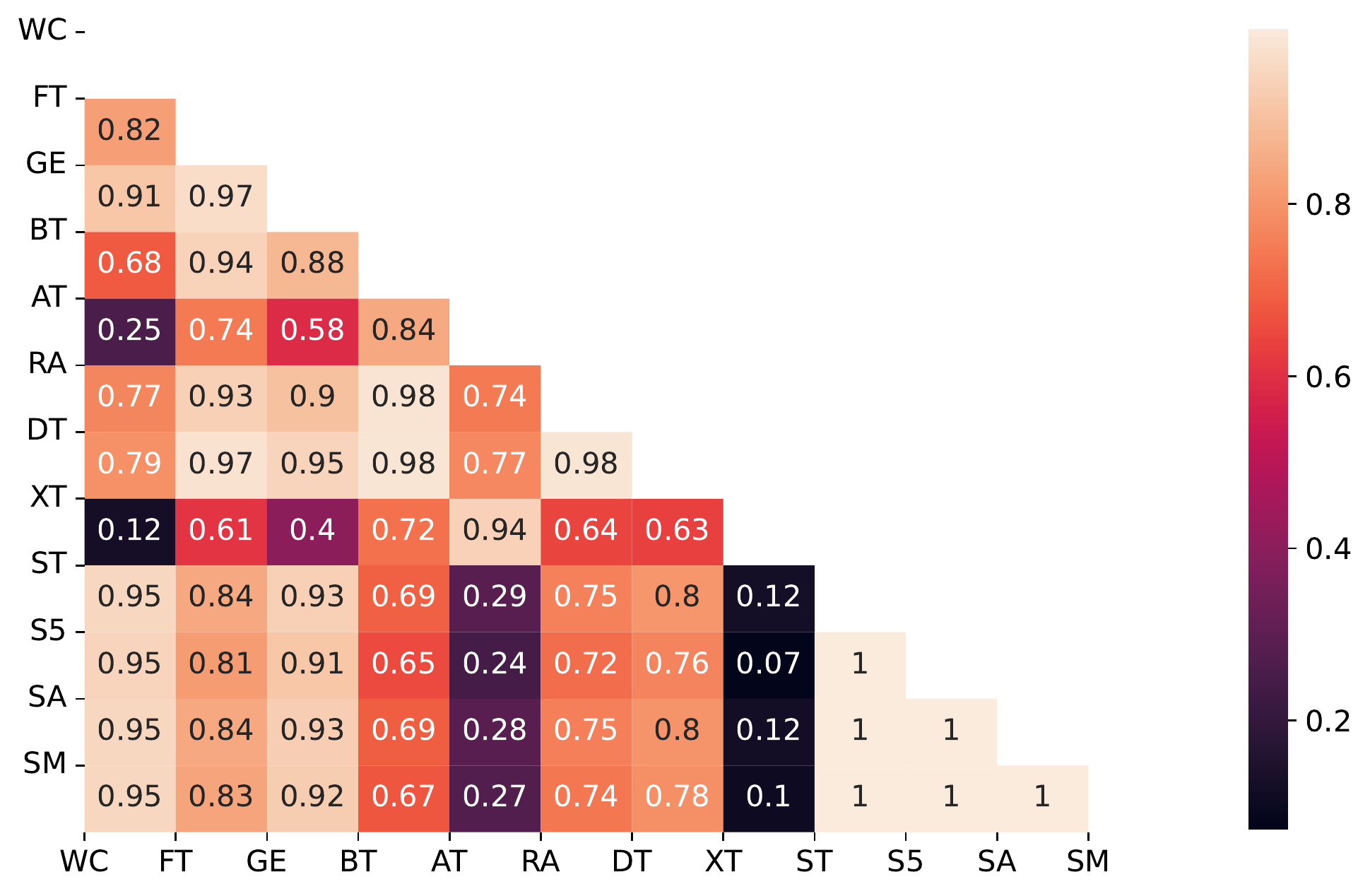}
% \caption{{\small Pearson correlation per pair of models wrt blocking~recall.}}
\caption{Pearson correlation of models wrt blocking recall  (Schema-Based).}
\label{fig:blk_heat_pear_sch}
\end{figure}

\begin{figure}[!t]
\centering
{\includegraphics[trim=0.12cm 0.12cm 0.12cm 0.12cm, clip, width=0.4\textwidth]{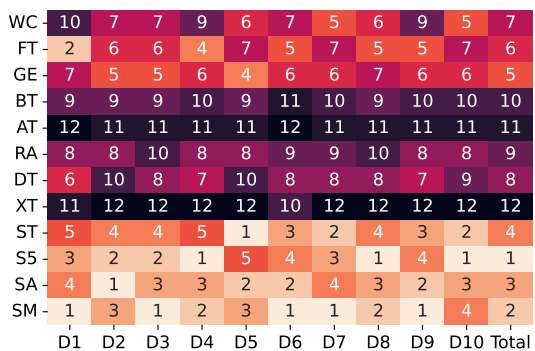}\label{subfig:match_unsup_heat_sb}}
\caption{The ranking position of each method per dataset in unsupervised settings for Matching. Lower is better (Schema-Based).}
\label{fig:matching_heat_sb}
\end{figure}

\begin{figure}[!t]
\centering
{\includegraphics[trim=0.12cm 0.12cm 0.12cm 0.12cm, clip, width=0.4\textwidth]{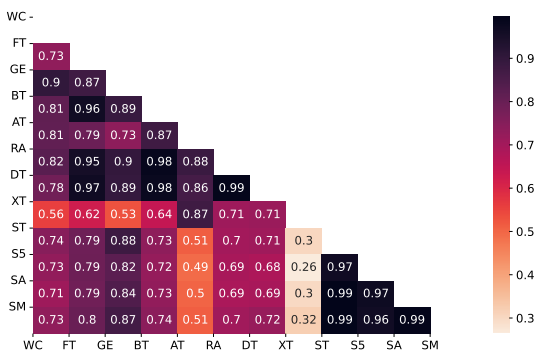}\label{subfig:sup_corr_sb}}

\caption{Pearson correlation with respect to F1 per pair of language models for unsupervised matching (Schema-Based).}
\label{fig:match_corr_sb}
\end{figure}

\begin{figure*}[!t]
\centering
\subfloat[D1 / Recall]{\includegraphics[trim=0.12cm 0.12cm 0.12cm 0.12cm, clip, width=0.25\textwidth, height=30mm]{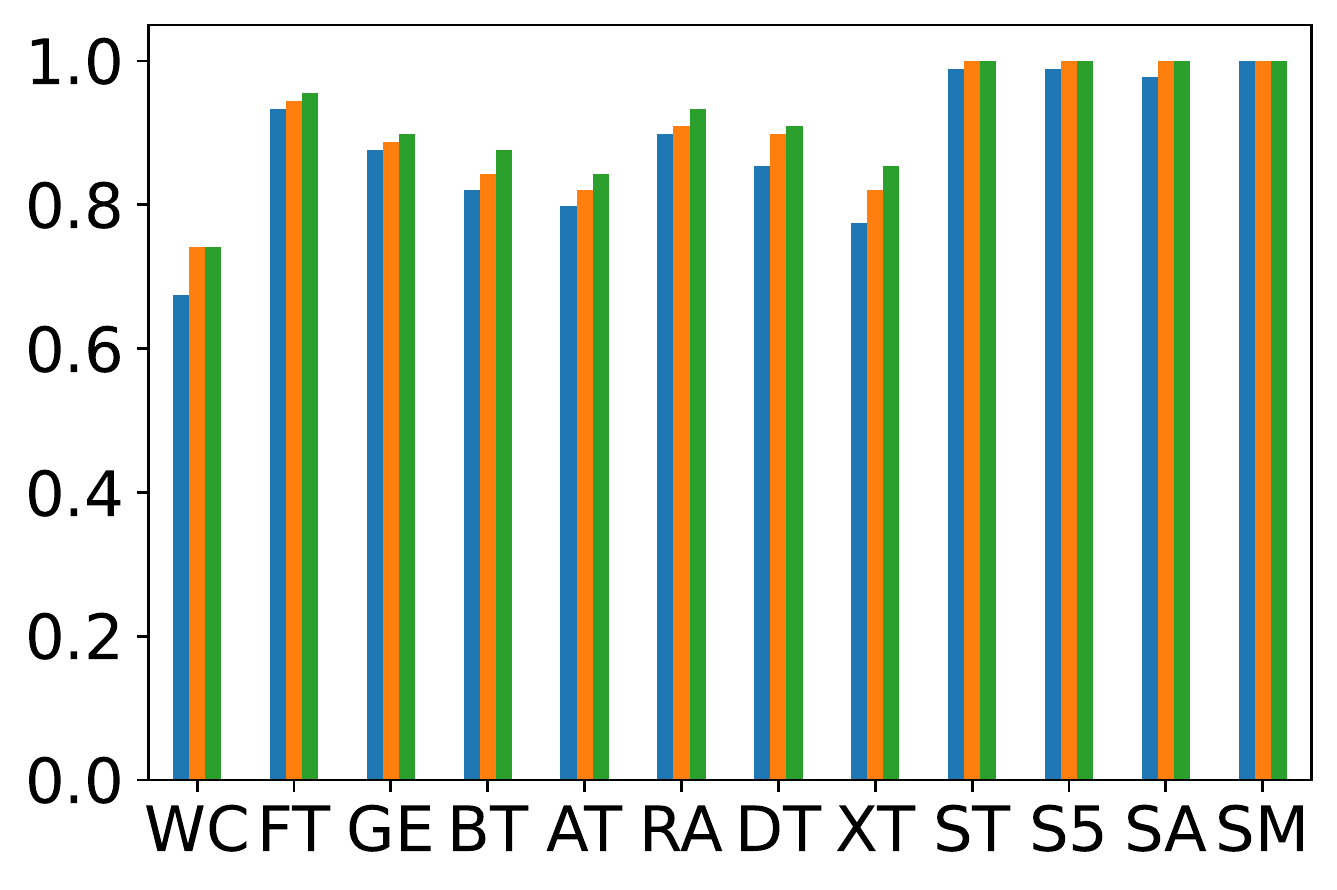}} 
\subfloat[D1 / Time]{\includegraphics[trim=0.12cm 0.12cm 0.12cm 0.12cm, clip, width=0.25\textwidth, height=30mm]{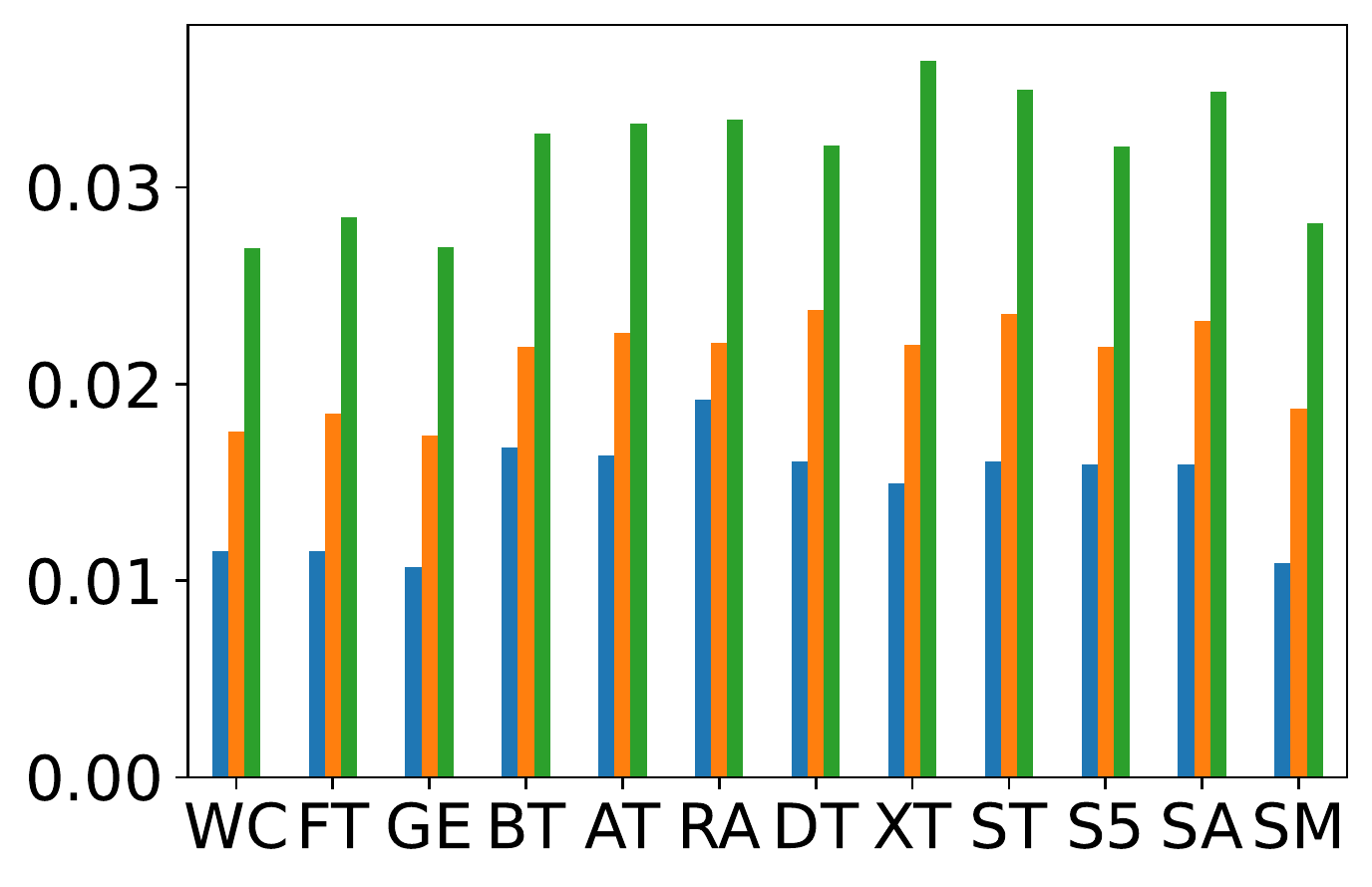}}
\subfloat[D2 / Recall]{\includegraphics[trim=0.12cm 0.12cm 0.12cm 0.12cm, clip, width=0.25\textwidth, height=30mm]{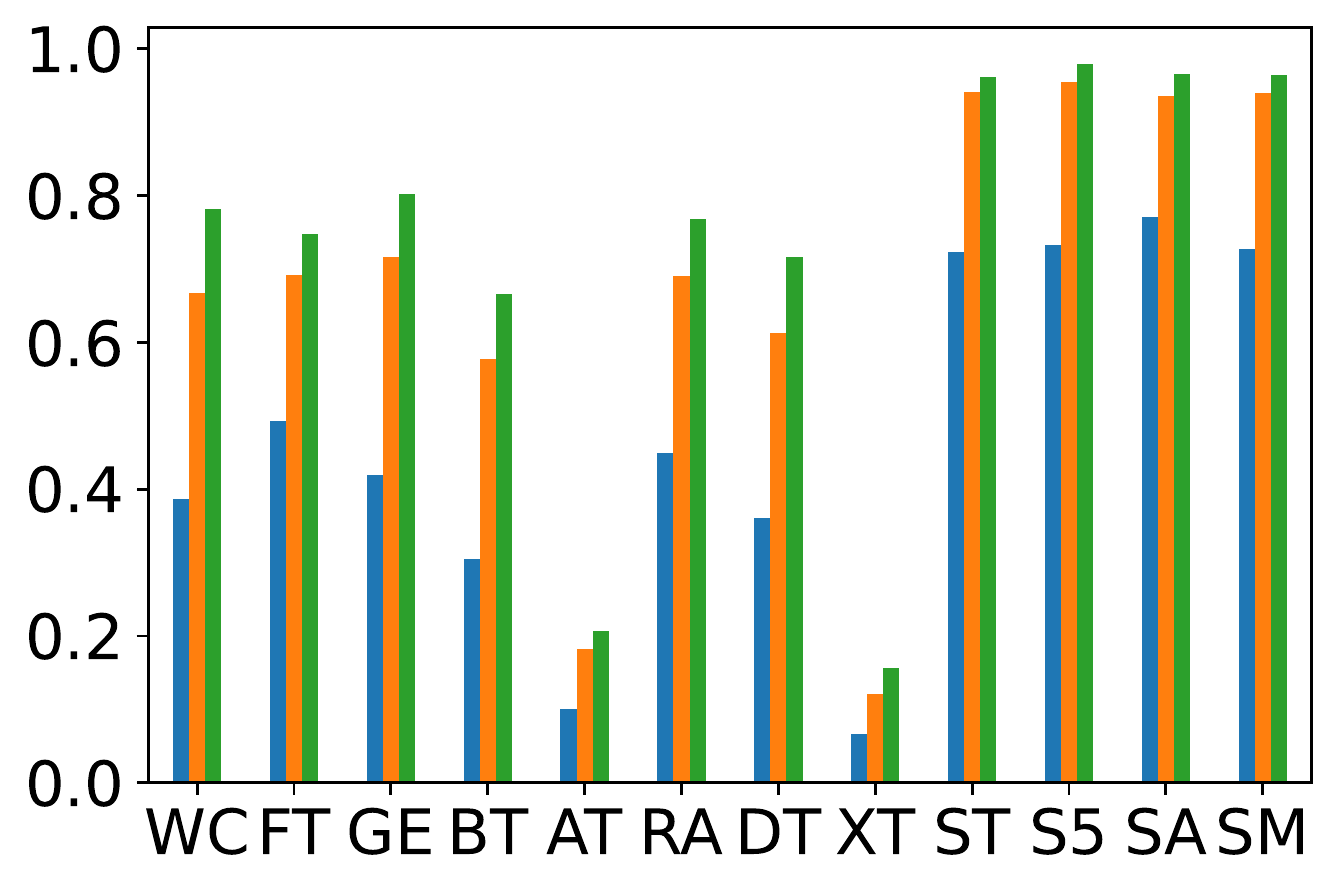}} 
\subfloat[D2 / Time]{\includegraphics[trim=0.12cm 0.12cm 0.12cm 0.12cm, clip, width=0.25\textwidth, height=30mm]{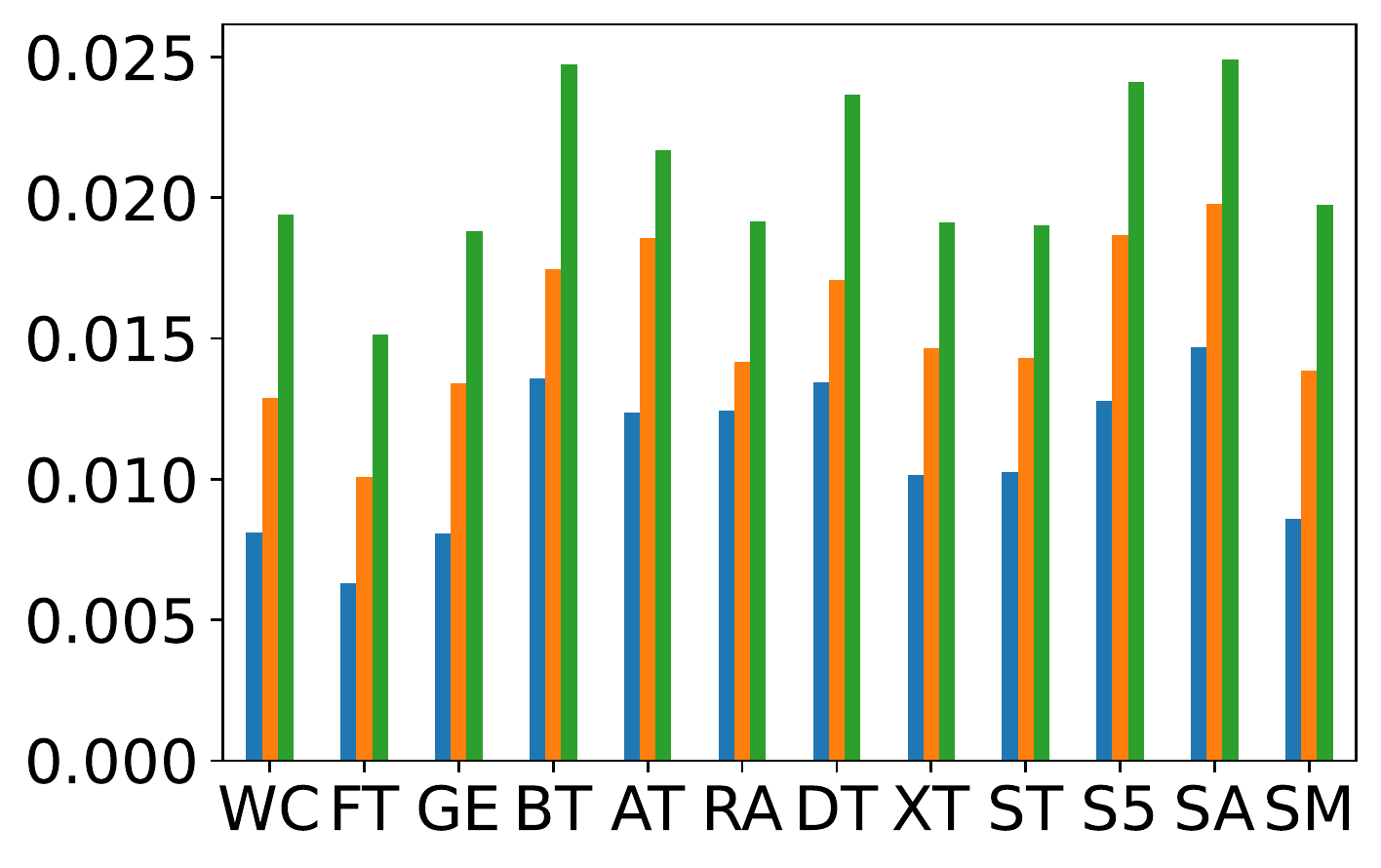}}
\newline
\subfloat[D3 / Recall]{\includegraphics[trim=0.12cm 0.12cm 0.12cm 0.12cm, clip, width=0.25\textwidth, height=30mm]{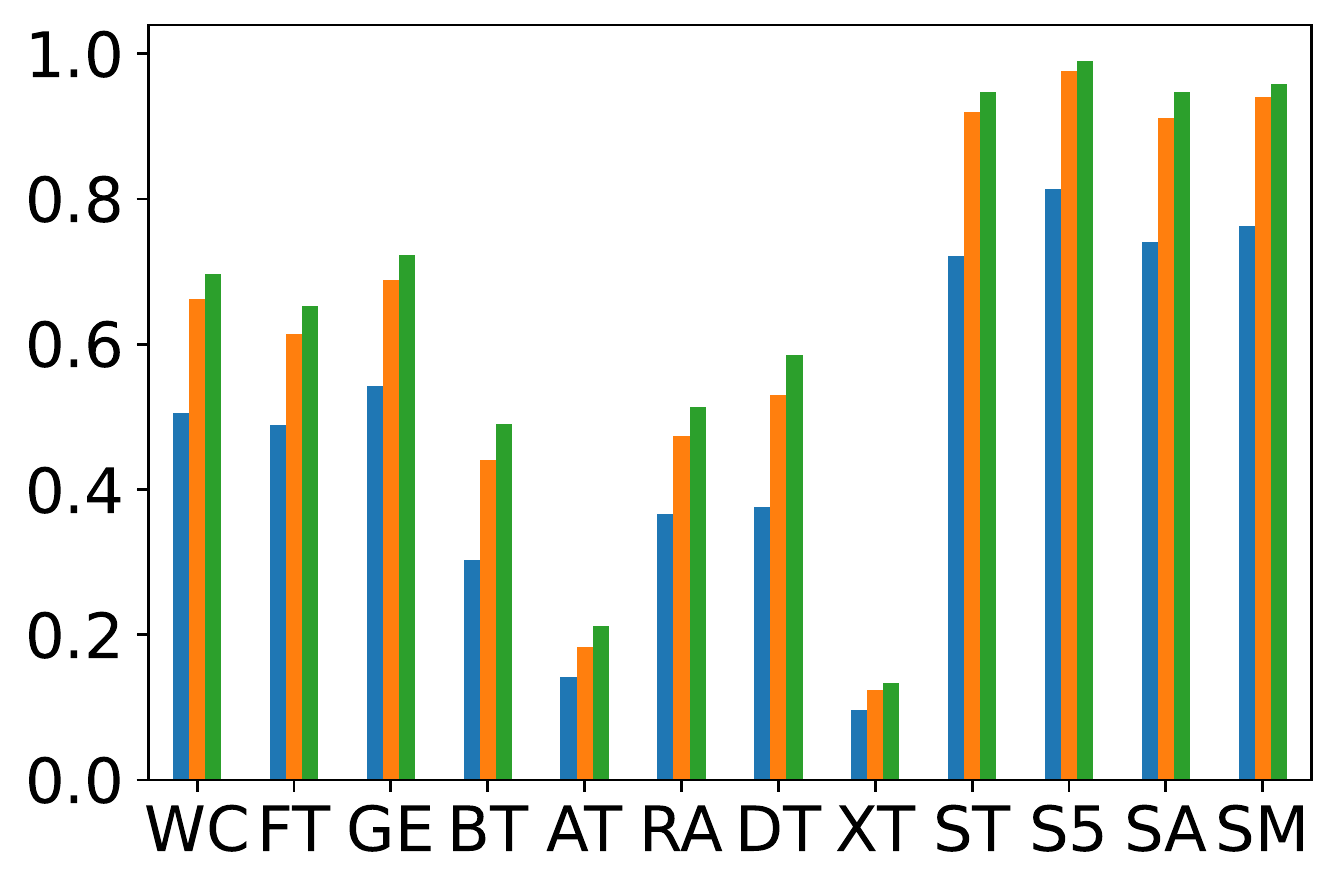}} 
\subfloat[D3 / Time]{\includegraphics[trim=0.12cm 0.12cm 0.12cm 0.12cm, clip, width=0.25\textwidth, height=30mm]{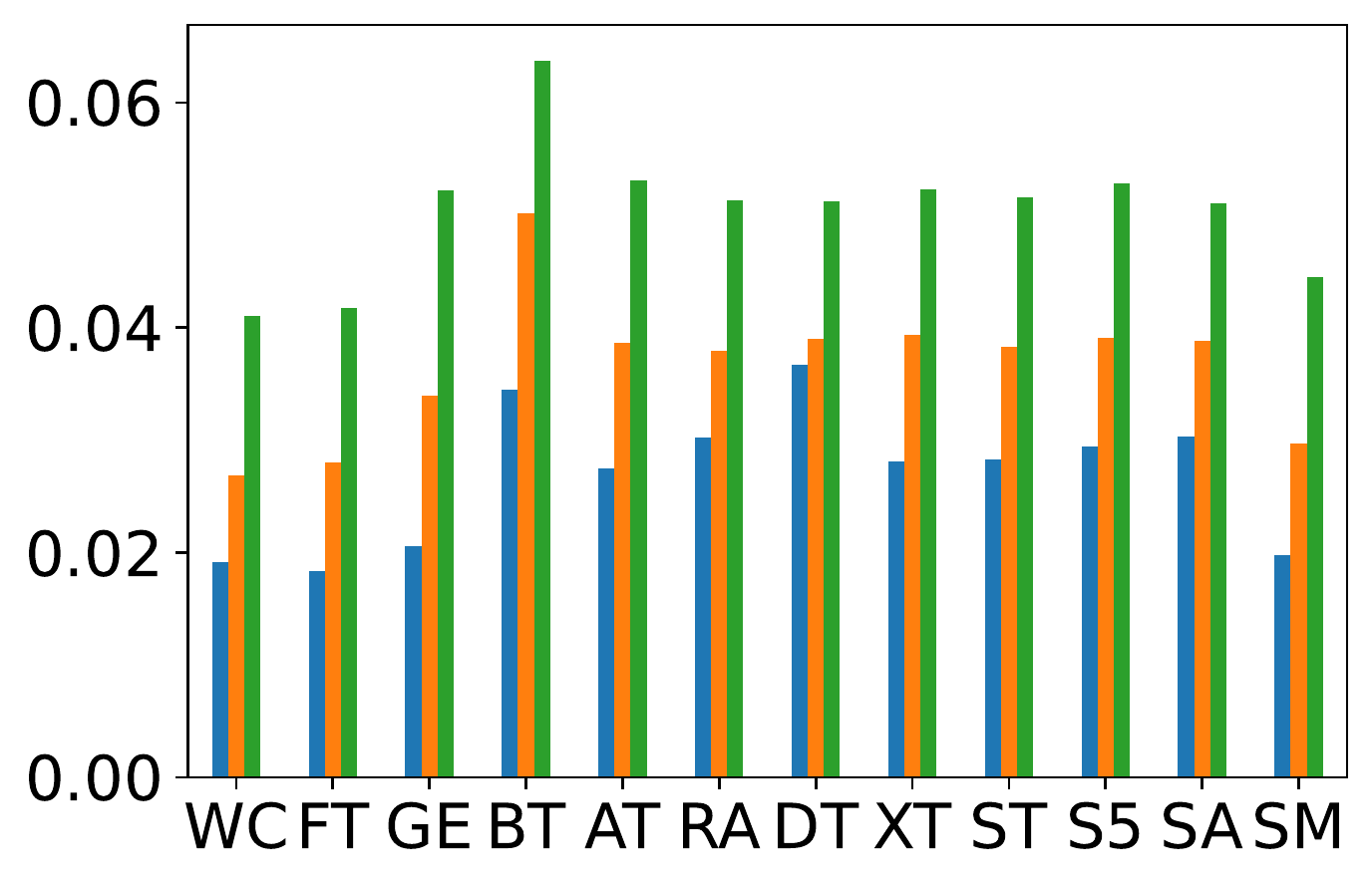}}
\subfloat[D4 / Recall]{\includegraphics[trim=0.12cm 0.12cm 0.12cm 0.12cm, clip, width=0.25\textwidth, height=30mm]{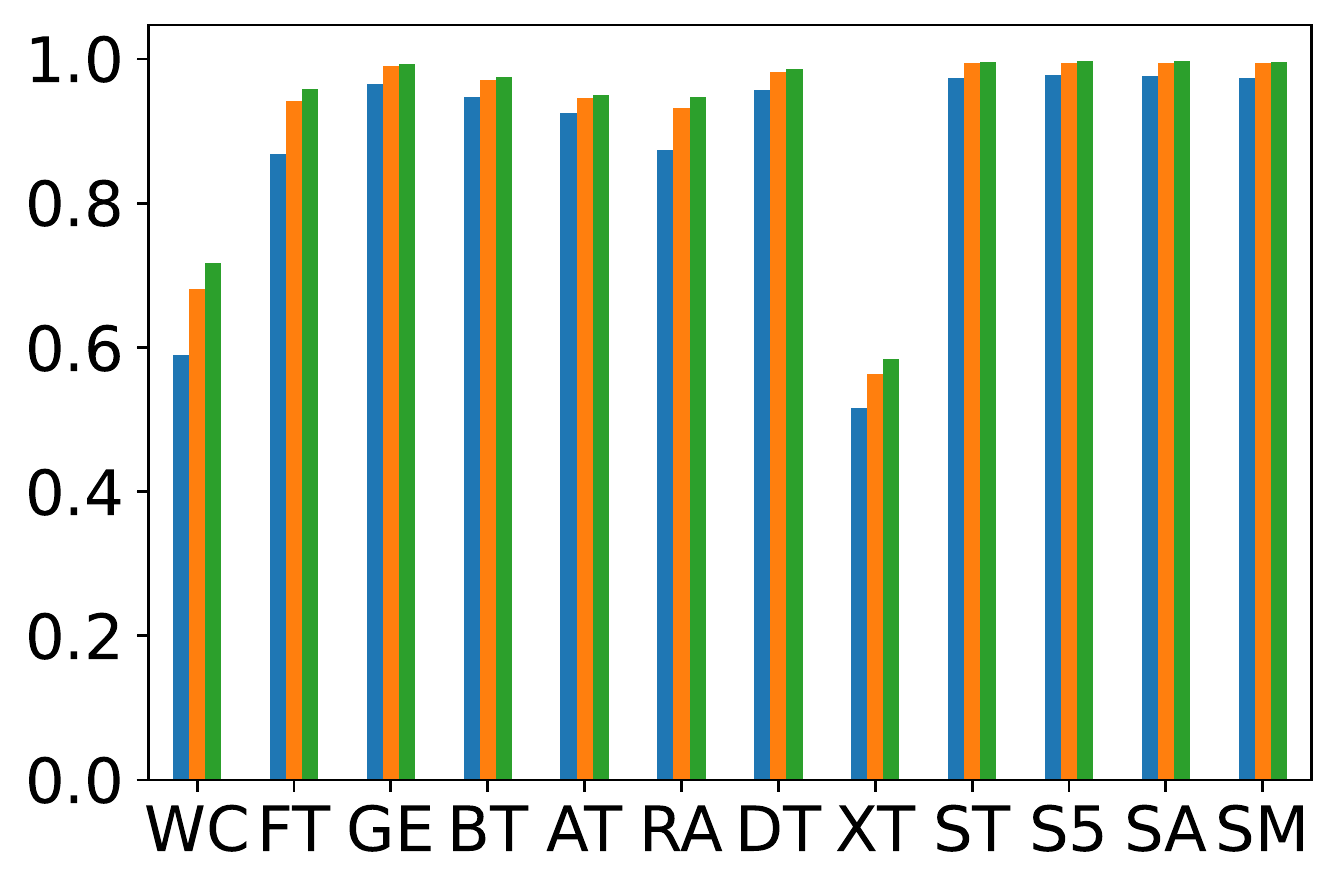}} 
\subfloat[D4 / Time]{\includegraphics[trim=0.12cm 0.12cm 0.12cm 0.12cm, clip, width=0.25\textwidth, height=30mm]{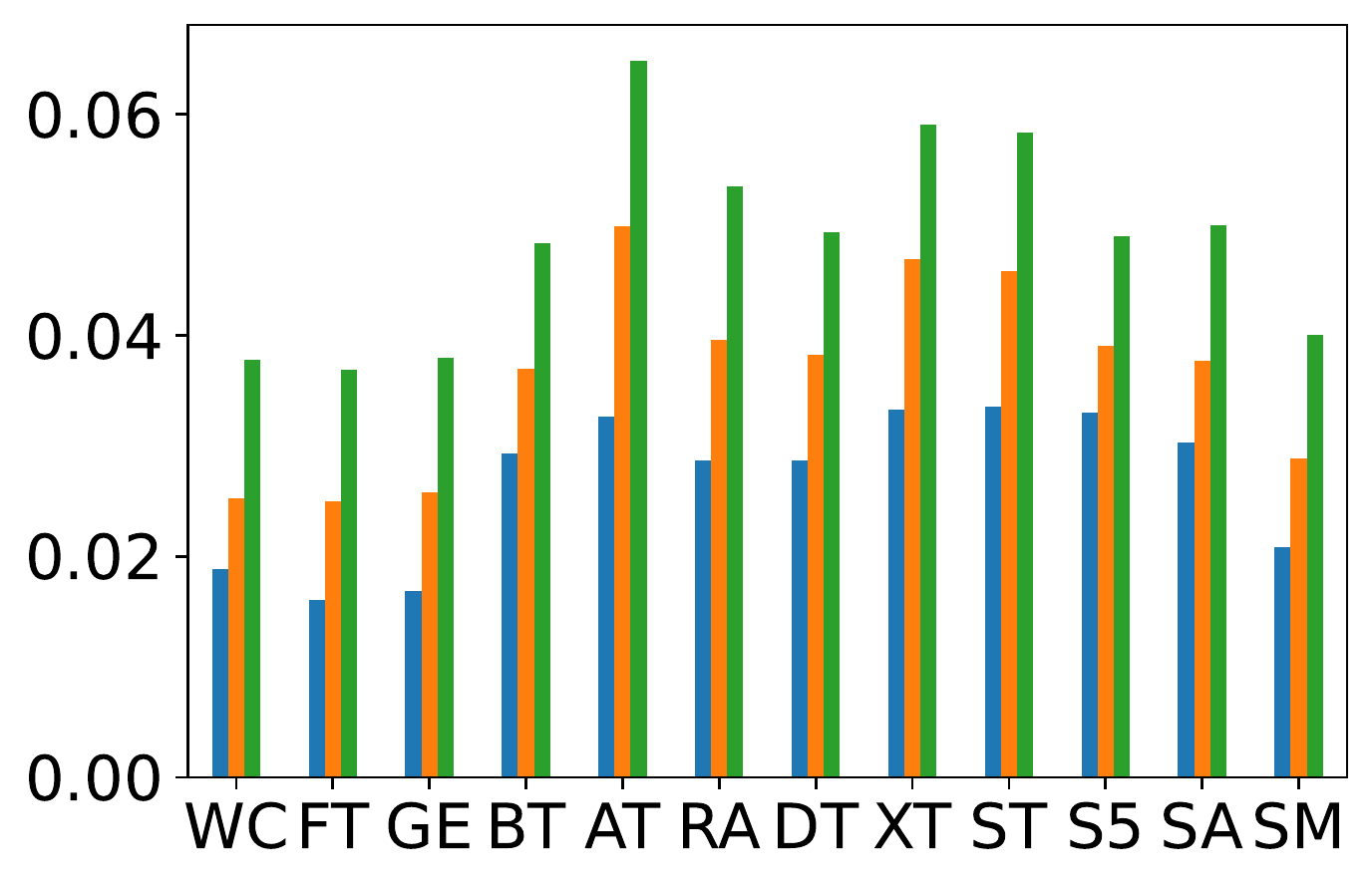}}
\newline
\subfloat[D5 / Recall]{\includegraphics[trim=0.12cm 0.12cm 0.12cm 0.12cm, clip, width=0.25\textwidth, height=30mm]{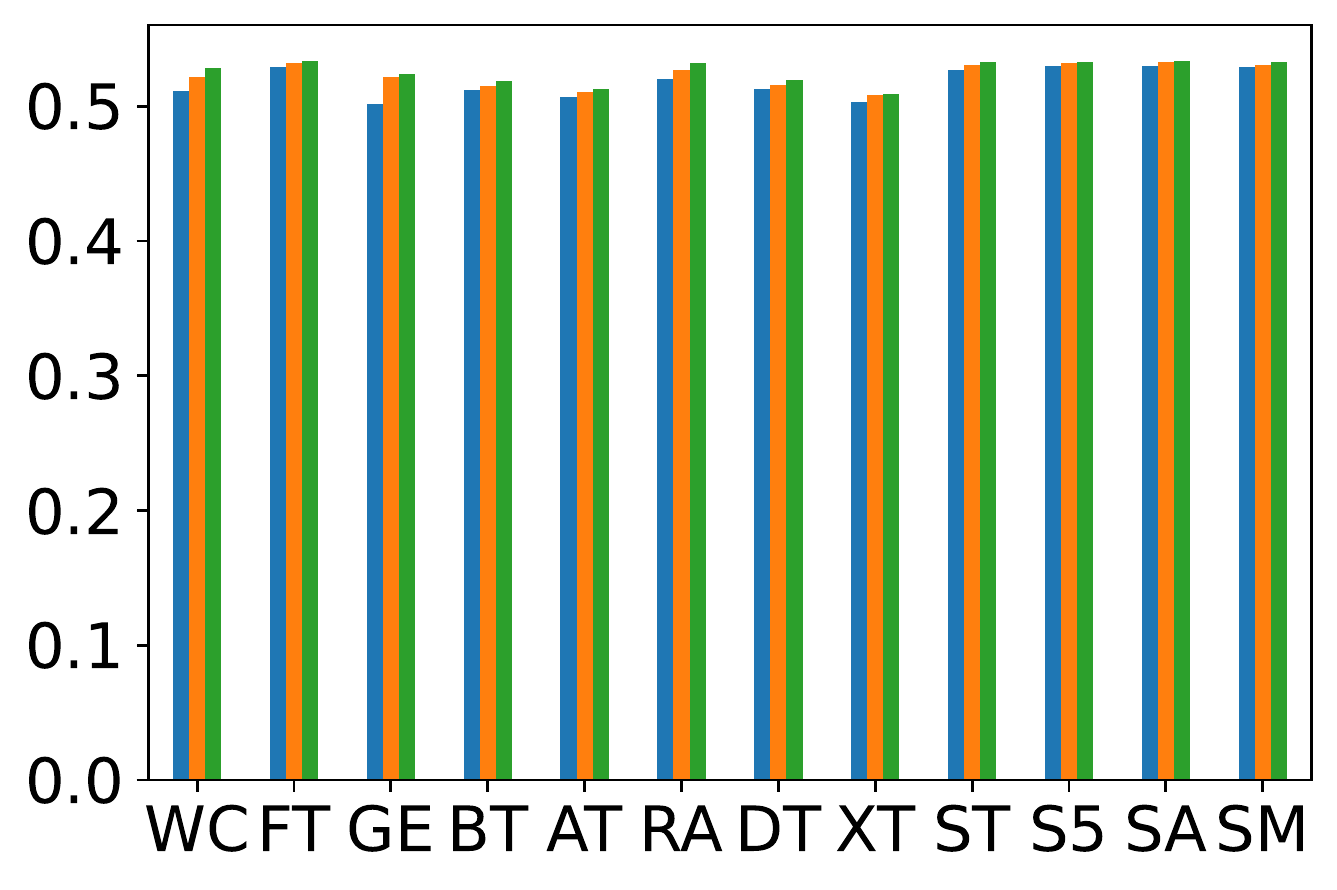}} 
\subfloat[D5 / Time]{\includegraphics[trim=0.12cm 0.12cm 0.12cm 0.12cm, clip, width=0.25\textwidth, height=30mm]{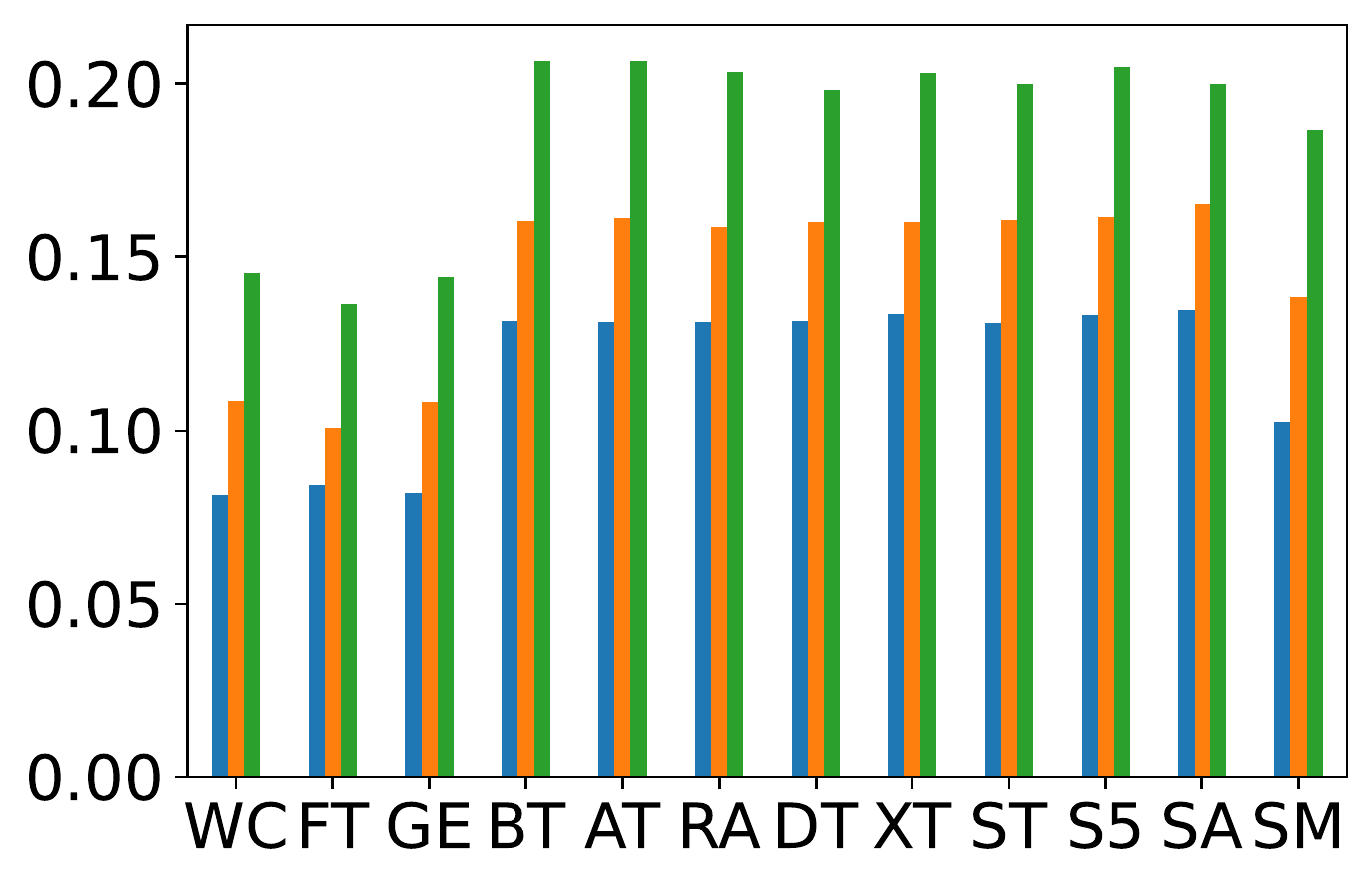}}
\subfloat[D6 / Recall]{\includegraphics[trim=0.12cm 0.12cm 0.12cm 0.12cm, clip, width=0.25\textwidth, height=30mm]{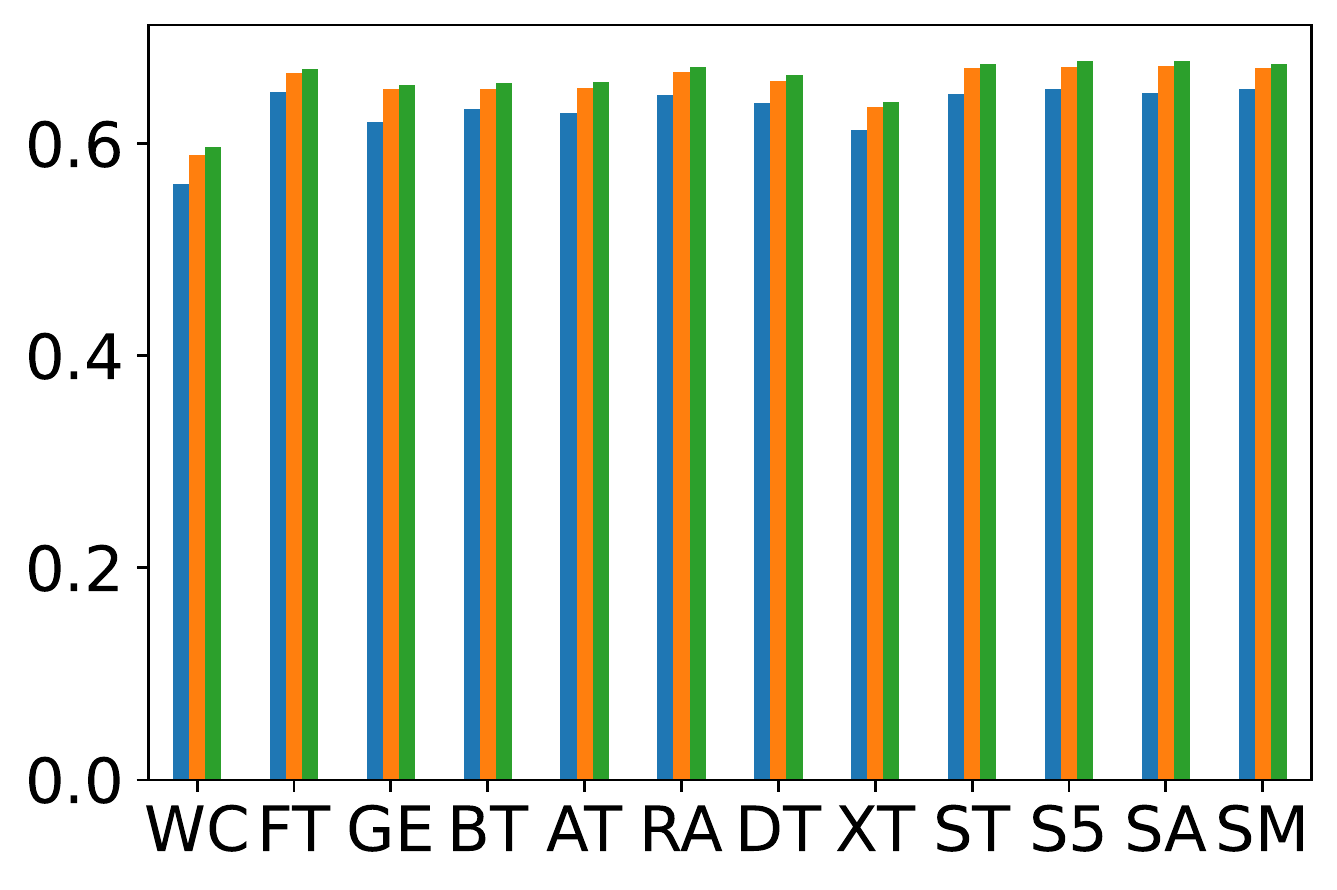}} 
\subfloat[D6 / Time]{\includegraphics[trim=0.12cm 0.12cm 0.12cm 0.12cm, clip, width=0.25\textwidth, height=30mm]{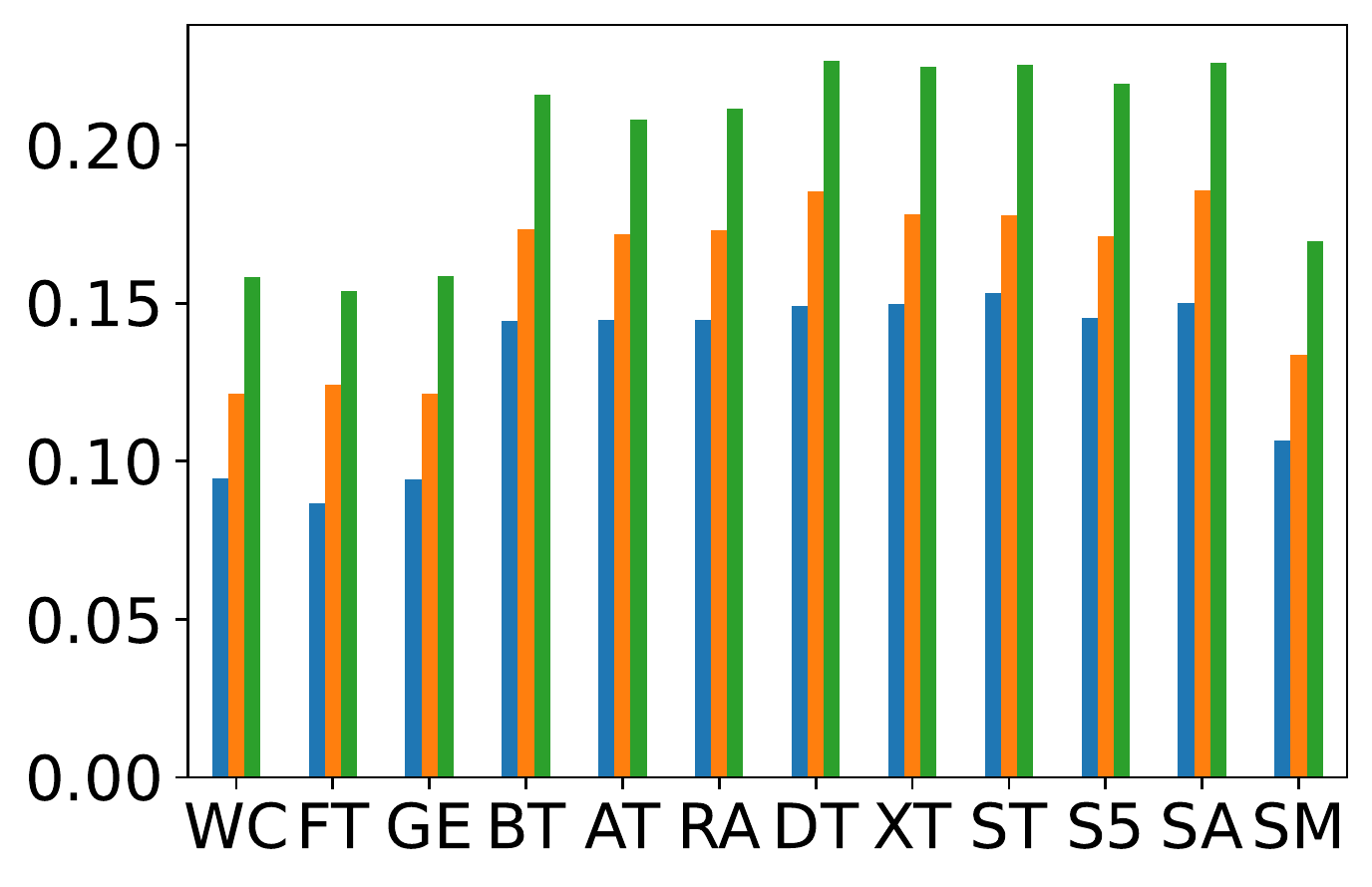}}
\newline
\subfloat[D7 / Recall]{\includegraphics[trim=0.12cm 0.12cm 0.12cm 0.12cm, clip, width=0.25\textwidth, height=30mm]{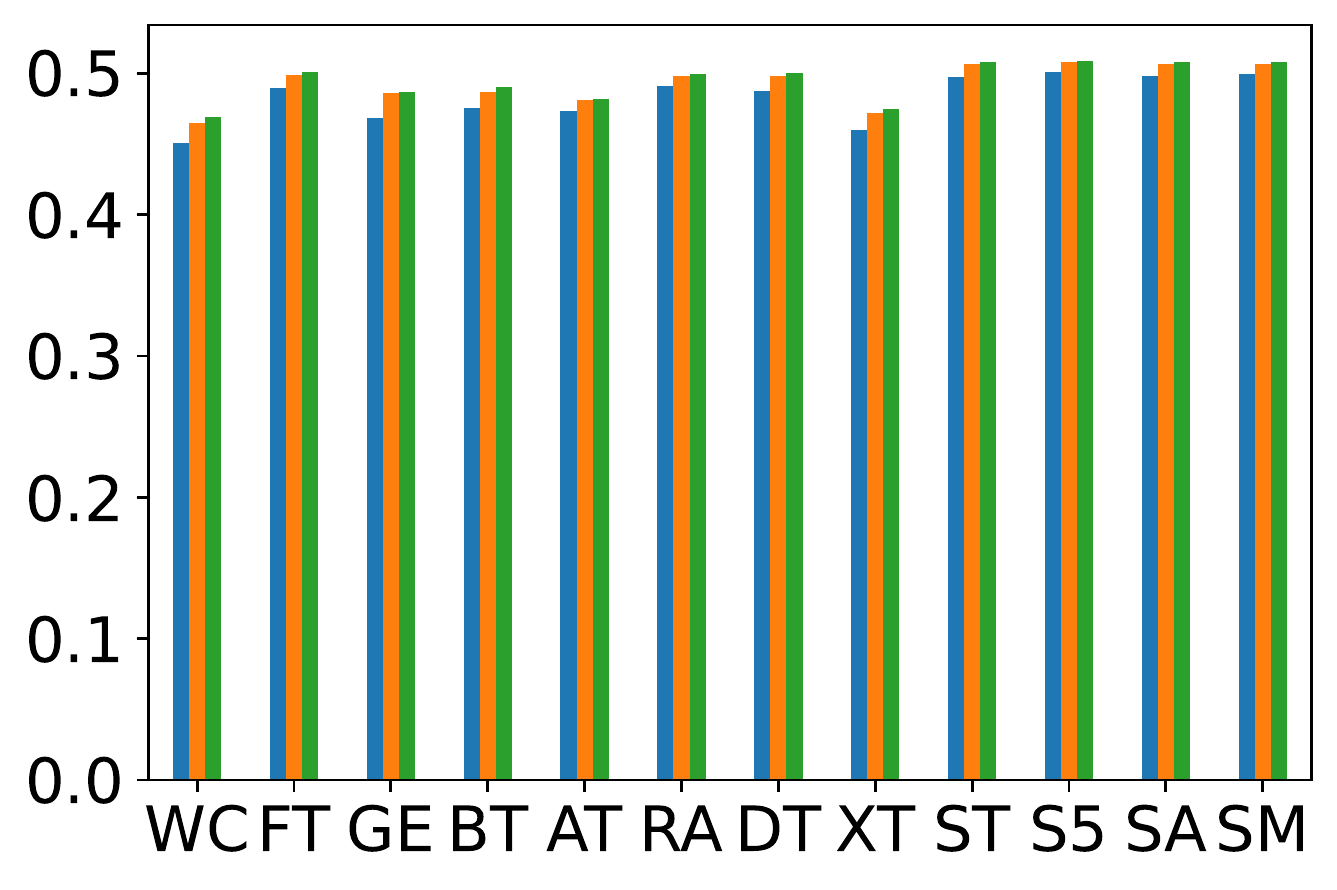}} 
\subfloat[D7 / Time]{\includegraphics[trim=0.12cm 0.12cm 0.12cm 0.12cm, clip, width=0.25\textwidth, height=30mm]{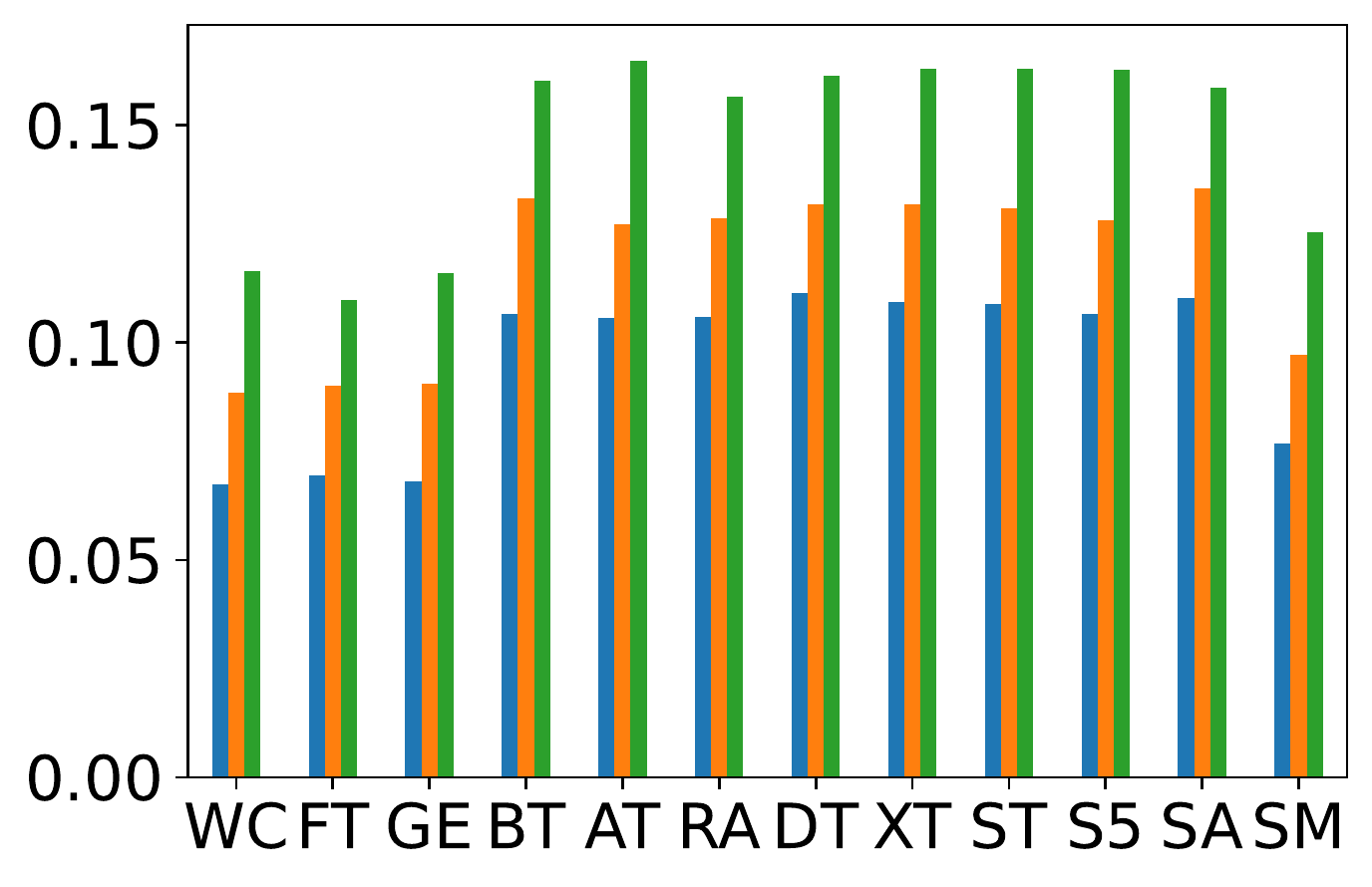}}
\subfloat[D8 / Recall]{\includegraphics[trim=0.12cm 0.12cm 0.12cm 0.12cm, clip, width=0.25\textwidth, height=30mm]{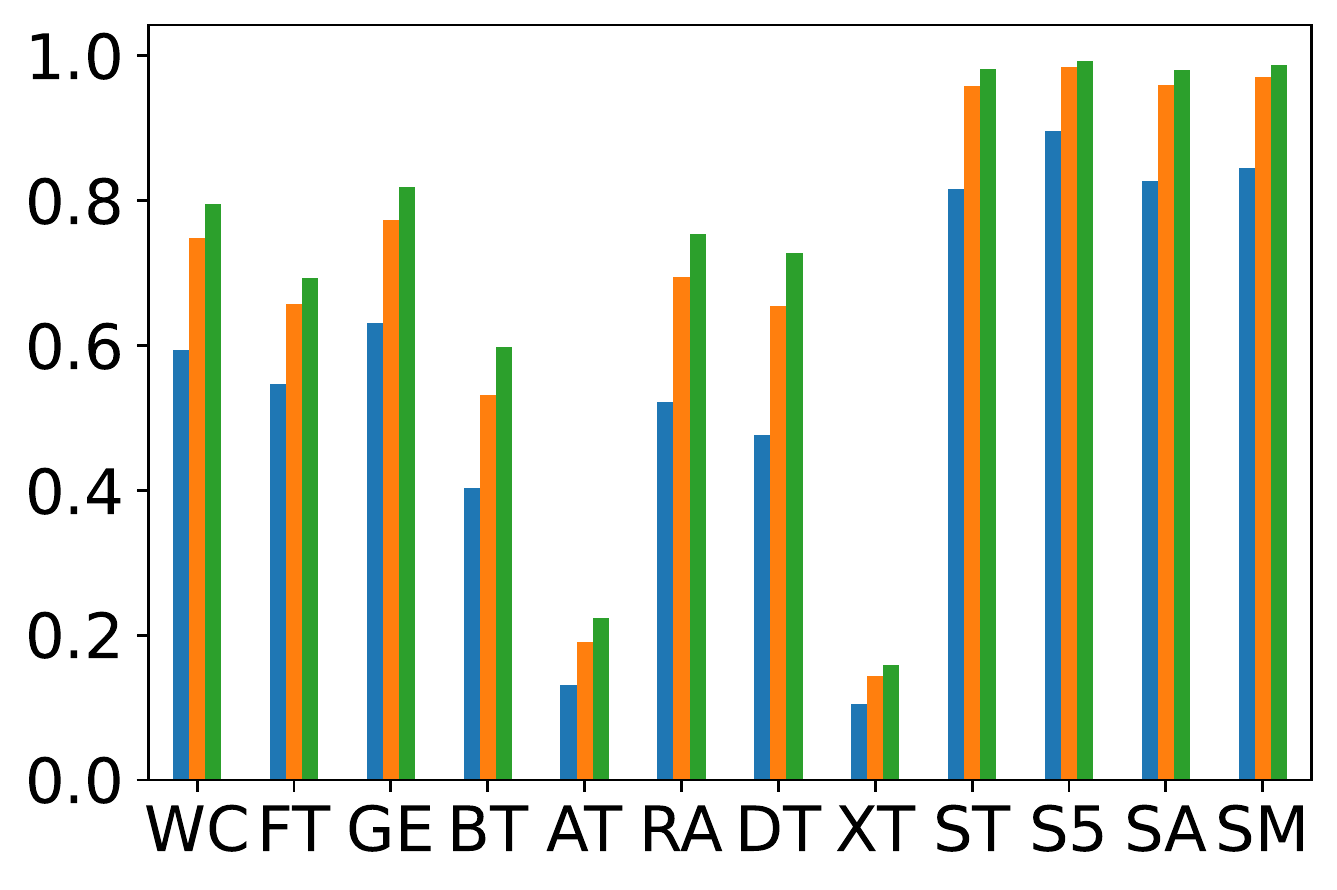}} 
\subfloat[D8 / Time]{\includegraphics[trim=0.12cm 0.12cm 0.12cm 0.12cm, clip, width=0.25\textwidth, height=30mm]{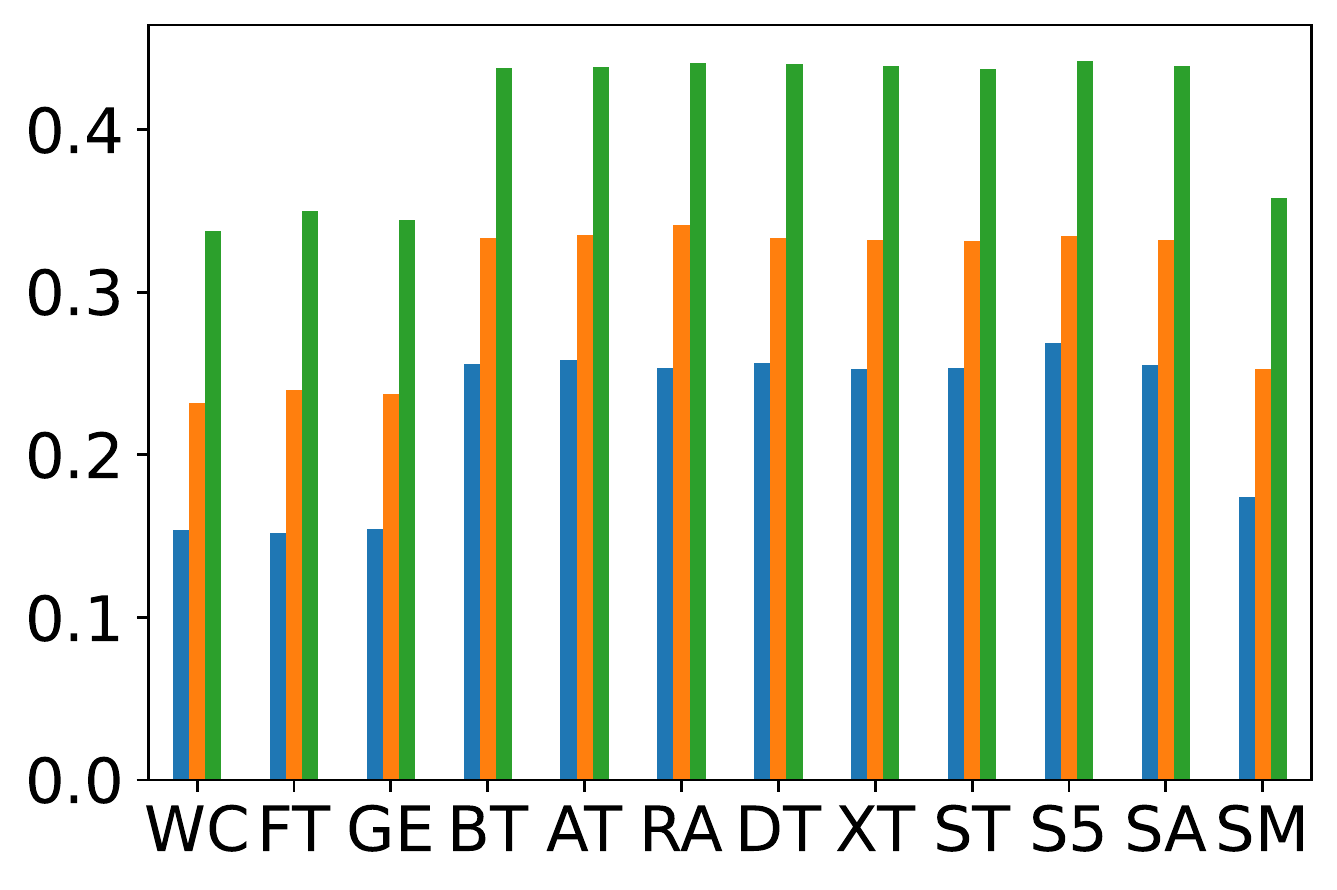}}
\newline
\subfloat[D9 / Recall]{\includegraphics[trim=0.12cm 0.12cm 0.12cm 0.12cm, clip, width=0.25\textwidth, height=30mm]{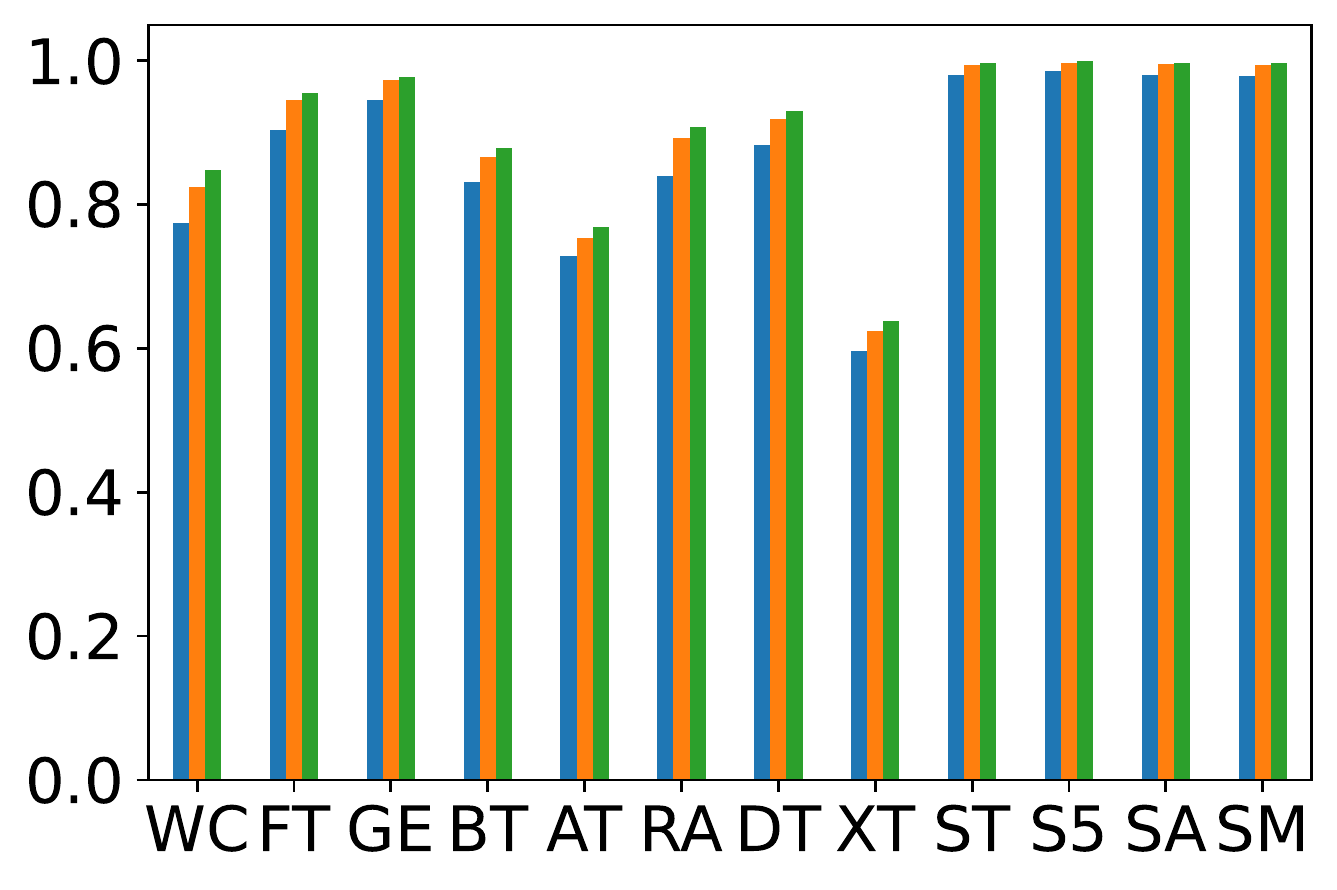}} 
\subfloat[D9 / Time]{\includegraphics[trim=0.12cm 0.12cm 0.12cm 0.12cm, clip, width=0.25\textwidth, height=30mm]{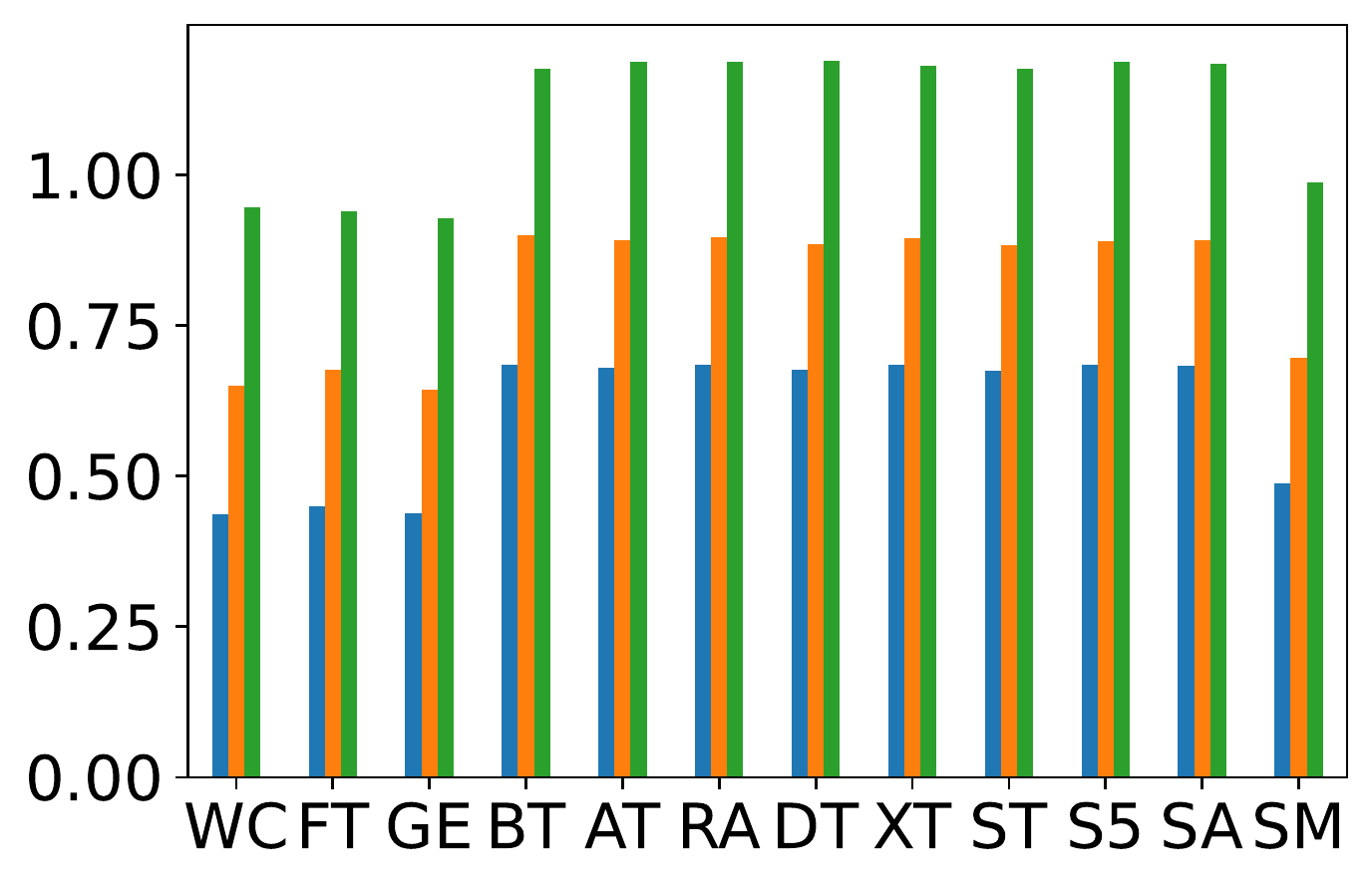}}
\subfloat[D10 / Recall]{\includegraphics[trim=0.12cm 0.12cm 0.12cm 0.12cm, clip, width=0.25\textwidth, height=30mm]{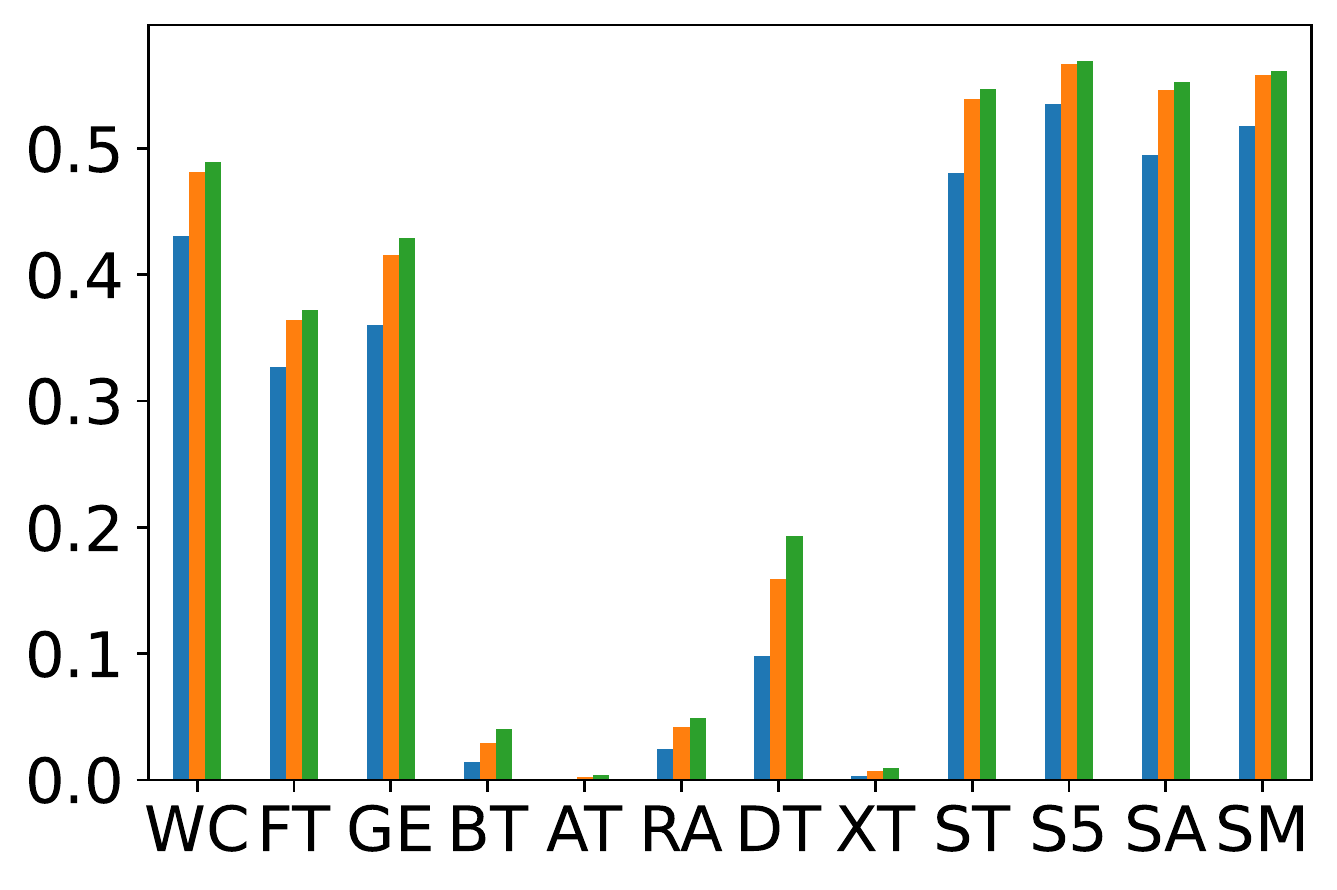}} 
\subfloat[D10 / Time]{\includegraphics[trim=0.12cm 0.12cm 0.12cm 0.12cm, clip, width=0.25\textwidth, height=30mm]{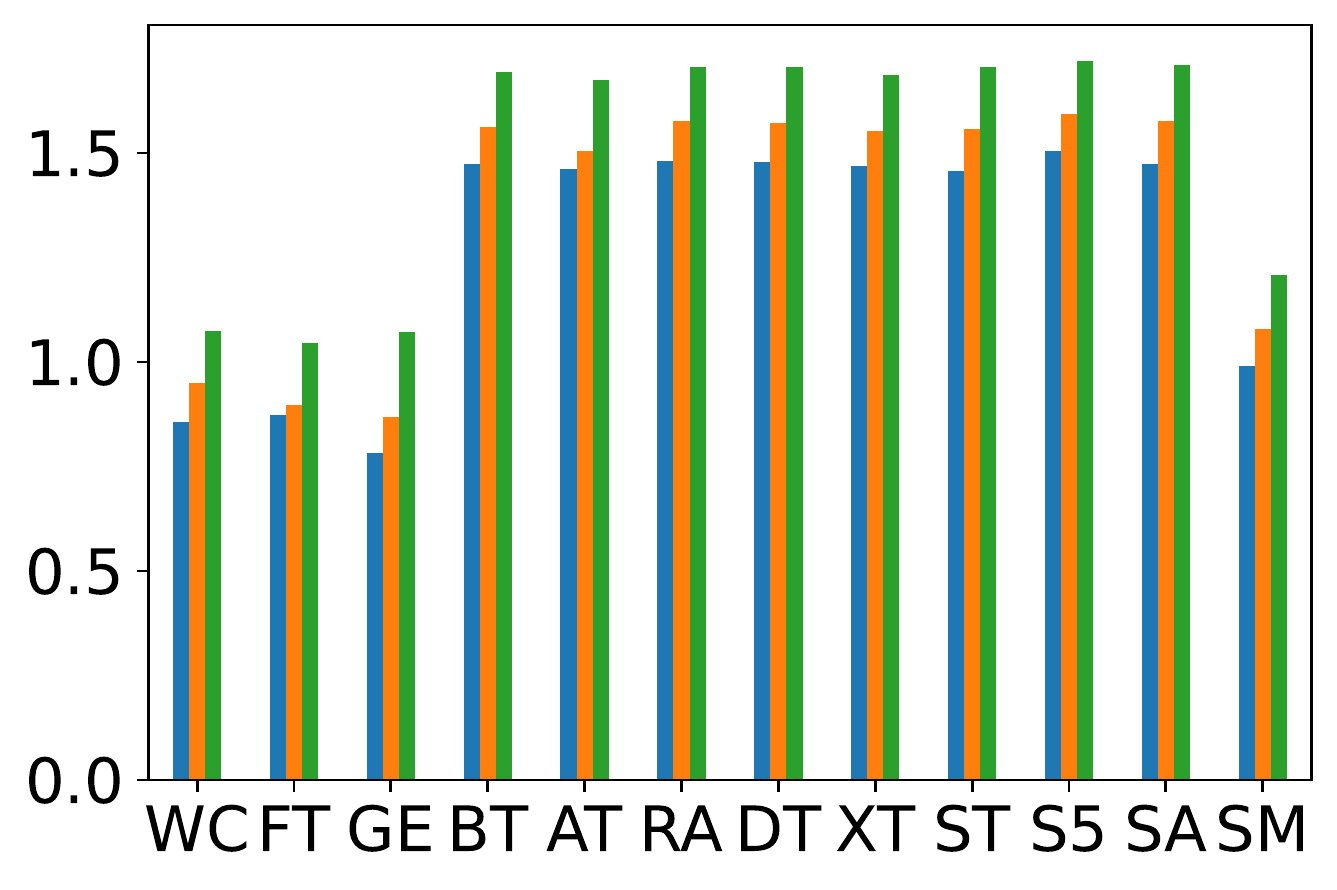}}
\caption{Recall and Blocking Time (sec) per method from real data per case. There are three values for $k \in \{${\color{myblue}1}, {\color{myorange}5}, {\color{mygreen}10} $\}$  (Schema-Based).}
\label{fig:blk_real_sch}
\end{figure*}

\begin{figure*}[!t]
\centering

% \subfloat{\includegraphics[width=0.6\linewidth]{matching_real_legend.pdf}}
% \setcounter{subfigure}{0}
% \newline

\subfloat[D1 / Scores]{\includegraphics[trim=0.12cm 0.12cm 0.12cm 0.12cm, clip, width=0.25\textwidth, height=30mm]{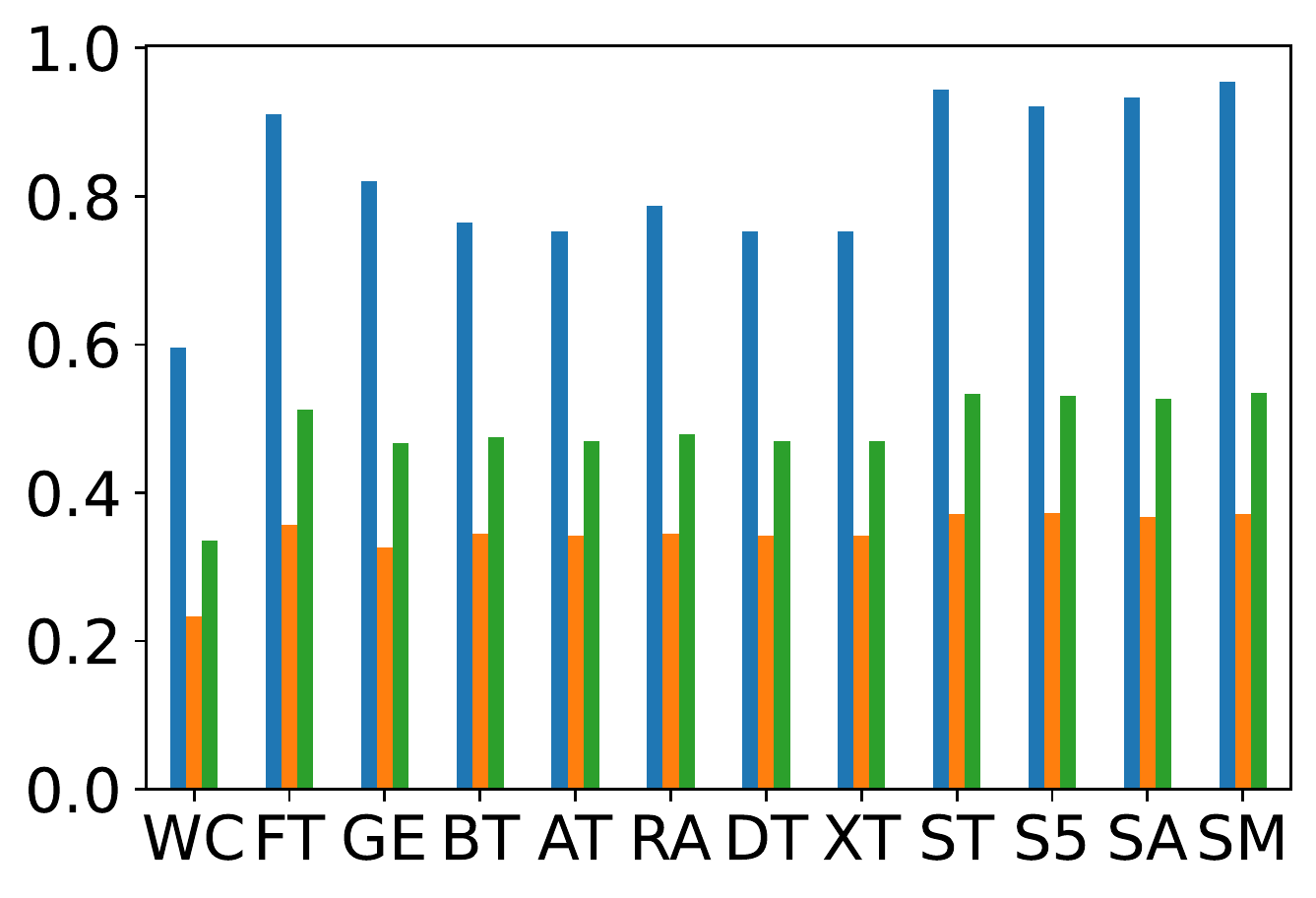}}
\subfloat[D1 / Time]{\includegraphics[trim=0.12cm 0.12cm 0.12cm 0.12cm, clip, width=0.25\textwidth, height=30mm]{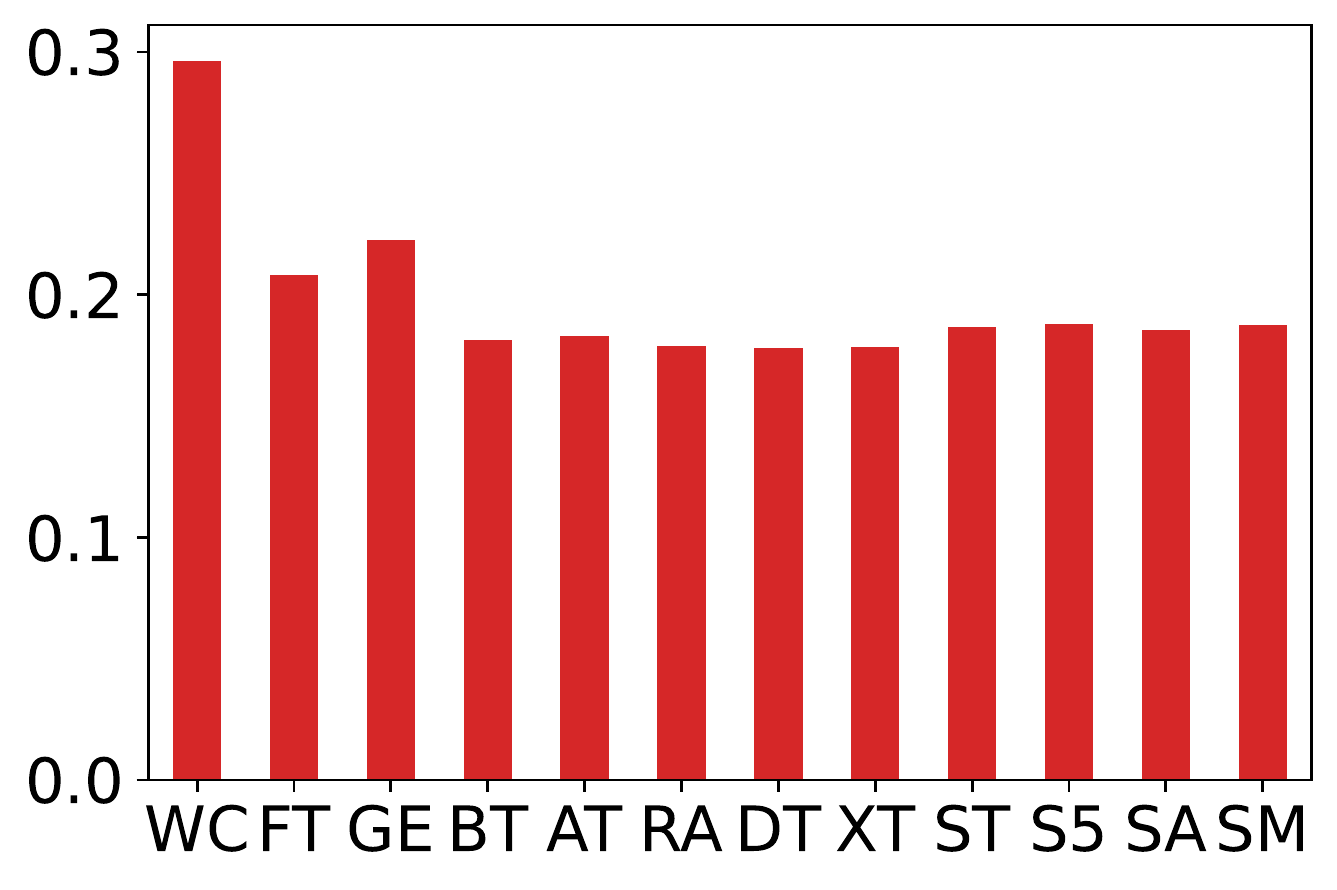}} 
%\hspace{5mm}%
\subfloat[D2 / Scores]{\includegraphics[trim=0.12cm 0.12cm 0.12cm 0.12cm, clip, width=0.25\textwidth, height=30mm]{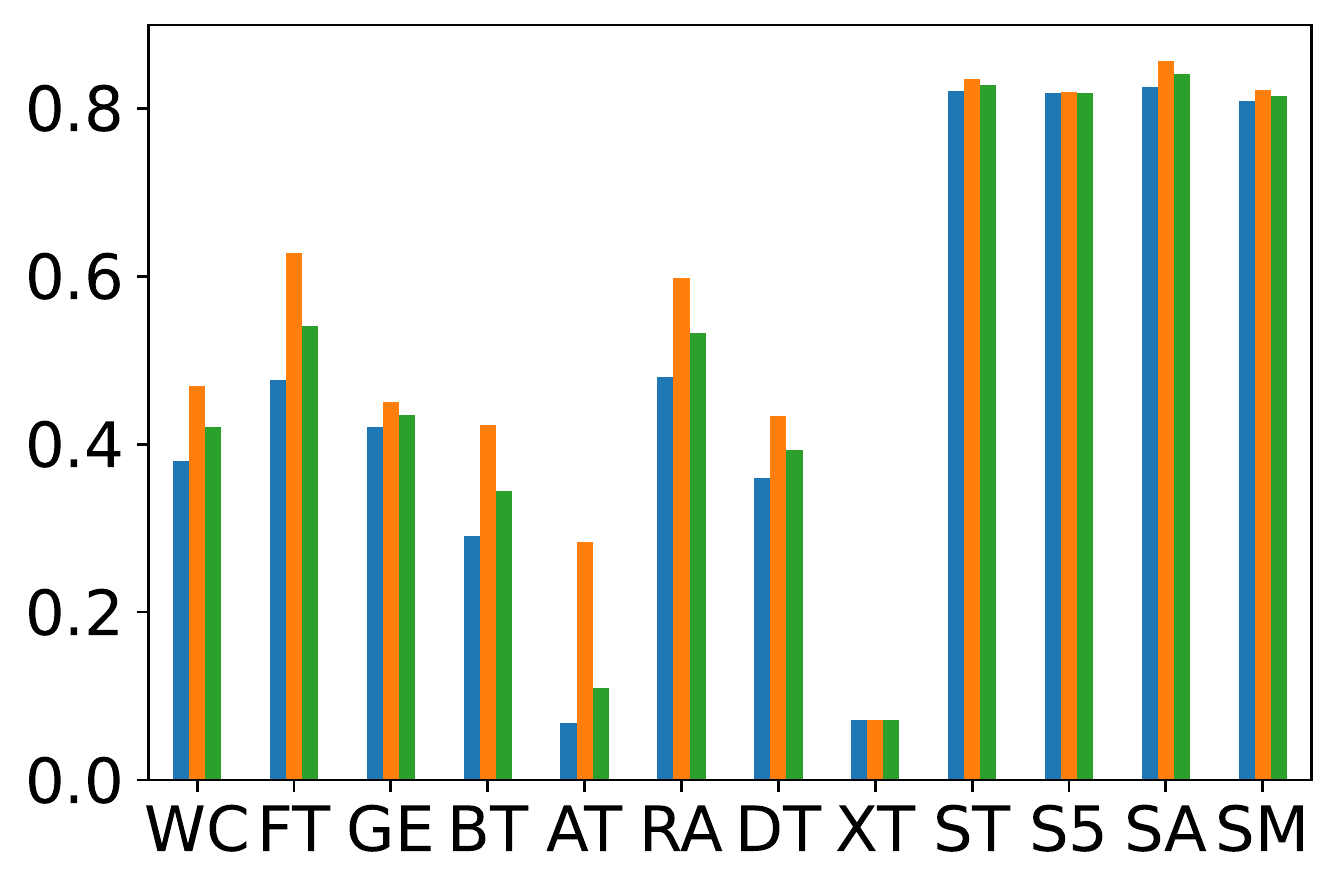}}
\subfloat[D2 / Time]{\includegraphics[trim=0.12cm 0.12cm 0.12cm 0.12cm, clip, width=0.25\textwidth, height=30mm]{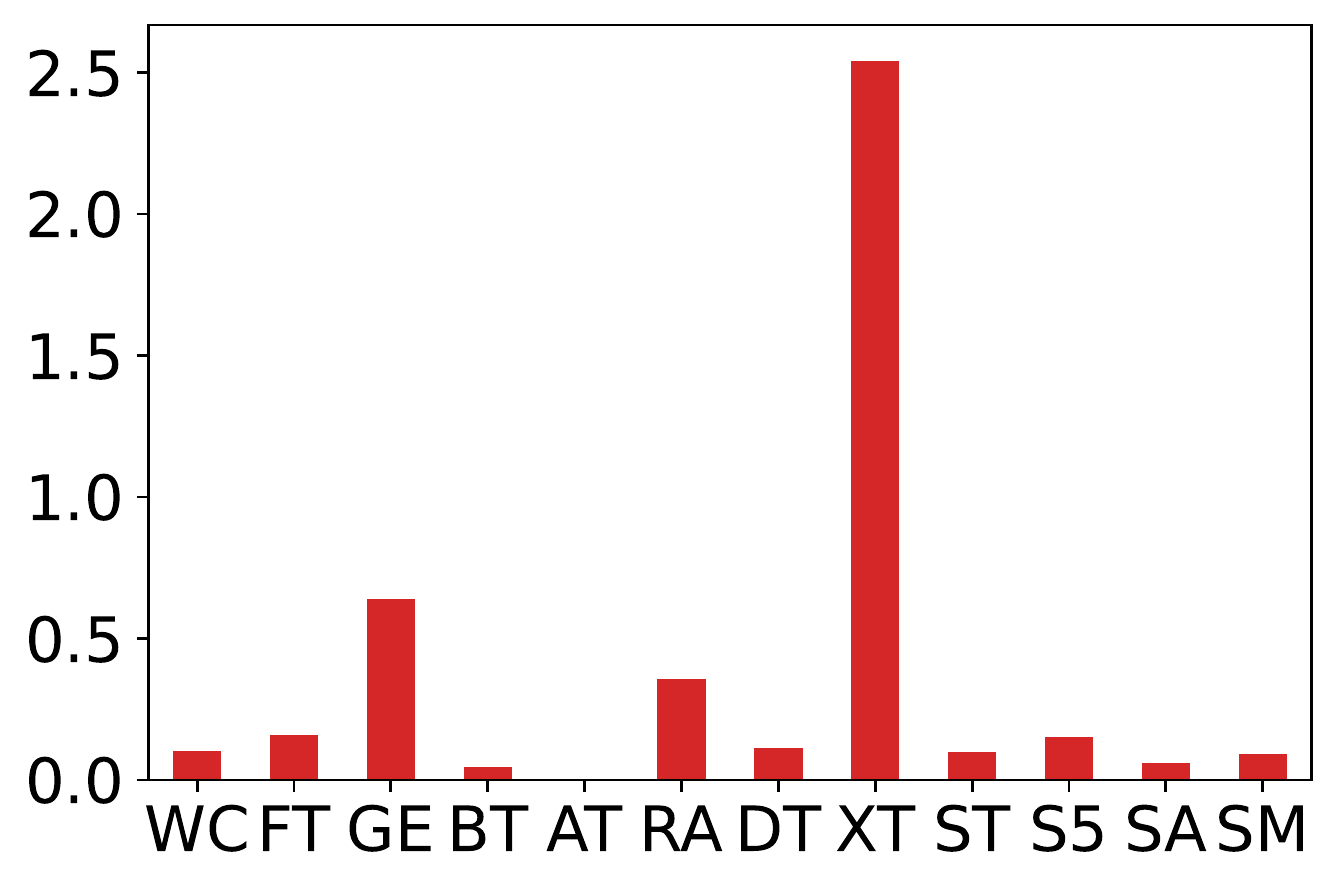}}
\newline
\subfloat[D3 / Scores]{\includegraphics[trim=0.12cm 0.12cm 0.12cm 0.12cm, clip, width=0.25\textwidth, height=30mm]{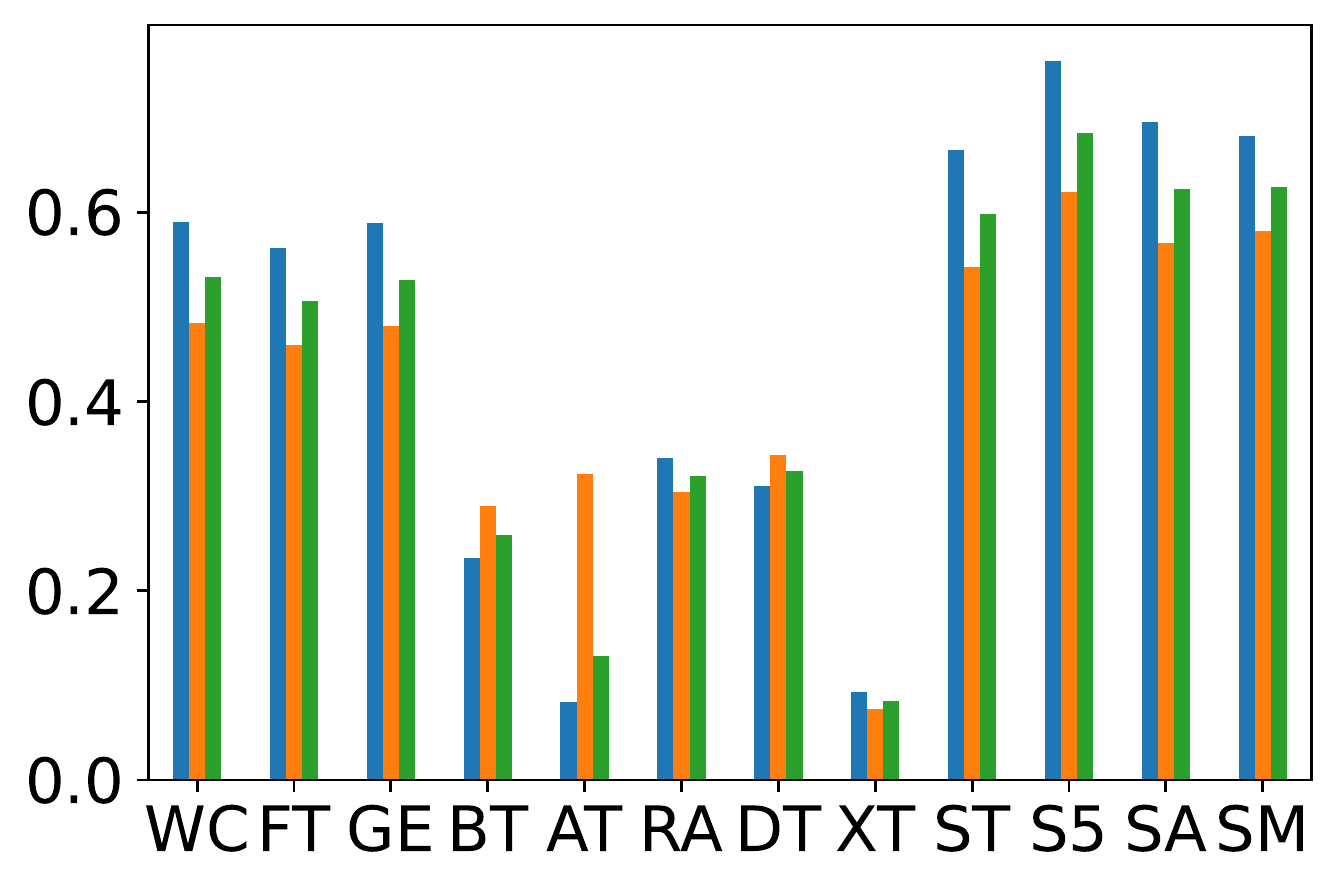}}
\subfloat[D3 / Time]{\includegraphics[trim=0.12cm 0.12cm 0.12cm 0.12cm, clip, width=0.25\textwidth, height=30mm]{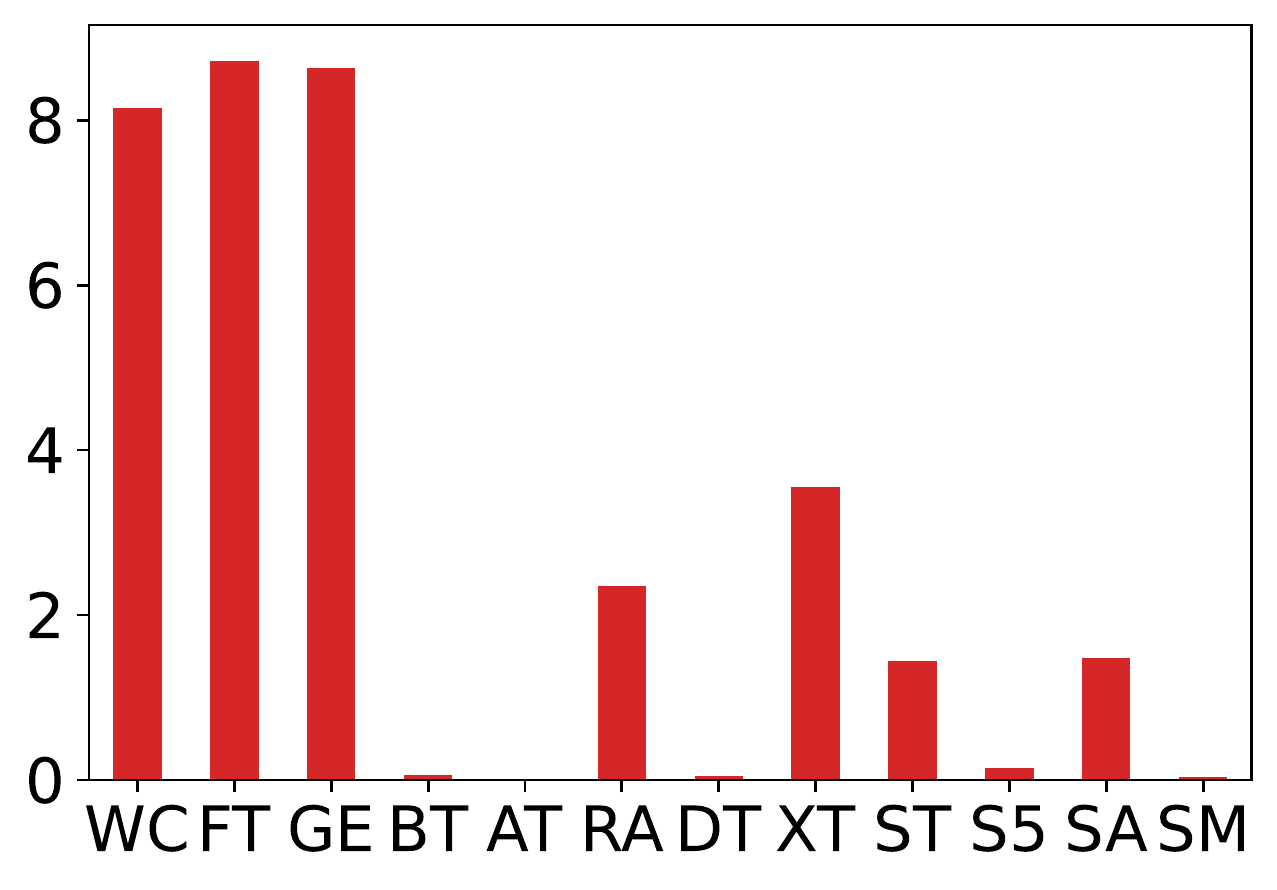}} 
%\hspace{5mm}%
\subfloat[D4 / Scores]{\includegraphics[trim=0.12cm 0.12cm 0.12cm 0.12cm, clip, width=0.25\textwidth, height=30mm]{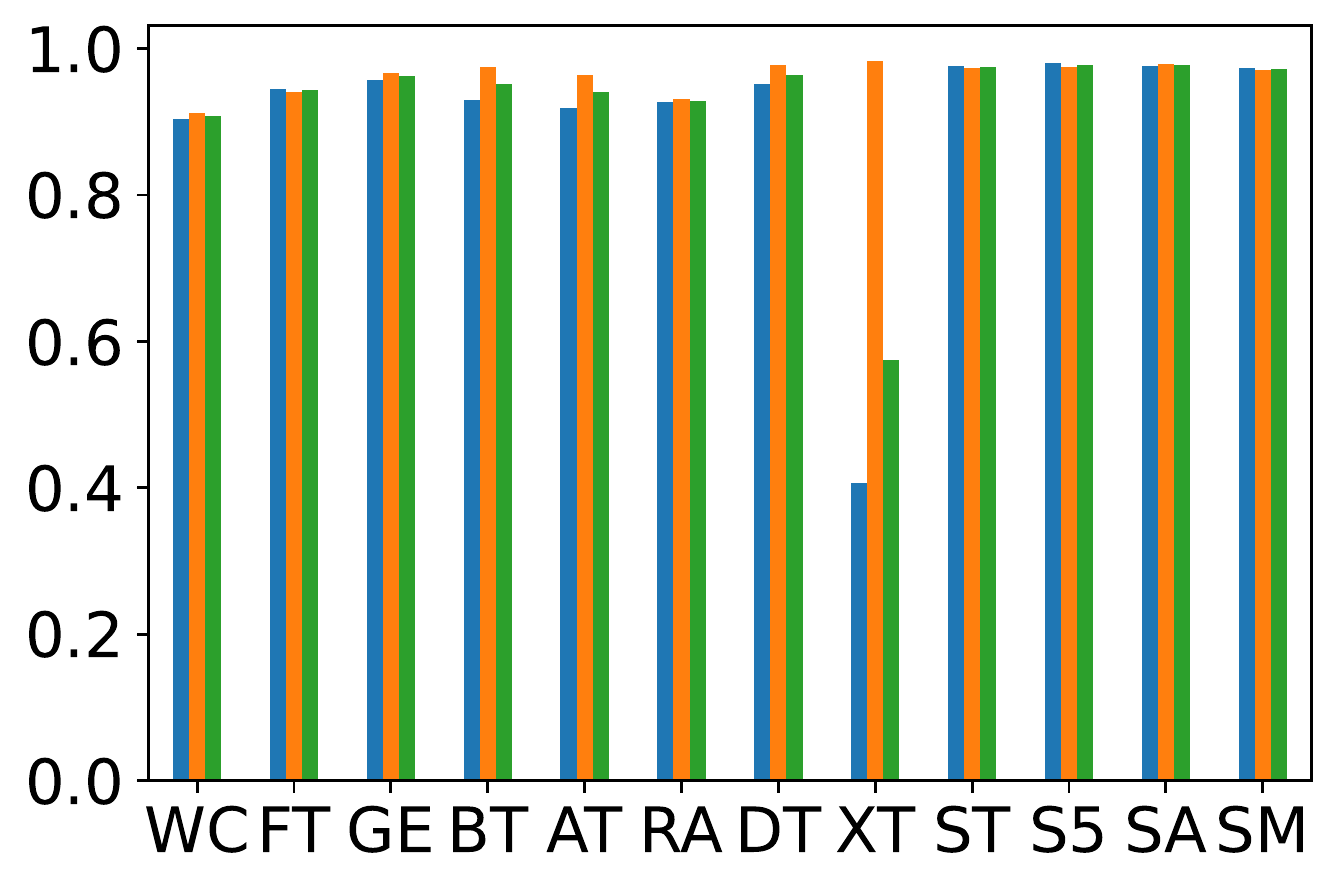}}
\subfloat[D4 / Time]{\includegraphics[trim=0.12cm 0.12cm 0.12cm 0.12cm, clip, width=0.25\textwidth, height=30mm]{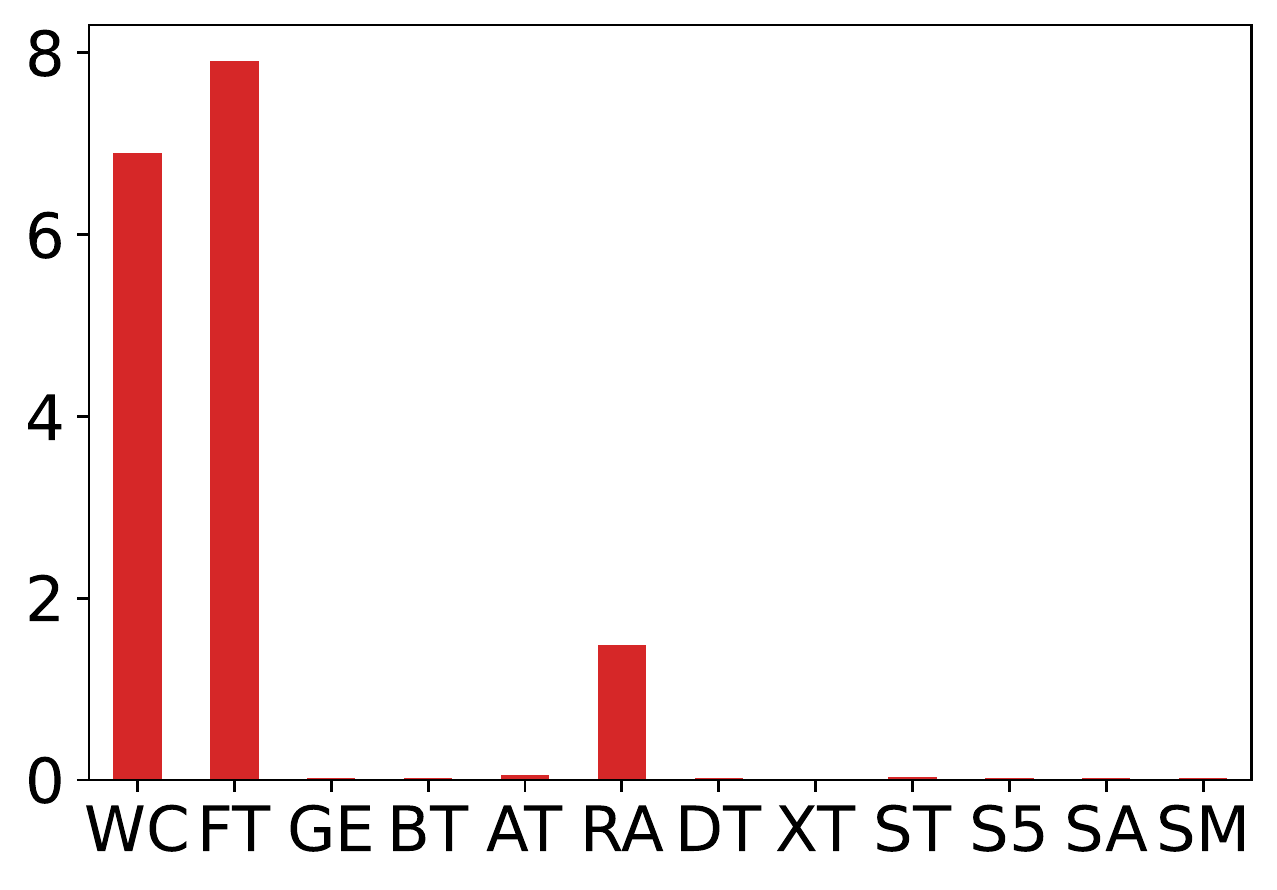}}
\newline
\subfloat[D5 / Scores]{\includegraphics[trim=0.12cm 0.12cm 0.12cm 0.12cm, clip, width=0.25\textwidth, height=30mm]{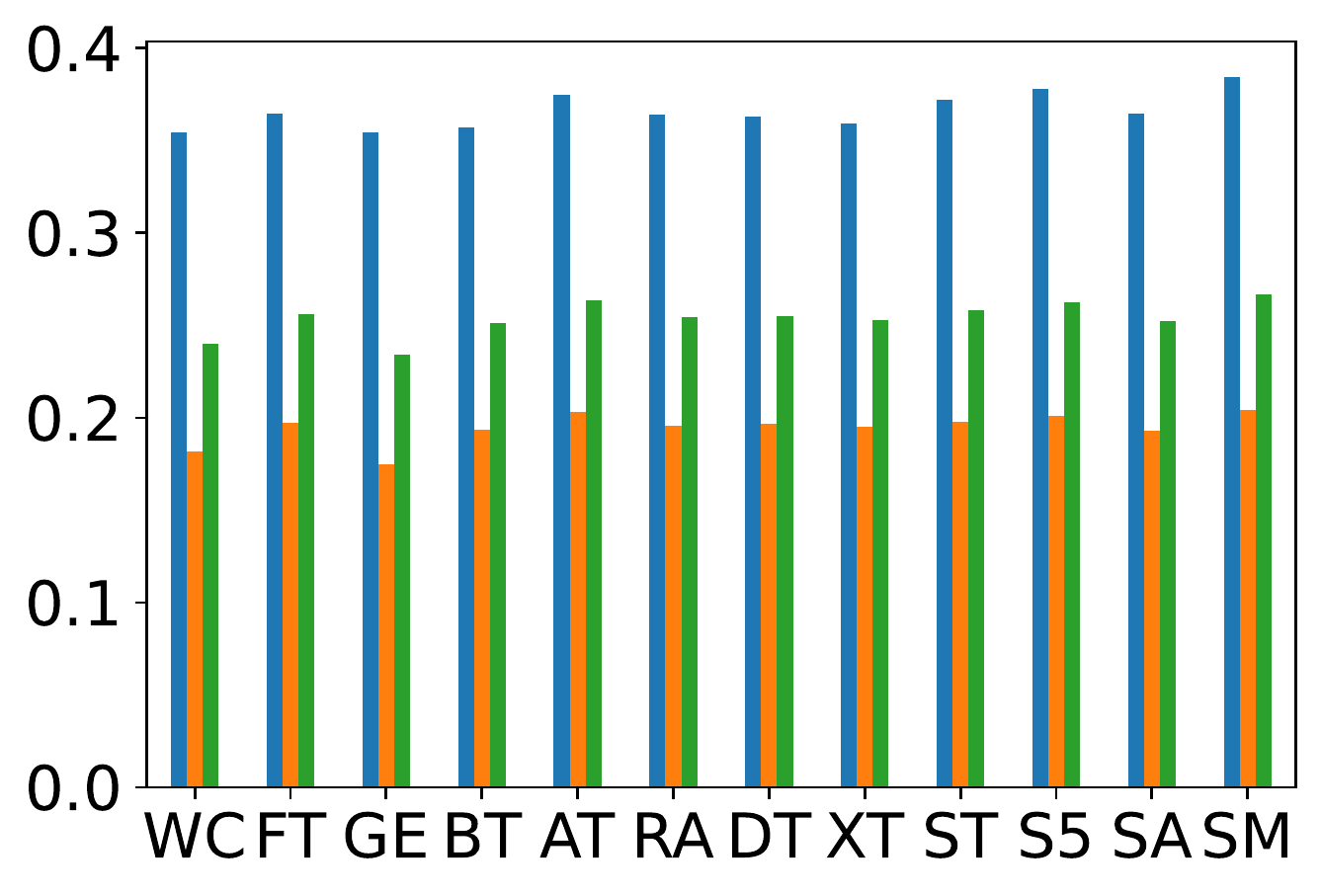}}
\subfloat[D5 / Time]{\includegraphics[trim=0.12cm 0.12cm 0.12cm 0.12cm, clip, width=0.25\textwidth, height=30mm]{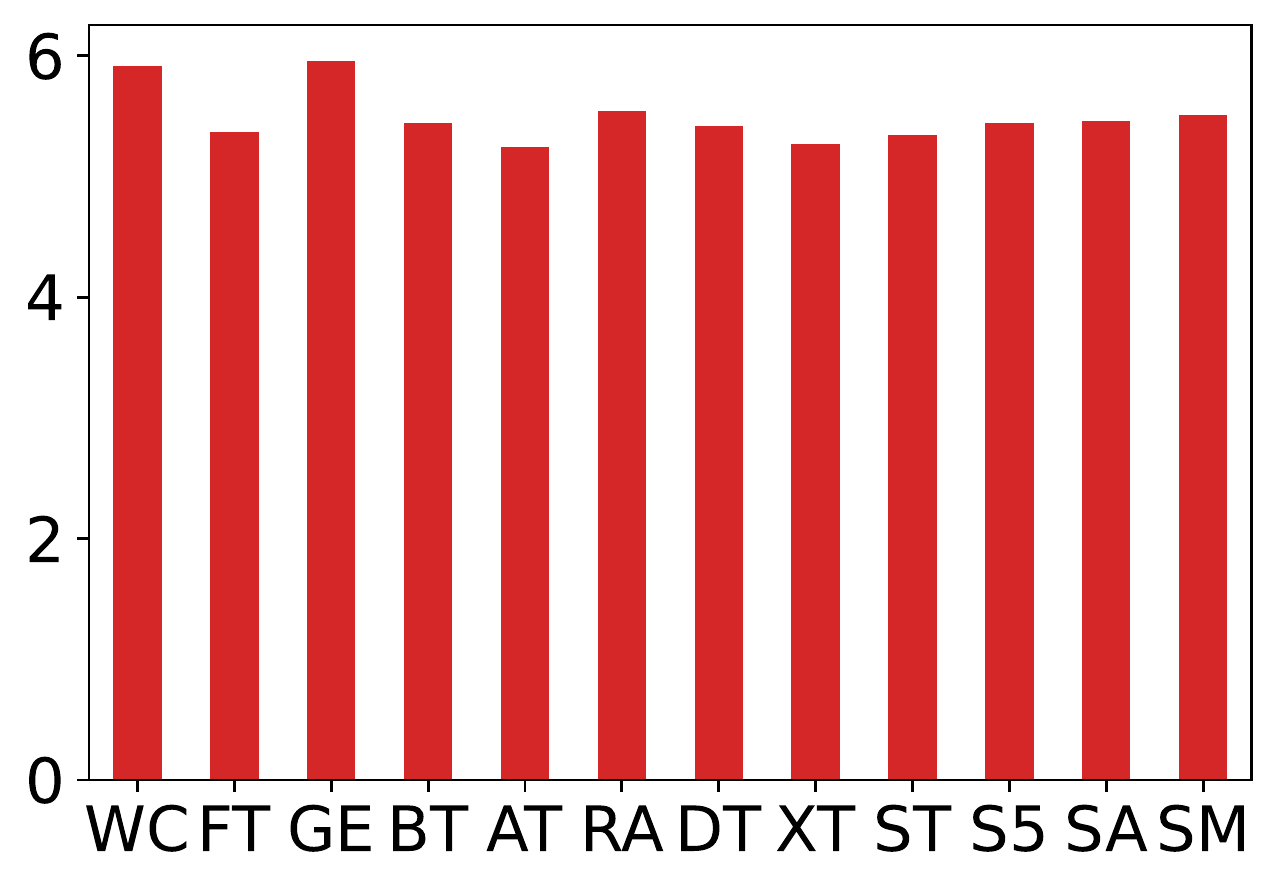}} 
%\hspace{5mm}%
\subfloat[D6 / Scores]{\includegraphics[trim=0.12cm 0.12cm 0.12cm 0.12cm, clip, width=0.25\textwidth, height=30mm]{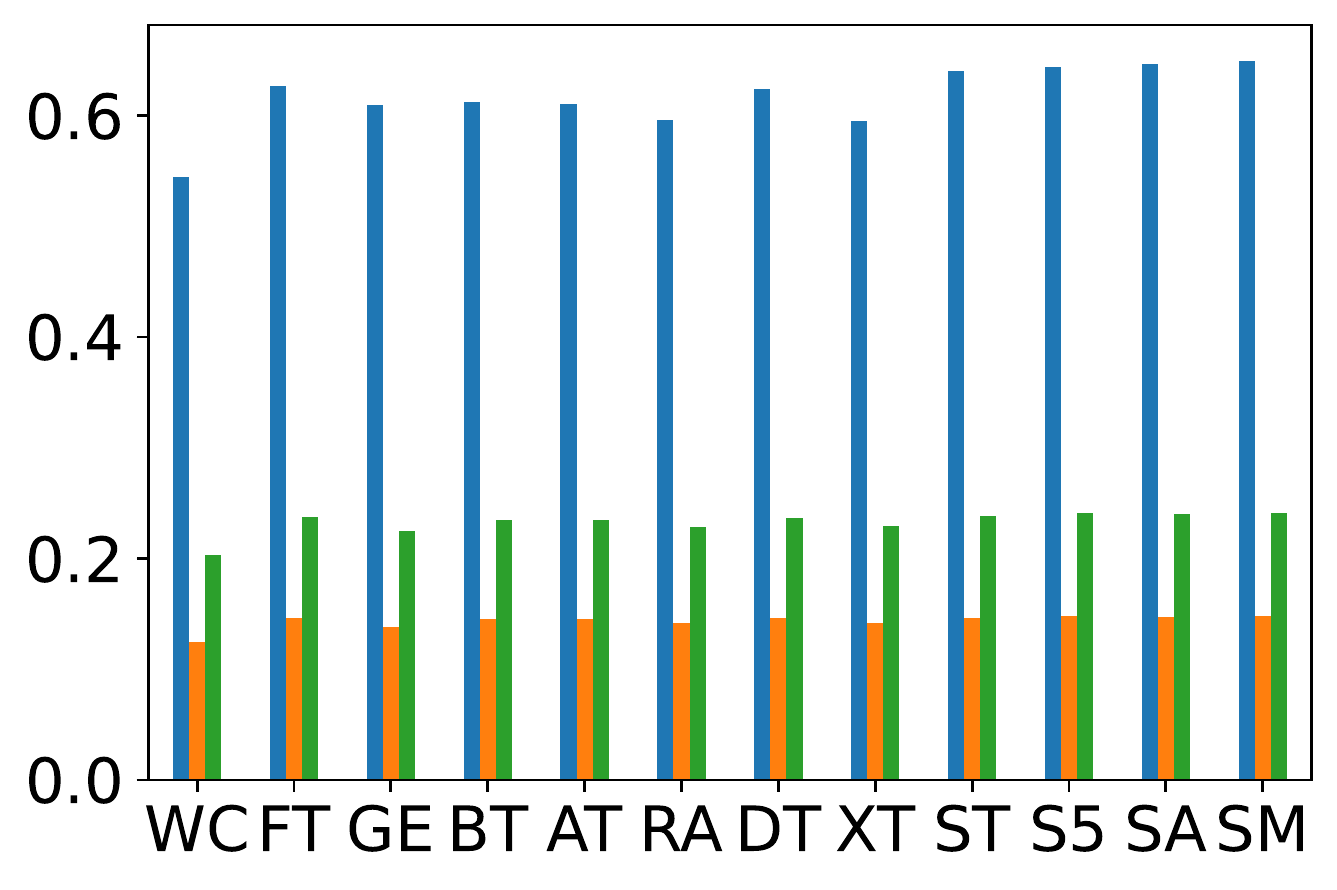}}
\subfloat[D6 / Time]{\includegraphics[trim=0.12cm 0.12cm 0.12cm 0.12cm, clip, width=0.25\textwidth, height=30mm]{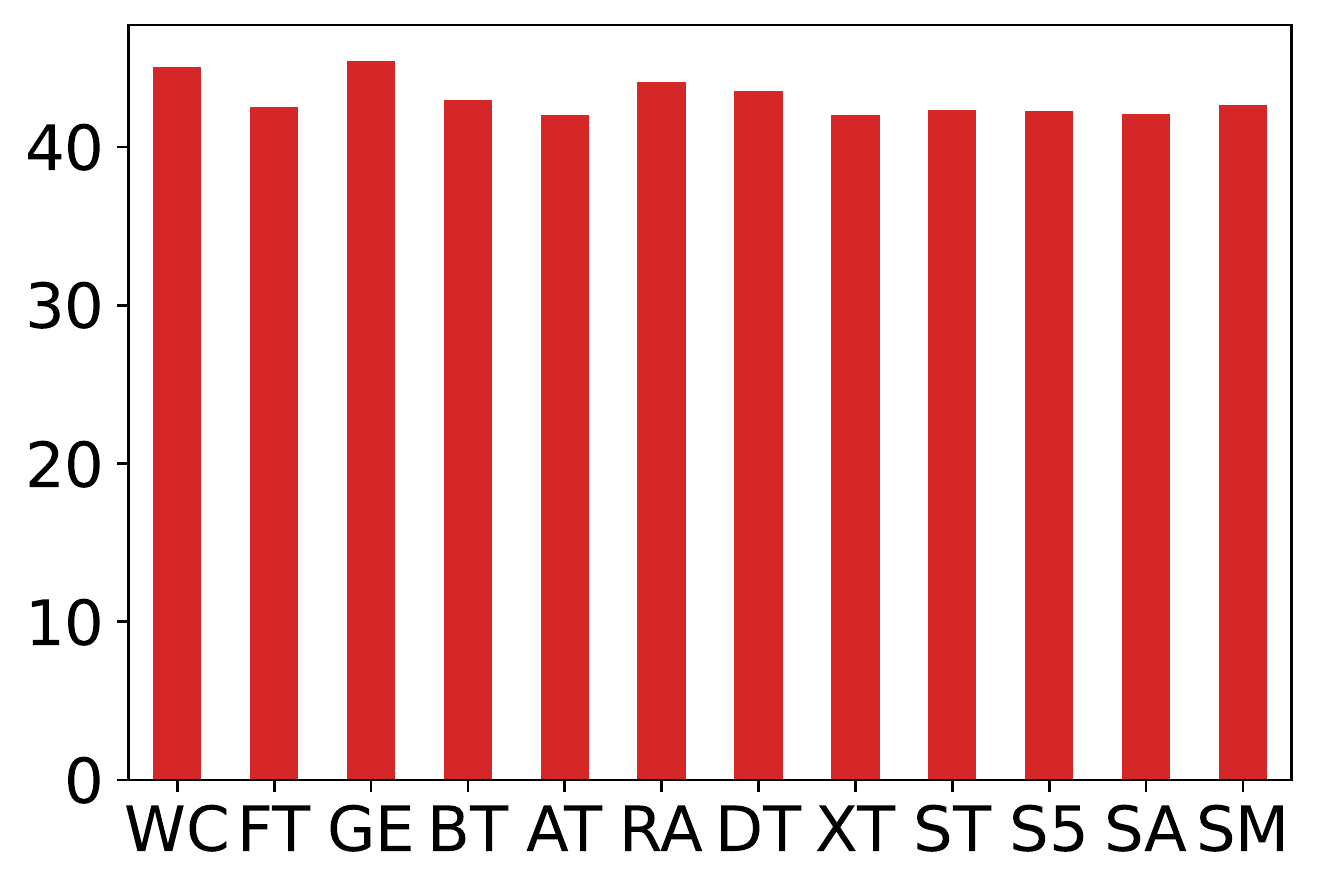}}
\newline
\subfloat[D7 / Scores]{\includegraphics[trim=0.12cm 0.12cm 0.12cm 0.12cm, clip, width=0.25\textwidth, height=30mm]{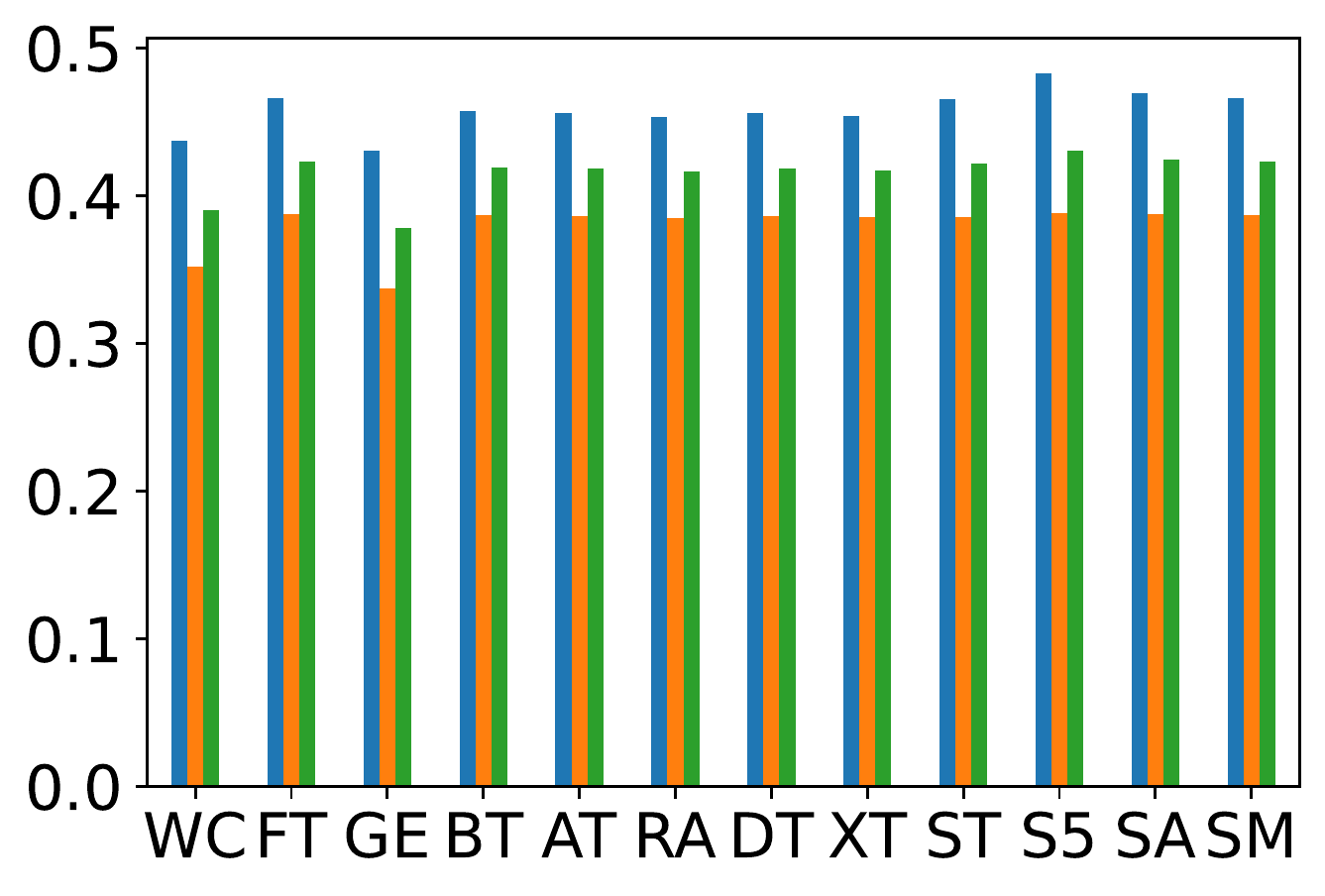}}
\subfloat[D7 / Time]{\includegraphics[trim=0.12cm 0.12cm 0.12cm 0.12cm, clip, width=0.25\textwidth, height=30mm]{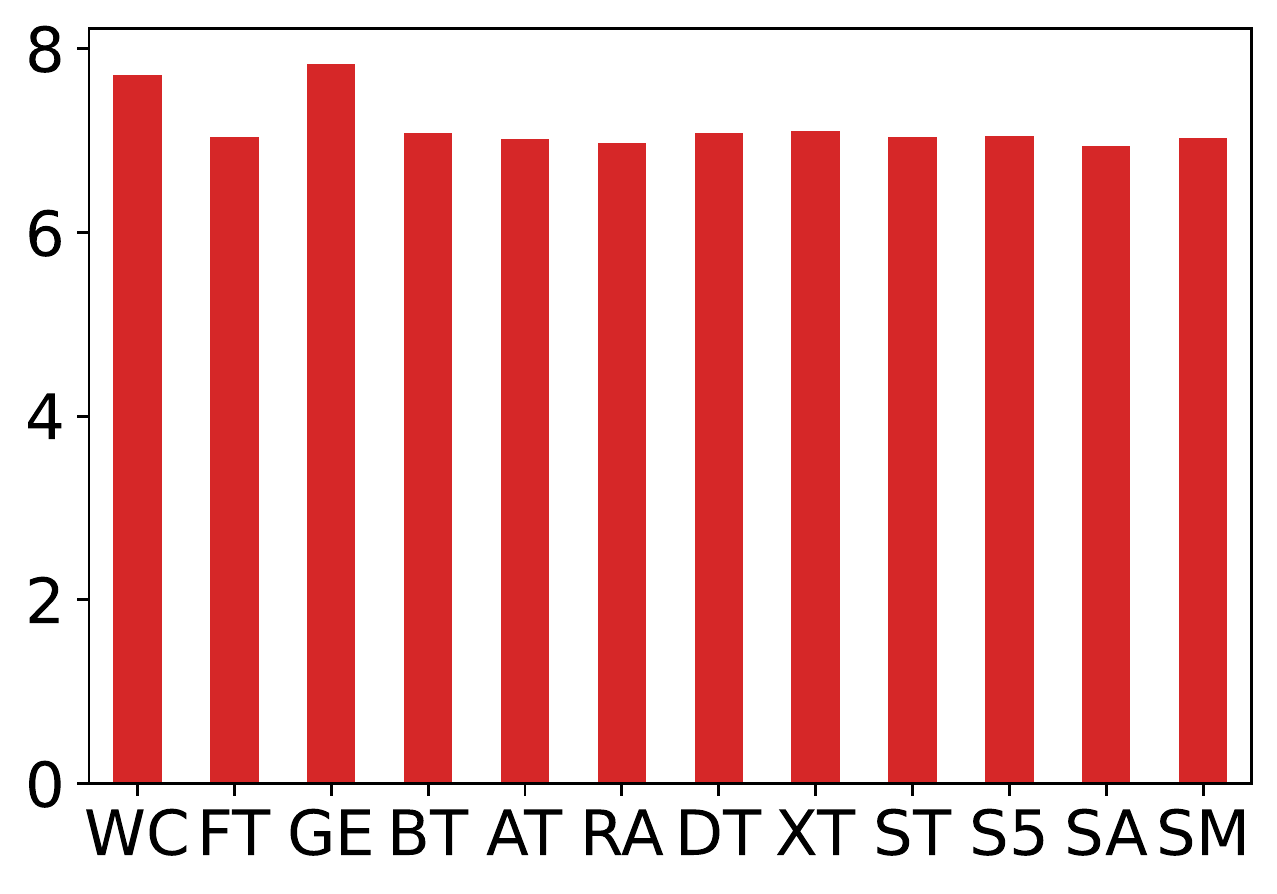}} 
%\hspace{5mm}%
\subfloat[D8 / Scores]{\includegraphics[trim=0.12cm 0.12cm 0.12cm 0.12cm, clip, width=0.25\textwidth, height=30mm]{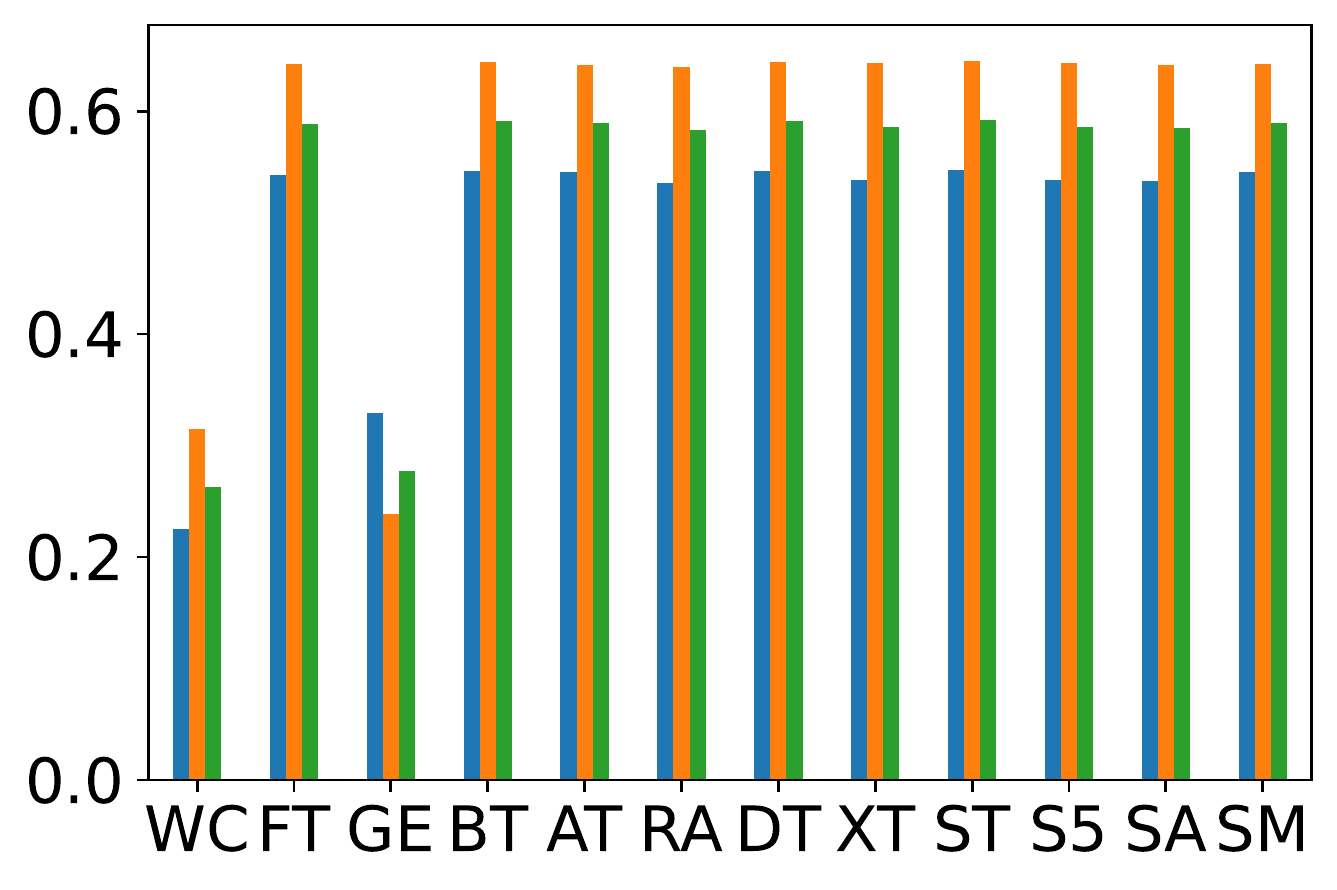}}
\subfloat[D8 / Time]{\includegraphics[trim=0.12cm 0.12cm 0.12cm 0.12cm, clip, width=0.25\textwidth, height=30mm]{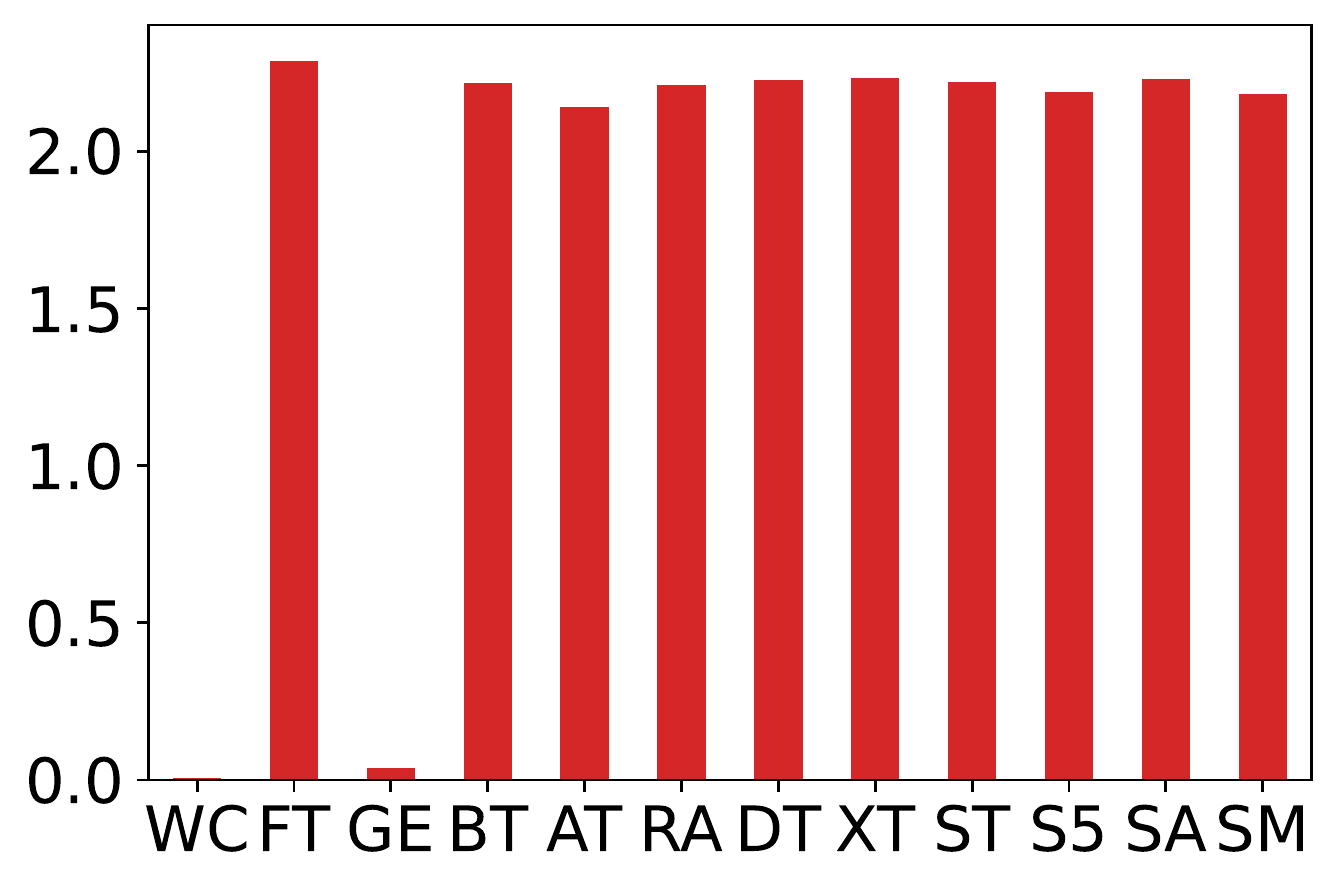}}
\newline
\subfloat[D9 / Scores]{\includegraphics[trim=0.12cm 0.12cm 0.12cm 0.12cm, clip, width=0.25\textwidth, height=30mm]{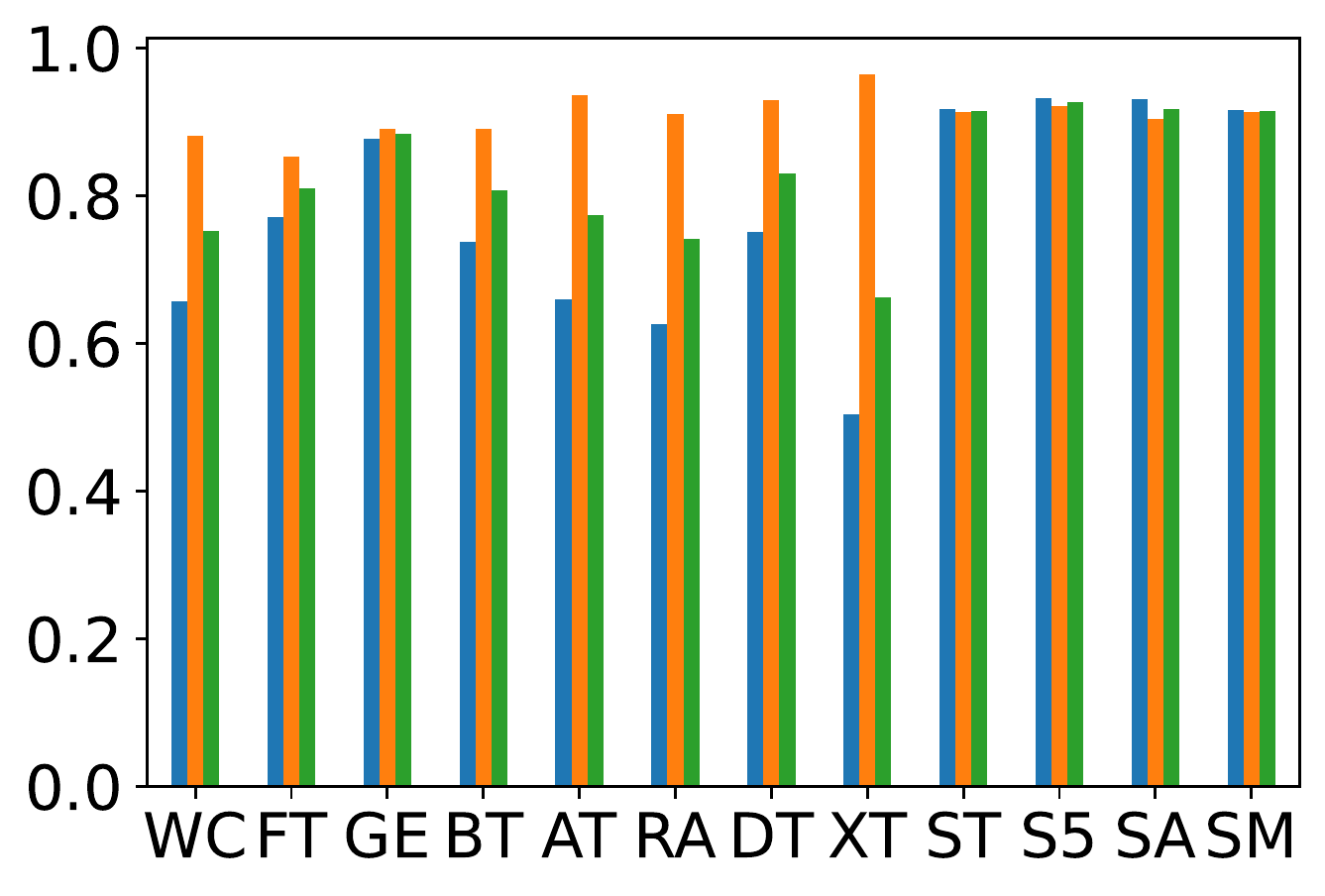}}
\subfloat[D9 / Time]{\includegraphics[trim=0.12cm 0.12cm 0.12cm 0.12cm, clip, width=0.25\textwidth, height=30mm]{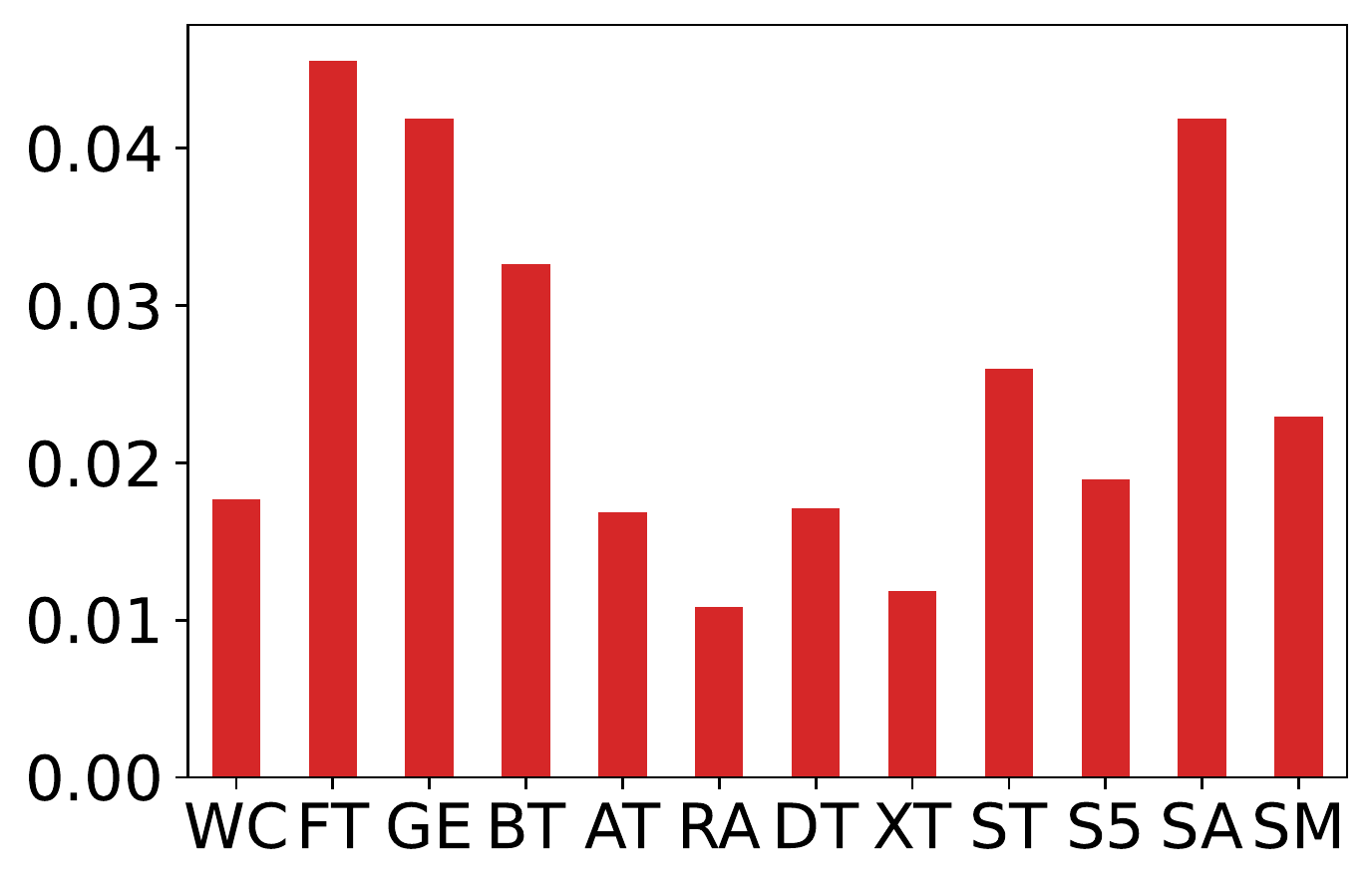}} 
%\hspace{5mm}%
\subfloat[D10 / Scores]{\includegraphics[trim=0.12cm 0.12cm 0.12cm 0.12cm, clip, width=0.25\textwidth, height=30mm]{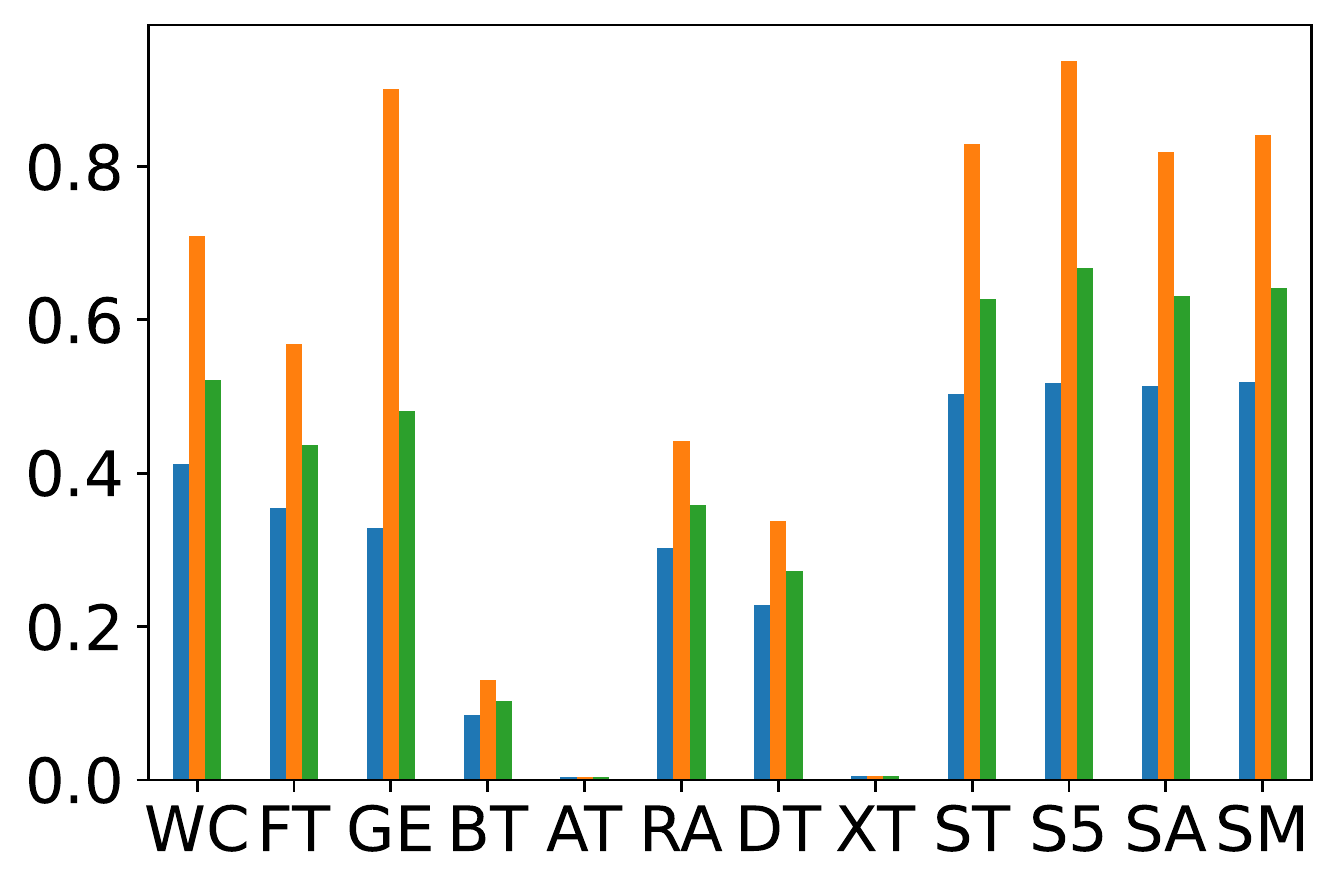}}
\subfloat[D10 / Time]{\includegraphics[trim=0.12cm 0.12cm 0.12cm 0.12cm, clip, width=0.25\textwidth, height=30mm]{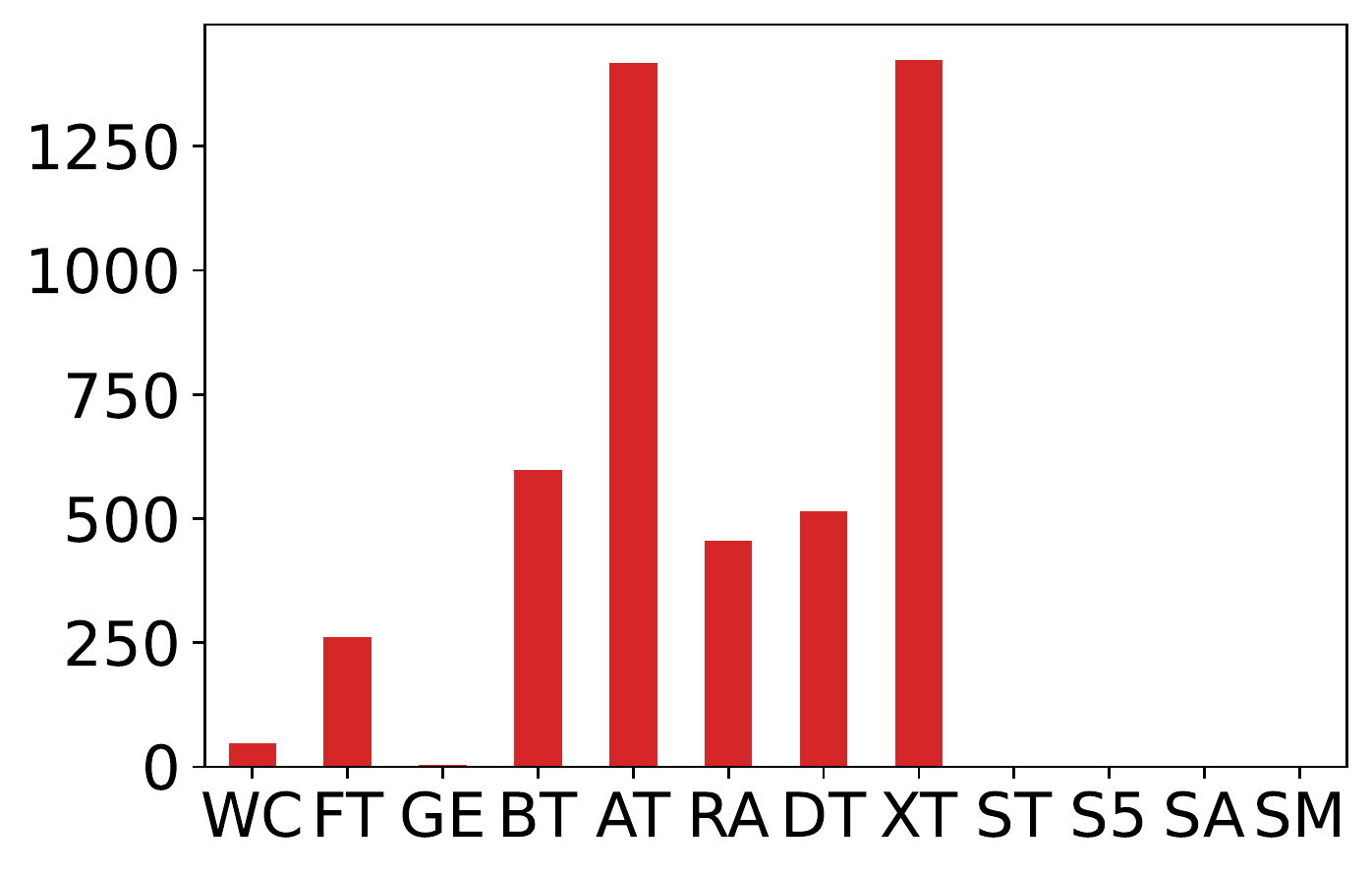}}
\newline
\caption{{\color{myblue}Precision}, {\color{myorange}Recall}, {\color{mygreen}F1} and {\color{myred}Matching Time} for Unsupervised Matching per dataset in Table \ref{tb:datasets}(a) (Schema-Based).}
\label{fig:match_unsup_sb}
\end{figure*}

In Figures \ref{fig:blk_heat_sch}, \ref{fig:blk_heat_pear_sch} and \ref{fig:blk_real_sch} we see results of Blocking on schema-based settings. The trend is similar to schema-agnostic, i.e. static models are faster, but not that efficient, while SBERT models are dominant. A similar trend is followed in Figures \ref{fig:match_unsup_sb}, \ref{fig:match_corr_sb} and \ref{fig:matching_heat_sb} for Unsupervised Matching.

\end{document}